\documentclass[fleqn,usenatbib,useAMS]{mnras}

\usepackage{graphicx}	
\usepackage{amsmath}	
\usepackage{multicol}        
\usepackage{bm}		
\usepackage{pdflscape}	
\usepackage{wasysym}
\usepackage{float}
\usepackage{tablefootnote}

\usepackage[T1]{fontenc}
\usepackage{ae,aecompl}

\usepackage{newtxtext,newtxmath}

\title[MNRAS \LaTeX\ guide for authors]{Flares and superflares on the southern active binary CC~Eri}

\author[M. Leitzinger]{M. Leitzinger,$^{1}$\thanks{E-mail:martin.leitzinger@uni-graz.at}
P. Odert,$^{1}$
R. Greimel,$^{2}$
P. Kab\'ath,$^{3}$
J. Lipt\'ak,$^{3,4}$
E.W. Guenther,$^{5}$
P. Heinzel,$^{3,6}$
\newauthor
P. Gajdo\v{s},$^{3,7}$
M. Ornik,$^{1}$
J. Wollmann,$^{3,4}$
M. Skarka,$^{3}$
J. Srba,$^{3}$
P. \v{S}koda,$^{3}$
J. Fr\'yda,$^{4}$
R. Brahm,$^{8,9,10}$
\newauthor
L. Vanzi,$^{11,12}$
J. Jan\'ik,$^{13}$
P. Pintr$^{14}$
\\
$^{1}$Institute of Physics/AGP, University of Graz, Universit\"atsplatz 5, A-8010 Graz, Austria\\
$^{2}$RG Science, Schanzelgasse 17, A-8010 Graz, Austria\\
$^{3}$Astronomical Institute, Czech Academy of Sciences, 25165 Ond\v{r}ejov, Czech Republic\\
$^{4}$Astronomical Institute of Charles University, V Hole\v{s}ovi\v{c}k\'{a}ch 2, 180 00 Prague, Czech Republic\\
$^{5}$Th\"uringer Landessternwarte Tautenburg, Sternwarte 5, 07778 Tautenburg, Germany\\
$^{6}$University of Wroc{\l}aw, Center of Scientific Excellence – Solar and Stellar Activity, Kopernika 11, 51-622 Wroc{\l}aw, Poland\\
$^{7}$Institute of Physics, Faculty of Science, Pavol Jozef \v{S}af\'arik University, Park Angelinum 9, 04001 Ko\v{s}ice, Slovakia\\
$^{8}$Facultad de Ingeniera y Ciencias, Universidad Adolfo Ib\'a\~nez, Av. Diagonal las Torres 2640, Pe\~nalol\'en, Santiago, Chile\\
$^{9}$Millennium Institute for Astrophysics, Chile\\
$^{10}$Data Observatory Foundation, Chile\\
$^{11}$Centre of Astro-Engineering, Pontificia Universidad Catolica de Chile, Av. Vicu\~na Mackenna 4860, Santiago, Chile\\
$^{12}$Department of Electrical Engineering, Pontificia Universidad Catolica de Chile, Av. Vicu\~na Mackenna 4860, Santiago, Chile\\
$^{13}$Department of Theoretical Physics and Astrophysics, Faculty of Science, Masaryk University, Kotl\'{a}\v{r}sk\'{a} 2, Brno, CZ-611 37, Czech Republic\\
$^{14}$Institute of Plasma Physics AS CR, v.v. i., TOPTEC centre, Sobotecka 1660, 511 01 Turnov, Czech Republic}

\date{Last updated 2020 June 10; in original form 2013 September 5}

\pubyear{2020}

\begin{document}
\label{firstpage}
\pagerange{\pageref{firstpage}--\pageref{lastpage}}
\maketitle

\begin{abstract}
Flares and CMEs are known to be the dominating high-energy phenomena on cool stars. Superflares were thoroughly investigated using broadband photometry predominantly from Kepler, K2, and TESS. Here we present a spectroscopic investigation of superflares on the very active spectroscopic binary CC~Eri. We focus on spectroscopic signatures of (super)-flares and line asymmetries with the goal to characterize superflares spectroscopically.  In 70 nights of spectroscopic observations obtained at the ESO~1.52m telescope with the Echelle spectrograph PUCHEROS+ hosted by the PLATOSpec consortium we identified 31 flares from which two are superflares already from the deduced g’-band energies. We also find a broad blue-wing asymmetry occurring in the impulsive phase of another superflare which shows great potential to be a prominence eruption. For the second most luminous flare we find the largest number of excess emissions during the impulsive and gradual flare. We identify the flare to be occurring close to the stellar limb which indicates that the flare was even more energetic than derived from its g'-band and spectral line energies. We identify more than sixty spectral lines in the spectral range of 4100 and 7200\AA{} showing excess emission during this flare. We detect continuum enhancements as well as photospheric line fillings during the flare. Generally we find that depending on the flare energy the number of spectral lines revealing excess emission increases, especially for the more energetic superflares. We therefore conclude that superflares are likely scaled-up versions of less energetic normal flares.
\end{abstract}




\newpage
 
\section{Introduction}

With the detection of exoplanets also the host star's activity became another important aspect why to more detailled investigate stellar activity of selected stars. Stellar flares are a subject of interest since the first half of the last century \citep[][]{Joy1949, Luyten1949} and stellar coronal mass ejections (CMEs) much later since the 1990ies \citep[][]{Houdebine1990, Guenther1997}. The more energetic superflares (E$>$10$^{33}$erg) have been identified as such going back to the early 2000s \citep{Schaefer2000}. Superflares gained a broader attention when it was found that superflares are a rather common phenomenon on solar-like stars \citep{Maehara2012}. \citet{Maehara2012} present 148 superflares found on 148 G-type main-sequence stars and \citet{Shibayama2013} extended the sample and found 1547 superflares on 279 solar-like stars, both using Kepler data. Interestingly, these superflares occurred not only on fast-rotating stars, but some superflares were also found on slow-rotating stars. The characterization of selected superflare stars has been conducted by subsequent spectroscopic observations \citep{Notsu2015a, Notsu2015b} confirming the slow rotation and the existence of large spots on these superflare stars. \citet{Okamoto2021} presented a final and more complete census of superflares on solar-like stars using all of the Kepler primary mission data. This led to 2341 superflares on 265 solar-type stars (5100-6000~K) and 26 superflares on 15 Sun-like stars (5600-6000~K). \citet{Vasilyev2024} also analysed Kepler data and identified 2889 superflares on 2527 Sun-like stars. These authors deduce a superflare rate (E$>$10$^{34}$~erg) for Sun-like stars of one per hundred years. Furthermore their stellar superflare frequency-energy distribution was found to be consistent with the extrapolation of the solar one to higher energies which led the authors to the conclusion that both solar flares and stellar superflares share the same physical mechanism.\\
With the launch of the Transiting Exoplanet Survey Satellite (TESS) further studies on superflare occurrence were conducted revealing possible selection biases with respect to Kepler which was looking at a fixed field in Cygnus whereas TESS performs an all-sky survey. \citet{Tu2020} analyzed the first year of TESS data in the southern hemisphere and found 1216 superflares on 400 solar-type stars. More or less simultaneously \citet{Doyle2020} analyzed the same data set and presented 1980 flares, from which 92\% were superflares ($\sim$1820), from 209 solar-type stars. Furthermore, \citet{Tu2021} analyzed the second year of TESS observations covering the northern hemisphere. In this census, 1272 superflares on 311 solar-type stars have been found. As one can see, the census of superflares deduced from Kepler and TESS data are comparable.\\
The driver and sources of emission of superflares was and is also a subject of discussion. There are two scenarii discussed in the literature, either superflares are simply scaled up solar flares \citep[e.g.][]{Notsu2013} and their origin is reconnection of the stellar magnetic field in active regions, or superflares are triggered by close-in orbiting objects \citep[e.g. brown dwarfs, planets,][]{Rubenstein2000, Cuntz2000, Ip2004} by reconnecting planetary and stellar magnetic fields. Only recently planet induced flares have been reported for the very young G dwarf HIP67522 \citep{Ilin2025}. Prior to this detection \citet{Klein2022} report on excess emission in the HeI D3 line at 5876\AA{} close to the orbital period of AU~Mic~b. The HeI D3 line at 5876\AA{} is temperature sensitive and is known to reveal excess emission during flares. Only for few of the superflare stars companions have been identified \citep[see e.g.][]{Tu2020, Tu2021}, whereas the majority of light curves reveals spot modulation. However, a clear relation between spottedness and flare and especially superflare occurrence has yet not been established \citep[see e.g.][]{Roettenbacher2018}. \citet{Heinzel2018} investigate the possible white-light contribution of superflare loops and find that the contribution may be significant due to the large spatial scale of stellar flaring loops and may even dominate in the gradual superflare phase.\\
The generation scenario of superflares is also debated \citep[see e.g.][]{Mullan2018}, leading to the question if spectroscopic signatures of solar flares and superflares may be probably different. For large flares these authors find a steeper\footnote{Usual flare frequency/energy distributions derived e.g. from Kepler or TESS show the flare energy on the x-axis and the flare frequency on the y-axis. In \citet{Mullan2018} the axis are switched and therefore a shallower power law slope is translated into a steeper power-law slope when following the more usual representation of flare frequency/energy distributions.} power-law slope (as also observed in cool stars) and suggest that a change in the power-law slope may take place in cool stars when the electric conductivity decreases. \\
The widely used superflare definition is an arbitrarily chosen bolometric energy threshold (one order of magnitude larger than the largest solar flares), rather than being a definition dependent on the physical mechanism. Maybe there are differences in the physical mechanism of superflares when compared to normal flares and maybe this produces signatures visible in spectra? Moreover such signatures maybe present only in even more energetic flares than the ones being at 10$^{33}$~erg. Up to now there have been only very few superflares captured spectroscopically. Either those have been captured with discontinuous spectral regions \citep[e.g.][blue and red, with a gap in between]{Lalitha2013}, being focused in the blue part of the optical spectrum only \citep[e.g.][]{Hawley1991}, not having the spectral resolving power to resolve spectral lines with an FWHM smaller the typical broad Balmer profiles \citep[][]{Kowalski2013}, not focusing on spectral line analysis of other than the strongest chromospheric lines \citep[e.g.][]{Notsu2024}, or having only a small spectral range \citep[][]{Namekata2021, Leitzinger2024}. All of those studies had partly different scientific questions to answer, therefore the various observational setups. In two cases \citep[][]{Paulson2006, Muheki2020b} the observational setups are comparable to the ones of the present study, i.e. capable of answering the same scientific questions, from which the study by \citet[][]{Muheki2020b} also presents several flares, possibly also one being a superflare.\\
Another aspect of flares/superflares is their flaring plasma flows and possibly accompanying eruptive components such as eruptive filaments/prominences which may evolve into CMEs. In the optical range, signatures of eruptive filaments/prominences can be investigated only by time resolved spectroscopy. In the recent past there have been dedicated observational programs to search for signatures of erupting filaments/prominences and CMEs using a number of different instruments and approaches \citep[for reviews on that topic see][and references therein]{Moschou2019, Leitzinger2022c, Osten2023, Tian2023, Vida2024}. The so far most frequently used method is the search for Doppler shifted emission/absorption, seen most prominently in the Balmer lines. Only few events showing a high probability of being indeed eruptive events have been reported so far \citep{Houdebine1990, Guenther1997, Vida2016, Namekata2021, Inoue2023, Leitzinger2024, Namekata2024}. Many more candidate events (events with low projected velocities) have been reported  \citep[e.g.][]{Fuhrmeister2018, Vida2019, Muheki2020a, Muheki2020b, Notsu2024}. Dedicated searches which revealed only few events have been presented as well \citep{Leitzinger2014, Korhonen2017}. Archival data have been used extensively as well which interestingly revealed only few events, although large amounts of data have been searched through \citep[e.g.][]{Vida2019,Leitzinger2020, Koller2021, Lu2022}. Further detections of single events revealing projected bulk velocities below the stars escape velocity have been reported as well \citep[e.g.][]{Gunn1994, FuhrmeisterSchmitt2004, Fuhrmeister2024,Lu2025, Cao2025}. The method of Doppler-shifted emission has been used also at shorter wavelengths to search for stellar CMEs \citep[e.g.][]{Leitzinger2011a, Argiroffi2019, Chen2022, Inoue2024}.\\
Beside the method of Doppler-shifted emission/absorption several other methods have been applied in the literature to search for stellar CMEs. The search for radio bursts being related to CMEs, as known from the Sun, i.e. type II and type IV bursts was attempted several times using single-channel receivers \citep[e.g.][]{Jackson1990, AbdulAziz1995, Abranin1997, Abranin1998} and later with multi-channel receivers \citep[e.g.][]{Leitzinger2009, Boiko2012, Konovalenko2012, Crosley2016, Crosley2018a, Crosley2018b, Villadsen2019, Bloot2024} but succeeded to our knowledge only two times with type IV bursts \citep[][]{Zic2020, Mohan2024} but so far not for type II bursts.\\
The method of X-ray absorption has been applied in several cases in the past \citep[e.g.][]{Haisch1983, Favata1999} and only recently by \citet[][]{Wang2023aa}.\\
Another solar phenomenon being related to CMEs, coronal dimmings, has been successfully searched for on stars \citep[][]{Veronig2021} revealing more than 20 candidate events having great potential of being indeed associated with stellar CMEs. A multi-wavelength study aiming at the simultaneous detection of CMEs/erupting filaments/prominences in the optical and X-ray domains has been conducted targeting at the solar analogue EK~Dra, with one event revealing a clear optical and a possible X-ray signature (coronal dimming) as well \citep[][]{Namekata2024}.\\
In this study we aim for a spectroscopic investigation of superflares and possible accompanying erupting filaments/prominences in the optical. We specifically aim to characterize superflares spectroscopically in comparison to less energetic normal flares. To increase the probability of detecting superflares we select, based on the results presented in \citet{Greimel2024}, one of the most frequent flaring stars from TESS, namely the southern spectroscopic binary CC~Eri. This study is a follow-up study of \citet{Odert2025} who aim for the same scientific goals but targeting a different star (AU~Mic).

\section{Target object, observations, and data reduction}

\subsection{Target star}
CC~Eri (HD16157) is a non-eclipsing double-line spectroscopic binary \citep[SB2,][]{Strassmeier1993}. The system is tidally locked and therefore the orbital period is the rotation period of the K-star component making it one of the fastest rotators in the solar neighbourhood. In Table~\ref{tab:1} we list parameters of this system from literature and from the present study. CC~Eri was observed so far only once with TESS in sector 3. It was several times the target of X-ray investigations. CC~Eri was already detected in X-rays in the early 1980ies revealing a logLx of 29.3 \citep[2-20keV band ][]{Tsikoudi1982}. \citet{Pallavicini1988} derived from EXOSAT observations a logLx of 29.6 in the 0.04-2keV band. Numerous flares \citep{Pan1995,CrespoChacon2007,Pandey2008, Karmakar2024} and superflares \citep{Evans2008,Karmakar2016} have been detected on CC~Eri. Furthermore CC~Eri was a target of multi-wavelength studies \citep[in X-rays EUV, optical spectropolarimetry, as well as radio, ][]{Osten2002,Slee2004,Budding2006}. Flares have been also found at optical and UV wavelengths \citep{Byrne1992,Amado2000}.

\begin{table}
 \caption{Parameters of the target star CC~Eri. Estimated XUV and H$\alpha$ flare rates were determined using the power laws from \citet{Audard2000} and \citet{Leitzinger2020}. The observed H$\alpha$ flare rate was determined from the observations of the present study.}
 \label{tab:1}
 \begin{tabular}{lcc}
  \hline
                                            &            CC~Eri        & reference\\
  \hline
  spectral type                             &          K7Ve+M3Ve       & $^{1}$ \\[2pt] 
  distance                                  &           11.55~pc        & $^{2}$\\[2pt]
  age                                       &        $\sim$9~Gyr       & $^{3}$\\[2pt]
  mass ratio                                &      $\sim$2             & $^{4}$ \\[2pt]
  mass$_{M}$                                &   0.31~M$_{\astrosun}$   & $^{5,1}$ \\[2pt]
  mass$_{K}$                                &   0.57~M$_{\astrosun}$   & $^{5,1}$ \\[2pt]
  radius$_{M}$                              &   0.41~R$_{\astrosun}$   & $^{5,1}$ \\[2pt]
  radius$_{K}$                              &   0.64~R$_{\astrosun}$   & $^{5,1}$ \\[2pt]
  logLx                                     &  29.2-29.8 erg~s$^{-1}$  & $^{6,7,8}$ \\[2pt]
  P$_{\mathrm{orbit, rot}}$                 &          1.56~d          & $^{1}$\\[2pt]
  inclination                               &    40 ... 50$^{\circ}$   & $^{7,10}$\\[2pt]
  Estimated flare rate XUV (E$>$10$^{32}$~erg)        &    10 ... 39~d$^{-1}$       & $^{11}$\\[2pt]
  Estimated flare rate H$\alpha$ (E$>$10$^{32}$~erg)  &    0.04 ... 6.3~d$^{-1}$      & $^{11}$\\[2pt]
  Observed flare rate H$\alpha$ (E$>$10$^{32}$~erg)  &    0.11~d$^{-1}$      & $^{11}$\\[2pt]
  Flare rate TESS                           &    4.9~d$^{-1}$        & $^{11}$\\[2pt]
    \hline
 \end{tabular}
  
  \begin{scriptsize} 
 $^{1}$\citet{Amado2000}; $^{2}$\citet{GaiaCollaboration2023}; $^{3}$\citet{Demircan2006}; $^{4}$\citet{Evans1959}; $^{5}$\citet{Eker2008}; $^{6}$\citet{Schmitt1987}; $^{7}$\citet{Pallavicini1988, Pallavicini1990}; $^{8}$\citet{Pan1995}; $^{9}$\citet{BoppEvans1973}; $^{10}$\citet{Glebocki1995} ; $^{11}$ this study
 \end{scriptsize}

 \end{table}

\subsection{TESS}
CC~Eri was observed by TESS in one sector only so far (sector 3) from 22nd of September to 17th of October 2018. The data are available as 2 minute cadence data. We use the SAP flux data product to determine the broadband flare rate of CC~Eri (see section~\ref{resTess}).
\subsection{The ESO1.52m telescope}
To obtain spectroscopic time series we utilize the ESO152m telescope located on LaSilla, Chile. The ESO152m telescope is hosted by the PLATOSpec consortium\footnote{The PLATOSpec consortium is led by the Astronomical Institute Ond\v{r}ejov of the Czech Academy of Sciences and consisting of Th\"uringer Landessternwarte Tautenburg and Universidad Catolica de Chile as major partners, Universidad Adolfo Ibanez, Masaryk University, and the Institute of Plasma Physics, Czech Academy of Sciences, as minor partners, and University of Graz as collaborating partner.}.  It was refurbished in 2022 and can be operated remotely. It was equipped from 2022-2024 with an interim  Echelle spectrograph PUCHEROS+ \citep{Antonucci2025}, an upgraded version of PUCHEROS \citep{Vanzi2012}, offering a wavelength range of $\sim$400-700nm and a spectral resolving power of R=18000. The telescope hosts two additional finder telescopes which are used for simultaneously photometry and low-resolution spectroscopy. On one finder telescope OndCam is installed, a CMOS camera equipped with Sloan filters. The other finder telescope is equipped with GrazCam, also a CMOS camera equipped with the low-resolution transmission grating SA200 as well as Sloan and Johnson filters. For a more detailed description see \citet{Odert2025}.
\subsection{Observations: Echelle spectroscopy}
The CC~Eri observations started on 02-08-2023 and lasted until 11-01-2024 (see Table~\ref{tab:A1}). In this period 70 data sets of each few hours were recorded. These 70 data sets add up to 2659 spectra or to a total on-source time of 230.3~hours. The predominantly used integration time was 300s, only few times integration times of 480 and 600s were used to compensate for different observing conditions.
\begin{figure*}
 \includegraphics[width=18cm]{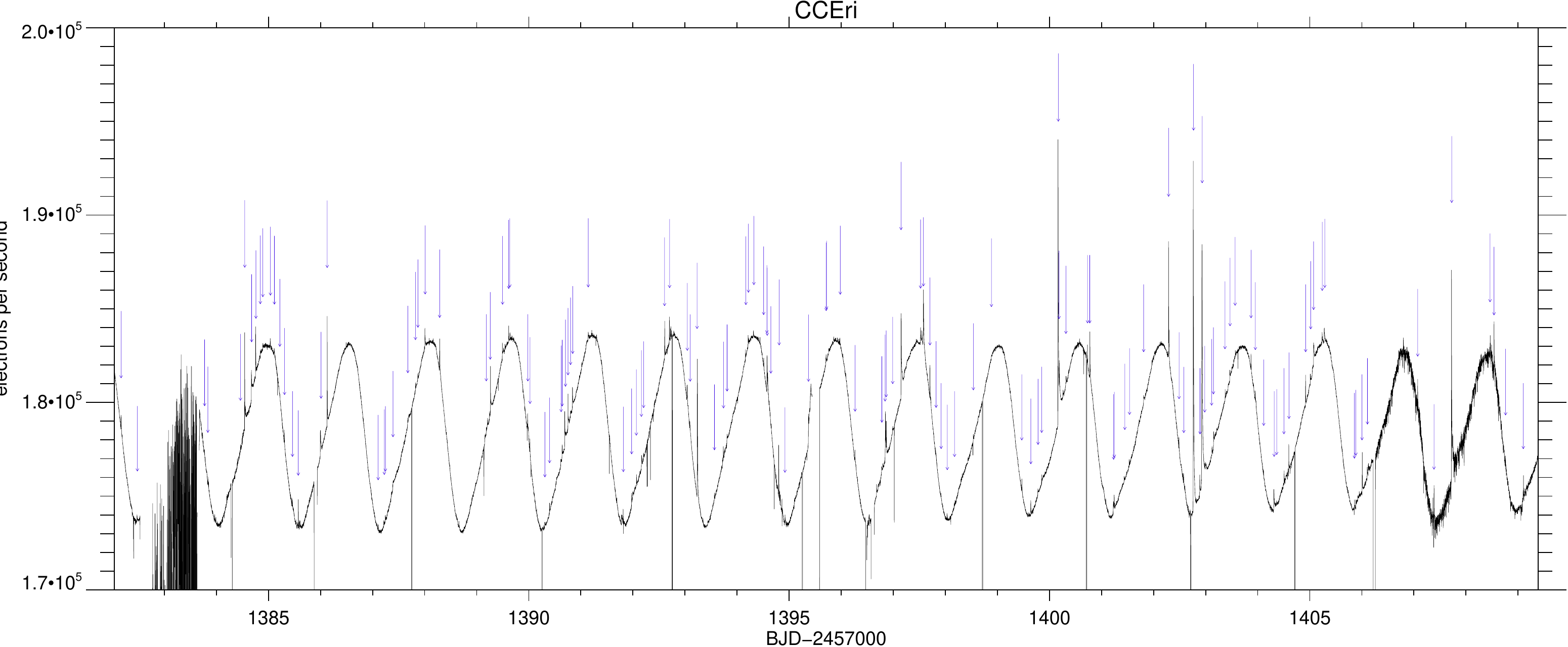}
  \caption{Only TESS light curve (sector 3) of CC~Eri. Flares are indicated by blue arrows.}
 \label{fig:example}
\end{figure*}
\subsection{Observations: Photometry and low resolution spectroscopy}
The simultaneous photometry was predominantly carried out using Sloan g'-filter, only in two nights the r'-filter was used and in seven nights g'- and r'-filter were used alternately. In six out of 70 nights no coordinated photometry was available (see Table~\ref{tab:A1}). Typical integration times are in the range of a few seconds.\\
SA200 observations have been performed coordinated only in three nights (see Table~\ref{tab:A1}).
\subsection{Data reduction and preparation}
The raw data frames are reduced with the Ceres+ pipeline, an upgraded version of the Ceres pipeline \citep{Brahm2017}. The pipeline provides a number of data products. We use the non-normalized deblazed and wavelength calibrated data as starting point. We apply barycentric wavelength correction, conversion to air wavelengths, as well as cross-correlation using the first spectrum of the time series to avoid wavelength drift. By doing so we shift the spectra to the rest wavelength of the dominating K-star component. Then the spectra are normalized by a linear fit to the spectra. These prepared spectra are used then for analysis.\\
For the flare analysis spectra need to be prepared which represent the stellar quiescent phase. For this purpose we use spectra before the impulsive phase which show no or minimum variability, as estimated from the H$\alpha$ light curves. For flares where no impulsive phase could have been captured we use spectra after the gradual phase to build a quiescent spectrum. For one event (flare no.~7, 2023-12-06), as there were no spectra before or after the impulsive /gradual flare phase, we utilized spectra from a different night but at the same orbital phase (see section~\ref{lineasymcme}), to build a quiescent spectrum.\\
The photometric data from OndCam are reduced with a dedicated pipeline \citep{Fryda2023} performing Flat-Fielding, Bias and Dark correction. To produce light curves we use aperture photometry. As comparison star one we use CD-44 791 classified as a giant and as comparison star two we use HD15899 classified as a K2/3 III star.\\
The SA200 data are only background subtracted. To do so we select a region below and above the target spectrum, average the spectra and subtract those from the target spectrum. Then the 2-dimensional target spectrum is collapsed in slit direction to obtain a 1-dimensional spectrum. The dispersion of the grating can be calculated according to the formula given in the SA200 manual\footnote{\url{https://www.shelyak.com/wp-content/uploads/Star-Analyser-200-Instructions-v1.2.pdf}}. The dispersion calculation involves the size of the aperture, the size of the detector pixels, lines per mm of the grating and the distance of the grating to the detector. Applying these values gives finally a dispersion of 7\AA{}/pixel. To apply a wavelength solution we need the starting wavelength $\lambda_{0}$. To obtain $\lambda_{0}$ we apply the detector response function of GrazCam to the GAIA spectrum of CC~Eri\footnote{taken from VizieR (\url{https://vizier.cds.unistra.fr}) catalogue I/355/spectra} \citep{GaiaCollaboration2023}. Then by comparing the modified GAIA spectrum to the SA200 spectrum we are able to match both spectra and determine then $\lambda_{0}$. We furthermore correct the SA200 spectrum for the detector response function. To flux calibrate the SA200 spectrum we fit both the GAIA and the SA200 spectra with Planck functions and by using the ratio of the Planck functions we are then able to flux calibrate the SA200 spectrum (see section~\ref{sec:lowres}).

\section{Results}
\subsection{TESS flare rate}
\label{resTess}
 From the TESS light curve we determine the TESS flare rate given in Table~\ref{tab:1}. Both components of CC~Eri lie in one TESS pixel, therefore one can analyse only the combined light curve in TESS. As usually every automated flare algorithm has its threshold of detecting low energy flares and there is only one TESS light curve of CC~Eri (see Fig.~\ref{fig:example}) we determined the TESS flare rate by eye. To do so we only subtract the spot modulated light curve, which we determined by folding a Savitzky-Golay filter with the TESS light curve of CC~Eri. We count 134 flares, which gives a flare rate of 4.9 flares per day. Small flares in the TESS light curve yield energies (in the TESS band) in the order of 10$^{31}$~erg whereas the flare with the largest peak (flare at $\sim$1402.8~BJD) yields an energy of $\sim$2$\times$10$^{33}$~erg.
 
 \subsection{Flares from spectroscopy}
 \label{resflarespec}
As CC~Eri is a non-eclipsing binary there are no spectral line profile variations due to occultation, but the system is classified as a SB2 system, i.e. a system which is orbiting so close that it can not be resolved and therefore both, the emission from the dK and the dM component, can be seen in the spectra, especially in stronger spectral lines such as the Balmer lines. In such spectral lines one sees the shift according to the radial velocity of the system. We decided not to disentangle both stars into separate spectra and therefore start the analysis with the binary star spectra.\\
\begin{figure}
\includegraphics[width=8.58cm]{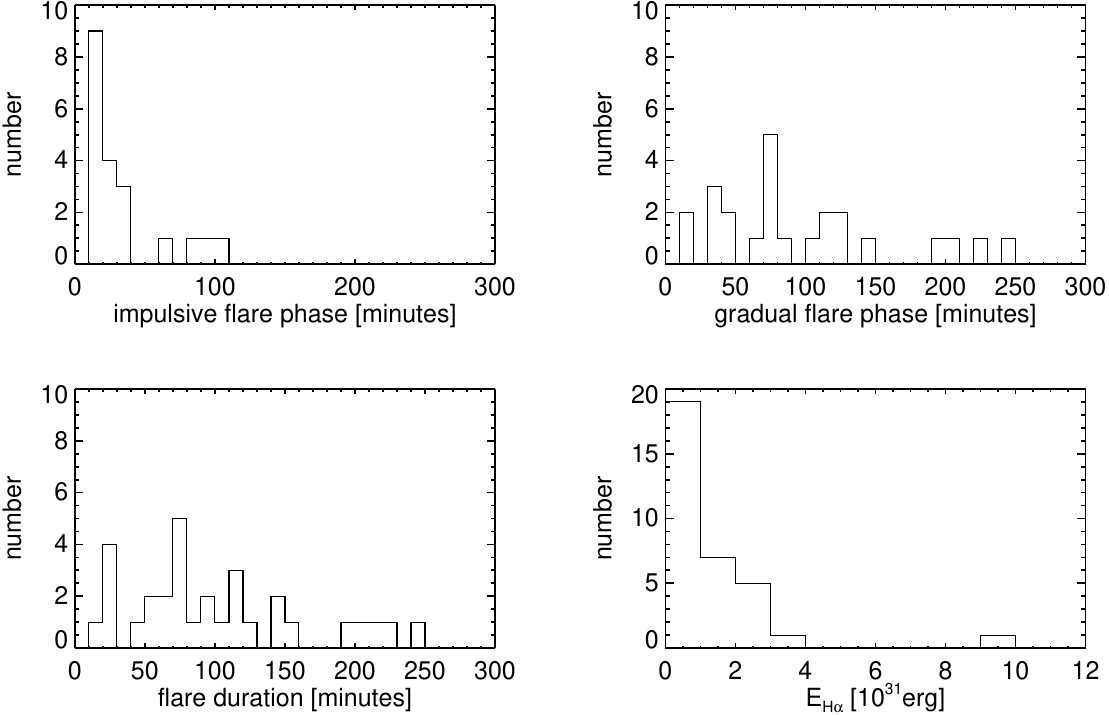}
\caption{Histograms of timing and energy information of the flares. Upper left panel: Histogram of the length of the H$\alpha$ impulsive phase. Upper right panel: Histogram of the length of the H$\alpha$ gradual phase. Lower left panel: Histogram of the H$\alpha$ full flare duration. Lower right panel: Histogram of H$\alpha$ flare energies.}
 \label{fig:flarehisto}
\end{figure}
Although we have coordinated g'-band photometry we define flares from Balmer light curves. This is related to the fact that not every Balmer flare has also a broad-band component, as low energy flares will not be detectable in a broad-band light curve. Therefore we produce equivalent-width (EW) Balmer light curves. To do so we follow the approach presented in \citet{Odert2025} by defining the continuum as an average of a blue  and red region of the corresponding spectral line. Thereby the EW is definded as follows EW = $\int$(C - F)/C d$\lambda$. Using this definition gives negative EWs for emission and positive EWs for absorption lines. Using this convention we generate EW light curves in the Balmer lines, H$\alpha$, H$\beta$, and H$\gamma$ (Fig.~\ref{fig:EWlc1}, ~\ref{fig:EWlc2}, ~\ref{fig:EWlc3}, ~\ref{fig:EWlc4}, ~\ref{fig:EWlc5}, ~\ref{fig:EWlc6}, ~\ref{fig:EWlc7}, ~\ref{fig:EWlc8} for the remaining flares). For the most energetic flare of the sample we also extract the H$\delta$, HeI(5876\AA), and sodium D1+D2 EW light curves (see Fig.~\ref{fig:exampleflare}). The line windows necessary to calculate the EW are set that way that those cover always both spectral components of CC~Eri. The definition of a flare is usually a rapid increase in flux and a slow decay. This evolution is more prominent in broad-band light curves than in Balmer EW light curves. However, a definite scheme how to decide what is a flare and what not is still difficult. In the present study we apply the following rules which a flare should fulfill: a) impulsive and/or gradual phase should be visible, b) the increase in EW during the impulsive phase must be $>$3~$\sigma$, c) the duration of the gradual phase should be greater equal the duration of the impulsive phase, d) a simultaneously detected 
\begin{figure}
 \includegraphics[width=8cm]{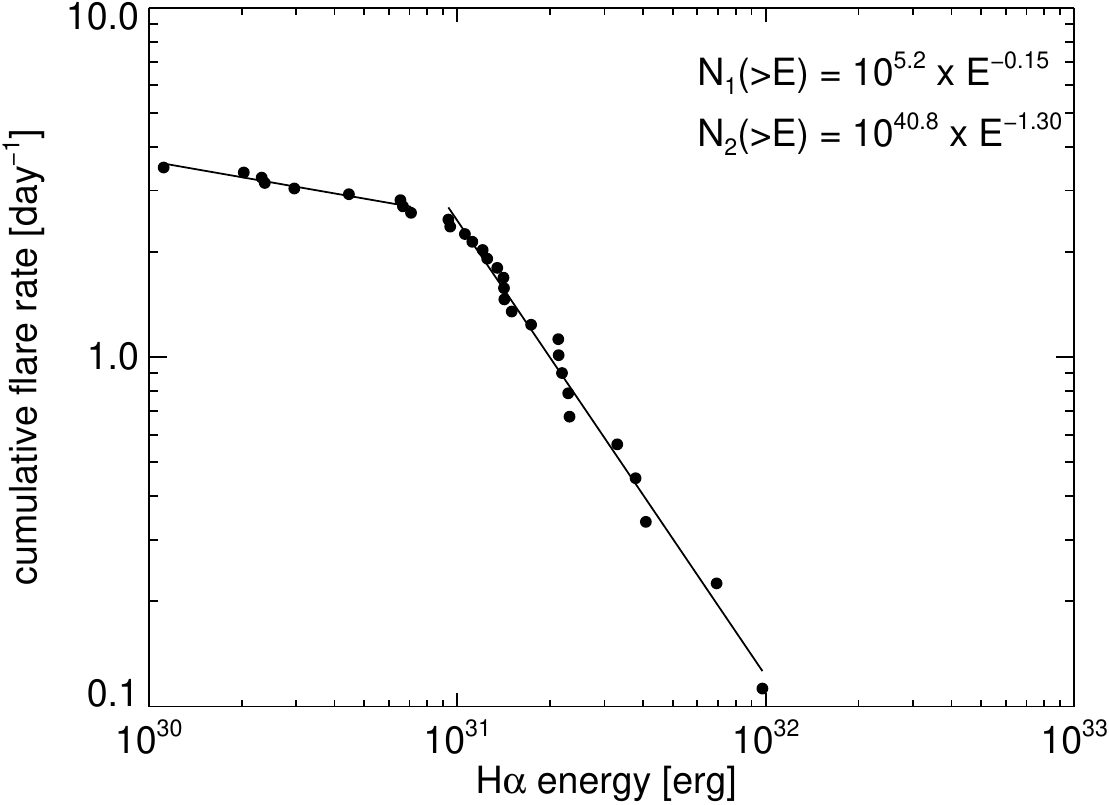}
  \caption{Cumulative flare frequency distribution of the H$\alpha$ flares of CC~Eri of the present study.}
 \label{fig:FFD}
\end{figure}
g'-band flare classifies also the Balmer EW light curve as flare light curve, d) at least one of the Balmer line cores needs to be enhanced. The majority of these rules are applicable only to a fully covered flare light curve. The flare light curves we detected consist sometimes of a gradual or impulsive phase only, or the impulsive/gradual phase is not fully covered.\\
 \begin{figure*}
 \includegraphics[width=18cm]{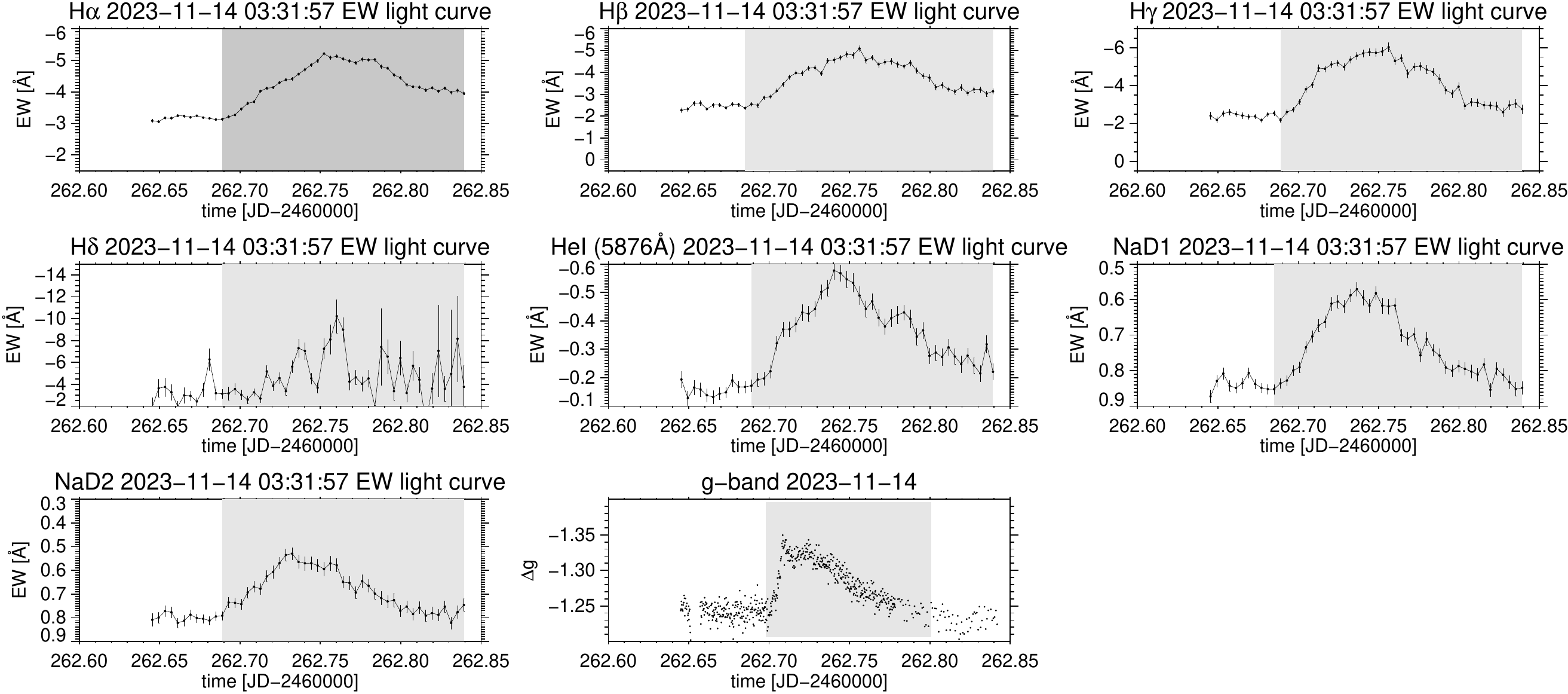}
 \caption{H$\alpha$, H$\beta$, H$\gamma$, H$\delta$, HeI D3 (5876\AA), NaD2+D1 and g'-band (from left to right and top to bottom) light curves of CC~Eri in the night of 2023-11-14 (flare no.~4) where the most energetic H$\alpha$ flare occurred during the monitoring of CC~Eri. The shaded area denotes the part of the EW light curve belonging to the flare.}
 \label{fig:exampleflare}
\end{figure*}
Applying these rules to the Balmer line EW light curves results in 31 flares. In Table~\ref{tab:B1} we list the 31 detected flares together with their H$\alpha$, H$\beta$, H$\gamma$, and g'-band (if available), start-, peak-, and stop-times, durations of the flare, impulsive, and the gradual flare phase (if available) in H$\alpha$, H$\beta$, and H$\gamma$, decay time of the g'-band flares, as well as the H$\alpha$, H$\beta$, H$\gamma$, and g'-band energies. The temporal flare parameters (e.g. start times, durations etc.) are not obtained from fitting the spectra but taken directly from the spectrum and therefore the accuracy of the values is limited by the cadence of the data. To calculate energies we follow the approach given in \citet{Odert2025}. Quiescent fluxes of the Balmer lines were taken from GAIA DR3 I/355/spectra.\\ 
If the flare is not fully covered then these parameters are lower limits only. We group the flares in three categories, in ''distinct`` flares, ''weak`` flares, and ''gradual phase only`` . In the group of ''distinct`` flares we group flares which have a g'-band counterpart. In addition we added also the flares from 2023-10-27 and 2024-01-10, as these are by far the flares with the largest peak enhancements having no g'-band counterpart. For flare 2023-10-27 there was no simultaneous photometry that night, because of its peak enhancement we would expect a g'-band counterpart. For flare 2024-01-10 only a partially simultaneous g'-band  light curve is available which starts during the impulsive phase of the H$\alpha$ flare. The first few g'-band data show a decay therefore we also assume that there could have been a coordinated g'-band flare which has not been fully captured. In the group of weak flares we collect flares without a g'-band counterpart and which have no pronounced flare peak enhancements. And finally in the group ``gradual phase only'' we group only flares where we have not observed the impulsive phase. This categorization gives an equal distribution of one third each.\\
In  Fig.\ref{fig:flarehisto} we show the histograms of H$\alpha$ flare parameters, namely the length of H$\alpha$ impulsive- and gradual phase (upper panels) as well as H$\alpha$ total duration and energy (lower panels). For the H$\alpha$ impulsive flare phase we see that shorter impulsive flare phases are more common. The H$\alpha$ gradual flare phase histogram shows a trend to shorter gradual phases. For the H$\alpha$ flare duration we see also more short flares than long ones. And for the H$\alpha$ energy we clearly see that there are more less energetic H$\alpha$ flares than H$\alpha$ flares revealing large energies.\\
In the cumulative flare frequency distribution of the H$\alpha$ flares (see Fig.~\ref{fig:FFD}) we see, as expected, a trend of an increasing number of flares with smaller energies. The flatter part of the distribution is related to detection thresholds of the instrument, i.e. we miss flares with smaller energies. To account for this behaviour we fit the distribution with two separate power laws (both are given in the plot). The H$\alpha$ power law slopes yield -0.15 and -1.30. Comparing the steeper slope with the ones determined from \citet[][AD~Leo:-1.56; EV~Lac:-1.81]{Muheki2020a, Muheki2020b} and \citet[][AU~Mic:-1.75]{Odert2025} we see that CC~Eri reveals a shallower slope than the ones from AD~Leo, EV~Lac, or AU~Mic indicating that the distribution is dominated by more energetic flares than in the case of the other three stars. 

\subsection{Flares from photometry}
We detect eight g'-band flares coordinated with the optical spectroscopy. In Table~\ref{tab:B1} we list temporal characteristics of the g'-band flares as well as their energies. To determine the g'-band energy we need to define start and stop times of the flare, i.e. when the flare deviates from the background. To do so we us the flare model from \citet{Gryciuk2017} which assumes a gaussian impulsive phase and an exponential decay. Applying the model by fitting the g'-band light curves and using a quiet g'-band magnitude of g'=9.52 results in flare energies given in Table~\ref{tab:B1}. To calculate the g'-band magnitude we us the conversion between Sloan g' and Johnson B and V \citep[taken from APASS DR9][B=10.161, V=8.71]{Henden2016} given in \citet[g = V + 0.634$\times$(B-V) - 0.108][]{Bilir2005}.  We find flare durations of 10 - 250 minutes and energies of $\sim$1-20$\times$10$^{32}$~erg. Two flares stand out from the sample, namely flares 2023-11-08 and 2023-11-14. Both are alone from their g'-band energies superflares (1.6, 2.0$\times$10$^{33}$~erg). Both have durations of 150 and 160 minutes, respectively. The decay times are 40 and $\sim$70 minutes, respectively. All the remaining g'-band flares have decay times of $<$6 minutes.

\subsection{Low-resolution spectroscopy from SA200}
\label{sec:lowres}
\begin{figure}
 \includegraphics[width=8cm]{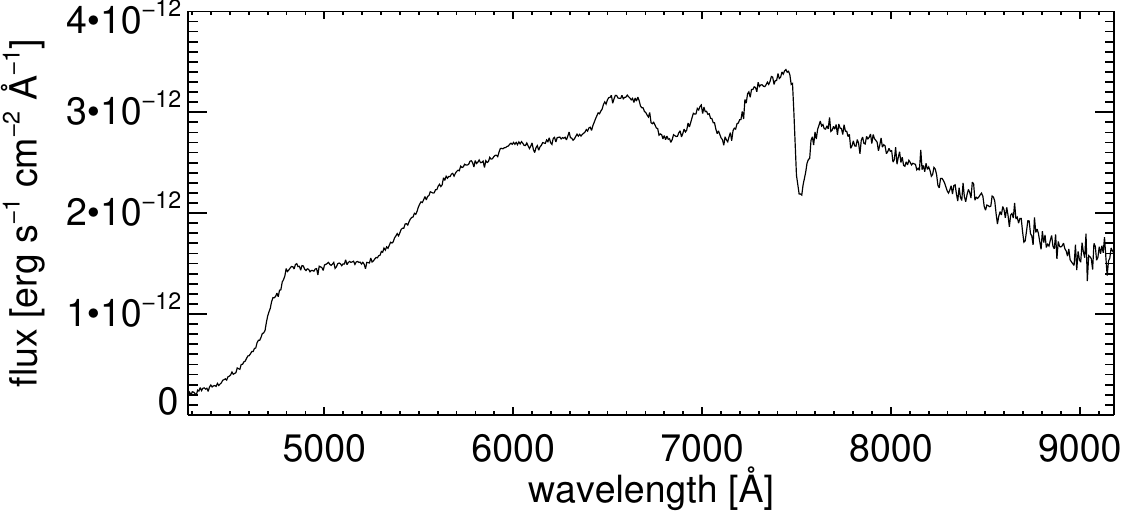}
 \caption{SA200 spectrum of CCEri obtained from all spectra of the night of 2023-11-28.}
 \label{fig:SA200}
\end{figure}

The ESO152 has two additional instruments mounted on the two finder telescopes. One of them is GrazCam which is capable of doing low-resolution spectroscopy using a transmission grating. During the CC~Eri campaign only three times the observations have been performed coordinated. Mainly therefore we did not catch any flare. However, in Fig.~\ref{fig:SA200} we show the SA200 spectrum of CC~Eri obtained during the night 2023-11-28. 

\subsection{Rotational modulation}
\begin{figure}
 \includegraphics[width=8cm]{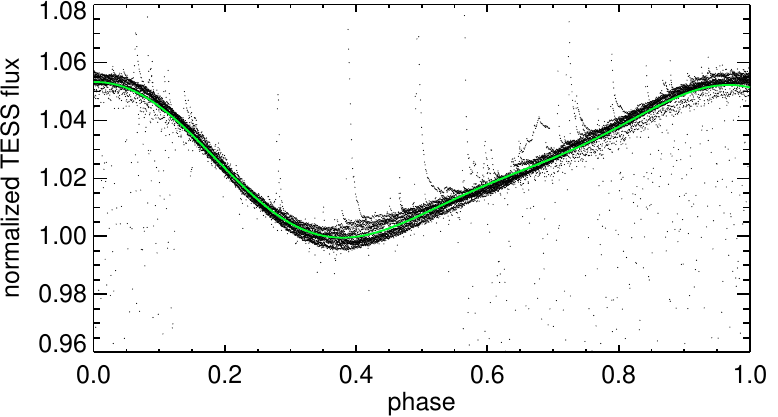}
 \includegraphics[width=8cm]{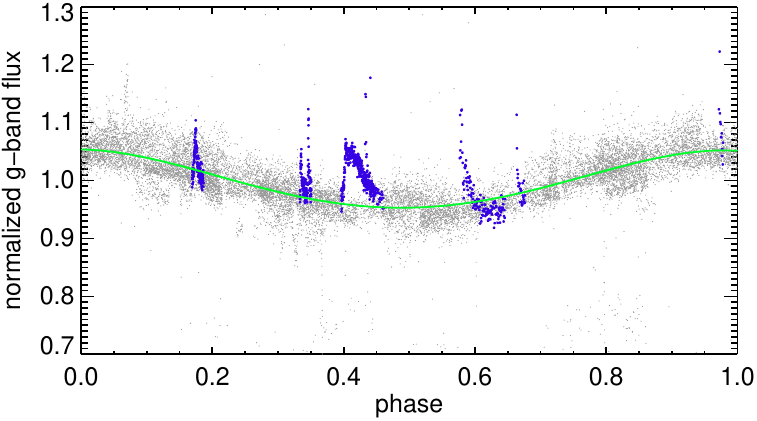}
 \includegraphics[width=8cm]{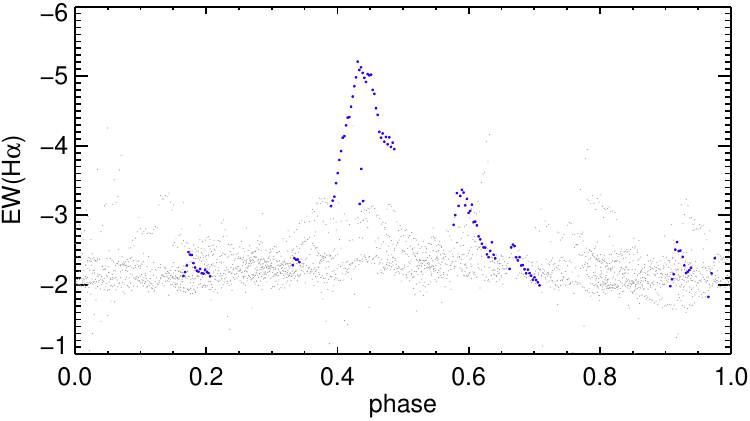}
 \caption{Phasefolded normalized flux light curves derived from TESS (upper panel) and g'-band (middle panel), and phasefolded light curve of EW(H$\alpha$) (lower panel). The g'band and TESS light curves are fitted with a double sinusoidal function (green solid line). The blue colored dots in the middle panel represent all detected g'-band flares (see Table~\ref{tab:B1}) and the blue dots in the lower panel represent all distinct flares detected in H$\alpha$ (see Table~\ref{tab:B1}).}
 \label{fig:phasefolded}
\end{figure}

The rotational modulation which is visible from the TESS light curve is supposed to originate from stellar spots. As CC~Eri is a tidally locked system, the same hemispheres are facing each other all the time. In the upper panel of Fig.~\ref{fig:phasefolded} we show the phasefolded TESS light curve. In the middle panel we show the phasefolded normalized g'-band flux of CC~Eri. The phasefolded data have been fitted with a double sinusodial function to determine the minimum and maximum of the light curve. In the lower panel of Fig.~\ref{fig:phasefolded} we show the normalized phasefolded H$\alpha$ flux which reveals a weak enhancement at phase 0.4. The phasefolded g'-band data reveal a clear minimum at this phase, which is an expected behavior \citep[see e.g.][]{Odert2025}. The amplitude of the g'-band phasefolded data is 0.096 whereas the amplitude of the phasefolded TESS data is 0.054, as determined from the sinusodial fits. Also in \citet{Odert2025} it was found that the g'-band amplitude was larger than the amplitude determined in TESS data. For CC~Eri we find a factor of $\sim$2 difference between g'-band and TESS which is roughly consistent with the factors found for AU~Mic in \citet[2 ... 8]{Odert2025}. Comparing the H$\alpha$ and g'-band phasefolded light curves we see at phase $\sim$0.4 flare no.~4 as significant event.

\subsection{Spectral line identification}
\label{lineident}
As we are interested in the excess emission of spectral lines during flares and superflares we need to perform a spectral line identification going beyond the prominent lines such as the Balmer, HeI, or MgI spectral lines. For this purpose we utilize the NIST\footnote{\url{https://physics.nist.gov/PhysRefData/ASD/lines_form.html}} data base and the spectral line catalogue of \citet{Moore1972} available at Vizier \citep{Coluzzi1993}. We also used the performed spectral line identifications in the literature of other stars during flaring \citep{Garcia-Alvarez2002, Paulson2006, Fuhrmeister2008, Fuhrmeister2011, Lalitha2013, Muheki2020b} as well as a study of a solar flare \citep{Johns-Krull1997} performed with Echelle spectroscopy. Furthermore we utilized the compilation of solar spectral atlases \citep{Malherbe2024} in the wavelength region of 3000 to 8800\AA{}. We apply the spectral line identification to spectral lines revealing excess emission during the impulsive and gradual flare phases on CC~Eri. All spectral excess emissions could be identified with this information. For flare no.~4 we identify the largest number of excess emissions. Furthermore we find 20 spectral lines (see Tab.~\ref{tab:D1}) revealing excess emissions only during the impulsive/gradual phase of flare no.~4. For a discussion on those spectral lines see section~\ref{superflares:special}.

\subsection{Spectral line asymmetries}

\begin{figure}
\centering
 \includegraphics[width=6.5cm]{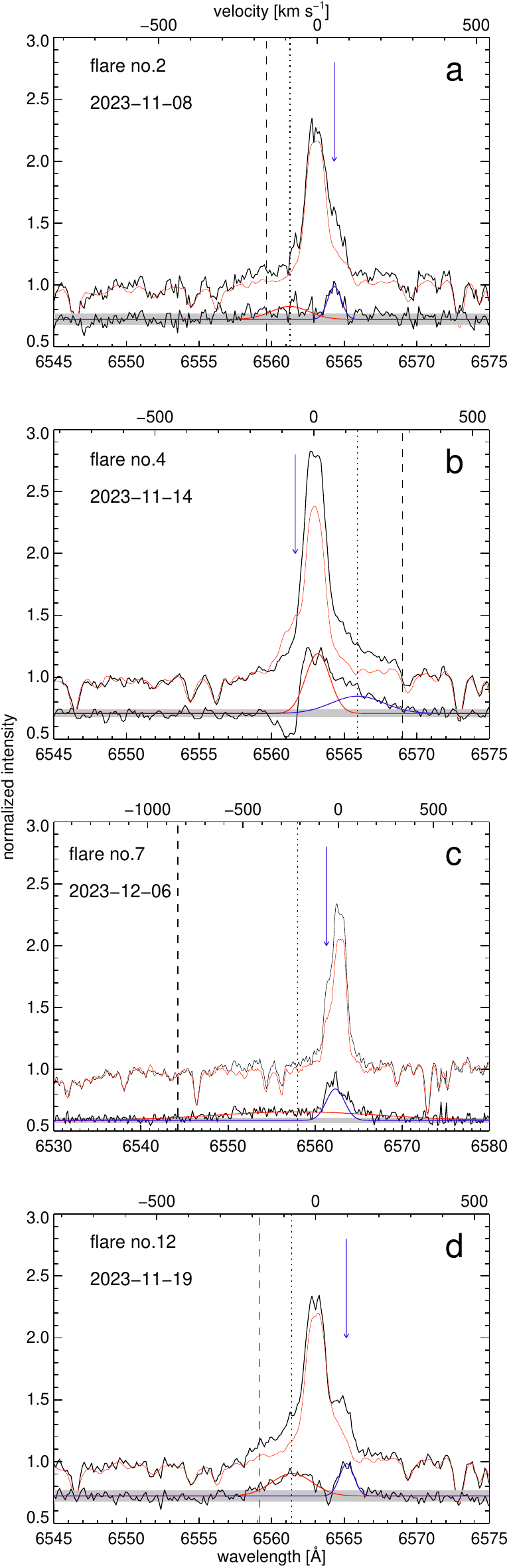}
 \caption{Detected asymmetries in H$\alpha$ during flares. For deduced parameters see Table~\ref{tab:C1}. In every subpanel we plot the spectrum showing the largest bulk velocity per night (solid black line). Overplotted is a pre- or post-flare spectrum (red solid line). Additionally we plot also in every subpanel below the actual spectrum the residual spectrum (shifted to a value of 0.7). The residual spectrum is overplotted with two gaussians accounting for the blue- or red asymmetries. Furthermore we have indicated the Doppler shift caused by the bulk velocities with a vertical dotted line and the one caused by the maximum velocity with a vertical dashed line. Also overplotted is a blue arrow indicating the position of the M-star component. }
 \label{fig:allasym}
\end{figure}

Flares are known to drive plasma motions visible at optical wavelengths. Furthermore also plasma ejections (eruptive filaments/prominences) being related to flares are visible in the optical. These phenomena can be investigated in optical spectra as motions leaving footprints in spectra in the form of extra emissions/absorptions around spectral lines or line asymmetries. Therefore we construct quiet spectra which are build from spectra which do not show signs of strong activity. By overplotting the spectra with the quiet spectrum of each night we are able to identify distinct asymmetries which may occur. We use the H$\alpha$ line as indicator of flare plasma motions and CMEs (see Fig.~\ref{fig:allasym} and Fig.~\ref{fig:asymapp}). In Table~\ref{tab:C1} we list all flares found during the observing campaign and their H$\alpha$ asymmetries (including their projected bulk and maximum velocities), if available. In Fig.~\ref{fig:allasym} and Fig.~\ref{fig:asymapp} we plot the spectra and residual spectra revealing the largest shifts in wavelength. \\
As one can see, in Fig.~\ref{fig:allasym} and Fig.~\ref{fig:asymapp} we overplotted two gaussians on the residual spectra to account for the spectral line asymmetry as well as possible line core enhancements. The residual spectra are generated by subtracting the aforementioned quiet spectrum from the impulsive or gradual flare phase spectrum. Using the gaussian fitting we are also able to deduce projected bulk and maximum velocities of the asymmetries (see Table~\ref{tab:C1}). Bulk velocities are determined from the gaussian fit component representing the asymmetry. As reference or central wavelength we use the K-star component. Therefore the  projected bulk velocity is calculated as the wavelength difference of the K-star component and the center of the gaussian fit representing the asymmetry. The maximum velocity is defined as the wavelength where the gaussian fit representing the asymmetry equals the 1-$\sigma$ error of the residuals. The errors of the projected bulk velocities are determined from the errors of the fitting routine. As the maximum velocity depends on the assumption where it is measured, we do not give errors and consider it a representative estimate rather than a measured value. In the case of CC~Eri and for the identification of spectral line asymmetries one has to take into account the binarity of the system. For the majority of flares we are able to identify the M-star component in H$\alpha$. If the spectral line asymmetry occurs close or overlayed on the M-star component it is difficult to identify the spectral line asymmetry. If the M-star component is on the opposite of the asymmetry (blue and red or red and blue) then the asymmetry can be clearly identified.
As mentioned above, we have grouped the detected flares in ''distinct``, ''weak`` and ''gradual phase only``. Seven out of nine (78\%) of the ''distinct`` flares group, eight out of 13 (62\%) of the ''weak`` flares group, and only three out of nine (33\%) of the ''gradual phase only`` group show H$\alpha$ asymmetries during flaring. In the following we comment on the asymmetries of the two g'-band superflares (flares no. 2 and 4) as well as on two further flares revealing distinct blue asymmetries (flares no.~7 and 12). All other, less pronounced, line asymmetries which we have detected are described in the appendix in section~\ref{asymap} where in Table~\ref{tab:C1} time and velocity characteristics of the asymmetries are given.\\
Flare no.~2 (2023-11-08): This superflare shows a blue wing enhancement during the impulsive flare phase (see panel (a) of Fig.~\ref{fig:allasym}). The blue wing enhancement starts in the first spectrum of the impulsive phase and lasts until four spectra after the peak. The total duration of the blue wing enhancement is 51~minutes. The projected bulk velocity is -86.2$\pm$9.4 and the projected maximum velocity is -161~km~s$^{-1}$. H$\beta$ (see panel (a) of Fig.~\ref{fig:asymapphbeta} in the appendix) and H$\gamma$ show asymmetries also after the peak as well. 
We can not comment on the red side of H$\alpha$ as the M-star component is located on the red side of the K-star component.\\
Flare no.~4 (2023-11-14): This superflare exhibits during its impulsive phase a red wing enhancement which increases in velocity during the gradual phase (see panel (b) of Fig.~\ref{fig:allasym}) and lasts until the last spectrum of the series. In H$\alpha$ the projected bulk velocity is 137.5$\pm$14.7 and the projected maximum velocity is 279.5~km~s$^{-1}$. H$\beta$ (see panel (b) of Fig.~\ref{fig:asymapphbeta} in the appendix) and also H$\gamma$ show a similar behaviour. Although the M-star component is on the blue side of the K-star component, we do not see any detectable blue asymmetry as in the case of flare no.~2.\\
Flare no.~7 (2023-12-06): This flare reveals a fast (the fastest of all detected asymmetries) blue wing enhancement in one spectrum of the impulsive phase (see panel (c) of Fig.~\ref{fig:allasym}). The projected bulk velocity is -213.5$\pm$49.3 and the maximum velocity reaches -842.1~km~s$^{-1}$. Also in H$\beta$ (see panel (c) of Fig.~\ref{fig:asymapphbeta} in the appendix) and H$\gamma$ these blue-wing enhancements can be seen.\\
Flare no.~12 (2023-11-19): In this spectral time series we see a rise in H$\alpha$ flux until the end of the time series lasting for 57~minutes. This rise is partly caused by blue wing enhancements (see panel (d) of Fig.~\ref{fig:allasym}) which increase towards the end of the time series. The projected bulk and maximum velocities are -74.7$\pm$13.7 and -143.9~km~s$^{-1}$, respectively. Also in H$\beta$ (see panel (d) of Fig.~\ref{fig:asymapphbeta} in the appendix) we see blue wing enhancements but not that distinct as in H$\alpha$.\\

\subsection{White-light component}
\label{results:whitelight}
Apart from the simple energy threshold of 10$^{33}$~erg which defines a superflare we also attempt the determination of a possible white-light component from the spectra. As unfortunately only few nights were observed coordinated with 
\begin{figure}
 \includegraphics[width=8.0cm]{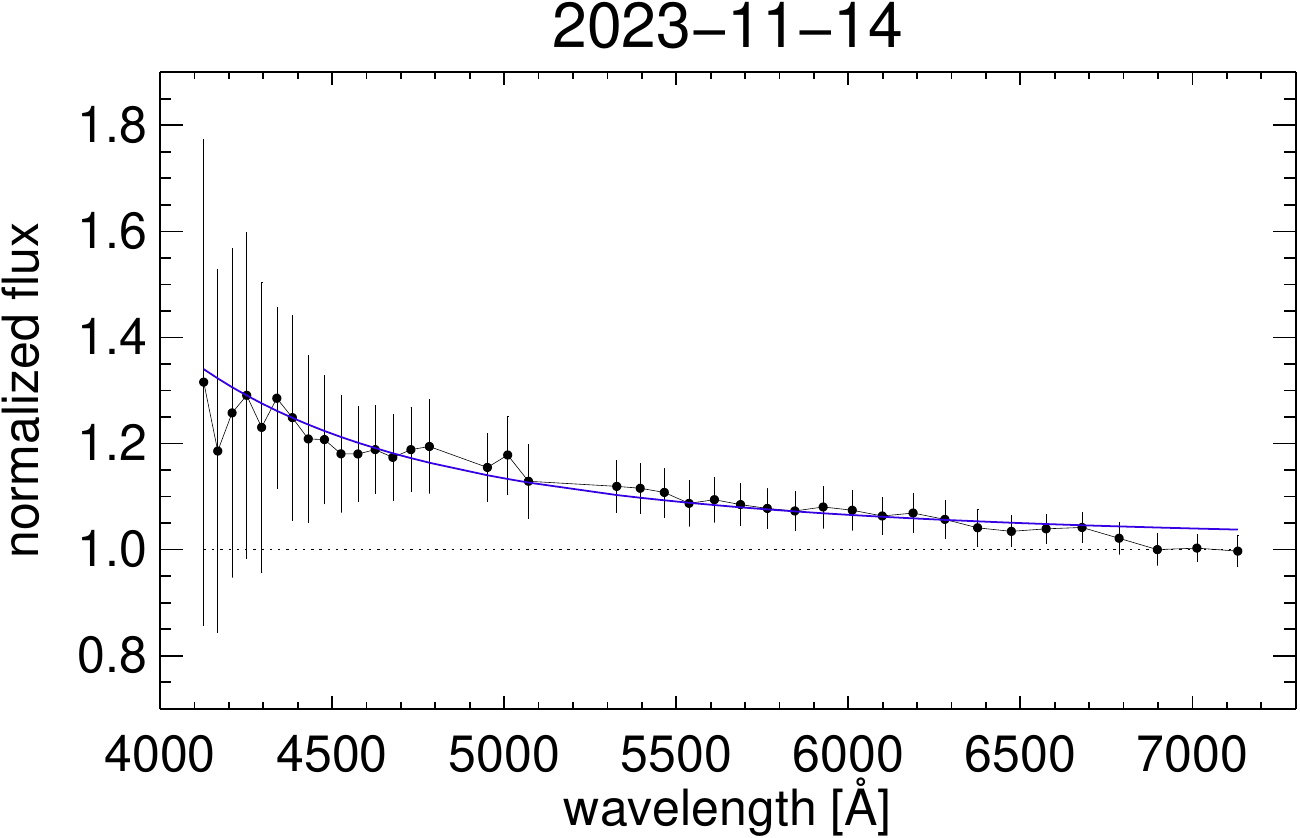}
 \includegraphics[width=8.0cm]{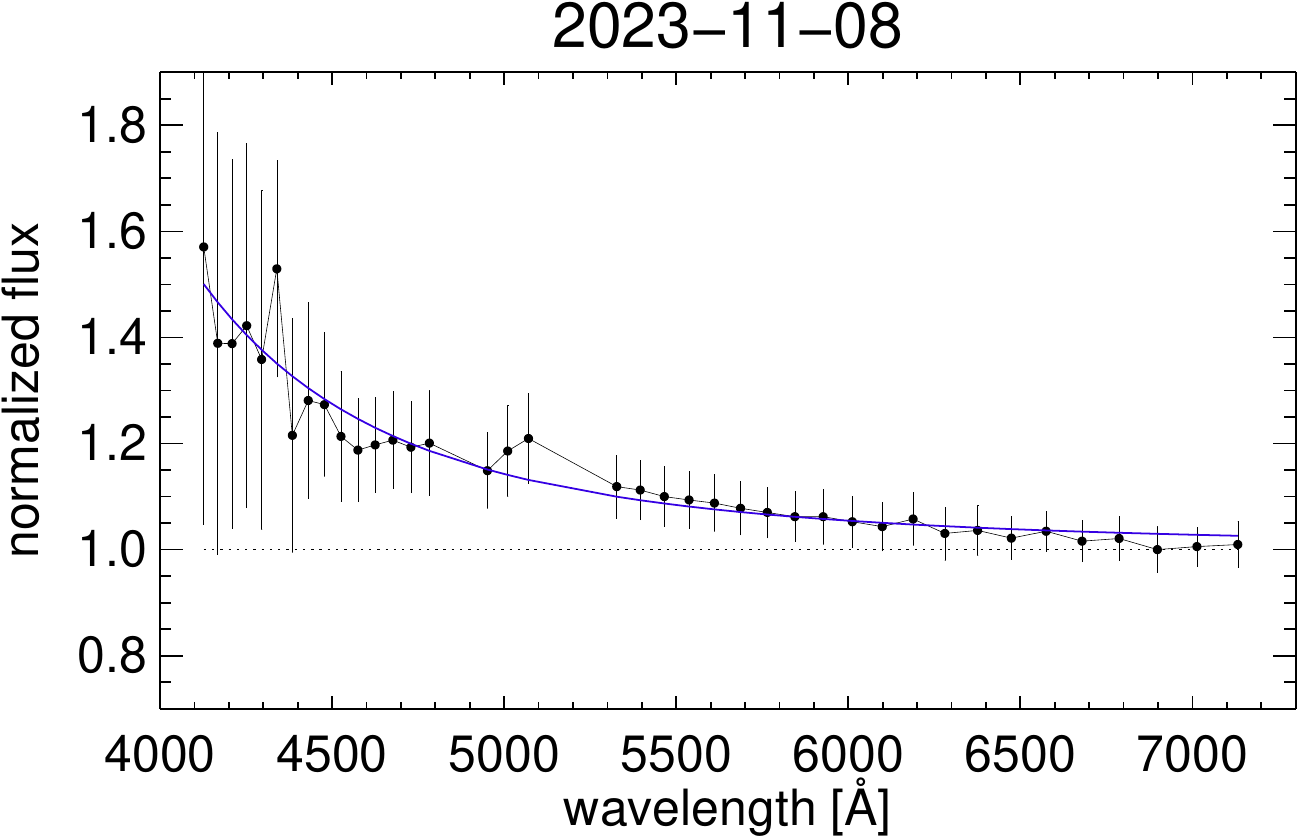}
 \caption{Median values of the central regions of the non-deblazed Echelle orders of the nights 2023-11-14 (upper panel) and 2023-11-08 (lower panel), i.e. flare no.~4 and no.~2. The horizontal dotted line represents the quiescent flux while the solid line represents the spectrum with the largest enhancement to the blue during the impulsive flare phase of each flare. Overplotted is a normalized Planck function.}
 \label{fig:whitelight20231114planck}
\end{figure}
GrazCam we alternatively use the method applied in \citet{Muheki2020b}. White-light emission of flares is continuum emission and originates from the heated footpoints of the flare loops anchored in the photosphere and/or, in the case of superflares, also from the flare loops \citep{Heinzel2018}. We use the non-deblazed Echelle spectra which still contain the continuum information from the star. To obtain a spectrum representing the continuum, we determined the median of each Echelle order. This spectrum is then normalized in the red wavelength range, in our case at the 3rd Echelle order. To make changes during flares/superflares visible we divide each spectrum by a representative quiescence spectrum obtained for each night. For flare no.~4 and no.~2 we find increased continua in the blue (see Fig.~\ref{fig:whitelight20231114planck}). 
In Fig.~\ref{fig:whitelight20231114planck} we show the continuum spectrum (black solid line) during the flare peak of flare no.~4 (upper panel)  and no.~2 (lower panel). Overplotted is a Planck function (blue solid line) determined from least-square fitting where we have set the effective temperature of CC~Eri to 3900~K. We obtain a flare temperature and area from Planck fitting (normalized Planck function depending on the stars and the flare effective temperature as well as a filling factor) of $\sim$~10000K and an area of 0.14\% (flare no.~4)  and $\sim$~31000~K and an area of 0.014\% (flare no.~2), respectively.\\
The error of the single spectral points is determined by gaussian error propagation of the ratio of the flare and quiet spectrum errors. These errors are given by the Ceres+ pipline and involve data error as well as read out noise. As the detector sensitivity in the blue decreases significantly we see much larger errors as on the red side of the spectrum. The CC~Eri spectra are affected by atmospheric dispersion during the night resulting in varying blue and red parts of the spectra during the night. As one can see from Fig.~\ref{fig:whitelight20231114comp} where we overplot a white-light light curve derived from the white-light spectra, as described above, the white-light light curve resembles the g'-band light curve which is not affected by atmospheric dispersion. For the generation of the white-light light curve we use the spectral range which resembles the g'-band, for better comparison. Apparently, the variation of the spectra due to atmospheric dispersion is much smaller than the variation due to the occurrence of the superflares shown in Fig.~\ref{fig:whitelight20231114comp}.
\begin{figure}
 \includegraphics[width=8.0cm]{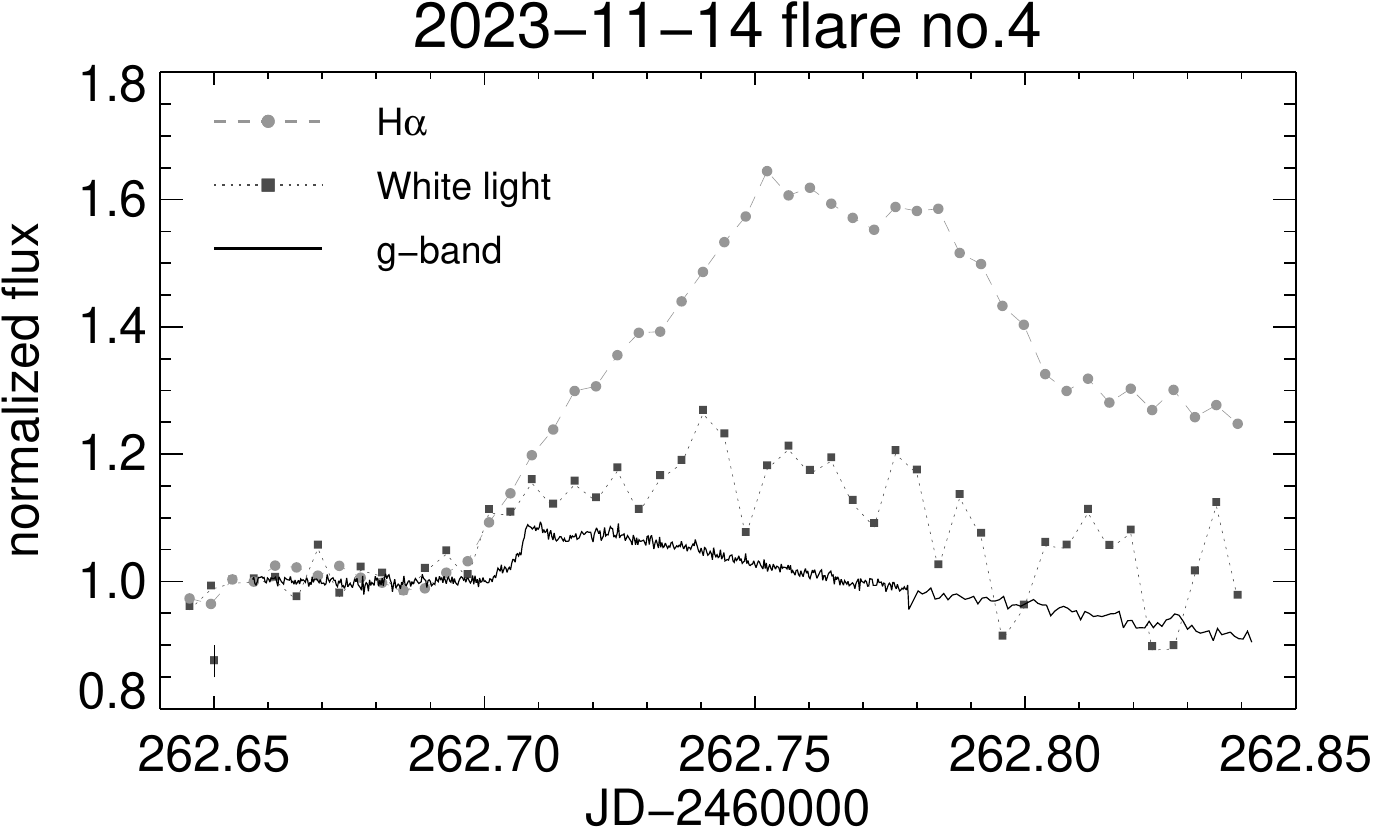}
 \includegraphics[width=8.0cm]{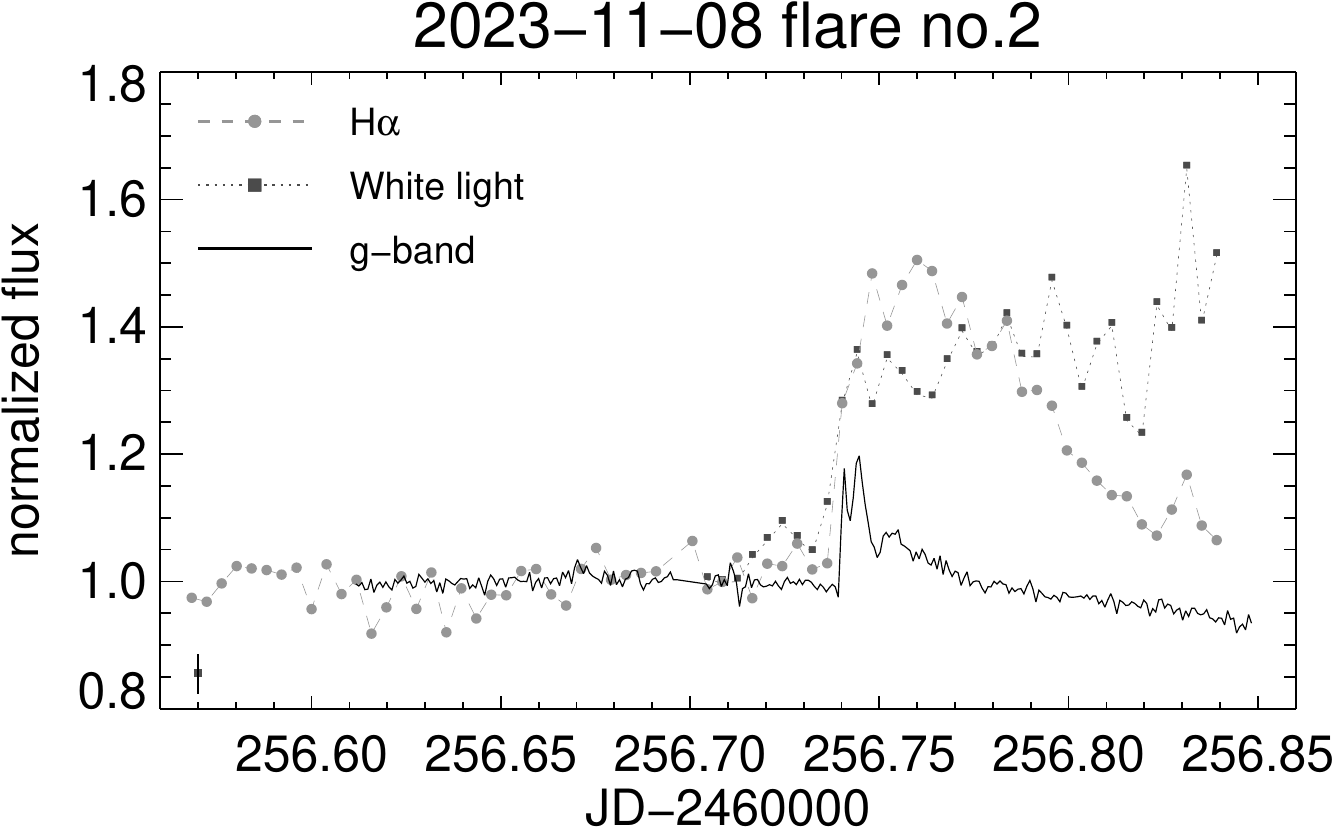}
 \caption{H$\alpha$ (light-grey dashed line and light-grey dots), g'-band photometry (solid black line), and white-light from spectroscopy (grey dotted line and grey filled squares) of the 2023-11-14 (upper panel) and 2023-11-08 (lower panel) superflare. In the lower left corner of each plot we plot the mean error of the white-light light curve.}
 \label{fig:whitelight20231114comp}
\end{figure}
For flares no.~2 and ~4 there is also coordinated g'-band photometry (see Fig.\ref{fig:exampleflare}) which we therefore can use to cross-check our deduced continuum spectra. In Fig.~\ref{fig:whitelight20231114comp} we show therefore three light curves of flares no.~2 and ~4. The H$\alpha$ light curve is represented by a dashed light-grey line, the g'-band light curve by a solid black line and the white-light light curve, deduced from the continua spectra. by a dotted dark-grey line. The g'-band light curve and the white-light light curve reveal a similar peak and light curve evolution. However, we have normalized the spectra, from which the white-light light curve is generated, in the red where also some flare-related emission might occur. This might cause some additional uncertainties, especially when comparing the white-light and g'-band light curves.\\

\subsection{Flare affected spectral lines}
\label{sec:speclines}
As our data consist of Echelle spectra we are able to investigate numerous spectral lines during flares and superflares. We examine every Echelle order during flares. CC~Eri shows the Balmer lines H$\alpha$, H$\beta$, H$\gamma$, and H$\delta$ in emission in quiescence. Also the cores of the NaD1+D2 spectral lines and the MgI triplet (5167, 5173, 5184\AA), as well as the temperature sensitive HeI D3(5876\AA ) line are seen in emission during quiescence. In Table~\ref{tab:D1}, \ref{tab:D2}, \ref{tab:D3}  we list all spectral lines which were enhanced during stellar flares. Ranking the flares according to their number of detected excess emission in spectral lines, flare no.~4 shows the largest number of detections in the impulsive phase, namely 63, followed by flare no.2 (40), and flare no.6 (25). Also in the gradual phase flare no. 4 shows the largest number of detections (29). The remaining gradual phases of flares from the ''distinct`` flare sample have detections $<$13. The weak flares show $<$11 detections in either impulsive or gradual phase. The sample of flares showing gradual phases only show $<$6 detections except for flare 31 which reveals 22 detections.\\
In the following we comment on flares which reveal excess emissions other than the Balmer lines during flares:\\ 
\textbf{Flare no.~1 (2023-10-27):}	In the impulsive flare phase we see excess emission in the Balmer, sodium, HeI(4471, 5876, 6678\AA{}), and MgI triplet lines. In the gradual phase the MgI triplet and the HeI(4471\AA{}) are not visible anymore, \\
\textbf{Flare no.~2 (2023-11-08):}	This is the flare with the second largest number of enhanced spectral lines during flaring. Also here the prominent Balmer, sodium, MgI triplet lines as well as a number of FeI and HeI lines are enhanced. We also see excess emission in few ionized lines such as e.g. the FeII(4549\AA{}) or the MgII(4481) lines. Also detected is the HeII(4686\AA{}) line which is known to be formed at higher temperatures as the HeI D3(5876\AA{}) line. Furthermore we see excess emission at 5316.61\AA{} which is also seen in flare no.~4. In the gradual phase we see a much lower number of enhanced spectral lines than in the impulsive phase. \\
\textbf{Flare no.~3 (2023-11-12):}	This flare shows only few spectral lines which are enhanced during the impulsive phase as well as the gradual phase including the Balmer lines and the sodium lines. The temperature sensitive HeI(5876\AA) spectral line is not detected in the impulsive nor in the gradual flare phase. On the other hand  we see excess emission in the HeII(4686\AA{}) line, detected in the gradual phase but not in the impulsive phase. \\
\textbf{Flare no.~4 (2023-11-14):}	This is the flare with the largest number of enhanced spectral lines. In the impulsive phase we see excess emission in the Balmer, sodium, MgI triplet, HeI and HeII lines. We also see further FeII lines which are not existent in the other flares of the study. The gradual phase shows besides the Balmer, sodium, MgI triplet and HeI lines also the HeII(4686\AA{}) line. Moreover we see 20 additional spectral lines revealing excess emission only visible during this flare.\\	
\textbf{Flare no.~5 (2023-11-17):}	This flare shows only few enhanced lines such as the Balmer, sodium and MgI triplet lines. \\	
\textbf{Flare no.~6 (2023-12-02):}	This flare, where we have captured only a fraction of the impulsive phase, shows several enhanced lines, such as Balmer, sodium, HeI and MgI triplet lines. Also seen is the temperature sensitive HeII(4686\AA{}) line.\\		
\textbf{Flare no.~7 (2023-12-06):}	This flare, where we have captured only a cutout around the probable peak shows excess emission in the Balmer lines, sodium and MgI triplet lines and also in the HeII(4686\AA{}) line. \\	
\textbf{Flare no.~8 (2023-12-11):}	This flare reveals several enhanced lines such as the Balmer, sodium, HeI and MgI triplet lines. We also see excess emission at 5317\AA{}, which was also seen in flares no.~2 and no.~4. The gradual phase on the contrary reveals only an enhanced H$\alpha$ spectral line.\\	
\textbf{Flare no.~9 (2024-01-10):}	This flare shows excess emission in the Balmer, sodium and MgI triplet lines during the impulsive and  gradual phase. No HeII(4686\AA{}) line is detected.\\	
The majority (flares no.~10, 11, 12, 14, 15, and 16) of the weak flare sample (see Table~\ref{tab:D2}) reveals excess emission in H$\alpha$ and partly also H$\beta$. The remaining weak flares (flares no.~17, 18, 19, 20, 21, and 22) show partly also excess emission in the sodium, the HeI(5876\AA{}) and in only one case the MgI triplet lines. No HeI(4471\AA{}) or HeII(4686\AA{}) lines are detected.\\
The same accounts for the sample of flares where we have captured only the gradual flare phase (see table~\ref{tab:D3}). The majority of events shows enhanced H$\alpha$ spectral lines with sometimes H$\beta$, and H$\gamma$, one component of the sodium lines, or some weak enhancement in other spectral lines (flares no.~23, 24, 25, 26, 28, 29, and 30). The last flare in the sample (flare no.~31) represents an exception in the sample of flares where we have detected a gradual phase only, as it shows by far the largest number of enhanced spectral lines in this sample. We find Balmer, sodium, few FeI, HeI lines, and also the MgI triplet lines.

\subsection{Flare energy relations}
\begin{figure}
\includegraphics[width=8.9cm]{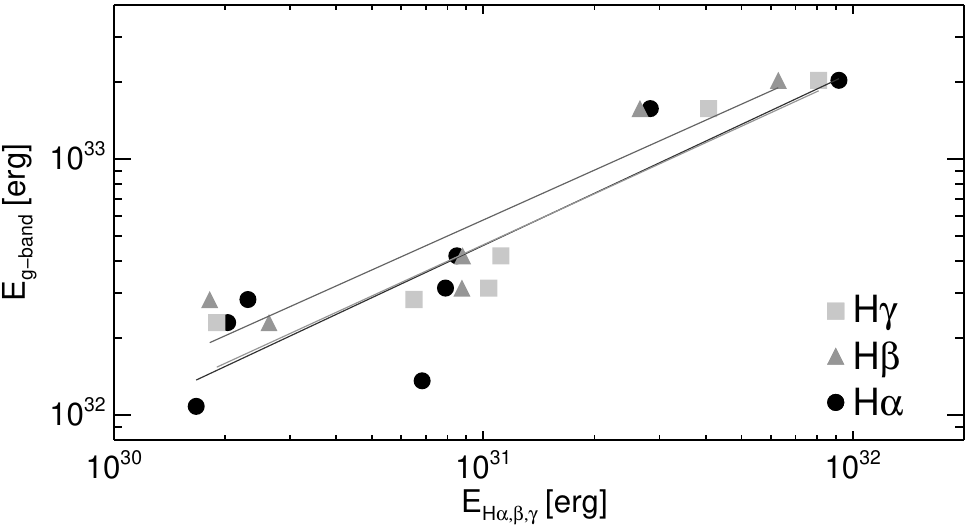}
 \caption{H$\alpha$, H$\beta$, and H$\gamma$ flare energies versus g'-band flare energies. The different grey colored lines are line fits to the corresponding H$\alpha$ (black), H$\beta$ (grey), and H$\gamma$ (light grey) energies. For parameters of these fits see Table~\ref{tab:flareparam}.}
 \label{fig:balmergband}
\end{figure}

\begin{figure}
\includegraphics[width=8.9cm]{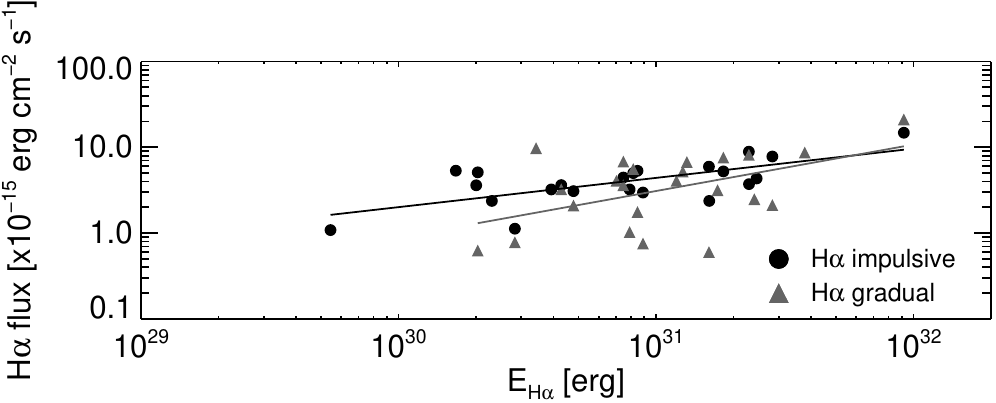}
\includegraphics[width=8.9cm]{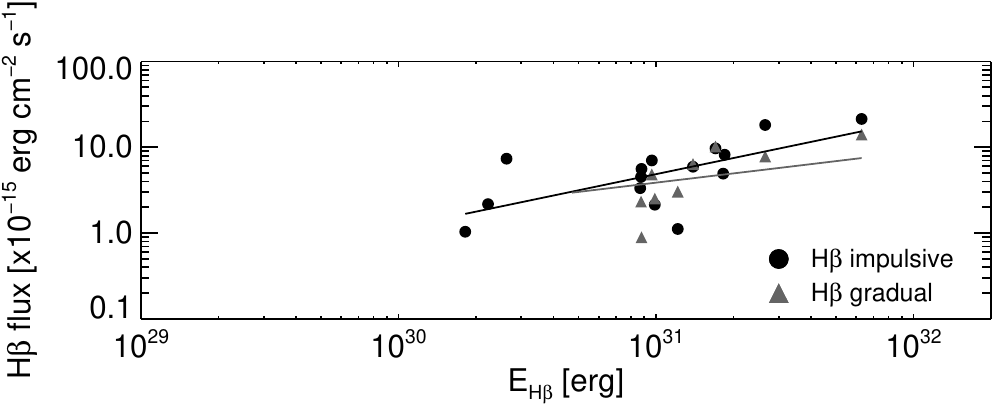}
\includegraphics[width=8.9cm]{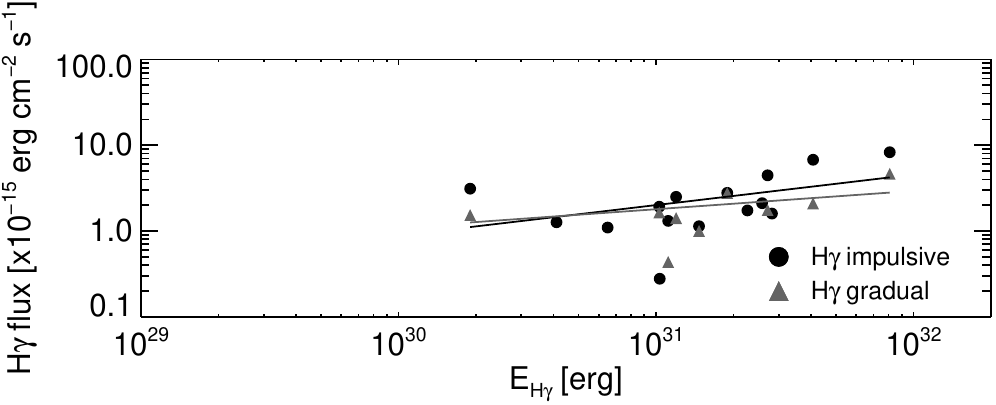}
 \caption{H$\alpha$, H$\beta$, and H$\gamma$ flare energies versus corresponding Balmer line mean fluxes for impulsive (black dots) and gradual (grey filled triangles) flare phases. Overplotted are line fits (impulsive flare phase: black solid line; gradual flare phase: grey solid line) to the data. For parameters of these fits see Table~\ref{tab:flareparam}.}
 \label{fig:balmerfluxenergy}
\end{figure}

\begin{figure}
 \includegraphics[width=8.18cm]{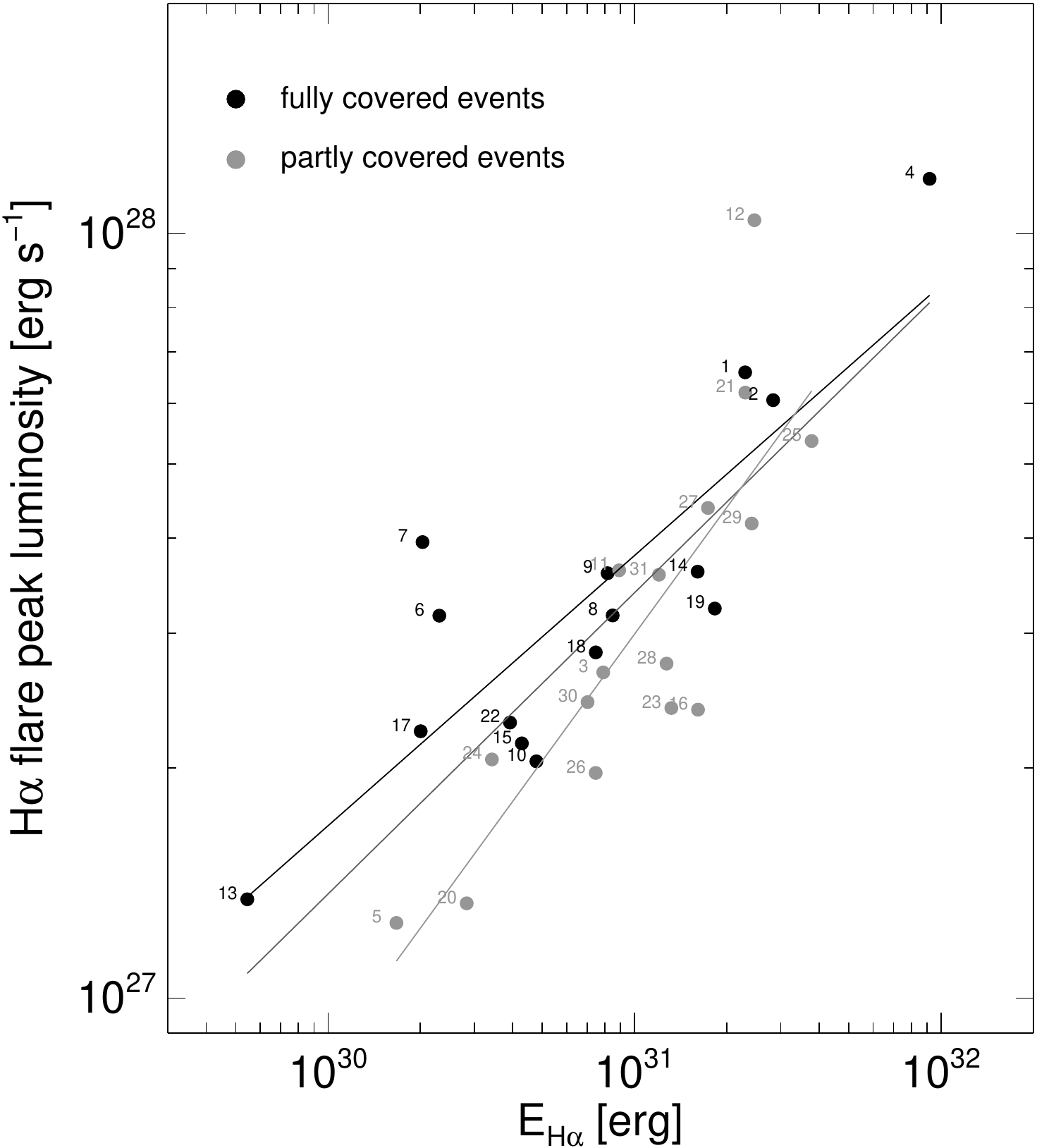}
 \caption{H$\alpha$ flare energy versus H$\alpha$ flare peak luminosity of the flare sample of the current study. Grey dots mark lower limits for E$_{H\alpha}$ or H$\alpha$ luminosity as there are some events where we have captured only the impulsive or the gradual phase. Overplotted are line fits to three different data groups: all flares (grey solid line), fully covered flares (black solid line), and not fully covered flares (light grey solid line). For parameters of these fits see Table~\ref{tab:flareparam}.}
 \label{fig:energylum}
\end{figure}

In this section we relate energies derived from different wavelength regimes and investigate their dependence. We find that the g'-band flare energies scale with H$\alpha$, H$\beta$, and H$\gamma$ flare energies. The larger the g'-band energies the larger are also the Balmer line energies during flares (see Fig.~\ref{fig:balmergband}). 
Investigating the dependence of the mean Balmer line fluxes and energies (see Fig.~\ref{fig:balmerfluxenergy}) analysed here for the impulsive and gradual flare phases reveals the same behaviour for all investigated Balmer lines. The larger the Balmer flare energy the larger also the Balmer line fluxes for both, the impulsive and gradual flare phases. We fit all relations with power laws. The fitting parameters are given in Table~\ref{tab:flareparam}.
We also investigate the relation of H$\alpha$ flare energy and H$\alpha$ flare peak luminosity. From figure~\ref{fig:energylum} one can see that the larger the H$\alpha$ flare energy the larger also the H$\alpha$ flare peak luminosity.

\begin{table}
	\centering
	\caption{Fitting parameters of the relation between H$\alpha$ flare energy and H$\alpha$ flare peak luminosity (see Fig.~\ref{fig:energylum}), the relation between g'-band flare and Balmer flare energies (see Fig.~\ref{fig:balmergband}), as well as fitting parameters of the relation between Balmer flare line fluxes and Balmer flare energies (see Fig.~\ref{fig:balmerfluxenergy}).}
	 \label{tab:flareparam}
	\begin{tabular}{lcc}
		\hline
                     &      a1       &      a2 \\
    \hline
                     &   log(E$_{\mathrm{H\alpha}}$)= a2 + a1$\times$log(L$_{\mathrm{H_{\alpha, peak}}}$)           &       \\                
   \hline
   all flares               &   0.395270 &  15.2655     \\ 
   fully covered flares     &   0.360562 &  16.3825     \\
   not fully covered flares &   0.506238 &  11.7837     \\
   \hline   
                     &    log(E$_{\mathrm{g'-band}}$)= a2 + a1$\times$log(E$_{\mathrm{H_{\alpha}}}$)                &        \\
   \hline                  
   H$\alpha$         &   0.676997    &   11.6759 \\
   H$\beta$          &   0.647505    &   12.6897 \\
   H$\gamma$         &   0.663003    &   12.1122 \\
   \hline  
                     &   log(F$_{\mathrm{H_{\alpha}}}$)= a2 + a1$\times$log(E$_{\mathrm{H_{\alpha}}}$)           &          \\
   \hline    
   H$\alpha_{\mathrm{imp}}$   &   0.36957422  & -10.812468   \\        
   H$\alpha_{\mathrm{grad}}$  &   0.76903796  & -23.436469   \\
   H$\beta_{\mathrm{imp}}$    &   0.40568967  & -11.842021   \\
   H$\beta_{\mathrm{grad}}$   &   0.27266524  & -7.8445791   \\
   H$\gamma_{\mathrm{imp}}$   &   0.24911670  & -7.2505008   \\
   H$\gamma_{\mathrm{grad}}$  &   0.23866840  & -7.1207789  \\     
   \hline
              
	\end{tabular}
\end{table}

\subsection{Flares and binarity}
\begin{figure}
 \includegraphics[width=8.18cm]{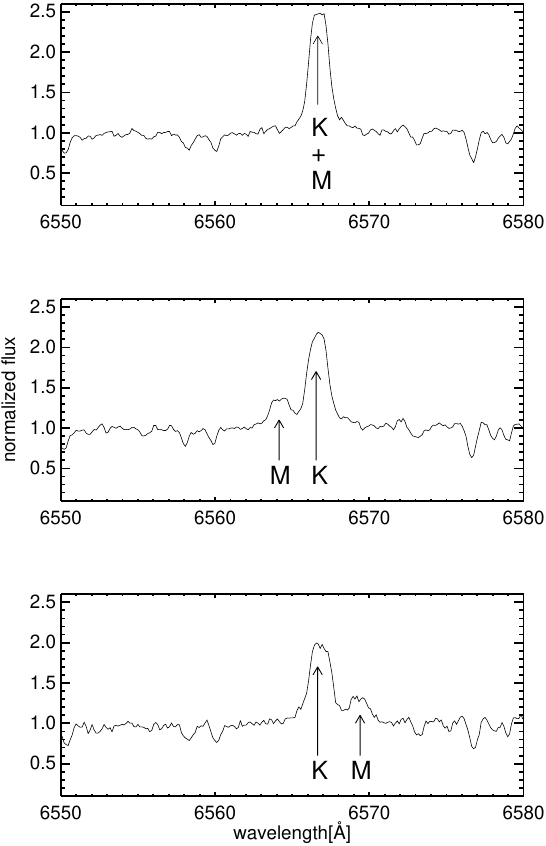}
 \caption{Upper panel: Merged H$\alpha$ line profile of K7 and M3 component. Middle panel: Separated H$\alpha$ line profile of K7 and M3 component, M3 component bluewards the K7 component. Lower panel: Separated H$\alpha$ line profile of K7 and M3 component, M3 component redwards the K7 component.}
 \label{fig:CCEriorbitspectrum}
\end{figure}

CC~Eri is a non-eclipsing binary which can not be spatially resolved. This means that light from both, the K7 and the M3 component, falls onto the target fibre of the spectrograph. This becomes obvious especially for the spectral lines where both components contribute significantly, such as the Balmer lines (see e.g. Fig.~\ref{fig:CCEriorbitspectrum}). From Table.~\ref{tab:1} one can deduce that the stellar disk of the K7 component is a factor of 2.4 larger than for the M3 component.

\begin{table}
	\centering
	\caption{Flares of the present study and their possible flaring source. Plus and minus signs indicate from which component the flare may have arisen. In the first colum we give the flare number, in the second the date, in the third and fourth the assignment to the binary components, in the fifth the corresponding scenario (see text), and in the sixth comments. }
	 \label{tab:flareorigin}
	\begin{tabular}{lccccc}
\hline
     & distinct flares & K7 & M3  & scenario & comments\\
\hline
1    &   2023-10-27    & +  &  - &     b     & \\
2    &   2023-11-08    & +  &  + &     b     & impulsive phase M \\
     &                 &    &    &           & gradual phase K\\
3    &   2023-11-12    & +  &  ? &     a     & K dominating\\
4    &   2023-11-14    & +  &  ? &     b     & K dominating\\
5    &   2023-11-17    & -  &  + &     a     & \\
6    &   2023-12-02    & +  &  ? &    b/c    & K dominating\\
7 	 &   2023-12-06    & +  &  ? &     b     & K dominating\\
8  	 &   2023-12-11    & +  &  - &     a     & \\
9  	 &   2024-01-10    & ?  &  + &    a/b    & M dominating\\
\hline
     &  weak flares & & &\\
\hline
10   &   2023-10-26    & ?  &  + &     a     & M dominating\\
11   &   2023-11-06    & +  &  - &     a     & \\
12   &   2023-11-19    & +  &  + &     a     & blue asymmetry\\
13   &   2023-11-23    & +  &  - &    a/b    & \\
14   &   2023-11-24    & +  &  ? &     a     & K dominating\\
15   &   2023-11-28    & +  &  + &    a/b    & 1st-M, 2nd-K\\
16   &   2023-12-02    & ?  &  ? &     a     & not assignable\\
17   &   2023-12-05    & -  &  + &     a     & \\
18   &   2023-12-21    & +  &  - &    a/b    & \\
19   &   2023-12-23    & +  &  ? &    a/b    & K dominating\\
20   &   2024-01-05    & +  &  - &     a     & \\
21   &   2024-01-07    & +  &  ? &    b/c    & K dominating\\
22   &   2024-01-08    & ?  &  ? &    a/b    & not assignable\\
\hline
     & gradual phase only & & &\\
\hline
23   &   2023-10-28    & ?  &  ? &    b/c    & not assignable\\
24   &   2023-11-13    & +  &  - &     a     & \\
25   &   2023-11-22    & ?  &  ? &     b/c   & not assignable\\
26   &   2023-11-30    & +  &  ? &    a/b    & K dominating\\
27   &   2023-12-12    & -  &  + &    a/b    & \\
28   &   2023-12-16    & +  &  ? &     a     & K dominating\\
29   &   2023-12-19    & +  &  ? &     a     & K dominating\\
30   &   2024-01-06    & +  &  - &     a     & \\
31   &   2024-01-11    & ?  &  ? &     b/c   & not assignable\\
\hline
\end{tabular}
\end{table}

Accordingly, the contributions to the Balmer lines are different. Dependent on orbital phase one either sees both components separated or merged. We decided to not disentangle the merged profiles as this is a challenging task \citep[see e.g.][]{Kriskovics2013} for constructing quiescent spectral profiles and for flaring this would require dedicated modelling and this is beyond the scope of the study. 
The short orbital period of the system makes it difficult to e.g. build average or quiescent spectra. Therefore we use only an average of few spectra prior the flare spectra (e.g. impulsive flare phase). This ensures that the separation of the K7 and M3 star component in the  average or quiescent spectra have a roughly similar separation as in the active spectra. 
In Fig.\ref{fig:CCEriorbitspectrum} three different orbital positions of the CC Eri system are shown. In the upper panel one can see an H$\alpha$ spectrum of CC~Eri, where the H$\alpha$ profiles of both components are merged. As the presented spectra originate from successive nights and we know that the difference of the spectrum in the upper panel to the middel one is $\sim$ 27~h the merged spectrum represents the K7 component behind the M3 component. In the middle panel one can see the M3 component bluewards the K7 component, which means that with respect to the prior night the M3 component has shifted to the blue, meaning that it has moved behind the K7 component and from behind the K7 component towards the observer. The configuration is here, M3 component on the left and K7 component on the right. In the lower panel one can see the opposite configuration, namely the M3 component redwards the K7 component. The configuration here is M3 component on the right and K7 component on the left. This is consistent with a time difference of $\sim$20~h between middel panel and lower panel, corresponding to roughly half of the orbital period.\\
The binary nature of CC~Eri raises the question if a flare occurs, from which component it originates or does it originate from the magnetic fields of both components. The binarity of the system is mainly obvious in the Balmer lines, as we see here two components (cf. Fig.~\ref{fig:CCEriorbitspectrum}). We assign the K-type component to the stronger Balmer component and the M-type component to the weaker Balmer component \citep[see also][]{Amado2000}. Thereby we can at least partly determine from which component the flare originates. We identify three different scenarii, a) both spectral profiles are well separated, b) the spectral profiles start to merge or separate, and c) the spectral profiles are merged. All of these scenarii are evident during the detected flares. Scenario a) enables the full determination of the flaring component, scenario b) the determination of the flaring component with a larger uncertainty, and scenario c) no determination of the flaring component. Applying this to the flares reveals the results shown in table~\ref{tab:flareorigin}. In section~\ref{Hamorph} in the appendix we describe the evolution of the H$\alpha$ components (K7 and M3) of every detected flare. This description is reflecting and supporting the content of Table~\ref{tab:flareorigin}. \\
\begin{figure}
 \includegraphics[width=8cm]{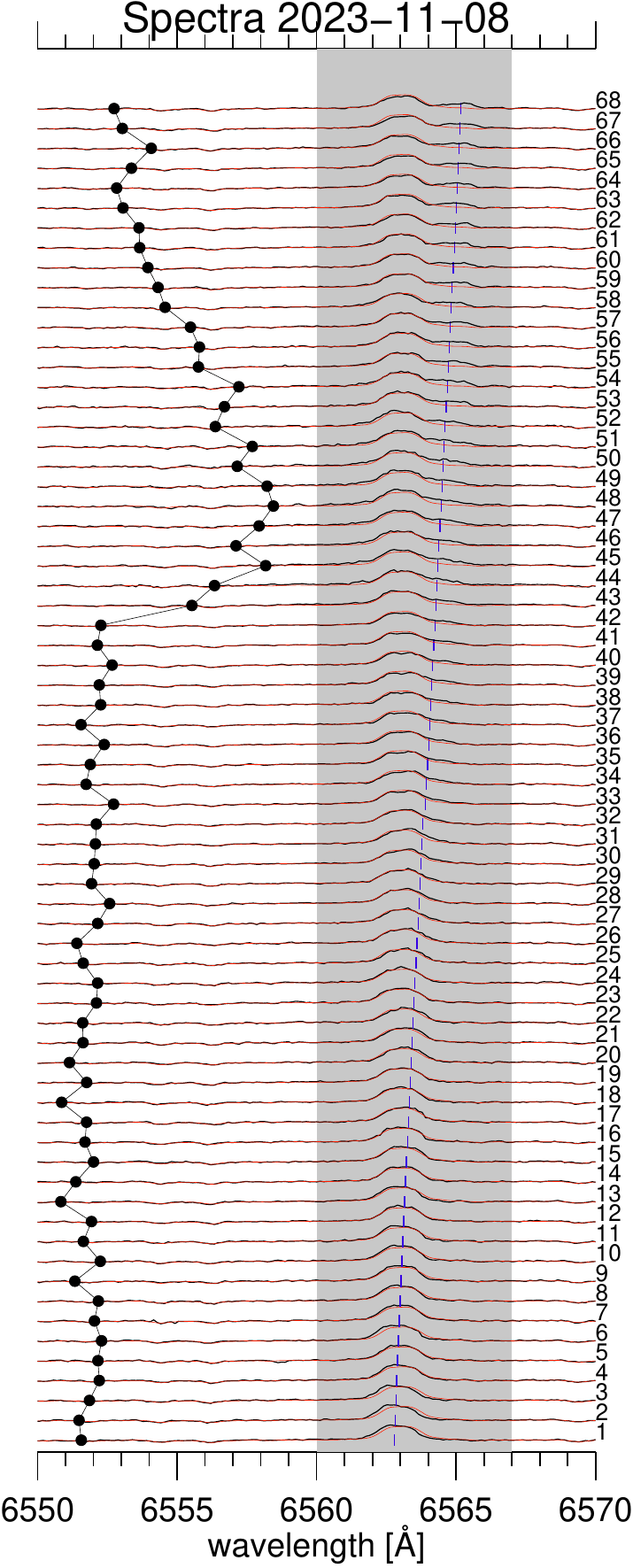}
 \caption{Spectral H$\alpha$ time series (from bottom to top) of the night of 2023-11-08 covering flare no.~2. Black solid lines denote the actual spectra whereas the red solid lines denote the quiescent spectrum. The blue vertical line denotes the position of the dM star component. Shown is also the H$\alpha$ light curve (black solid line, black dots) being displayed vertically.} 
 \label{fig:flaretimeseries}
\end{figure}
For the ``distinct flares'' sample we see that for two third of the flares (no.~1, 3, 4, 6, 7, and 8) the K-dwarf is the source of flaring. For one flare (no.~2) we see excess emission from both components and for two flares (no.~5, 9) the flare source is the M-dwarf.  For the ``weak flares'' more than half of the flares (flares no.~11, 13, 14, 18. 19, 20, and 21) show as source the K-dwarf, only one sixth (flares no.~10, 17) were originating from the M-dwarf, for two flares we could not assign a flaring source (16, 22), and the rest was originating from both stars (flares no.~12, 15). For the ``gradual phase only'' flares also around half show as source of flaring the K-dwarf (24, 26, 28, 29, and 30), one flare (27) shows the M-dwarf as source, and for three events (23, 25, and 31) we could not assign a flaring source. \\


\section{Discussion}
\subsection{Flare morphology from light curves}
\label{Discussion:flaremorph}
CC~Eri has revealed 31 flares (see Table~\ref{tab:B1}) during the optical spectroscopic monitoring with the Echelle spectrograph Pucheros+ on the ESO1.52m telescope. This data set represents the first documentation of the effect of flares and superflares on spectral lines on CC~Eri. We see a variety of different flares which we have grouped in section~\ref{resflarespec} in ''distinct``, ''weak``, and ''gradual phase only`` flares. The majority of flares shows H$\alpha$ impulsive phases with durations of $<$20~min and a trend towards shorter ($<$30~min) H$\alpha$ flare durations (see Fig.~\ref{fig:flarehisto}). The H$\alpha$ impulsive flare phase distribution peaks at 15 minutes with a secondary peak at 25 minutes. The maximum length of the H$\alpha$ impulsive flare phase in the histogram is flare no.~21 from 2024-01-07 (see upper left panel of Fig.~\ref{fig:EWlc6}) with a length of 104~min followed by the strongest flare in the sample, flare no.~4 from 2023-11-14 (see Fig.~\ref{fig:exampleflare}) with 91~min. \\
The H$\alpha$ gradual flare phases reveal values between 10 ad 250~min and show a trend towards shorter lengths ($<$130~min) of the gradual phases. The histogram peaks at 65~min . The longest H$\alpha$ gradual phase has occurred in flare no.~25 from 2023-11-22 (see upper left panel of Fig.~\ref{fig:EWlc7}) where nearly the whole light curve belongs to the gradual H$\alpha$ phase.\\
We find a mean H$\alpha$ flare duration of $\sim$120~min (median of flare duration is $\sim$114~min), a mean H$\alpha$ flare impulsive phase of $\sim$32~min (median is $\sim$21~min), and a mean H$\alpha$ flare gradual phase of $\sim$99~min (median is $\sim$81~min).  \citet{Notsu2024} analyzed dMe star flares and from their H$\alpha$ flare durations we find a mean of 132~min and for the flares on AU~Mic analyzed by \citet{Odert2025} we find a mean H$\alpha$ flare duration of 154~min, a mean H$\alpha$ flare rise time of 69~min, and a mean H$\alpha$ flare decay time of 99~min. All of these values relate well with our derived values. Relating these to values from solar H$\alpha$ flares the stellar values are significantly larger. 
\citet{Temmer2001} examined a statistical analysis of solar H$\alpha$ flares from a quarter of a century of solar observations (1975-1999). These authors present mean and median values for the same H$\alpha$ flare parameters which we deduced from our observations. The solar H$\alpha$ mean flare duration is $\sim$21~min, the solar H$\alpha$ mean impulsive phase gives $\sim$5~min, and the solar H$\alpha$ mean gradual phase gives $\sim$16~min. Comparing these values to the flares of CC~Eri reveals a difference by a factor of $\sim$6. What is consistent from this comparison is that the solar and stellar histograms share a similar behaviour, namely that the long duration flares are more rare than the short duration flares. The same is true for the impulsive and gradual phases. Of course our flare sample is much smaller than the one used for the solar analysis by \citet{Temmer2001} and therefore less significant, but also stellar flare-frequency distributions reveal that the less energetic flares are much more numerous than the energetic flares, and the energetic flares are usually the ones which have longer flare durations \citep[see e.g.][]{Odert2025}. Moreover, with integration times of stellar spectroscopic observations and limiting factors such as S/N it is obvious that such observations can not compete with solar observations. 
Furthermore, solar observations are spatially resolved observations in contrast to stellar observations, also this fact hinders the detection of small events. Therefore many short-duration events having low signal are not detectable within stellar observations. This is also a factor to take into account when comparing stellar and solar flare parameters.\\
The presented flares reveal different morphologies. For the impulsive phases we see the typical fast rise - long decay for flares no.~1, 2, 3 , 8, 9, 10, 11, 14, 16, 18, and 22 (see Fig.~\ref{fig:EWlc1}, Fig.~\ref{fig:EWlc2}, Fig.~\ref{fig:EWlc3}, Fig.~\ref{fig:EWlc4}, Fig.~\ref{fig:EWlc5}, and Fig.~\ref{fig:EWlc6}). For flares no.~23-31 (see Fig.~\ref{fig:EWlc6}, Fig.~\ref{fig:EWlc7}, and Fig.~\ref{fig:EWlc8}) we see gradual phases only, as we missed the impulsive phases.  There are also flares which show a slow rise in the impulsive phase (see upper left panel of Fig.~\ref{fig:flarehisto}) and those are flares no.~4,12,15,17, and 21 (see Fig.~\ref{fig:EWlc1}, Fig.~\ref{fig:EWlc3}, and Fig.~\ref{fig:EWlc4}).\\
We know that flares exhibit different morphologies in different wavelength ranges on the Sun and other stars. One example for that is the Neupert effect from the Sun and stars manifesting itself by the earlier peak of hard X-ray flares compared to the later peak seen in soft X-ray flares similar to optical and H$\alpha$ flares. This reveals that the wavelength dependent flare morphology relates to different thermal and non-thermal physical processes caused by the flare, such as fast electron beams, plasma motions, and heating processes. For e.g. Proxima Centauri this was demonstrated using X-ray (XMM-Newton) and optical (ESO VLT/UVES) observations \citep{Fuhrmeister2011}.\\ 
On the Sun the majority of H$\alpha$ flares reveal impulsive phases of $<$10~min with maximum values of 20 \citep{Temmer2001} to 60~min \citep{Wilson1982, Wilson1987}. For instance, a slow rise phase ($\sim$30~min) of a solar flare in X-rays as well as H$\alpha$ has been reported by \citet{Yurchyshyn2015} which was triggered by an emerging flux rope in a sunspot in between two umbrae. The impulsive phase of flares seen in H$\alpha$ was presented by \citet{Wuelser1989}. The H$\alpha$ light curves of flare kernels were presented and the shape of the light curves resemble typical flares with short rise and long decay phase as well as light curves with shapes similar to the H$\alpha$ light curve shown in Fig.\ref{fig:exampleflare}.\\
H$\alpha$ flare light curves with long rise phases are typical for RS~CVn type systems. Only recently \citet{Cao2025} present H$\alpha$ monitoring of the RS~CVn system UX~Ari. The flare they present shows a slow rise phase, which is way longer (in the order of days) than for flare no.~4 on CC~Eri. The authors assign the flare to the KIV rather than to a flare from both components.\\
Long rise phases in solar flares are not uncommon but rare. On RS~CVn systems long flare rise phases are evident but we can not evaluate if a long flare rise phase is a signature of a flare originating from the reconnection of magnetic fields from both components of the RS~CVn system (see also above).\\ 
Another aspect to consider is rotation and projection. The H$\alpha$ emission during a flare may stem from two different regions. Those are either the footpoints of the magnetic loops, so the area where the flare is anchored to the star, for H$\alpha$ this is the chromosphere, or the cool flare loops \citep{Heinzel2018, Wollmann2023}.  \\
When H$\alpha$ footpoints rotate onto the visible hemisphere the area of the footpoints, as they appear on the stellar limb, is seen in projection and therefore the footpoint area is smaller than its true unprojected value. As the footpoints rotate onto the stellar disk the area increases until it appears unprojected at the center of the stellar disk, and then the area again decreases due to projection as the footpoints approach again the stellar limb. Projection also affects flare loops, but in different ways as those are elongated structures \citep{Bicz2024}. In the case of CC~Eri, these projection effects can alter the true or unprojected flare light curve and let the light curve appear to have a longer rise phase as it had in the unprojected case.


\subsection{Line asymmetries as possible signatures of stellar coronal mass ejections}
\label{lineasymcme}
Spectral line asymmetries are a very common phenomenon seen during flares. As already mentioned in section~\ref{Discussion:flaremorph} either these asymmetries are related to plasma motions during flares within the flare region or those are related to filament/prominence eruptions.\\
During the flaring process plasma fills the magnetic loops and cools down in the decay phase of the flare. This coronal condensation (plasma cooling and downflow) is seen in H$\alpha$ on the Sun \citep[e.g.][]{Ichimoto1984} and stars \citep[e.g.][]{Namekata2022a, Wollmann2023, Leitzinger2024}, whereas chromospheric evaporation (filling of the flare loops) during the impulsive phase is seen on the Sun \citep[e.g.][]{Heinzel1994} but for stars this may be hard to detect as the evaporation signature in H$\alpha$ happens on short time scales (too fast cooling) which may be challenging to detect in stellar spectra. Chromospheric evaporation at low velocities has been found on the Sun even in the gradual flare phase and is termed gentle chromospheric evaporation \citep{Schmieder1987}.\\
However, blue wing asymmetries in H$\alpha$ during the impulsive phase of stellar flares with projected velocities being significantly below the stars escape velocity have been interpreted as chromospheric evaporation \citep[e.g.][]{Gunn1994}. On the Sun, blue asymmetries within the flare ribbons were analysed by \citet{Heinzel1994}.\\
For CC~Eri we find blue- and red-asymmetries in H$\alpha$ with bulk velocities being well below the stars escape velocity. In Table~\ref{tab:C1} we list all detections of H$\alpha$ line asymmetries found in the flare sample (see also Fig.~\ref{fig:allasym} and Fig.~\ref{fig:asymapp}). As mentioned in section~\ref{Discussion:flaremorph} coronal condensation (i.e. cool flare loops or coronal rain) is typically seen as red asymmetry. As we are of the opinion that in stellar observations chromospheric evaporation is very hard to detect as the time scales of this phenomenon are below our integration times due to fast heating, we attribute blue asymmetries to filament/prominence motions.\\
We have detected red asymmetries during flares no.~4, 18, 25, and 30 (see Fig.~\ref{fig:allasym} and Fig.~\ref{fig:asymapp}). The strongest red asymmetry is detected in flare no.~4 (see panel (b) of Fig.~\ref{fig:allasym}) during the gradual flare phase. Also flares no.~25 and 30 (see Fig.~\ref{fig:asymapp}) show red asymmetries during their gradual phase. As this is reminiscent to coronal condensation from flare loops in the gradual flare phase \citep{Wollmann2023}, we assign those three events to coronal condensation. The red asymmetry seen in flare no.~18 (see Fig.~\ref{fig:asymapp}) occurs during the impulsive flare phase. This is also seen in H$\beta$. We suggest that the asymmetry was possibly caused by chromospheric condensation, although on the Sun this phenomenon is rather slow, in the order few tens of km~s$^{-1}$ \citep[e.g.][]{Canfield1990, Graham2020}.\\
The mean bulk velocities of all events were found to be below 215~km~s$^{-1}$. Flare no.~7 (see panel (c) of Fig.~\ref{fig:allasym}) was captured by five spectra only, possibly showing the peak of the flare with two spectra each before and after the peak. From the H$\alpha$ light curve (see lower left panel of Fig.~\ref{fig:EWlc2}) one can see that the light curve starts at an EW value of $\sim$-3 which is above a quiescent level of $\sim$-2 (estimated by eye from the H$\alpha$ light curves in other nights). This means that the first data point in the light curve belongs already to the flare. To evaluate if a flare reveals asymmetries we need a quiescent spectrum. If there are several spectra prior to the impulsive phase of the flare then we build the quiescent spectrum from the mean of those few spectra prior to the flare. If we have only a decaying tail and some spectra afterwards, we build the quiescent spectrum from a mean of those spectra after the decaying tail. For flare no.~7 there are no quiescent spectra before or after the flare as we only obtained a snapshot of five spectra during the flare. For this event we use quiescent spectra from another night obtained at the same orbital phase, this is the case for night 2023-12-20. Here we use the average of five spectra which are representative for the quiescent phase at that orbital period. With this quiescent spectrum we find then in spectrum no.~2 the blue asymmetry (see panel (c) of Fig.~\ref{fig:allasym}). This asymmetry represents the fastest of all events in terms of maximum velocity with a bulk velocity of -213.5$\pm$49.3~km~s$^{-1}$ and a maximum speed of -842.1~km~s$^{-1}$, seen in H$\alpha$, and a bulk velocity of -394.7.0$\pm$63.5~km~s$^{-1}$ and a maximum speed of -953.6~km~s$^{-1}$, seen in H$\beta$ (see panel (c) Fig.~\ref{fig:asymapphbeta} in the appendix). We also see in H$\alpha$ a red asymmetry in the spectrum where the broad blue asymmetry is detected which even increases in flux in the successive spectrum. 
\begin{figure}
 \includegraphics[width=\linewidth]{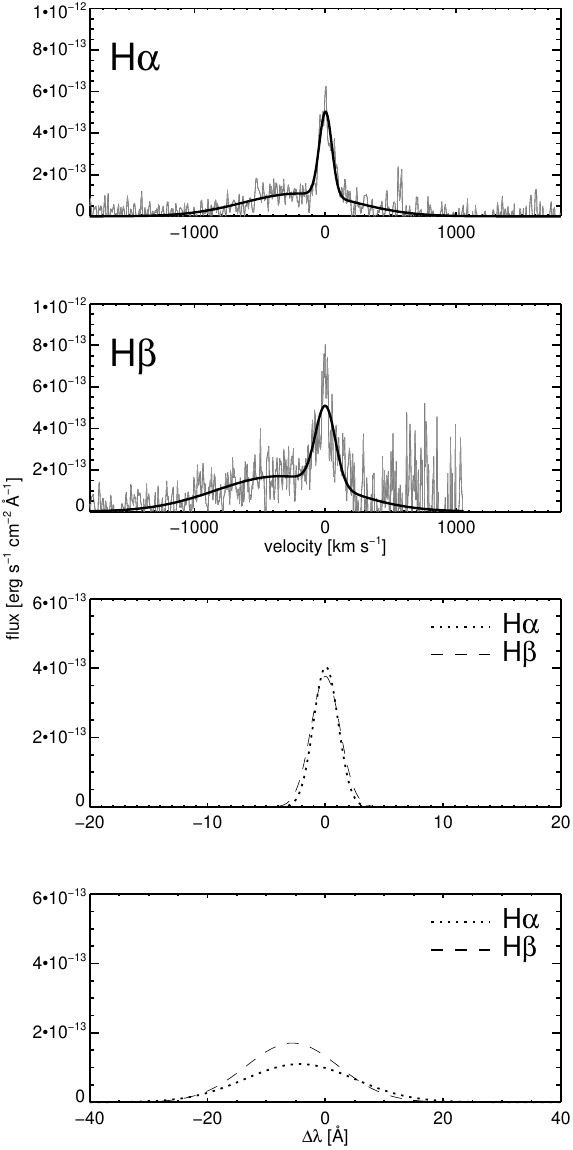}
 \caption{Upper panel: Residual H$\alpha$ spectrum (grey line) of the fast asymmetry of flare no.~7 overplotted with a double gauss fit (black line). Middle upper panel: Residual H$\beta$ spectrum (grey line) of the fast asymmetry of flare no.~6 overplotted with a double gauss fit (black line). Middel lower panel: Single gauss components of the respective line core (dotted line: H$\alpha$, dashed line: H$\beta$). Lower panel: Single gauss components of the respective blue asymmetry (dotted line: H$\alpha$, dashed line: H$\beta$)}.
 \label{fig:asymfastevent}
\end{figure}
We also see the asymmetry in H$\beta$ with the same spectral evolution as for H$\alpha$ (broad blue asymmetry followed by a distinct red asymmetry). In H$\gamma$ we see also a blue extra emission but at lower maximum velocity ($\sim$200~km~s$^{-1}$). Also here the successive spectrum reveals a distinct red asymmetry. The blue asymmetries are also the ones with the largest FWHM (FWHM$_{H_\alpha}$ = 17.0\AA, FWHM$_{H_\beta}$ = 11.4\AA) of all detected asymmetries detected on CC~Eri.\\
The wide bluewing asymmetry is pronounced only in the second spectrum of the five-spectrum long time series of that night. We see weak
and slow excess emission bluewards H$\alpha$, H$\beta$, and H$\gamma$ in the first spectrum, in the second spectrum the broad blue-wing asymmetry is evident in H$\alpha$ and H$\beta$, but significantly less broad in H$\gamma$. In the third spectrum the broad blue-wing asymmetry has drastically reduced to maximum velocities of $\sim$270~km~s$^{-1}$. At the same time we see in H$\alpha$,H$\beta$, and H$\gamma$ distinct red asymmetries at maximum velocities (200 ... 275~km~s$^{-1}$, from H$\alpha$ to H$\gamma$). In the last two spectra of the series the red asymmetry is still evident but decreases in velocity and flux.\\
The broadness of the event may be explained by a phenomenon known from the Sun, namely self-similar expansion seen in solar CMEs \citep[e.g.][]{Subramanian2014}. Thermal broadening and standard microturbulence can not explain such broad features having FWHMs in the order of few hundreds of km~s$^{-1}$ \citep[see also][]{Namekata2024}. As a CME moves away from the Sun it expands in a self-similar way, i.e. the ratio of the diameter of the fluxrope and the height of the fluxrope above the solar surface is roughly constant. This might explain the broadness of the asymmetry. The appearance of a broad asymmetry in only one spectrum is more challenging to explain as one would expect to see a signature of an erupting prominence for a longer time. One spectrum of this series has an integration time of 300s. With a mean bulk velocity of the H$\alpha$ and H$\beta$ signatures of 294~km~s$^{-1}$ this means that the core material has reached one fifth of the stellar radius of the dK star component of CC~Eri. On the other hand for the fast part of the signatures the mean maximum velocity derived from H$\alpha$ and H$\beta$ yields 969~km~s$^{-1}$, which results in a travelled distance of 290700~km which is a bit (0.65 R$\mathrm{_{K-comp}}$) more than one half radius of the dK star component of CC~Eri. Rather short-lived Doppler-shifted excess emissions have been also found in \citet{Houdebine1990} where the authors report the so far fastest Doppler-shifted excess emission interpreted as a flare mass ejection lasting for few minutes only. This scenario is also in agreement with our observations.\\
The Balmer decrement, i.e. the ratio of Balmer line fluxes of different lines, may also indicate if such blue asymmetries are caused by prominence or flare plasma. \citet{Heinzel1994} presented theoretical relations between prominence parameters and the emitted radiation based on 140 prominence NLTE models. In their Fig.~2 the behaviour of the Balmer decrements H$\alpha$/H$\beta$ and H$\gamma$/H$\beta$ is shown. H$\alpha$/H$\beta$ values for prominences in this plot are between 2 and 10. For solar flares Balmer decrements have been presented by \citet{Johns-Krull1997} yielding H$\alpha$/H$\beta$ values of $\sim$1.5 ... 2.5. From that one can see that the Balmer decrement H$\alpha$/H$\beta$ in prominences is larger than the one for flares. There is overlap in the value range of 2 ... 2.5 which is also roughly the range where H$\beta$ becomes optically thick \citep{Heinzel1994}. Accordingly we examined the H$\alpha$/H$\beta$ Balmer decrement for the flare and prominence line components for CC~Eri flare no.~7, i.e. the one spectrum showing the broad blue-wing emission. Therefore we fitted the residual H$\alpha$ and H$\beta$ spectra (see upper panels of Fig.~\ref{fig:asymfastevent}) of that flare and the wide blue-wing asymmetry with a double gaussian. Thereby we are able to determine the flux of the shifted broad component and the unshifted narrow component separately. The single gaussians are shown in the lower panels of Fig.~\ref{fig:asymfastevent}. From those we compute the Balmer decrement for the flare (lower middle panel of Fig.~\ref{fig:asymfastevent}) and the prominence case (lower panel of Fig.~\ref{fig:asymfastevent}). The residual spectra have been flux calibrated from a GAIA DR3 (I/355/spectra) CC~Eri spectrum. Doing so yields a H$\alpha$/H$\beta$(flare) = 0.91 and a H$\alpha$/H$\beta$(prominence) = 0.77, i.e. H$\alpha$/H$\beta$(prominence) $<$ H$\alpha$/H$\beta$(flare). We see that the Balmer decrement of the flare is comparable to that of the probable prominence. The integral of the residuum of the asymmetry and the flare is calculated using the gaussian fits as we can not otherwise separate the flare and prominence component. As one can see the H$\beta$ residual is much (by a factor of more than 3) more noisy ($\sigma_{H\beta}$ = 0.075) than its H$\alpha$ counterpart ($\sigma_{H\alpha}$ = 0.023).\\
However, the asymmetry of flare no.~7 has the largest potential of all detected asymmetries to be indeed the signature of an erupting prominence due to its large maximum speed. \\

\subsection{The effect of flares on spectral lines}
 \label{discussion:speclines}

\begin{figure*}
 \includegraphics[width=18cm]{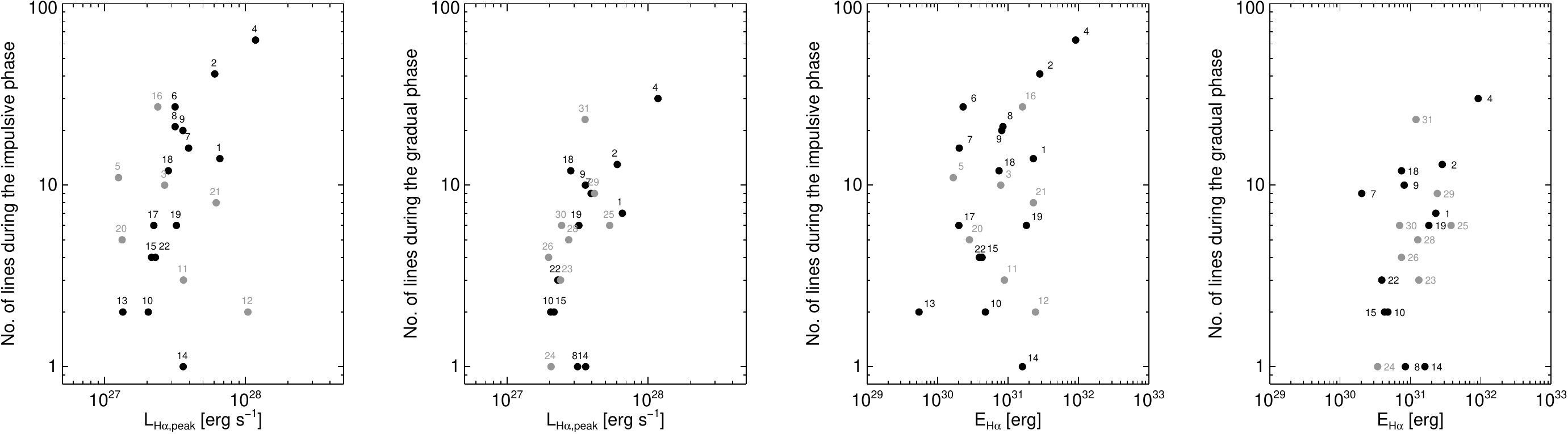}
 \caption{Left panel: Flare peak luminosity versus number of detected spectral lines during the impulsive flare phase. Left middle panel: Flare peak luminosity versus number of detected spectral lines during the gradual flare phase. Right middle panel: H$\alpha$ flare energy versus the number of detected spectral lines during the impulsive flare phase. Right panel: H$\alpha$ flare energy versus number of detected spectral lines during the gradual flare phase. Grey filled symbols mark lower limits for E$_{H\alpha}$ or H$\alpha$ luminosity as there are some events where we have captured only the impulsive or the gradual phase. Black filled symbols denote events which have been fully captured. The numbers denote the flare events according to table~\ref{tab:B1}.}
 \label{fig:energyrelations}
\end{figure*}

As the data used in this study are Echelle spectra we have the unique opportunity to look into many more spectral lines during flares and superflares beside the typical suspects H$\alpha$, H$\beta$, and H$\gamma$. In Tables ~\ref{tab:D1},~\ref{tab:D2}, and ~\ref{tab:D3} we list all spectral lines which we have found to be enhanced during flares on CC~Eri. We investigated, where available, impulsive and gradual flare phases separately. One can of course also try to find spectral line enhancements during flares and superflares in single spectra but then only for the most energetic flares one will see spectral line enhancements. First we did a single spectra analysis and found spectral line enhancements for the most energetic flares of our sample such as flare no.~4. Second, all spectra which belong to the H$\alpha$ impulsive or gradual phase were averaged and in the residual impulsive and gradual spectra line enhancements where searched for. In section~\ref{sec:speclines} we describe in detail which spectral lines are enhanced during which phase. One aspect of the present study is the characterization of superflares and flares in terms of their spectral line emission. Focusing on this aspect reveals the following. Comparing the number of enhanced spectral lines among the flares reveals a trend which might be expected. The more energetic the flare or the higher the peak luminosity the more spectral lines one can see enhanced. In Fig.~\ref{fig:energyrelations} we show the relation of the H$\alpha$ flare energy to the number of enhanced spectral lines during the impulsive and gradual flare phases, as well as the relation of H$\alpha$ flare peak luminosity to the number of enhanced spectral lines during the impulsive and gradual flare phases. We see a trend of more energetic or more luminous flares showing a larger number of enhanced spectral lines during flaring.
If we look into the plots in Fig.~\ref{fig:energyrelations} in detail we see the following. \\
The leftmost panel of Fig.~\ref{fig:energyrelations} shows H$\alpha$ peak luminosity versus the number of detected spectral lines during the impulsive flare phase. The largest H$\alpha$ flare peak luminosity related to the largest number of detected spectral lines during the impulsive flare phase shows flare no.~4, which is the flare in our sample with the largest peak luminosity and H$\alpha$ energy. The second largest number of detected spectral lines during the impulsive phase is not as maybe expected related to the second most luminous H$\alpha$ flare peak (flare no.~12) or second most energetic H$\alpha$ flare (flare no.~25), it is flare no.~2. Also this is not surprising if we take a look at the aformentioned flares no.~12 and 25. Both events have not been entirely captured (see Fig.~\ref{fig:EWlc3} and Fig.~\ref{fig:EWlc7}). Flare no.~12 shows only an impulsive phase which is characterized by a slow increase in H$\alpha$ flux and we can not be sure if we have captured the true flare peak and the quiescent EW level is already enhanced in comparison to other quiet EW levels (see lower panel of Fig.~\ref{fig:phasefolded}). For flare~no.~25 we have captured the gradual phase only. The reason why this flare, although we have only captured its gradual phase, reveals the second largest H$\alpha$ flare energy is related to its duration, the gradual phase lasts for more than four hours which is the longest duration of all flares of the sample. For the determination of excess emission in spectral lines during the flares of our sample we average all spectra belonging to the impulsive phase as well as all spectra belonging to the gradual phase. If, as in the case of flares no.~12 and 25, the impulsive and gradual phases evolve slowly then one has to take into account that due to the slow increase and decay the excess emission due to flaring may be averaged out or weakened. Usually the excess emission arises shortly before and after the flare peak which we have seen in the single spectra analysis (see  Fig.~\ref{fig:singlespectranalysis1},~\ref{fig:singlespectranalysis2},~\ref{fig:singlespectranalysis3}) of the most energetic flare of the sample (flare no.~4).\\
Another flare which shows a significant H$\alpha$ flare peak luminosity (4th largest) and also H$\alpha$ flare energy (fifth largest) but a rather low number of spectral lines with excess emission is flare no.~21. Also this flare has not been entirely captured (see Fig.~\ref{fig:EWlc6}), here only the impulsive phase, or at least a part of it was captured. The same arguments as given above account also for this flare and explain the low number of spectral lines with excess emission. We see from the leftmost panel in Fig.~\ref{fig:energyrelations} that all ``distinct'' flares (flares no.1 .. 9) show at least $>$ 10 spectral lines with excess emission during the impulsive flare phase. 
Taking a look at the number of spectral lines being enhanced during the gradual flare phase in dependence on H$\alpha$ flare peak luminosity (left middle panel of Fig.~\ref{fig:energyrelations}) the picture is the following. In general we see a lower number of spectral lines with excess emission compared to the impulsive phase. Again flare no.~4 reveals the largest number of spectral lines with excess emission during the gradual flare phase followed by flare no.~31 which is a flare where we have captured the gradual phase only (see Fig.~\ref{fig:EWlc8}). Flares 2, 9, and 18 show $\ge$ 10 spectral lines, the rest, also flares from the ``distinct flares'' sample (flares no.~1, 7, and 8) show $<$ 10 spectral lines revealing excess emission. Surprisingly, flare no.~31 shows the second highest number of spectral lines with excess emissions during the gradual flare phase. From its H$\alpha$ peak flux it is comparable to flare no.~25, although the duration of its H$\alpha$ gradual phase is shorter. Moreover also the level of the post flare phase is still enhanced as it is higher (-2.5) as the typical post flares EW level (-2) in other nights.\\
In Fig.~\ref{fig:energylum} we plot the H$\alpha$ flare energy versus the H$\alpha$ flare peak luminosity. As one can see both are related and reveal that the higher the H$\alpha$ flare energy the higher also the H$\alpha$ flare peak luminosity \citep[see also][]{Odert2025}. Therefore one can expect that the interpretation of the two rightmost panels of Fig.~\ref{fig:energyrelations} is similar to the interpretation of the leftmost of the same figure. The rightmost panels of Fig.~\ref{fig:energyrelations} show in the left panel H$\alpha$ flare energy versus number of spectral lines showing excess emission during the impulsive flare phase whereas the right panel shows the same for the gradual flare phase. We see the same flares revealing $\ge$ 10 spectral lines showing excess emission in the impulsive and gradual phases as for the two leftmost panels in Fig.~\ref{fig:energyrelations}.\\
From this we can summarize that in principle the larger the H$\alpha$ flare peak luminosity or also the larger the flare H$\alpha$ energy the larger the number of spectral lines being affected in terms of excess emission.\\
Taking a look at the temperature sensitive spectral lines which we have detected during the flares, e.g. the HeI D3(5876\AA) and HeII (4686\AA) spectral lines, where HeI is effectively photoionized to HeII by the XUV radiation coming from surrounding hot loops, also indicates that the more energetic flares show the emergence of these lines. This is the case for flare no.~4 and no.~8. For flare no.~4 the HeII line is also seen in excess emission in the gradual flare phase. Flare no.~6 shows also the HeII line but much weaker than for flares no.~2 and 4. Furthermore we see the HeII line also in flare no.~3. This flare is according to its Balmer light curves not eye-catching but it shows a coordinated g'-band flare (see Fig.~\ref{fig:EWlc1}. However, this g'-band flare shows a symmetrical light curve, i.e. the impulsive and gradual flare phases have a similar duration. For all other g'-band flares we find that the gradual phase is longer than the impulsive one. The flare shows the HeII line not during the impulsive phase but during the gradual flare phase. This contradicts the results for flare no.~4 as this also shows excess emission in the HeII(4686\AA) line but here also in the impulsive phase. An explanation here may be that the HeII(4686\AA) line can be also excited collisionally \citep[see e.g.][]{Chugai2023}. This would presume that during the gradual phase regions of increased plasma density should have been existent. During the gradual phase plasma cooling takes place and the plasma moves down the flare loop. This motion probably may not cause such regions which would provide sufficient collisions to excite the HeII(4686\AA) line. This flare shows marginal red asymmetries during the gradual phase which are typically interpreted as coronal condensation, i.e. the downward motion of plasma in flaring loops.
\subsection{Flares versus Superflares or what makes Superflares special}
\label{superflares:special}
To address possible differences between normal flares and superflares we need to estimate the bolometric energies of our flare sample. Judging from the g'-band data alone, we have two flares in the sample which show g'-band energies $\ge$ 10$^{33}$~erg, meaning that their bolometric output must be even higher. These are flares no.~4 and 2. For the remaining five g'-band flares (flares no.~3, 5, 6, 7, and 8) we also need to derive their bolometric energy. Based on the approach presented in \citet[][equations 1, 5, and 6]{Shibayama2013} we estimate the bolometric energy of our g'-band flares. To account for the binarity of CC~Eri we adapt 
\begin{figure*}
 \includegraphics[width=16.58cm]{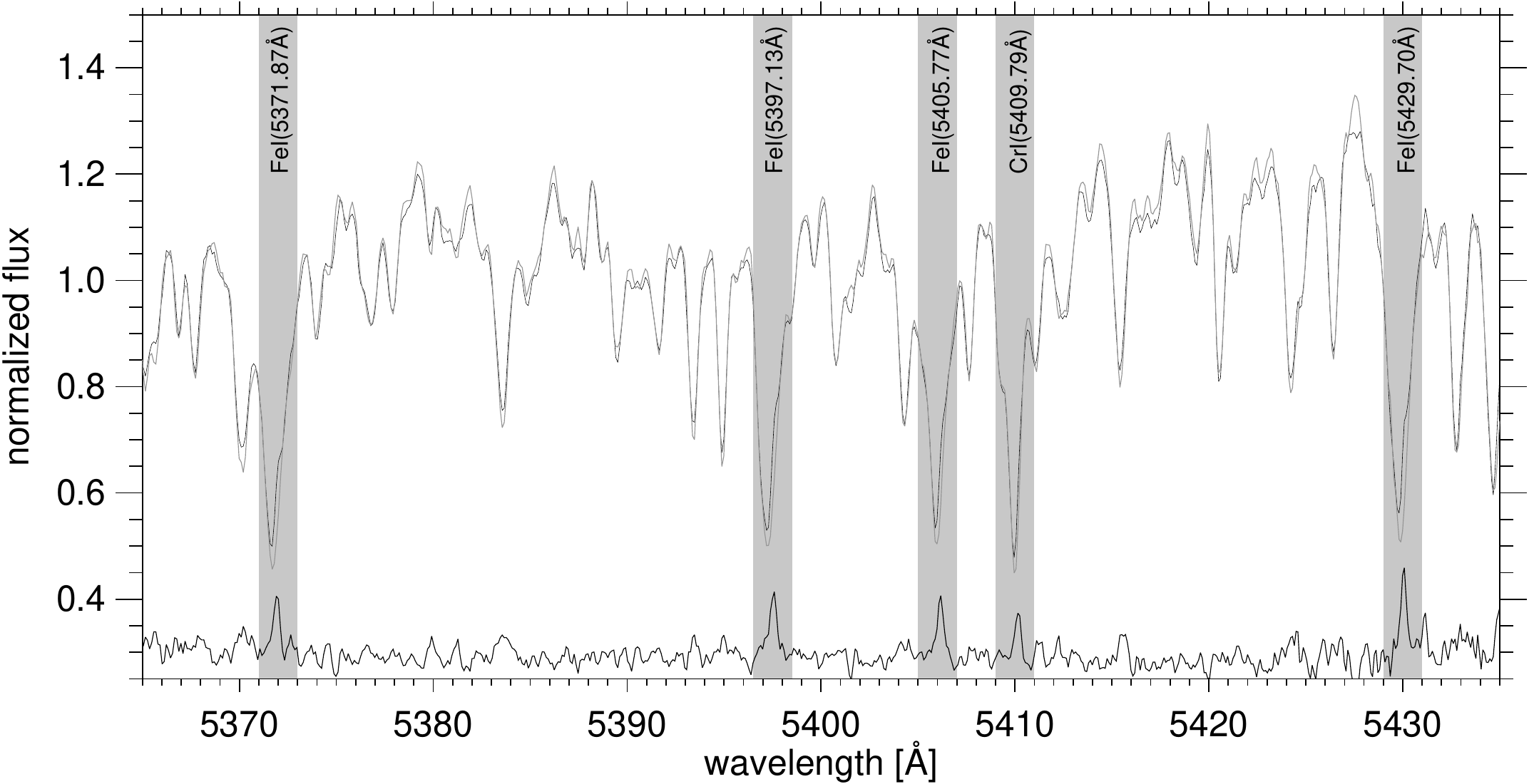}
 \caption{Impulsive phase (black solid line) and preflare (grey solid line) spectra of one Echelle order of flare no.~4. Indicated (by vertical grey area) are five spectral lines revealing shifted excess emission appearing in the red wing of the corresponding absorption lines due to a flaring region being located close to the stellar limb. Also shown is the residual (impulsive - preflare phase) shifted by a value of 0.3.}
 \label{fig:order19}
\end{figure*}
equation 5 in \citet{Shibayama2013} to Eq.~\ref{flareareacceri}, assuming the flare originates from the K-component.
\begin{equation}
\frac{A_{F}}{A_{\ast, A}} = C'\frac{F_{\ast, A}' + F_{\ast, B}'(\frac{R_{\ast, B}}{R_{\ast, A}})^2}{F_{F}' - F_{\ast, A}'}    \label{flareareacceri}
\end{equation}
Here, A$_{F}$ is the flare area, A$_{\ast, A}$ the area of component A (for CC~Eri this is the dK7 star), \textit{F$_{\ast, A}$'} the flux of component A, \textit{F$_{\ast, B}$'} the flux of component B, \textit{F$_{F}$'} the flux of the flare, \textit{R$_{\ast, B}$} and \textit{R$_{\ast, A}$} the radii of components A and B, and C' the observed normalized flare light curve. The flux is defined as $\int R_{\lambda} B_{\lambda, T} d\lambda$, where R$_{\lambda}$ is the transmission function (in our case the g'-band transmission function) and B$_{\lambda, T}$ is the Planck function. For the effective temperature of a dK7 star we adopt 4100~K, for a dM3 star 3430~K\footnote{\url{https://www.pas.rochester.edu/~emamajek/EEM_dwarf_UBVIJHK_colors_Teff.txt}}, and for the flare 9000~K. Using equations 1 and 6 from \citet{Shibayama2013} yields bolometric energies of $>$10$^{33}$~erg for every g'-band flare (E$_{\mathrm{bol, no.2}}$ = 6.3$\times$10$^{34}$~erg, E$_{\mathrm{bol, no.3}}$ = 3.3$\times$10$^{33}$~erg, E$_{\mathrm{bol, no.4}}$ = 5.5$\times$10$^{34}$~erg, E$_{\mathrm{bol, no.5}}$ = 3.4$\times$10$^{33}$~erg and 3.6$\times$10$^{33}$~erg, E$_{\mathrm{bol, no.6}}$ = 1$\times$10$^{34}$~erg, E$_{\mathrm{bol, no.7}}$ = 8$\times$10$^{33}$~erg, and E$_{\mathrm{bol, no.8}}$ = 7.1$\times$10$^{33}$~erg).  For the flares without g'-band data we utilize flare energy relations from the literature \citep{Butler1988, Hawley1991, Osten2015} similar as \citet{Odert2025} we found that applying those to our remaining flares (flares with H$\gamma$ flare energies $>$3$\times$10$^{30}$) yields 14 additional probable superflares (flares no.~1, 9, 11, 12, 17, 18, 19, 29, 21, 25, 26, 29, 30, 31). in addition to that, with the findings from \citet{Namekata2024} and a threshold of E$_{H_{\alpha}}>$2$\times$10$^{31}$~erg \citep{Odert2025} we find two more flares which may be probable superflares (flares no.~27 and 28). Summarizing, we find seven distinct superflares, 16 possible superflares, and eight normal flares.\\
As mentioned in section~\ref{discussion:speclines} the temperature sensitive HeII(4686\AA) spectral line has been found pronounced only in the impulsive phases of flares no.~4 and 2 (strongest superflares in our sample) and rather weak in the impulsive phase of flare no.~6 (third strongest superflare). \citet{Muheki2020b} present Echelle spectroscopic observations of flares on the young and active dMe star EV~Lac. These authors performed a spectral line identification of two of their most energetic flares. According to the determined H$\alpha$ energies (F2=12.2$\times$10$^{31}$~erg, F3=10.6$\times$10$^{31}$~erg) these flares are likely superflares. Beside other spectral lines they also found excess emission in the HeII(4686\AA) line in their two most energetic flares. \\
For flare no.~4 we also find the CaI (4456.61,4226.73\AA), CrI (4274.80, 5409.79\AA), FeI (4459.12, 4427.31, 4405.75, 4271.65\AA), FeII (5234.62, 4384.68, 4303.17, 4233.17, 4178.86, 4173.45\AA), MnI (4783.42\AA), TiI (4443.80\AA), and TiII(4395.03\AA) revealing excess emission.\\
Beside these excess emissions in the above mentioned spectral lines which are probably related to a more energetic flare regime and probably to the superflare nature, the white-light emission evident in a rise of the continuum, typically increasing towards the blue and UV, was measured for the two strongest superflares no.~4 and 2. In section~\ref{results:whitelight} we present the white-light spectra of flare no.~4 and no~2. Both flares reveal a clear signature of excess continuum enhancement in the blue. For both flares we have selected the flare peak of the corresponding H$\alpha$ light curves in Fig.~\ref{fig:whitelight20231114planck} as representatives. For flares no.~4 and no.~2 we have identified more characteristics compared to the remaining superflares of the sample. All g'-band flares fulfill all energy thresholds (E$_{bol}$>10$^{33}$, E$_{H_{\gamma}}$>3$\times$10$^{30}$) of being a superflare. The bolometric energies of $>$10$^{34}$~erg show that those are powerful superflares, i.e. being an order of magnitude above the superflare energy threshold of E$_{bol}$>10$^{33}$. Then we can clearly identify a white-light continuum. And third, for flare no.~4, we see several spectral lines which are not existent in the remaining flares/superflares.\\
Flares no.~2 and no.~4 stand out from the remaining superflares by being roughly an order of magnitude more energetic as derived from their g'-band light curves. Both have comparable flare energies, a comparable length of the H$\alpha$ gradual phase (114.0, 125.3~min), but different lengths of the H$\alpha$ impulsive phase (34.2, 91.1~min) as well as different decay times (40.0, 69.9~min) as derived from the g'-band light curves. Furthermore, flare no.~2 reveals 40 spectral lines showing excess emission during the flare whereas flare no.~4 reveals 63 spectral lines. So the question arises why flare no.~4 shows a factor of 1.6 more spectral lines showing excess emissions during the flare.\\
During the spectral line identification of flare no.~4 we recognized extra emissions in absorption lines appearing in the red wing of those lines (see Fig.~\ref{fig:order19}). This behaviour could be only detected in flare no.~4. We have detected such shifts in 14 absorption lines (CaI7148.15, MgI5528.40, FeI5455.61, FeI5446.92, FeI5429.70, CrI5409.79, FeI5405.77, FeI5397.13, FeI5371.87, FeI5227.19, CrI5208.44, FeI4957.30, MgI4702.98, and FeI4427.31). In Fig.~\ref{fig:vsini} we plot these detected wavelength shifts along the wavelength of the 14 absorption lines. The median of the shifts is 0.27\AA. 
\begin{figure}
 \includegraphics[width=\linewidth]{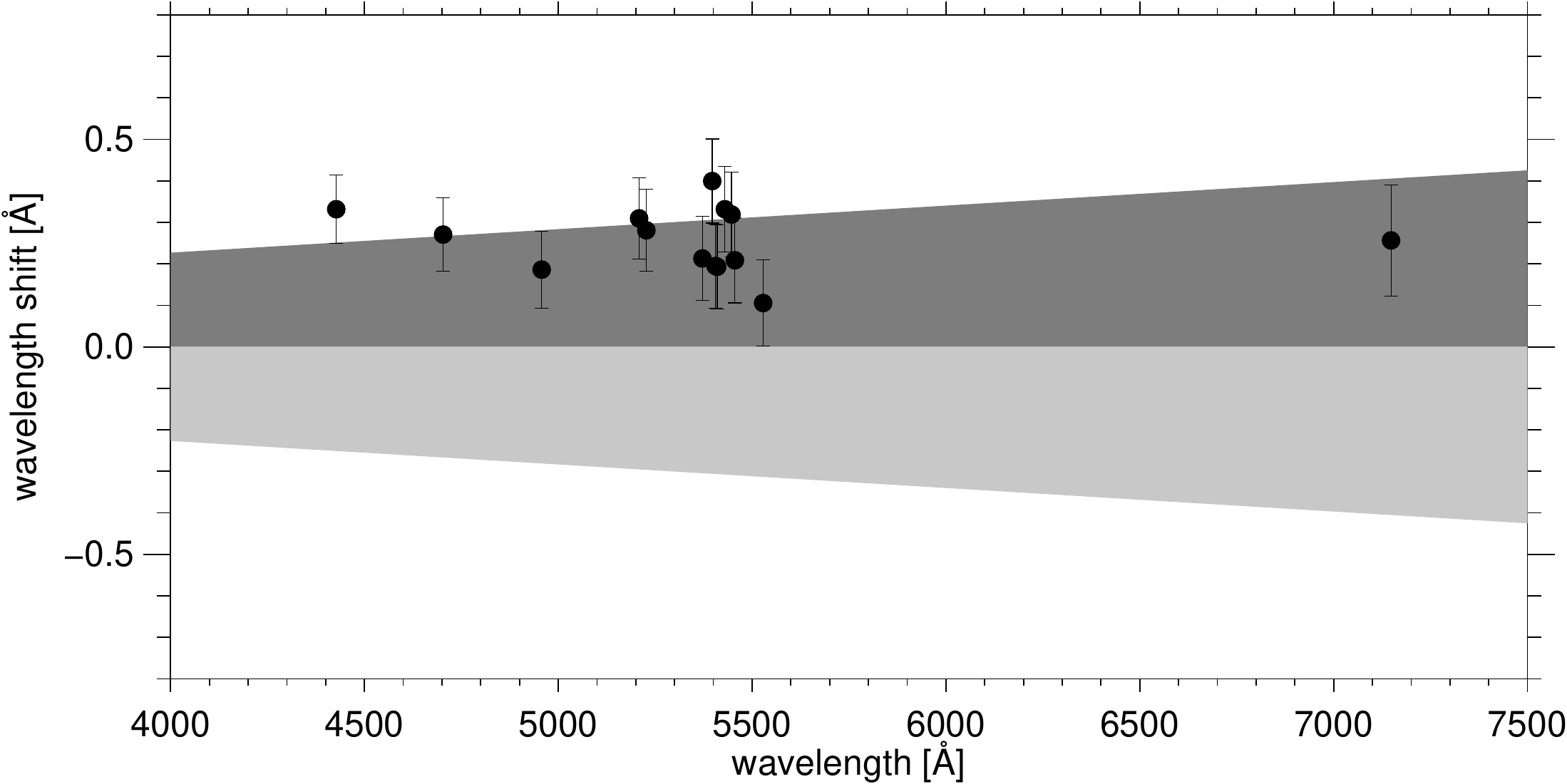}
 \caption{Wavelength shift of the flare emissions for the absorption lines showing excess emission in the red wing. Grey and light grey area denotes the $\pm \textit{v} sin(\textit{i})$ range. }
 \label{fig:vsini}
\end{figure}
We also plot the wavelength range (grey and light grey area) corresponding to the vsini of CC~Eri \citep[17~km~s$^{-1}$][]{Donati1997}. As one can see, although there is scatter in the data, the data points are gathering at the vsini value of CC~Eri, indicating that the region of flare emission must be located close or at the stellar limb. The rotational period of the CC~Eri system is about one and a half days. As the system has a bound rotation this is also the rotation period of the dK and dM star. For a rotation of half of the stellar hemisphere it takes around nine hours. We see the excess emissions in the red wings of the absorption lines partly also in the gradual flare phase. The H$\alpha$ flare duration of flare no.~4 is around three and a half hours. Also therefore it is reasonable to take into account such a scenario. Taking all this into account, this possibly explains also the difference in the number of detected excess emissions of flare no.~2 and 4. If flare no.~4 occurs at the stellar limb then the emitting flare region is seen in projection, i.e. it appears smaller than its true value and this further means that the energy values we have derived are lower limits.\\
In the appendix we show the full spectral evolution in various spectral lines of flare no.~4 (see Fig.~\ref{fig:singlespectranalysis0}, Fig.~\ref{fig:singlespectranalysis1}, Fig.~\ref{fig:singlespectranalysis2}, and Fig.~\ref{fig:singlespectranalysis3}). In Fig.~\ref{fig:singlespectranalysis0} we show especially the evolution in the Balmer lines. As one can see the small Balmer component (M-star component) is at the beginning of flare no.~4 on the blue side of the large Balmer component (K-star component). This corresponds to an orbital configuration of the M-star being at phase$>$0.75 (phase=0 corresponds to M-star in front of K-star)  where it has its maximum blue shift in the spectrum. Then from this position the M-star moves in front of the K-star until both are seen in line (phase=0). So, flare no.~4 occurs in this orbital period. We see the excess emissions in the spectral lines red shifted. This means, as mentioned above, that the flaring region must be close to the K-star's limb (phase$<$0.25). As the excess flux is red-shifted (moving away from the observer) it must occur on the far side from the M-star, and therefore it is unlikely that the M-star is the flaring source for flare no.~4.\\
As said, spectroscopic investigations of energetic or even superflares are rare, especially fully covering the region between 4100 and 7500\AA{} with spectral resolving powers sufficiently large to resolve spectral lines with smaller FWHM. One prominent example is the great flare on AD Leo presented by \citet{Hawley1991} where the authors analyzed simultaneously obtained optical (3560-4440\AA) and UV spectra (1150-2000 and 1900-3100\AA), as well as coordinated U,B,V,R.H$\alpha$, and H$\beta$ photometry. The total flare energy determined from these wavelength regimes gives 10$^{34}$~erg and the authors find that the flare when compared to other well observed stellar flares was similar in terms of time evolution and line broadening of the emissions features as well as the Balmer decrement. \citet{Paulson2006} report on a flare from the old Barnard's star, an M4 dwarf. The authors observed the flare with a nearly continuously spectral coverage of 3600-10800\AA{} and a spectral resolving power of $\sim$60000. The authors detect nearly 160 spectral lines revealing excess emission during the flare. The authors confirm all spectral lines found also in \citet[][]{Hawley1991} and argue that the spectral lines not detected by \citet{Hawley1991} are related to the fact of higher spectral resolution and higher S/N in their data. No energy is given for the flare on Barnard's star. Optical spectroscopy of 14 non-white-light flares on AD~Leo have been reported by \citet{CrespoChacon2007} using the IDS spectrograph on the INT of the ING on LaPalma. The instrument covers the wavelength region of  3554-5176\AA{} in the blue and  5527-7137\AA{} in the red. Prominent chromospheric lines, such as the Balmer, sodium D1 and D2, CaII H+K as well as the HeI4207 lines were detected. Optical spectroscopy of a giant flare on the dM star  CN~Leo was presented by \citet{Fuhrmeister2008} using UVES installed on one of ESO's VLTs. Both, the blue (3050-3860\AA) and the red (6400-10080\AA) arms of UVES have been used with a resolving power of $\sim$40000. The authors detect in total 1143 spectral lines revealing excess emission from which only 154 were located in the red arm. Our nearest neighbour Proxima Centauri was observed spectroscopically in the optical \citep{Fuhrmeister2011}, again with UVES on the VLT (3290-4500\AA{} in the blue arm and 6400-10080\AA{} in the red arm). In total 474 spectral lines revealing excess emissions from which less than 10\% originated in the red arm are presented. \citet{Lalitha2013} report on multi-wavelength observations of flares on the very young late G/early K star AB~Dor. Optical spectroscopic (UVES/VLT 3720-4945\AA{} at R$\sim$40000 and 5695-9465\AA{}, at R$\sim$60000) investigation of three flares including spectral line analysis are shown. All three events were clustered events, i.e. occurring closely together. Event two occurred in the decaying tail of event one where event three occurred more separated and can therefore maybe counted as a separate event. The X-ray energies of events one and two as well as event three are in the order of 10$^{33}$, making them superflares. 
\citet{Kowalski2013} reported on 20 M-dwarf flares observed with photometry and spectroscopy. The spectral range was 3400-9200\AA{} and the resolving power was R<1000 which allows spectral line analysis of broad spectral lines, such as e.g. the Balmer lines. Here especially Balmer lines, CaII H+K and HeI4471 lines were measured. From the given U-band energies already two flares from YZ~CMi are superflares in the order of E$_{U}\sim$10$^{34}$~erg, i.e. with a bolometric energy being even larger than that. Another impressive data set of spectroscopic time series observations on dM stars was presented by \citet[][]{Notsu2024}. With a spectral coverage of 3800-10000\AA{}, the resolving power of $\sim$32000 makes these data special, as spectral line analysis of spectral lines with smaller FWHM is feasible. \citet[][]{Notsu2024} focused on spectral line asymmetries in Balmer lines occurring during flares rather than the effect on other numerous spectral lines being visible during flares. The authors focus on prominent chromospheric spectral lines such as the Balmer, sodium D1 and D2, CaII H+K, CaII8542, and HeI5876 lines. From the given energies in this study we see that from g'-band energy one flare seems to be a superflare, which is the flare from the night of the 27th of August 2020 on EV~Lac. In the study of \citet{Muheki2020b} on flares of EV~Lac their two most energetic flares, being likely superflares, show also a different number of spectral lines revealing excess emission. The more energetic flare (F2) of their study also shows a larger number of excess emissions compared to their slightly less energetic flare F3. Both flares differ by a factor of 1.44 in number of detected excess emissions during flaring.\\
From the above studies only two fully cover the wavelength region between 4100 and 7500\AA{}, the studies by \citet{Paulson2006} on a flare on Barnard's star showing a continuum rise (i.e. a white-light flare) and \citet{Muheki2020b} on flares on EV~Lac one of which also showing a continuum increase towards the blue. \citet{Paulson2006} find many more lines revealing excess emission ($\sim$180) than listed in \citet[36, ][]{Muheki2020b} and in the present study (63). The reasons for that are related to flares with different energies, then instruments with different spectral resolving powers R$\sim$60000 \citep[][]{Paulson2006}, R$\sim$35000 \citep[][]{Muheki2020b} and R$\sim$18000 (present study), different S/N regimes, and different wavelength coverage (\citet{Paulson2006} start already at $\sim$3800\AA). However, we and \citet{Muheki2020b} see partly the same lines, we see a subset of those presented by \citet{Paulson2006} but can not resolve the remaining ones or the flares presented in the present study and in \citet{Muheki2020b} did not produced them.\\
\citet{Johns-Krull1997} analyzed a solar M7.7 flare occurring on the 6th of March 1993 with an Echelle spectrograph. The spectrograph slit was placed on the flaring knot. Beside the analysis of chromospheric lines being enhanced during the flare and stating the detection of line aymmetries, the authors list all photospheric lines which showed filling-in due to the flare. We cross-checked the results from this solar study with the filling ins of our absorption lines. From their 91 detections of filled absorption lines we found 18 in flare no.~4 of the present study and 10 in flare 2 presented in \citet{Muheki2020b}. \citet{Johns-Krull1997} state that the solar M7.7 flare was a flare without white-light emission. Flare no.~4 of the present study is a flare showing white-light emission as well is flare 2 presented in \citet{Muheki2020b}.\\
In the present study we have detected 31 flares from which eight are superflares, from which two are energetic superflares ($\sim$ an order of magnitude larger than the remaining superflares), and the rest are normal flares.
This means that we can relate normal flares to superflares in terms of excess emission in spectral lines, because all have been observed with the same observational setup, same spectral resolving power, same wavelength region and comparable S/N. The results for the analysis of excess emissions in spectral lines during the flares have been shown in Fig.~\ref{fig:energyrelations}, where one can see that the more energetic flares reveal more spectral lines during flares revealing excess emission. If we now take especially a look at the photospheric lines with excess emission, the normal flares show very few or none. This is consistent with flares revealing a flare signature in chromospheric lines only (e.g. Balmer lines etc.) corresponding to the flare process not being able to generate electron beams sufficiently strong to penetrate deep enough to affect the photosphere, i.e. causing excess emission of photospheric lines or continuum enhancements. On the other hand, we can not exclude, in the case of normal flares, that there is continuum enhancement or photospheric line filling, which we can not measure due to e.g. the S/N being too low causing the possible signatures being hidden in the data noise. Also photospheric line filling has been seen on the Sun during flares without a visible white-light component \citep[see e.g.][]{Johns-Krull1997}. However, for flares no.~4 and 2 of the present study, respresenting more energetic superflares, we have seen the continua being enhanced towards the blue, photospheric lines showing excess emission, the number of spectral lines showing excess emission being significantly higher than for normal flares, and we see typical flare plasma motions (such as red asymmetries reminiscent of coronal condensations being a typical flare plasma motion seen on the Sun and stars). Because of all these findings we suggest that superflares, at least the ones we have detected in the present study, are scaled up normal flares rather than being a distinct phenomenon of stellar activity. This conclusion does not rule out that more energetic superflares, i.e. more energetic as the ones we have detected in the present study ($>$10$^{34}$~erg), possibly may be distinct phenomena which may have a more different physical background than the superflares we have detected on CC~Eri.

\section{Conclusion}
We have examined an analysis of spectral monitoring of the non-eclipsing active southern binary CC~Eri to characterize its flaring and superflaring activity spectroscopically as well as to characterize flare related plasma motions and plasma motions possibly related to eruptive phenomena such as eruptive filaments/prominences. We have identified 31 flares from the spectroscopic data. Eight flares are also visible in the coordinated g'-band photometry. From the 31 flares we identify two superflares (flares no.~2 and 4) from their g'-band energies alone and seven (flares no.~2, 3, 4, 5a, 5b, 6, 7, and 8) when accounting for their bolometric energy. We find distinct white-light emission (enhanced continua) for flares no.~2 and 4. We are able to identify the component of the binary system from which flares originate only for some flares, as we have events with flare signatures also in both components. Spectral line asymmetries are evident in the stronger flares. Fast asymmetries, i.e. which exceed the stars escape velocity were not detected, but we have detected one event which maximum projected velocities exceed the stars escape velocity and which is a potential candidate of being a stellar prominence eruption. Spectral line and excess emission identification during the impulsive and gradual flare phases reveals that the strongest or most energetic flare, flare no.~4, reveals excess emissions which are not seen in the remaining flares/superflares (partly in flare no.~2). For flare no.~4 we have identified several  excess emissions occurring in the red wing of absorption lines which we interpret as the flaring region being close to the stellar limb.
Taking into account the relations between number of excess emissions and H$\alpha$ flare energies/luminosities, flare no.~4 being a limb event, i.e. being in reality even more energetic, flares no.~2 and 4 revealing continuum enhancements and excess emissions in several photospheric lines, and showing plasma motions during these flares which are commonly observed on the Sun and stars, we conclude that the most energetic superflares of our study are likely scaled up versions of normal flares. 

\section*{Acknowledgements}
\addcontentsline{toc}{section}{Acknowledgements}
The authors of this study thank the anonymous referee for helpful comments which improved the quality of the manuscript. This research was funded in whole, or in part, by the Austrian Science Fund (FWF) [10.55776/I5711, 10.55776/P37256, 10.55776/PAT4657624]. For the purpose of open access, the author has applied a CC BY public copyright licence to any Author Accepted Manuscript version arising from this submission. PK, JL, PH, and JW acknowledge GACR grant 22-30516K. The EXOWORLD project is supported by the European Union under the Horizon Europe Programme Marie Sk{\l}odowska-Curie Actions Staff Exchanges. PK and JL acknowledge travel funding for mobility to Chile from Horizon 2020 Project ID: 101086149 EXOWORLD. PH was supported by the program 'Excellence Initiative - Research University' for years 2020–2026 at University of Wroc{\l}aw, project No. BPIDUB.4610.96.2021.KG. This work was generously supported by the Th\"uringer Ministerium f\"ur Wirtschaft, Wissenschaft und Digitale Gesellschaft and the Th\"uringer Aufbaubank. LV acknowledges projects ANID Fondecyt n. 1211162, ANID Quimal ASTRO20-0025 and ANID BASAL FB210003. The research of P.G. was supported by the Slovak Research and Development Agency under contract No. APVV-24-0160 and the internal grant No. VVGS-2023-2784 of the P. J. {\v S}af{\'a}rik University in Ko{\v s}ice, funded by the EU NextGenerationEU through the Recovery and Resilience Plan for Slovakia under the project No. 09I03-03-V05-00008.
We thank the observers (in alphabetical order) who contributed to collecting the data presented in this paper: Luca Antonucci, Zuzana Balk\'oov\'a, Raine Karjalainen, Lud\v{e}k \v{R}ezba,  Veronika Schaffenroth, Jan Sloup, Ji\v{r}i Srba, Michaela V\'itkov\'a, Elizaveta Vostretcova, Ji\v{r}i \v{Z}\'ak, and Eva \v{Z}\v{d}\'arsk\'a. We also thank Jan Fuchs for software support.
This paper includes data collected by the \textit{TESS} mission, which are publicly available from the Mikulski Archive for Space Telescopes (MAST). Funding for the \textit{TESS} mission is provided by the NASA's Science Mission Directorate. This work has made use of data from the European Space Agency (ESA) mission {\it Gaia} (\url{https://www.cosmos.esa.int/gaia}), processed by the {\it Gaia} Data Processing and Analysis Consortium (DPAC, \url{https://www.cosmos.esa.int/web/gaia/dpac/consortium}). Funding for the DPAC has been provided by national institutions, in particular the institutions participating in the {\it Gaia} Multilateral Agreement. 

\section*{Data Availability}

 The data collected for the work presented in this article are available upon request in their raw, but also in their CERES+ reduced versions. We can provide photometric and also spectroscopic data if requested. The \textit{TESS} light curve data are available at the Mikulski Archive for Space Telescopes (MAST; \url{https://mast.stsci.edu}). The \textit{Gaia} DR3 spectrum is available at VizieR (\url{https://vizier.cds.unistra.fr}) in Table I/355/spectra.



\bibliographystyle{mnras}
\bibliography{Mybibfile} 

\begin{thebibliography}{}
\makeatletter
\relax
\def\mn@urlcharsother{\let\do\@makeother \do\$\do\&\do\#\do\^\do\_\do\%\do\~}
\def\mn@doi{\begingroup\mn@urlcharsother \@ifnextchar [ {\mn@doi@}
  {\mn@doi@[]}}
\def\mn@doi@[#1]#2{\def\@tempa{#1}\ifx\@tempa\@empty \href
  {http://dx.doi.org/#2} {doi:#2}\else \href {http://dx.doi.org/#2} {#1}\fi
  \endgroup}
\def\mn@eprint#1#2{\mn@eprint@#1:#2::\@nil}
\def\mn@eprint@arXiv#1{\href {http://arxiv.org/abs/#1} {{\tt arXiv:#1}}}
\def\mn@eprint@dblp#1{\href {http://dblp.uni-trier.de/rec/bibtex/#1.xml}
  {dblp:#1}}
\def\mn@eprint@#1:#2:#3:#4\@nil{\def\@tempa {#1}\def\@tempb {#2}\def\@tempc
  {#3}\ifx \@tempc \@empty \let \@tempc \@tempb \let \@tempb \@tempa \fi \ifx
  \@tempb \@empty \def\@tempb {arXiv}\fi \@ifundefined
  {mn@eprint@\@tempb}{\@tempb:\@tempc}{\expandafter \expandafter \csname
  mn@eprint@\@tempb\endcsname \expandafter{\@tempc}}}

\bibitem[\protect\citeauthoryear{{Abdul-Aziz} et~al.,}{{Abdul-Aziz}
  et~al.}{1995}]{AbdulAziz1995}
{Abdul-Aziz} H.,  et~al., 1995, \aaps, \href
  {https://ui.adsabs.harvard.edu/abs/1995A&AS..114..509A} {114, 509}

\bibitem[\protect\citeauthoryear{{Abranin} et~al.,}{{Abranin}
  et~al.}{1997}]{Abranin1997}
{Abranin} E.~P.,  et~al., 1997, \mn@doi [\apss] {10.1023/A:1001192915841},
  \href {https://ui.adsabs.harvard.edu/abs/1997Ap&SS.257..131A} {257, 131}

\bibitem[\protect\citeauthoryear{{Abranin} et~al.,}{{Abranin}
  et~al.}{1998}]{Abranin1998}
{Abranin} E.~P.,  et~al., 1998, \mn@doi [Astronomical and Astrophysical
  Transactions] {10.1080/10556799808232093}, \href
  {https://ui.adsabs.harvard.edu/abs/1998A&AT...17..221A} {17, 221}

\bibitem[\protect\citeauthoryear{{Amado}, {Doyle}, {Byrne}, {Cutispoto},
  {Kilkenny}, {Mathioudakis}  \& {Neff}}{{Amado} et~al.}{2000}]{Amado2000}
{Amado} P.~J.,  {Doyle} J.~G.,  {Byrne} P.~B.,  {Cutispoto} G.,  {Kilkenny} D.,
   {Mathioudakis} M.,   {Neff} J.~E.,  2000, \aap, \href
  {https://ui.adsabs.harvard.edu/abs/2000A&A...359..159A} {359, 159}

\bibitem[\protect\citeauthoryear{{Antonucci} et~al.,}{{Antonucci}
  et~al.}{2025}]{Antonucci2025}
{Antonucci} L.,  et~al., 2025, \mnras, submitted

\bibitem[\protect\citeauthoryear{{Argiroffi} et~al.,}{{Argiroffi}
  et~al.}{2019}]{Argiroffi2019}
{Argiroffi} C.,  et~al., 2019, \mn@doi [Nature Astronomy]
  {10.1038/s41550-019-0781-4}, \href
  {https://ui.adsabs.harvard.edu/abs/2019NatAs...3..742A} {3, 742}

\bibitem[\protect\citeauthoryear{{Audard}, {G{\"u}del}, {Drake}  \&
  {Kashyap}}{{Audard} et~al.}{2000}]{Audard2000}
{Audard} M.,  {G{\"u}del} M.,  {Drake} J.~J.,   {Kashyap} V.~L.,  2000, \mn@doi
  [\apj] {10.1086/309426}, \href
  {http://adsabs.harvard.edu/abs/2000ApJ...541..396A} {541, 396}

\bibitem[\protect\citeauthoryear{{Bicz}, {Falewicz}, {Heinzel}, {Pietras}  \&
  {Pre{\'s}}}{{Bicz} et~al.}{2024}]{Bicz2024}
{Bicz} K.,  {Falewicz} R.,  {Heinzel} P.,  {Pietras} M.,   {Pre{\'s}} P.,
  2024, \mn@doi [\apjl] {10.3847/2041-8213/ad6c06}, \href
  {https://ui.adsabs.harvard.edu/abs/2024ApJ...972L..11B} {972, L11}

\bibitem[\protect\citeauthoryear{{Bilir}, {Karaali}  \& {Tun{\c{c}}el}}{{Bilir}
  et~al.}{2005}]{Bilir2005}
{Bilir} S.,  {Karaali} S.,   {Tun{\c{c}}el} S.,  2005, \mn@doi [Astronomische
  Nachrichten] {10.1002/asna.200510358}, \href
  {https://ui.adsabs.harvard.edu/abs/2005AN....326..321B} {326, 321}

\bibitem[\protect\citeauthoryear{{Bloot} et~al.,}{{Bloot}
  et~al.}{2024}]{Bloot2024}
{Bloot} S.,  et~al., 2024, \mn@doi [\aap] {10.1051/0004-6361/202348065}, \href
  {https://ui.adsabs.harvard.edu/abs/2024A&A...682A.170B} {682, A170}

\bibitem[\protect\citeauthoryear{{Boiko}, {Konovalenko}, {Koliadin}  \&
  {Melnik}}{{Boiko} et~al.}{2012}]{Boiko2012}
{Boiko} A.~I.,  {Konovalenko} A.~A.,  {Koliadin} V.~L.,   {Melnik} V.~N.,
  2012, Advances in Astronomy and Space Physics, \href
  {https://ui.adsabs.harvard.edu/abs/2012AASP....2..121B} {2, 121}

\bibitem[\protect\citeauthoryear{{Bopp} \& {Evans}}{{Bopp} \&
  {Evans}}{1973}]{BoppEvans1973}
{Bopp} B.~W.,  {Evans} D.~S.,  1973, \mn@doi [\mnras]
  {10.1093/mnras/164.4.343}, \href
  {https://ui.adsabs.harvard.edu/abs/1973MNRAS.164..343B} {164, 343}

\bibitem[\protect\citeauthoryear{{Brahm}, {Jord{\'a}n}  \& {Espinoza}}{{Brahm}
  et~al.}{2017}]{Brahm2017}
{Brahm} R.,  {Jord{\'a}n} A.,   {Espinoza} N.,  2017, \mn@doi [\pasp]
  {10.1088/1538-3873/aa5455}, \href
  {https://ui.adsabs.harvard.edu/abs/2017PASP..129c4002B} {129, 034002}

\bibitem[\protect\citeauthoryear{{Budding} et~al.,}{{Budding}
  et~al.}{2006}]{Budding2006}
{Budding} E.,  et~al., 2006, \mn@doi [\apss] {10.1007/s10509-006-9086-z}, \href
  {https://ui.adsabs.harvard.edu/abs/2006Ap&SS.304...13B} {304, 13}

\bibitem[\protect\citeauthoryear{{Butler}, {Rodono}  \& {Foing}}{{Butler}
  et~al.}{1988}]{Butler1988}
{Butler} C.~J.,  {Rodono} M.,   {Foing} B.~H.,  1988, \aap, \href
  {https://ui.adsabs.harvard.edu/abs/1988A&A...206L...1B} {206, L1}

\bibitem[\protect\citeauthoryear{{Byrne}, {Agnew}, {Cutispoto}, {Kilkenny},
  {Neff}  \& {Panagi}}{{Byrne} et~al.}{1992}]{Byrne1992}
{Byrne} P.~B.,  {Agnew} D.~J.,  {Cutispoto} G.,  {Kilkenny} D.~W.,  {Neff}
  J.~E.,   {Panagi} P.~M.,  1992, in {Byrne} P.~B.,  {Mullan} D.~J.,  eds, ,
  Vol.~397, Surface Inhomogeneities on Late-Type Stars.
p.~255, \mn@doi{10.1007/3-540-55310-X_161}

\bibitem[\protect\citeauthoryear{{Canfield}, {Penn}, {Wulser}  \&
  {Kiplinger}}{{Canfield} et~al.}{1990}]{Canfield1990}
{Canfield} R.~C.,  {Penn} M.~J.,  {Wulser} J.-P.,   {Kiplinger} A.~L.,  1990,
  \mn@doi [\apj] {10.1086/169345}, \href
  {http://adsabs.harvard.edu/abs/1990ApJ...363..318C} {363, 318}

\bibitem[\protect\citeauthoryear{{Cao} \& {Gu}}{{Cao} \& {Gu}}{2025}]{Cao2025}
{Cao} D.,  {Gu} S.,  2025, \mn@doi [\aap] {10.1051/0004-6361/202452857}, \href
  {https://ui.adsabs.harvard.edu/abs/2025A&A...695L...2C} {695, L2}

\bibitem[\protect\citeauthoryear{{Chen}, {Tian}, {Li}, {Wang}, {Lu}, {Xu},
  {Hou}  \& {Wu}}{{Chen} et~al.}{2022}]{Chen2022}
{Chen} H.,  {Tian} H.,  {Li} H.,  {Wang} J.,  {Lu} H.,  {Xu} Y.,  {Hou} Z.,
  {Wu} Y.,  2022, \mn@doi [\apj] {10.3847/1538-4357/ac739b}, \href
  {https://ui.adsabs.harvard.edu/abs/2022ApJ...933...92C} {933, 92}

\bibitem[\protect\citeauthoryear{{Chugai} \& {Utrobin}}{{Chugai} \&
  {Utrobin}}{2023}]{Chugai2023}
{Chugai} N.~N.,  {Utrobin} V.~P.,  2023, \mn@doi [Astronomy Letters]
  {10.1134/S1063773723350013}, \href
  {https://ui.adsabs.harvard.edu/abs/2023AstL...49..639C} {49, 639}

\bibitem[\protect\citeauthoryear{{Coluzzi}}{{Coluzzi}}{1993}]{Coluzzi1993}
{Coluzzi} R.,  1993, Bulletin d'Information du Centre de Donnees Stellaires,
  \href {https://ui.adsabs.harvard.edu/abs/1993BICDS..43....7C} {43, 7}

\bibitem[\protect\citeauthoryear{{Crespo-Chac{\'o}n}, {Micela}, {Reale},
  {Caramazza}, {L{\'o}pez-Santiago}  \& {Pillitteri}}{{Crespo-Chac{\'o}n}
  et~al.}{2007}]{CrespoChacon2007}
{Crespo-Chac{\'o}n} I.,  {Micela} G.,  {Reale} F.,  {Caramazza} M.,
  {L{\'o}pez-Santiago} J.,   {Pillitteri} I.,  2007, \mn@doi [\aap]
  {10.1051/0004-6361:20077601}, \href
  {https://ui.adsabs.harvard.edu/abs/2007A&A...471..929C} {471, 929}

\bibitem[\protect\citeauthoryear{{Crosley} \& {Osten}}{{Crosley} \&
  {Osten}}{2018a}]{Crosley2018b}
{Crosley} M.~K.,  {Osten} R.~A.,  2018a, \mn@doi [\apj]
  {10.3847/1538-4357/aaaec2}, \href
  {https://ui.adsabs.harvard.edu/abs/2018ApJ...856...39C} {856, 39}

\bibitem[\protect\citeauthoryear{{Crosley} \& {Osten}}{{Crosley} \&
  {Osten}}{2018b}]{Crosley2018a}
{Crosley} M.~K.,  {Osten} R.~A.,  2018b, \mn@doi [\apj]
  {10.3847/1538-4357/aacf02}, \href
  {https://ui.adsabs.harvard.edu/abs/2018ApJ...862..113C} {862, 113}

\bibitem[\protect\citeauthoryear{{Crosley} et~al.,}{{Crosley}
  et~al.}{2016}]{Crosley2016}
{Crosley} M.~K.,  et~al., 2016, \mn@doi [\apj] {10.3847/0004-637X/830/1/24},
  \href {https://ui.adsabs.harvard.edu/abs/2016ApJ...830...24C} {830, 24}

\bibitem[\protect\citeauthoryear{{Cuntz}, {Saar}  \& {Musielak}}{{Cuntz}
  et~al.}{2000}]{Cuntz2000}
{Cuntz} M.,  {Saar} S.~H.,   {Musielak} Z.~E.,  2000, \mn@doi [\apjl]
  {10.1086/312609}, \href
  {https://ui.adsabs.harvard.edu/abs/2000ApJ...533L.151C} {533, L151}

\bibitem[\protect\citeauthoryear{{Demircan}, {Eker}, {Karata{\c{s}}}  \&
  {Bilir}}{{Demircan} et~al.}{2006}]{Demircan2006}
{Demircan} O.,  {Eker} Z.,  {Karata{\c{s}}} Y.,   {Bilir} S.,  2006, \mn@doi
  [\mnras] {10.1111/j.1365-2966.2005.09948.x}, \href
  {https://ui.adsabs.harvard.edu/abs/2006MNRAS.366.1511D} {366, 1511}

\bibitem[\protect\citeauthoryear{{Donati}, {Semel}, {Carter}, {Rees}  \&
  {Collier Cameron}}{{Donati} et~al.}{1997}]{Donati1997}
{Donati} J.~F.,  {Semel} M.,  {Carter} B.~D.,  {Rees} D.~E.,   {Collier
  Cameron} A.,  1997, \mn@doi [\mnras] {10.1093/mnras/291.4.658}, \href
  {https://ui.adsabs.harvard.edu/abs/1997MNRAS.291..658D} {291, 658}

\bibitem[\protect\citeauthoryear{{Doyle}, {Ramsay}  \& {Doyle}}{{Doyle}
  et~al.}{2020}]{Doyle2020}
{Doyle} L.,  {Ramsay} G.,   {Doyle} J.~G.,  2020, \mn@doi [\mnras]
  {10.1093/mnras/staa923}, \href
  {https://ui.adsabs.harvard.edu/abs/2020MNRAS.494.3596D} {494, 3596}

\bibitem[\protect\citeauthoryear{{Eker} et~al.,}{{Eker}
  et~al.}{2008}]{Eker2008}
{Eker} Z.,  et~al., 2008, \mn@doi [\mnras] {10.1111/j.1365-2966.2008.13670.x},
  \href {https://ui.adsabs.harvard.edu/abs/2008MNRAS.389.1722E} {389, 1722}

\bibitem[\protect\citeauthoryear{{Evans}}{{Evans}}{1959}]{Evans1959}
{Evans} D.~S.,  1959, \mn@doi [\mnras] {10.1093/mnras/119.5.526}, \href
  {https://ui.adsabs.harvard.edu/abs/1959MNRAS.119..526E} {119, 526}

\bibitem[\protect\citeauthoryear{{Evans} et~al.,}{{Evans}
  et~al.}{2008}]{Evans2008}
{Evans} P.~A.,  et~al., 2008, The Astronomer's Telegram, \href
  {https://ui.adsabs.harvard.edu/abs/2008ATel.1787....1E} {1787, 1}

\bibitem[\protect\citeauthoryear{{Favata} \& {Schmitt}}{{Favata} \&
  {Schmitt}}{1999}]{Favata1999}
{Favata} F.,  {Schmitt} J.~H.~M.~M.,  1999, \mn@doi [\aap]
  {10.48550/arXiv.astro-ph/9909041}, \href
  {https://ui.adsabs.harvard.edu/abs/1999A&A...350..900F} {350, 900}

\bibitem[\protect\citeauthoryear{Fr\'{y}da}{Fr\'{y}da}{2023}]{Fryda2023}
Fr\'{y}da J.,  2023, Master's thesis, Charles University, \url
  {https://dspace.cuni.cz/handle/20.500.11956/182197?locale-attribute=en}

\bibitem[\protect\citeauthoryear{{Fuhrmeister} \& {Schmitt}}{{Fuhrmeister} \&
  {Schmitt}}{2004}]{FuhrmeisterSchmitt2004}
{Fuhrmeister} B.,  {Schmitt} J.~H.~M.~M.,  2004, \mn@doi [\aap]
  {10.1051/0004-6361:20035644}, \href
  {http://adsabs.harvard.edu/abs/2004A%26A...420.1079F} {420, 1079}

\bibitem[\protect\citeauthoryear{{Fuhrmeister}, {Liefke}, {Schmitt}  \&
  {Reiners}}{{Fuhrmeister} et~al.}{2008}]{Fuhrmeister2008}
{Fuhrmeister} B.,  {Liefke} C.,  {Schmitt} J.~H.~M.~M.,   {Reiners} A.,  2008,
  \mn@doi [\aap] {10.1051/0004-6361:200809379}, \href
  {https://ui.adsabs.harvard.edu/abs/2008A&A...487..293F} {487, 293}

\bibitem[\protect\citeauthoryear{{Fuhrmeister}, {Lalitha}, {Poppenhaeger},
  {Rudolf}, {Liefke}, {Reiners}, {Schmitt}  \& {Ness}}{{Fuhrmeister}
  et~al.}{2011}]{Fuhrmeister2011}
{Fuhrmeister} B.,  {Lalitha} S.,  {Poppenhaeger} K.,  {Rudolf} N.,  {Liefke}
  C.,  {Reiners} A.,  {Schmitt} J.~H.~M.~M.,   {Ness} J.~U.,  2011, \mn@doi
  [\aap] {10.1051/0004-6361/201117447}, \href
  {https://ui.adsabs.harvard.edu/abs/2011A&A...534A.133F} {534, A133}

\bibitem[\protect\citeauthoryear{{Fuhrmeister} et~al.,}{{Fuhrmeister}
  et~al.}{2018}]{Fuhrmeister2018}
{Fuhrmeister} B.,  et~al., 2018, \mn@doi [\aap] {10.1051/0004-6361/201732204},
  \href {https://ui.adsabs.harvard.edu/abs/2018A&A...615A..14F} {615, A14}

\bibitem[\protect\citeauthoryear{{Fuhrmeister} et~al.,}{{Fuhrmeister}
  et~al.}{2024}]{Fuhrmeister2024}
{Fuhrmeister} B.,  et~al., 2024, \mn@doi [\aap] {10.1051/0004-6361/202451697},
  \href {https://ui.adsabs.harvard.edu/abs/2024A&A...691A.208F} {691, A208}

\bibitem[\protect\citeauthoryear{{Gaia Collaboration} et~al.,}{{Gaia
  Collaboration} et~al.}{2023}]{GaiaCollaboration2023}
{Gaia Collaboration} et~al., 2023, \mn@doi [\aap]
  {10.1051/0004-6361/202243940}, \href
  {https://ui.adsabs.harvard.edu/abs/2023A&A...674A...1G} {674, A1}

\bibitem[\protect\citeauthoryear{{Garc{\'\i}a-Alvarez}, {Jevremovi{\'c}},
  {Doyle}  \& {Butler}}{{Garc{\'\i}a-Alvarez}
  et~al.}{2002}]{Garcia-Alvarez2002}
{Garc{\'\i}a-Alvarez} D.,  {Jevremovi{\'c}} D.,  {Doyle} J.~G.,   {Butler}
  C.~J.,  2002, \mn@doi [\aap] {10.1051/0004-6361:20011743}, \href
  {https://ui.adsabs.harvard.edu/abs/2002A&A...383..548G} {383, 548}

\bibitem[\protect\citeauthoryear{{Glebocki} \& {Stawikowski}}{{Glebocki} \&
  {Stawikowski}}{1995}]{Glebocki1995}
{Glebocki} R.,  {Stawikowski} A.,  1995, \actaa, \href
  {https://ui.adsabs.harvard.edu/abs/1995AcA....45..725G} {45, 725}

\bibitem[\protect\citeauthoryear{{Graham}, {Cauzzi}, {Zangrilli}, {Kowalski},
  {Sim{\~o}es}  \& {Allred}}{{Graham} et~al.}{2020}]{Graham2020}
{Graham} D.~R.,  {Cauzzi} G.,  {Zangrilli} L.,  {Kowalski} A.,  {Sim{\~o}es}
  P.,   {Allred} J.,  2020, \mn@doi [\apj] {10.3847/1538-4357/ab88ad}, \href
  {https://ui.adsabs.harvard.edu/abs/2020ApJ...895....6G} {895, 6}

\bibitem[\protect\citeauthoryear{{Greimel}, {Kab\'ath}, {Leitzinger}, {Odert}
  \& {Liptak}}{{Greimel} et~al.}{2025}]{Greimel2024}
{Greimel} R.~J.,  {Kab\'ath} P.,  {Leitzinger} M.,  {Odert} P.,   {Liptak} J.,
  2025, to be submitted to \pasp

\bibitem[\protect\citeauthoryear{{Gryciuk}, {Siarkowski}, {Sylwester},
  {Gburek}, {Podgorski}, {Kepa}, {Sylwester}  \& {Mrozek}}{{Gryciuk}
  et~al.}{2017}]{Gryciuk2017}
{Gryciuk} M.,  {Siarkowski} M.,  {Sylwester} J.,  {Gburek} S.,  {Podgorski} P.,
   {Kepa} A.,  {Sylwester} B.,   {Mrozek} T.,  2017, \mn@doi [\solphys]
  {10.1007/s11207-017-1101-8}, \href
  {https://ui.adsabs.harvard.edu/abs/2017SoPh..292...77G} {292, 77}

\bibitem[\protect\citeauthoryear{{Guenther} \& {Emerson}}{{Guenther} \&
  {Emerson}}{1997}]{Guenther1997}
{Guenther} E.~W.,  {Emerson} J.~P.,  1997, \aap, \href
  {http://adsabs.harvard.edu/abs/1997A%26A...321..803G} {321, 803}

\bibitem[\protect\citeauthoryear{{Gunn}, {Doyle}, {Mathioudakis}, {Houdebine}
  \& {Avgoloupis}}{{Gunn} et~al.}{1994}]{Gunn1994}
{Gunn} A.~G.,  {Doyle} J.~G.,  {Mathioudakis} M.,  {Houdebine} E.~R.,
  {Avgoloupis} S.,  1994, \aap, \href
  {http://adsabs.harvard.edu/abs/1994A%26A...285..489G} {285, 489}

\bibitem[\protect\citeauthoryear{{Haisch}, {Linsky}, {Bornmann}, {Stencel},
  {Antiochos}, {Golub}  \& {Vaiana}}{{Haisch} et~al.}{1983}]{Haisch1983}
{Haisch} B.~M.,  {Linsky} J.~L.,  {Bornmann} P.~L.,  {Stencel} R.~E.,
  {Antiochos} S.~K.,  {Golub} L.,   {Vaiana} G.~S.,  1983, \mn@doi [\apj]
  {10.1086/160866}, \href
  {https://ui.adsabs.harvard.edu/abs/1983ApJ...267..280H} {267, 280}

\bibitem[\protect\citeauthoryear{{Hawley} \& {Pettersen}}{{Hawley} \&
  {Pettersen}}{1991}]{Hawley1991}
{Hawley} S.~L.,  {Pettersen} B.~R.,  1991, \mn@doi [\apj] {10.1086/170474},
  \href {https://ui.adsabs.harvard.edu/abs/1991ApJ...378..725H} {378, 725}

\bibitem[\protect\citeauthoryear{{Heinzel} \& {Shibata}}{{Heinzel} \&
  {Shibata}}{2018}]{Heinzel2018}
{Heinzel} P.,  {Shibata} K.,  2018, \mn@doi [\apj] {10.3847/1538-4357/aabe78},
  \href {https://ui.adsabs.harvard.edu/abs/2018ApJ...859..143H} {859, 143}

\bibitem[\protect\citeauthoryear{{Heinzel}, {Karlicky}, {Kotrc}  \&
  {Svestka}}{{Heinzel} et~al.}{1994}]{Heinzel1994}
{Heinzel} P.,  {Karlicky} M.,  {Kotrc} P.,   {Svestka} Z.,  1994, \mn@doi
  [\solphys] {10.1007/BF00680446}, \href
  {http://adsabs.harvard.edu/abs/1994SoPh..152..393H} {152, 393}

\bibitem[\protect\citeauthoryear{{Henden}, {Templeton}, {Terrell}, {Smith},
  {Levine}  \& {Welch}}{{Henden} et~al.}{2016}]{Henden2016}
{Henden} A.~A.,  {Templeton} M.,  {Terrell} D.,  {Smith} T.~C.,  {Levine} S.,
  {Welch} D.,  2016, {VizieR Online Data Catalog: AAVSO Photometric All Sky
  Survey (APASS) DR9 (Henden+, 2016)}, VizieR On-line Data Catalog: II/336.
  Originally published in: 2015AAS...22533616H

\bibitem[\protect\citeauthoryear{{Houdebine}, {Foing}  \& {Rodono}}{{Houdebine}
  et~al.}{1990}]{Houdebine1990}
{Houdebine} E.~R.,  {Foing} B.~H.,   {Rodono} M.,  1990, \aap, \href
  {http://adsabs.harvard.edu/abs/1990A%26A...238..249H} {238, 249}

\bibitem[\protect\citeauthoryear{{Ichimoto} \& {Kurokawa}}{{Ichimoto} \&
  {Kurokawa}}{1984}]{Ichimoto1984}
{Ichimoto} K.,  {Kurokawa} H.,  1984, \mn@doi [\solphys] {10.1007/BF00156656},
  \href {http://adsabs.harvard.edu/abs/1984SoPh...93..105I} {93, 105}

\bibitem[\protect\citeauthoryear{{Ilin}, {Vedantham}, {Poppenh{\"a}ger},
  {Bloot}, {Callingham}, {Brandeker}  \& {Chakraborty}}{{Ilin}
  et~al.}{2025}]{Ilin2025}
{Ilin} E.,  {Vedantham} H.~K.,  {Poppenh{\"a}ger} K.,  {Bloot} S.,
  {Callingham} J.~R.,  {Brandeker} A.,   {Chakraborty} H.,  2025, \mn@doi
  [\nat] {10.1038/s41586-025-09236-z}, \href
  {https://ui.adsabs.harvard.edu/abs/2025Natur.643..645I} {643, 645}

\bibitem[\protect\citeauthoryear{{Inoue}, {Maehara}, {Notsu}, {Namekata},
  {Honda}, {Namizaki}, {Nogami}  \& {Shibata}}{{Inoue}
  et~al.}{2023}]{Inoue2023}
{Inoue} S.,  {Maehara} H.,  {Notsu} Y.,  {Namekata} K.,  {Honda} S.,
  {Namizaki} K.,  {Nogami} D.,   {Shibata} K.,  2023, \mn@doi [\apj]
  {10.3847/1538-4357/acb7e8}, \href
  {https://ui.adsabs.harvard.edu/abs/2023ApJ...948....9I} {948, 9}

\bibitem[\protect\citeauthoryear{{Inoue} et~al.,}{{Inoue}
  et~al.}{2024}]{Inoue2024}
{Inoue} S.,  et~al., 2024, \mn@doi [\apjl] {10.3847/2041-8213/ad5667}, \href
  {https://ui.adsabs.harvard.edu/abs/2024ApJ...969L..12I} {969, L12}

\bibitem[\protect\citeauthoryear{{Ip}, {Kopp}  \& {Hu}}{{Ip}
  et~al.}{2004}]{Ip2004}
{Ip} W.-H.,  {Kopp} A.,   {Hu} J.-H.,  2004, \mn@doi [\apjl] {10.1086/382274},
  \href {https://ui.adsabs.harvard.edu/abs/2004ApJ...602L..53I} {602, L53}

\bibitem[\protect\citeauthoryear{{Jackson}, {Kundu}  \& {Kassim}}{{Jackson}
  et~al.}{1990}]{Jackson1990}
{Jackson} P.~D.,  {Kundu} M.~R.,   {Kassim} N.,  1990, \mn@doi [\solphys]
  {10.1007/BF00156801}, \href
  {https://ui.adsabs.harvard.edu/abs/1990SoPh..130..391J} {130, 391}

\bibitem[\protect\citeauthoryear{{Johns-Krull}, {Hawley}, {Basri}  \&
  {Valenti}}{{Johns-Krull} et~al.}{1997}]{Johns-Krull1997}
{Johns-Krull} C.~M.,  {Hawley} S.~L.,  {Basri} G.,   {Valenti} J.~A.,  1997,
  \mn@doi [\apjs] {10.1086/313030}, \href
  {http://adsabs.harvard.edu/abs/1997ApJS..112..221J} {112, 221}

\bibitem[\protect\citeauthoryear{{Joy} \& {Humason}}{{Joy} \&
  {Humason}}{1949}]{Joy1949}
{Joy} A.~H.,  {Humason} M.~L.,  1949, \mn@doi [\pasp] {10.1086/126150}, \href
  {https://ui.adsabs.harvard.edu/abs/1949PASP...61..133J} {61, 133}

\bibitem[\protect\citeauthoryear{{Karmakar} \& {Pandey}}{{Karmakar} \&
  {Pandey}}{2016}]{Karmakar2016}
{Karmakar} S.,  {Pandey} J.~C.,  2016, in 19th Cambridge Workshop on Cool
  Stars, Stellar Systems, and the Sun (CS19). Cambridge Workshop on Cool Stars,
  Stellar Systems, and the Sun.
p.~144, \mn@doi{10.5281/zenodo.59648}

\bibitem[\protect\citeauthoryear{{Karmakar}, {Pandey}, {Rawat}, {Singh}  \&
  {Shedge}}{{Karmakar} et~al.}{2024}]{Karmakar2024}
{Karmakar} S.,  {Pandey} J.~C.,  {Rawat} N.,  {Singh} G.,   {Shedge} R.,  2024,
  \mn@doi [Bulletin de la Societe Royale des Sciences de Liege]
  {10.25518/0037-9565.11701}, \href
  {https://ui.adsabs.harvard.edu/abs/2024BSRSL..93..333K} {93, 333}

\bibitem[\protect\citeauthoryear{{Klein} et~al.,}{{Klein}
  et~al.}{2022}]{Klein2022}
{Klein} B.,  et~al., 2022, \mn@doi [\mnras] {10.1093/mnras/stac761}, \href
  {https://ui.adsabs.harvard.edu/abs/2022MNRAS.512.5067K} {512, 5067}

\bibitem[\protect\citeauthoryear{{Koller}, {Leitzinger}, {Temmer}, {Odert},
  {Beck}  \& {Veronig}}{{Koller} et~al.}{2021}]{Koller2021}
{Koller} F.,  {Leitzinger} M.,  {Temmer} M.,  {Odert} P.,  {Beck} P.~G.,
  {Veronig} A.,  2021, \mn@doi [\aap] {10.1051/0004-6361/202039003}, \href
  {https://ui.adsabs.harvard.edu/abs/2021A&A...646A..34K} {646, A34}

\bibitem[\protect\citeauthoryear{{Konovalenko} et~al.,}{{Konovalenko}
  et~al.}{2012}]{Konovalenko2012}
{Konovalenko} A.~A.,  et~al., 2012, in European Planetary Science Congress
  2012. pp EPSC2012--902

\bibitem[\protect\citeauthoryear{{Korhonen}, {Vida}, {Leitzinger}, {Odert}  \&
  {Kov{\'a}cs}}{{Korhonen} et~al.}{2017}]{Korhonen2017}
{Korhonen} H.,  {Vida} K.,  {Leitzinger} M.,  {Odert} P.,   {Kov{\'a}cs} O.~E.,
   2017, in {Nandy} D.,  {Valio} A.,   {Petit} P.,  eds,  Vol. 328, Living
  Around Active Stars. pp 198--203 (\mn@eprint {arXiv} {1612.06643}),
  \mn@doi{10.1017/S1743921317003969}

\bibitem[\protect\citeauthoryear{{Kowalski}, {Hawley}, {Wisniewski}, {Osten},
  {Hilton}, {Holtzman}, {Schmidt}  \& {Davenport}}{{Kowalski}
  et~al.}{2013}]{Kowalski2013}
{Kowalski} A.~F.,  {Hawley} S.~L.,  {Wisniewski} J.~P.,  {Osten} R.~A.,
  {Hilton} E.~J.,  {Holtzman} J.~A.,  {Schmidt} S.~J.,   {Davenport} J. R.~A.,
  2013, \mn@doi [\apjs] {10.1088/0067-0049/207/1/15}, \href
  {https://ui.adsabs.harvard.edu/abs/2013ApJS..207...15K} {207, 15}

\bibitem[\protect\citeauthoryear{{Kriskovics }, {Vida}, {K{\H{o}}v{\'a}ri},
  {Garcia-Alvarez}  \& {Ol{\'a}h}}{{Kriskovics } et~al.}{2013}]{Kriskovics2013}
{Kriskovics } L.,  {Vida} K.,  {K{\H{o}}v{\'a}ri} Z.,  {Garcia-Alvarez} D.,
  {Ol{\'a}h} K.,  2013, \mn@doi [Astronomische Nachrichten]
  {10.1002/asna.201211974}, \href
  {https://ui.adsabs.harvard.edu/abs/2013AN....334..976K} {334, 976}

\bibitem[\protect\citeauthoryear{{Lalitha}, {Fuhrmeister}, {Wolter}, {Schmitt},
  {Engels}  \& {Wieringa}}{{Lalitha} et~al.}{2013}]{Lalitha2013}
{Lalitha} S.,  {Fuhrmeister} B.,  {Wolter} U.,  {Schmitt} J.~H.~M.~M.,
  {Engels} D.,   {Wieringa} M.~H.,  2013, \mn@doi [\aap]
  {10.1051/0004-6361/201321419}, \href
  {https://ui.adsabs.harvard.edu/abs/2013A&A...560A..69L} {560, A69}

\bibitem[\protect\citeauthoryear{{Leitzinger} \& {Odert}}{{Leitzinger} \&
  {Odert}}{2022}]{Leitzinger2022c}
{Leitzinger} M.,  {Odert} P.,  2022, \mn@doi [Serbian Astronomical Journal]
  {10.2298/SAJ2205001L}, \href
  {https://ui.adsabs.harvard.edu/abs/2022SerAJ.205....1L} {205, 1}

\bibitem[\protect\citeauthoryear{{Leitzinger}, {Odert}, {Hanslmeier},
  {Konovalenko}, {Vanko}, {Khodachenko}, {Lammer}  \& {Rucker}}{{Leitzinger}
  et~al.}{2009}]{Leitzinger2009}
{Leitzinger} M.,  {Odert} P.,  {Hanslmeier} A.,  {Konovalenko} A.~A.,  {Vanko}
  M.,  {Khodachenko} M.~L.,  {Lammer} H.,   {Rucker} H.~O.,  2009, in
  {E.~Stempels} ed.,  American Institute of Physics Conference Series Vol.
  1094, American Institute of Physics Conference Series. pp 680--683,
  \mn@doi{10.1063/1.3099205}

\bibitem[\protect\citeauthoryear{{Leitzinger}, {Odert}, {Ribas}, {Hanslmeier},
  {Lammer}, {Khodachenko}, {Zaqarashvili}  \& {Rucker}}{{Leitzinger}
  et~al.}{2011}]{Leitzinger2011a}
{Leitzinger} M.,  {Odert} P.,  {Ribas} I.,  {Hanslmeier} A.,  {Lammer} H.,
  {Khodachenko} M.~L.,  {Zaqarashvili} T.~V.,   {Rucker} H.~O.,  2011, \mn@doi
  [\aap] {10.1051/0004-6361/201015985}, \href
  {http://adsabs.harvard.edu/abs/2011A%26A...536A..62L} {536, A62}

\bibitem[\protect\citeauthoryear{{Leitzinger}, {Odert}, {Greimel}, {Korhonen},
  {Guenther}, {Hanslmeier}, {Lammer}  \& {Khodachenko}}{{Leitzinger}
  et~al.}{2014}]{Leitzinger2014}
{Leitzinger} M.,  {Odert} P.,  {Greimel} R.,  {Korhonen} H.,  {Guenther} E.~W.,
   {Hanslmeier} A.,  {Lammer} H.,   {Khodachenko} M.~L.,  2014, \mn@doi
  [\mnras] {10.1093/mnras/stu1161}, \href
  {http://adsabs.harvard.edu/abs/2014MNRAS.443..898L} {443, 898}

\bibitem[\protect\citeauthoryear{{Leitzinger} et~al.,}{{Leitzinger}
  et~al.}{2020}]{Leitzinger2020}
{Leitzinger} M.,  et~al., 2020, \mn@doi [\mnras] {10.1093/mnras/staa504}, \href
  {https://ui.adsabs.harvard.edu/abs/2020MNRAS.493.4570L} {493, 4570}

\bibitem[\protect\citeauthoryear{{Leitzinger}, {Odert}  \&
  {Greimel}}{{Leitzinger} et~al.}{2024}]{Leitzinger2024}
{Leitzinger} M.,  {Odert} P.,   {Greimel} R.,  2024, \mn@doi [\mnras]
  {10.1093/mnras/stae1404}, \href
  {https://ui.adsabs.harvard.edu/abs/2024MNRAS.532.1486L} {532, 1486}

\bibitem[\protect\citeauthoryear{{Lu} et~al.,}{{Lu} et~al.}{2022}]{Lu2022}
{Lu} H.-p.,  et~al., 2022, \mn@doi [\aap] {10.1051/0004-6361/202142909}, \href
  {https://ui.adsabs.harvard.edu/abs/2022A&A...663A.140L} {663, A140}

\bibitem[\protect\citeauthoryear{{Lu} et~al.,}{{Lu} et~al.}{2025}]{Lu2025}
{Lu} H.-P.,  et~al., 2025, \mn@doi [\apjl] {10.3847/2041-8213/ad93cc}, \href
  {https://ui.adsabs.harvard.edu/abs/2025ApJ...978L..32L} {978, L32}

\bibitem[\protect\citeauthoryear{{Luyten}}{{Luyten}}{1949}]{Luyten1949}
{Luyten} W.~J.,  1949, \mn@doi [\pasp] {10.1086/126169}, \href
  {https://ui.adsabs.harvard.edu/abs/1949PASP...61..179L} {61, 179}

\bibitem[\protect\citeauthoryear{{Maehara} et~al.,}{{Maehara}
  et~al.}{2012}]{Maehara2012}
{Maehara} H.,  et~al., 2012, \mn@doi [\nat] {10.1038/nature11063}, \href
  {http://adsabs.harvard.edu/abs/2012Natur.485..478M} {485, 478}

\bibitem[\protect\citeauthoryear{{Malherbe}}{{Malherbe}}{2024}]{Malherbe2024}
{Malherbe} J.-M.,  2024, \mn@doi [arXiv e-prints] {10.48550/arXiv.2404.16902},
  \href {https://ui.adsabs.harvard.edu/abs/2024arXiv240416902M} {p.
  arXiv:2404.16902}

\bibitem[\protect\citeauthoryear{{Mohan}, {Mondal}, {Wedemeyer}  \&
  {Gopalswamy}}{{Mohan} et~al.}{2024}]{Mohan2024}
{Mohan} A.,  {Mondal} S.,  {Wedemeyer} S.,   {Gopalswamy} N.,  2024, \mn@doi
  [arXiv e-prints] {10.48550/arXiv.2402.00185}, \href
  {https://ui.adsabs.harvard.edu/abs/2024arXiv240200185M} {p. arXiv:2402.00185}

\bibitem[\protect\citeauthoryear{{Moore}}{{Moore}}{1972}]{Moore1972}
{Moore} C.~E.,  1972, {A multiplet table of astrophysical interest - Pt.1:
  Table of multiplets - Pt.2: Finding list of all lines in the table of
  multiplets}

\bibitem[\protect\citeauthoryear{{Moschou}, {Drake}, {Cohen},
  {Alvarado-G{\'o}mez}, {Garraffo}  \& {Fraschetti}}{{Moschou}
  et~al.}{2019}]{Moschou2019}
{Moschou} S.-P.,  {Drake} J.~J.,  {Cohen} O.,  {Alvarado-G{\'o}mez} J.~D.,
  {Garraffo} C.,   {Fraschetti} F.,  2019, \mn@doi [\apj]
  {10.3847/1538-4357/ab1b37}, \href
  {https://ui.adsabs.harvard.edu/abs/2019ApJ...877..105M} {877, 105}

\bibitem[\protect\citeauthoryear{{Muheki}, {Guenther}, {Mutabazi}  \&
  {Jurua}}{{Muheki} et~al.}{2020a}]{Muheki2020b}
{Muheki} P.,  {Guenther} E.~W.,  {Mutabazi} T.,   {Jurua} E.,  2020a, \mn@doi
  [\mnras] {10.1093/mnras/staa3152}, \href
  {https://ui.adsabs.harvard.edu/abs/2020MNRAS.499.5047M} {499, 5047}

\bibitem[\protect\citeauthoryear{{Muheki}, {Guenther}, {Mutabazi}  \&
  {Jurua}}{{Muheki} et~al.}{2020b}]{Muheki2020a}
{Muheki} P.,  {Guenther} E.~W.,  {Mutabazi} T.,   {Jurua} E.,  2020b, \mn@doi
  [\aap] {10.1051/0004-6361/201936904}, \href
  {https://ui.adsabs.harvard.edu/abs/2020A&A...637A..13M} {637, A13}

\bibitem[\protect\citeauthoryear{{Mullan} \& {Paudel}}{{Mullan} \&
  {Paudel}}{2018}]{Mullan2018}
{Mullan} D.~J.,  {Paudel} R.~R.,  2018, \mn@doi [\apj]
  {10.3847/1538-4357/aaa960}, \href
  {https://ui.adsabs.harvard.edu/abs/2018ApJ...854...14M} {854, 14}

\bibitem[\protect\citeauthoryear{{Namekata} et~al.,}{{Namekata}
  et~al.}{2021}]{Namekata2021}
{Namekata} K.,  et~al., 2021, \mn@doi [Nature Astronomy]
  {10.1038/s41550-021-01532-8}, \href
  {https://ui.adsabs.harvard.edu/abs/2022NatAs...6..241N} {6, 241}

\bibitem[\protect\citeauthoryear{{Namekata} et~al.,}{{Namekata}
  et~al.}{2022}]{Namekata2022a}
{Namekata} K.,  et~al., 2022, \mn@doi [\apjl] {10.3847/2041-8213/ac4df0}, \href
  {https://ui.adsabs.harvard.edu/abs/2022ApJ...926L...5N} {926, L5}

\bibitem[\protect\citeauthoryear{{Namekata} et~al.,}{{Namekata}
  et~al.}{2024}]{Namekata2024}
{Namekata} K.,  et~al., 2024, \mn@doi [\apj] {10.3847/1538-4357/ad0b7c}, \href
  {https://ui.adsabs.harvard.edu/abs/2024ApJ...961...23N} {961, 23}

\bibitem[\protect\citeauthoryear{{Notsu} et~al.,}{{Notsu}
  et~al.}{2013}]{Notsu2013}
{Notsu} Y.,  et~al., 2013, \mn@doi [\apj] {10.1088/0004-637X/771/2/127}, \href
  {https://ui.adsabs.harvard.edu/abs/2013ApJ...771..127N} {771, 127}

\bibitem[\protect\citeauthoryear{{Notsu}, {Honda}, {Maehara}, {Notsu},
  {Shibayama}, {Nogami}  \& {Shibata}}{{Notsu} et~al.}{2015a}]{Notsu2015a}
{Notsu} Y.,  {Honda} S.,  {Maehara} H.,  {Notsu} S.,  {Shibayama} T.,  {Nogami}
  D.,   {Shibata} K.,  2015a, \mn@doi [\pasj] {10.1093/pasj/psv001}, \href
  {https://ui.adsabs.harvard.edu/abs/2015PASJ...67...32N} {67, 32}

\bibitem[\protect\citeauthoryear{{Notsu}, {Honda}, {Maehara}, {Notsu},
  {Shibayama}, {Nogami}  \& {Shibata}}{{Notsu} et~al.}{2015b}]{Notsu2015b}
{Notsu} Y.,  {Honda} S.,  {Maehara} H.,  {Notsu} S.,  {Shibayama} T.,  {Nogami}
  D.,   {Shibata} K.,  2015b, \mn@doi [\pasj] {10.1093/pasj/psv002}, \href
  {https://ui.adsabs.harvard.edu/abs/2015PASJ...67...33N} {67, 33}

\bibitem[\protect\citeauthoryear{{Notsu} et~al.,}{{Notsu}
  et~al.}{2024}]{Notsu2024}
{Notsu} Y.,  et~al., 2024, \mn@doi [\apj] {10.3847/1538-4357/ad062f}, \href
  {https://ui.adsabs.harvard.edu/abs/2024ApJ...961..189N} {961, 189}

\bibitem[\protect\citeauthoryear{{Odert} et~al.,}{{Odert}
  et~al.}{2025}]{Odert2025}
{Odert} P.,  et~al., 2025, \mn@doi [\mnras] {10.1093/mnras/stae2752}, \href
  {https://ui.adsabs.harvard.edu/abs/2025MNRAS.537..537O} {537, 537}

\bibitem[\protect\citeauthoryear{{Okamoto}, {Notsu}, {Maehara}, {Namekata},
  {Honda}, {Ikuta}, {Nogami}  \& {Shibata}}{{Okamoto}
  et~al.}{2021}]{Okamoto2021}
{Okamoto} S.,  {Notsu} Y.,  {Maehara} H.,  {Namekata} K.,  {Honda} S.,  {Ikuta}
  K.,  {Nogami} D.,   {Shibata} K.,  2021, \mn@doi [\apj]
  {10.3847/1538-4357/abc8f5}, \href
  {https://ui.adsabs.harvard.edu/abs/2021ApJ...906...72O} {906, 72}

\bibitem[\protect\citeauthoryear{{Osten}}{{Osten}}{2023}]{Osten2023}
{Osten} R.~A.,  2023, in {Vidotto} A.~A.,  {Fossati} L.,   {Vink} J.~S.,  eds,
  Vol. 370, Winds of Stars and Exoplanets. pp 25--36,
  \mn@doi{10.1017/S1743921322003714}

\bibitem[\protect\citeauthoryear{{Osten} \& {Wolk}}{{Osten} \&
  {Wolk}}{2015}]{Osten2015}
{Osten} R.~A.,  {Wolk} S.~J.,  2015, \mn@doi [\apj]
  {10.1088/0004-637X/809/1/79}, \href
  {http://adsabs.harvard.edu/abs/2015ApJ...809...79O} {809, 79}

\bibitem[\protect\citeauthoryear{{Osten}, {Brown}, {Wood}  \& {Brady}}{{Osten}
  et~al.}{2002}]{Osten2002}
{Osten} R.~A.,  {Brown} A.,  {Wood} B.~E.,   {Brady} P.,  2002, \mn@doi [\apjs]
  {10.1086/323666}, \href
  {https://ui.adsabs.harvard.edu/abs/2002ApJS..138...99O} {138, 99}

\bibitem[\protect\citeauthoryear{{Pallavicini}, {Monsignori-Fossi}, {Landini}
  \& {Schmitt}}{{Pallavicini} et~al.}{1988}]{Pallavicini1988}
{Pallavicini} R.,  {Monsignori-Fossi} B.~C.,  {Landini} M.,   {Schmitt}
  J.~H.~M.~M.,  1988, \aap, \href
  {https://ui.adsabs.harvard.edu/abs/1988A&A...191..109P} {191, 109}

\bibitem[\protect\citeauthoryear{{Pallavicini}, {Tagliaferri}  \&
  {Stella}}{{Pallavicini} et~al.}{1990}]{Pallavicini1990}
{Pallavicini} R.,  {Tagliaferri} G.,   {Stella} L.,  1990, \aap, \href
  {https://ui.adsabs.harvard.edu/abs/1990A&A...228..403P} {228, 403}

\bibitem[\protect\citeauthoryear{{Pan} \& {Jordan}}{{Pan} \&
  {Jordan}}{1995}]{Pan1995}
{Pan} H.~C.,  {Jordan} C.,  1995, \mn@doi [\mnras] {10.1093/mnras/272.1.11},
  \href {https://ui.adsabs.harvard.edu/abs/1995MNRAS.272...11P} {272, 11}

\bibitem[\protect\citeauthoryear{{Pandey} \& {Singh}}{{Pandey} \&
  {Singh}}{2008}]{Pandey2008}
{Pandey} J.~C.,  {Singh} K.~P.,  2008, \mn@doi [\mnras]
  {10.1111/j.1365-2966.2008.13342.x}, \href
  {https://ui.adsabs.harvard.edu/abs/2008MNRAS.387.1627P} {387, 1627}

\bibitem[\protect\citeauthoryear{{Paulson}, {Allred}, {Anderson}, {Hawley},
  {Cochran}  \& {Yelda}}{{Paulson} et~al.}{2006}]{Paulson2006}
{Paulson} D.~B.,  {Allred} J.~C.,  {Anderson} R.~B.,  {Hawley} S.~L.,
  {Cochran} W.~D.,   {Yelda} S.,  2006, \mn@doi [\pasp] {10.1086/499497}, \href
  {https://ui.adsabs.harvard.edu/abs/2006PASP..118..227P} {118, 227}

\bibitem[\protect\citeauthoryear{{Roettenbacher} \& {Vida}}{{Roettenbacher} \&
  {Vida}}{2018}]{Roettenbacher2018}
{Roettenbacher} R.~M.,  {Vida} K.,  2018, \mn@doi [\apj]
  {10.3847/1538-4357/aae77e}, \href
  {https://ui.adsabs.harvard.edu/abs/2018ApJ...868....3R} {868, 3}

\bibitem[\protect\citeauthoryear{{Rubenstein} \& {Schaefer}}{{Rubenstein} \&
  {Schaefer}}{2000}]{Rubenstein2000}
{Rubenstein} E.~P.,  {Schaefer} B.~E.,  2000, \mn@doi [\apj] {10.1086/308326},
  \href {https://ui.adsabs.harvard.edu/abs/2000ApJ...529.1031R} {529, 1031}

\bibitem[\protect\citeauthoryear{{Schaefer}, {King}  \&
  {Deliyannis}}{{Schaefer} et~al.}{2000}]{Schaefer2000}
{Schaefer} B.~E.,  {King} J.~R.,   {Deliyannis} C.~P.,  2000, \mn@doi [\apj]
  {10.1086/308325}, \href
  {https://ui.adsabs.harvard.edu/abs/2000ApJ...529.1026S} {529, 1026}

\bibitem[\protect\citeauthoryear{{Schmieder}, {Forbes}, {Malherbe}  \&
  {Machado}}{{Schmieder} et~al.}{1987}]{Schmieder1987}
{Schmieder} B.,  {Forbes} T.~G.,  {Malherbe} J.~M.,   {Machado} M.~E.,  1987,
  \mn@doi [\apj] {10.1086/165344}, \href
  {http://adsabs.harvard.edu/abs/1987ApJ...317..956S} {317, 956}

\bibitem[\protect\citeauthoryear{{Schmitt}, {Pallavicini}, {Monsignori-Fossi}
  \& {Harnden}}{{Schmitt} et~al.}{1987}]{Schmitt1987}
{Schmitt} J.~H.~M.~M.,  {Pallavicini} R.,  {Monsignori-Fossi} B.~C.,
  {Harnden} F.~R. J.,  1987, \aap, \href
  {https://ui.adsabs.harvard.edu/abs/1987A&A...179..193S} {179, 193}

\bibitem[\protect\citeauthoryear{{Shibayama} et~al.,}{{Shibayama}
  et~al.}{2013}]{Shibayama2013}
{Shibayama} T.,  et~al., 2013, \mn@doi [\apjs] {10.1088/0067-0049/209/1/5},
  \href {https://ui.adsabs.harvard.edu/abs/2013ApJS..209....5S} {209, 5}

\bibitem[\protect\citeauthoryear{{Slee}, {Budding}, {Carter}, {Mengel}, {Waite}
   \& {Donati}}{{Slee} et~al.}{2004}]{Slee2004}
{Slee} O.~B.,  {Budding} E.,  {Carter} B.~D.,  {Mengel} M.~W.,  {Waite} I.,
  {Donati} J.~F.,  2004, \mn@doi [\pasa] {10.1071/AS03034}, \href
  {https://ui.adsabs.harvard.edu/abs/2004PASA...21...72S} {21, 72}

\bibitem[\protect\citeauthoryear{{Strassmeier}, {Hall}, {Fekel}  \&
  {Scheck}}{{Strassmeier} et~al.}{1993}]{Strassmeier1993}
{Strassmeier} K.~G.,  {Hall} D.~S.,  {Fekel} F.~C.,   {Scheck} M.,  1993,
  \aaps, \href {https://ui.adsabs.harvard.edu/abs/1993A&AS..100..173S} {100,
  173}

\bibitem[\protect\citeauthoryear{{Subramanian}, {Arunbabu}, {Vourlidas}  \&
  {Mauriya}}{{Subramanian} et~al.}{2014}]{Subramanian2014}
{Subramanian} P.,  {Arunbabu} K.~P.,  {Vourlidas} A.,   {Mauriya} A.,  2014,
  \mn@doi [\apj] {10.1088/0004-637X/790/2/125}, \href
  {https://ui.adsabs.harvard.edu/abs/2014ApJ...790..125S} {790, 125}

\bibitem[\protect\citeauthoryear{{Temmer}, {Veronig}, {Hanslmeier}, {Otruba}
  \& {Messerotti}}{{Temmer} et~al.}{2001}]{Temmer2001}
{Temmer} M.,  {Veronig} A.,  {Hanslmeier} A.,  {Otruba} W.,   {Messerotti} M.,
  2001, \mn@doi [\aap] {10.1051/0004-6361:20010908}, \href
  {https://ui.adsabs.harvard.edu/abs/2001A&A...375.1049T} {375, 1049}

\bibitem[\protect\citeauthoryear{{Tian}, {Xu}, {Chen}, {Zhang}, {Lu}, {Chen},
  {Yang}  \& {Wu}}{{Tian} et~al.}{2023}]{Tian2023}
{Tian} H.,  {Xu} Y.,  {Chen} H.,  {Zhang} J.,  {Lu} H.,  {Chen} Y.,  {Yang} Z.,
    {Wu} Y.,  2023, \mn@doi [SCIENTIA SINICA Technologica]
  {10.1360/SST-2022-0212}, \href
  {https://ui.adsabs.harvard.edu/abs/2023ScSnT..53.2021T} {53, 2021}

\bibitem[\protect\citeauthoryear{{Tsikoudi}}{{Tsikoudi}}{1982}]{Tsikoudi1982}
{Tsikoudi} V.,  1982, \mn@doi [\apj] {10.1086/160417}, \href
  {https://ui.adsabs.harvard.edu/abs/1982ApJ...262..263T} {262, 263}

\bibitem[\protect\citeauthoryear{{Tu}, {Yang}, {Zhang}  \& {Wang}}{{Tu}
  et~al.}{2020}]{Tu2020}
{Tu} Z.-L.,  {Yang} M.,  {Zhang} Z.~J.,   {Wang} F.~Y.,  2020, \mn@doi [\apj]
  {10.3847/1538-4357/ab6606}, \href
  {https://ui.adsabs.harvard.edu/abs/2020ApJ...890...46T} {890, 46}

\bibitem[\protect\citeauthoryear{{Tu}, {Yang}, {Wang}  \& {Wang}}{{Tu}
  et~al.}{2021}]{Tu2021}
{Tu} Z.-L.,  {Yang} M.,  {Wang} H.~F.,   {Wang} F.~Y.,  2021, \mn@doi [\apjs]
  {10.3847/1538-4365/abda3c}, \href
  {https://ui.adsabs.harvard.edu/abs/2021ApJS..253...35T} {253, 35}

\bibitem[\protect\citeauthoryear{{Vanzi} et~al.,}{{Vanzi}
  et~al.}{2012}]{Vanzi2012}
{Vanzi} L.,  et~al., 2012, \mn@doi [\mnras] {10.1111/j.1365-2966.2012.21382.x},
  \href {https://ui.adsabs.harvard.edu/abs/2012MNRAS.424.2770V} {424, 2770}

\bibitem[\protect\citeauthoryear{{Vasilyev} et~al.,}{{Vasilyev}
  et~al.}{2024}]{Vasilyev2024}
{Vasilyev} V.,  et~al., 2024, \mn@doi [Science] {10.1126/science.adl5441},
  \href {https://ui.adsabs.harvard.edu/abs/2024Sci...386.1301V} {386, 1301}

\bibitem[\protect\citeauthoryear{{Veronig}, {Odert}, {Leitzinger}, {Dissauer},
  {Fleck}  \& {Hudson}}{{Veronig} et~al.}{2021}]{Veronig2021}
{Veronig} A.~M.,  {Odert} P.,  {Leitzinger} M.,  {Dissauer} K.,  {Fleck} N.~C.,
    {Hudson} H.~S.,  2021, \mn@doi [Nature Astronomy]
  {10.1038/s41550-021-01345-9}, \href
  {https://ui.adsabs.harvard.edu/abs/2021NatAs...5..697V} {5, 697}

\bibitem[\protect\citeauthoryear{{Vida} et~al.,}{{Vida}
  et~al.}{2016}]{Vida2016}
{Vida} K.,  et~al., 2016, \mn@doi [\aap] {10.1051/0004-6361/201527925}, \href
  {http://adsabs.harvard.edu/abs/2016A%26A...590A..11V} {590, A11}

\bibitem[\protect\citeauthoryear{{Vida}, {Leitzinger}, {Kriskovics}, {Seli},
  {Odert}, {Kov{\'a}cs}, {Korhonen}  \& {van Driel-Gesztelyi}}{{Vida}
  et~al.}{2019}]{Vida2019}
{Vida} K.,  {Leitzinger} M.,  {Kriskovics} L.,  {Seli} B.,  {Odert} P.,
  {Kov{\'a}cs} O.~E.,  {Korhonen} H.,   {van Driel-Gesztelyi} L.,  2019,
  \mn@doi [\aap] {10.1051/0004-6361/201834264}, \href
  {https://ui.adsabs.harvard.edu/abs/2019A&A...623A..49V} {623, A49}

\bibitem[\protect\citeauthoryear{{Vida} et~al.,}{{Vida}
  et~al.}{2024}]{Vida2024}
{Vida} K.,  et~al., 2024, \mn@doi [Universe] {10.3390/universe10080313}, \href
  {https://ui.adsabs.harvard.edu/abs/2024Univ...10..313V} {10, 313}

\bibitem[\protect\citeauthoryear{{Villadsen} \& {Hallinan}}{{Villadsen} \&
  {Hallinan}}{2019}]{Villadsen2019}
{Villadsen} J.,  {Hallinan} G.,  2019, \mn@doi [\apj]
  {10.3847/1538-4357/aaf88e}, \href
  {https://ui.adsabs.harvard.edu/abs/2019ApJ...871..214V} {871, 214}

\bibitem[\protect\citeauthoryear{{Wang}}{{Wang}}{2023}]{Wang2023aa}
{Wang} J.,  2023, \mn@doi [Research in Astronomy and Astrophysics]
  {10.1088/1674-4527/acd590}, \href
  {https://ui.adsabs.harvard.edu/abs/2023RAA....23i5019W} {23, 095019}

\bibitem[\protect\citeauthoryear{{Wilson}}{{Wilson}}{1982}]{Wilson1982}
{Wilson} R.~M.,  1982, {Statistical aspects of the 1980 solar flars. 1: Data
  base, frequency distributions, and overview remarks}

\bibitem[\protect\citeauthoryear{{Wilson}}{{Wilson}}{1987}]{Wilson1987}
{Wilson} R.~M.,  1987, NASA Technical Papers, \href
  {https://ui.adsabs.harvard.edu/abs/1987NASTP2714.....W} {2714}

\bibitem[\protect\citeauthoryear{{Wollmann}, {Heinzel}  \&
  {Kab{\'a}th}}{{Wollmann} et~al.}{2023}]{Wollmann2023}
{Wollmann} J.,  {Heinzel} P.,   {Kab{\'a}th} P.,  2023, \mn@doi [\aap]
  {10.1051/0004-6361/202244544}, \href
  {https://ui.adsabs.harvard.edu/abs/2023A&A...669A.118W} {669, A118}

\bibitem[\protect\citeauthoryear{{Wuelser} \& {Marti}}{{Wuelser} \&
  {Marti}}{1989}]{Wuelser1989}
{Wuelser} J.-P.,  {Marti} H.,  1989, \mn@doi [\apj] {10.1086/167567}, \href
  {https://ui.adsabs.harvard.edu/abs/1989ApJ...341.1088W} {341, 1088}

\bibitem[\protect\citeauthoryear{{Yurchyshyn}, {Kumar}, {Cho}, {Lim}  \&
  {Abramenko}}{{Yurchyshyn} et~al.}{2015}]{Yurchyshyn2015}
{Yurchyshyn} V.,  {Kumar} P.,  {Cho} K.~S.,  {Lim} E.~K.,   {Abramenko} V.~I.,
  2015, \mn@doi [\apj] {10.1088/0004-637X/812/2/172}, \href
  {https://ui.adsabs.harvard.edu/abs/2015ApJ...812..172Y} {812, 172}

\bibitem[\protect\citeauthoryear{{Zic} et~al.,}{{Zic} et~al.}{2020}]{Zic2020}
{Zic} A.,  et~al., 2020, \mn@doi [\apj] {10.3847/1538-4357/abca90}, \href
  {https://ui.adsabs.harvard.edu/abs/2020ApJ...905...23Z} {905, 23}

\makeatother
\end{thebibliography}



\appendix

\newpage

\section{Observation log}

\begin{table*}
	\centering
	\caption{Observation log of the CC Eri spectroscopic monitoring at ESO152. First column gives the local date, second column the start date and time in UT of the first spectrum, the third column the start date and time in UT of the last spectrum, the fourth column the number of spectra, the fifth column the exposure time in seconds of the spectra and the number of spectra having this exposure time, and the sixth column yields the filters used for coordinated photometry. if more than one filter is given then this means that those were used alternately.}
	 \label{tab:A1}
	\begin{tabular}{lcccccc}
		\hline
   night     &    t$_{\mathrm{obs}}$ start first spec & t$_{\mathrm{obs}}$ start last spec & Nspec & t$_{\mathrm{exp}}$ & coordinated photometry & coordinated low-res\\
             &                    [UT]                &                [UT]                &       &        [s]         &                        &   spectroscopy     \\ 
  \hline
2023-08-02	 &	            2023-08-02 06:52:55	   &	      2023-08-02 09:08:22	      &  14	  &	       600/12	    &	      - 	         &         +            \\
2023-09-20	 &	            2023-09-21 06:36:33	   &	      2023-09-21 07:30:03	      &  9	  &	       300/9	    &	      - 	         &         -            \\
2023-09-21	 &	            2023-09-22 07:29:44    &	      2023-09-22 08:56:14	      &	 16	  &	       300/16	    &	      r’	         &         -            \\
2023-09-22	 &	            2023-09-23 08:33:08	   &	      2023-09-23 08:38:55	      &	 2	  &	       300/2	    &	      - 	         &         -            \\
2023-09-23	 &	            2023-09-24 04:31:29	   &	      2023-09-24 09:04:32        &	 30	  &	       300/30	    &	      g’	         &         -            \\
2023-09-24	 &	            2023-09-25 03:45:29	   &	      2023-09-25 08:41:06 	      &	 34	  &	   300/27, 480/7	&	      - 	         &         -            \\
2023-09-25	 &	            2023-09-26 04:58:06	   &	      2023-09-26 08:45:18 	      &	 37	  &	       300/37	    &	      -  	         &         -            \\
2023-09-27	 &	            2023-09-28 03:35:39	   &	      2023-09-28 06:03:14	      &	 19	  &	       300/19	    &	     g’+r’	         &         -            \\
2023-09-28	 &	            2023-09-29 02:21:36	   &	      2023-09-29 03:27:28	      &	 12	  &	   300/11,480/1	    &	     g’+r’	         &         -            \\
2023-09-29	 &	            2023-09-30 04:33:28	   &	      2023-09-30 08:37:06	      &	 30	  &	       300/30	    &	     g’+r’	         &         -            \\
2023-10-03	 &	            2023-10-04 07:34:25	   &	      2023-10-04 08:27:53	      &	 11	  &	       300/11	    &	      g’	         &         -            \\
2023-10-04	 &	            2023-10-05 06:15:19	   &	      2023-10-05 08:41:38	      &	 25	  &	       300/25	    &	      g’	         &         +            \\
2023-10-07	 &	            2023-10-08 05:12:10	   &	      2023-10-08 06:14:00	      &	 11	  &	   300/10, 480/1	&	     g’+r’	         &         -            \\
2023-10-10	 &	            2023-10-11 05:52:36	   &	      2023-10-11 05:58:22	      &	 2	  &	       300/2   	    &	   g+r+clear	     &         -            \\
2023-10-11	 &	            2023-10-12 03:55:40	   &	      2023-10-12 08:48:08	      &	 29	  &	   300/17, 480/12	&	      g’	         &         -            \\
2023-10-25	 &	            2023-10-26 02:24:44	   &	      2023-10-26 09:08:59	      &	 46	  &	       300/46	    &	     g’+r’	         &         -            \\
2023-10-26	 &	            2023-10-27 05:45:51	   &	      2023-10-27 08:37:07	      &	 34	  &	   300/27, 480/7	&	      -  	         &         -            \\
2023-10-27	 &             	2023-10-28 03:08:50    &	      2023-10-28 08:43:14	      &	 56	  &	       300/56 	    &	   g’+r’	         &         -            \\
2023-10-29	 &	            2023-10-30 04:47:34	   &	      2023-10-30 06:33:46	      &	 19	  &	       300/19	    &	     g’	             &         -            \\
2023-11-03	 &	            2023-11-04 00:35:12	   &	      2023-11-04 08:30:42	      &	 79	  &	       300/79	    &	     g’	             &         -            \\
2023-11-05	 &	            2023-11-06 01:41:47	   &	      2023-11-06 07:10:00	      &	 55	  &	       300/55	    &	     g’	             &         -            \\
2023-11-06	 &	            2023-11-07 00:31:08	   &       	  2023-11-07 08:48:17	      &	 82	  &	       300/82	    &	     g’	             &         -            \\
2023-11-07	 &	            2023-11-08 01:40:54	   &	      2023-11-08 08:10:50	      &	 68	  &  	   300/68	    &	     g’	             &         -            \\
2023-11-11	 &	            2023-11-12 04:03:31	   &	      2023-11-12 08:20:00	      &	 46	  &	       300/46	    &	     g’	             &         -            \\
2023-11-12	 &	            2023-11-13 02:52:35	   &	      2023-11-13 08:59:21	      &	 64	  &	       300/64	    &  	     g’	             &         -            \\
2023-11-13	 &	            2023-11-14 03:31:57	   &	      2023-11-14 08:11:04	      &	 50	  &	       300/50	    &	     g’	             &         -            \\
2023-11-14	 &	            2023-11-15 01:50:04	   &	      2023-11-15 06:30:46	      &	 49	  &	       300/49	    &	     g’	             &         -            \\
2023-11-16	 &            	2023-11-17 02:53:43	   &	      2023-11-17 06:36:03	      &	 39	  &	       300/39	    &	     g’	             &         -            \\
2023-11-17	 &	            2023-11-18 02:29:40	   &	      2023-11-18 05:05:24	      &	 24	  &	   300/20, 600/4	&	     g’	             &         -            \\
2023-11-18	 &	            2023-11-19 01:34:20	   &	      2023-11-19 06:02:16	      &	 48	  &	       300/48	    &	     g’	             &         -            \\
2023-11-19	 &	            2023-11-20 01:19:27	   &	      2023-11-20 06:26:19	      &	 59	  &	       300/59	    &	     g’	             &         -            \\
2023-11-20	 &	            2023-11-21 00:57:27	   &	      2023-11-21 06:55:35	      &	 51	  &	   300/36, 600/15	&	     g’     	     &         -            \\
2023-11-21	 &	            2023-11-22 01:30:32	   &	      2023-11-22 06:01:00	      &	 47	  &	       300/47	    &	     g’ 	         &         -            \\
2023-11-22	 &	            2023-11-23 01:06:03	   &	      2023-11-23 06:19:40	      &	 55	  &	       300/55	    &	     g’	             &         -            \\
2023-11-23	 &	            2023-11-24 01:17:40	   &	      2023-11-24 06:14:57	      &	 52	  &	       300/52	    &	     g’	             &         -            \\
2023-11-24	 &	            2023-11-25 01:34:38	   &	      2023-11-25 05:52:05	      &	 26	  &	   300/19, 600/7	&	     g’	             &         -            \\
2023-11-27	 &	            2023-11-28 01:07:00	   &	      2023-11-28 06:32:48	      &	 56	  &	       300/56	    &	  g’+SA200	         &         -            \\
2023-11-28	 &             	2023-11-29 01:02:42    &	      2023-11-29 06:24:57	      &	 55	  &	       300/55	    &	     g’	             &         +            \\
2023-11-29	 &	            2023-11-30 02:52:01	   &	      2023-11-30 07:12:07	      &	 39	  &	   300/30, 600/9	&	     g’	             &         -            \\
2023-11-30	 &	            2023-12-01 04:05:14	   &	      2023-12-01 06:22:41	      &	 25	  &	       300/25	    &	     g’	             &         -            \\
2023-12-01	 &	            2023-12-02 01:02:26	   &	      2023-12-02 06:20:35	      &	 42	  &	   300/26, 600/16	&	     g’	             &         -            \\
2023-12-02	 &	            2023-12-03 02:10:06	   &	      2023-12-03 06:21:15	      &	 45	  &	       300/45	    &	     g’	             &         -            \\
2023-12-03	 &	            2023-12-04 02:02:19	   &	      2023-12-04 06:58:34	      &	 52	  &	       300/52	    &	     g’	             &         -            \\
2023-12-04	 &           	2023-12-05 02:17:24    &	      2023-12-05 05:36:46	      &	 36	  &	       300/36	    &	     g’	             &         -            \\
2023-12-05	 &	            2023-12-06 02:16:31	   &	      2023-12-06 02:39:18	      &	 5	  &	       300/5	    &	     g’	             &         -            \\
2023-12-06	 &	            2023-12-07 02:16:33	   &	      2023-12-07 06:21:33	      &	 44	  &	       300/44	    &	     g’	             &         -            \\
2023-12-07	 &	            2023-12-08 02:35:43	   &	      2023-12-08 06:00:54	      &	 37	  &	       300/37	    &	     g’	             &         -            \\
2023-12-08	 &	            2023-12-09 02:51:49	   &	      2023-12-09 05:02:49	      &	 24	  &	       300/24	    &	     r’	             &         -            \\
2023-12-09	 &	            2023-12-10 02:35:56	   &	      2023-12-10 04:41:13	      &	 23	  &	       300/23	    &	     g’	             &         -            \\
2023-12-10	 &	            2023-12-11 02:17:19	   &	      2023-12-11 05:41:09	      &	 35	  &	       300/35	    &	     g’	             &         -            \\
2023-12-11	 &            	2023-12-12 01:11:39	   &	      2023-12-12 06:45:49	      &	 59	  &	       300/59	    &	     g’	             &         -            \\
2023-12-13	 &          	2023-12-14 03:33:21	   &	      2023-12-14 05:44:21	      &	 24	  &	       300/24	    &	     g’	             &         -            \\
2023-12-14	 &	            2023-12-15 02:16:05	   &	      2023-12-15 05:52:41	      &	 39	  &	       300/39	    &	     g’	             &         -            \\
2023-12-15	 &	            2023-12-16 00:46:46	   &	      2023-12-16 06:05:58	      &	 53	  &	       300/53	    &	     g’	             &         -            \\
2023-12-16	 &	            2023-12-17 01:19:30	   &	      2023-12-17 06:13:12	      &	 51	  &	       300/51	    &	     g’	             &         -            \\
2023-12-17	 &	            2023-12-18 01:08:48	   &     	  2023-12-18 05:46:08	      &	 48	  &	       300/48	    &	     g’	             &         -            \\
2023-12-18	 &	            2023-12-19 01:09:36	   &	      2023-12-19 06:21:54	      &	 47	  &	   300/37, 600/10	&	     g’	             &         -            \\

  \hline
	\end{tabular}
\end{table*}

\begin{table*}
	\centering
	\contcaption{}
	\begin{tabular}{lcccccc}
		\hline
   night     &    t$_{\mathrm{obs}}$ start first spec & t$_{\mathrm{obs}}$ start last spec & Nspec & t$_{\mathrm{exp}}$ & coordinated photometry  & coordinated low-res\\
            &                                        &                                    &       &        [s]         &                         &   spectroscopy     \\ 

  \hline
2023-12-19	&	2023-12-20 01:03:07	&	2023-12-20 06:38:40	&	48	&	300/46, 600/2	&	g’	& -\\
2023-12-20	&	2023-12-21 01:08:03	&	2023-12-21 06:15:45	&	53	&	300/53	&	g’	& -	\\
2023-12-21	&	2023-12-22 05:22:17	&	2023-12-22 06:19:14	&	11	&	300/11	&	g’	& -	\\
2023-12-22	&	2023-12-23 01:20:34	&	2023-12-23 06:44:30	&	52	&	300/44, 600/8	&	g’	& -	\\
2024-01-03	&	2024-01-04 01:42:03	&	2024-01-04 05:29:52	&	41	&	300/41	&	g’	& -	\\
2024-01-04	&	2024-01-05 02:40:31	&	2024-01-05 05:20:03	&	29	&	300/29	&	g’	& -	\\
2024-01-05	&	2024-01-06 01:59:46	&	2024-01-06 05:11:17	&	34	&	300/34	&	g’	& -	\\
2024-01-06	&	2024-01-07 01:39:19	&	2024-01-07 05:12:51	&	38	&	300/38	&	g’	& -	\\
2024-01-07	&	2024-01-08 02:08:28	&	2024-01-08 04:59:59	&	35	&	300/35	&	g’	& -	\\
2024-01-08	&	2024-01-09 01:45:10	&	2024-01-09 05:01:55	&	40	&	300/40	&	g’	& -	\\
2024-01-09	&	2024-01-10 01:25:42	&	2024-01-10 04:56:52	&	37	&	300/37	&	g’	& -	\\
2024-01-10	&	2024-01-11 01:02:02	&	2024-01-11 05:27:22	&	40	&	300/35, 600/5	&	g’	& -	\\
2024-01-11	&	2024-01-12 01:46:32	&	2024-01-12 04:54:32	&	34	&	300/34	&	g’	& -	\\

  \hline
	\end{tabular}
\end{table*}



\clearpage

\section{Flares and their parameters}

\begin{table*}
	\centering
	\caption{Observation log of the CC Eri spectroscopic monitoring at ES152. First column gives the local date, second column the start date and time in UT of the first spectrum, the third column the start date and time in UT of the last spectrum, the fourth column the number of spectra, the fifth column the exposure time in seconds of the spectra and the number of spectra having this exposure time, and the sixth column yields the filters used for coordinated photometry. if more than one filter is given then this means that those were used alternately.}
		 \label{tab:B1}
	\begin{tabular}{lcccccccccc}
		\hline
  
 Flare   &   night     &   line    &   flare start      &     flare peak     &    flare stop       &  t$_{\mathrm{flare}}$  & t$_{\mathrm{imp}}$  & t$_{\mathrm{grad}}$ & t$_{\mathrm{decay}}$ &        E  \\
         &             &           &                    &                    &                     &                        &                     &                     &                      &           \\
   no.   &    [UT]     &           & [JD-2460000]       &    [JD-2460000]    &  [JD-2460000]       &          [min ]        &        [min]        &         [min]       &      [min]           &       [10$^{31}$~erg]\\ 
  \hline
 
  distinct flares &&&&&&&&&& \\

  \hline

1   & 2023-10-27  & H$\alpha$ & 244.79727 & 244.81329 & 244.85738 & 86.6  & 23.1 & 63.5 &        -        & 2.3$\pm$0.2\\
1   & 2023-10-27  & H$\beta$  & 244.79727 & 244.81329 & 244.85738 & 86.6  & 23.1 & 63.5 &        -        & 1.7$\pm$0.3\\
1   & 2023-10-27  & H$\gamma$ & 244.80127 & 244.81329 & 244.85738 & 80.8  & 17.3 & 63.5 &        -        & 2.7$\pm$0.5\\
2   & 2023-11-08  & H$\alpha$ & 256.73623 & 256.75998 & 256.83913 & 148.2 & 34.2 & 114.0 &       -        & 2.8$\pm$0.5\\
2   & 2023-11-08  & H$\beta$  & 256.73228 & 256.74810 & 256.83913 & 153.9 & 22.8 & 131.1&        -        & 2.7$\pm$0.6\\
2   & 2023-11-08  & H$\gamma$ & 256.73623 & 256.76393 & 256.83913 & 148.2 & 39.9 & 108.3&        -        & 4.1$\pm$0.9\\
    & 2023-11-08  & g-band    & 256.73916 & 256.74074 & 256.84833 & 160.0 & 1.1  & 154.9&     40.0        & 211.5$\pm$5.9\\
3   & 2023-11-12  & H$\alpha$ & 260.77820 & 260.79008 & 260.84550 & 96.9  & 17.1 & 79.8 &        -        & 0.8$\pm$0.3\\
3   & 2023-11-12  & H$\beta$  & 260.78612 & 260.79404 & 260.84550 & 85.5  & 11.4 & 74.1 &        -        & 0.9$\pm$0.3\\
3   & 2023-11-12  & H$\gamma$ & 260.77820 & 260.79404 & 260.84550 & 96.9  & 22.8 & 74.1 &        -        & 1.0$\pm$0.5\\
    & 2023-11-12  & g-band    & 260.78388 & 260.79207 & 260.81075 & 40.0  & 11.8 & 26.9 &     6.4         & 23.7$\pm$0.9\\
4   & 2023-11-14  & H$\alpha$ & 262.68898 & 262.75226 & 262.83929 & 216.5 & 91.1 & 125.3&        -        & 9.2$\pm$0.4\\
4   & 2023-11-14  & H$\beta$  & 262.68501 & 262.75622 & 262.83929 & 222.2 & 102.6& 119.6&        -        & 6.3$\pm$0.5\\
4   & 2023-11-14  & H$\gamma$ & 262.68898 & 262.75622 & 262.83929 & 216.5 & 96.8 & 119.6&        -        & 8.1$\pm$0.8\\
    & 2023-11-14  & g-band    & 262.69802 & 262.71183 & 262.80076 & 150.0 & 19.9 & 128.1&     69.9        & 206.1$\pm$2.1\\
5   & 2023-11-17  & H$\alpha$ & 265.71398 & 265.72189 & 265.73375 & 28.5  & 11.4 & 17.1 &        -        & 0.2$\pm$0.1\\
5   & 2023-11-17  & H$\alpha$ & 265.73771 & 265.74561 & 265.77330 & 51.3  & 11.4 & 39.9 &        -        & 0.4$\pm$0.1\\
    & 2023-11-17  & g-band    & 265.71997 & 265.72251 & 265.73753 & 30.0  & 3.7  & 21.6 &     4.0         & 14.0$\pm$2.1\\
    & 2023-11-17  & g-band    & 265.73859 & 265.74028 & 265.74729 & 10.0  & 2.4  & 10.1 &     1.8         & 15.2$\pm$2.4\\
6   & 2023-12-02  & H$\alpha$ & 280.74595 & 280.76083 & 280.76083 & 21.4  & 21.4 & 0.0  &        -        & 0.2$\pm$0.2\\
6   & 2023-12-02  & H$\beta$  & 280.74595 & 280.76083 & 280.76083 & 21.4  & 21.4 & 0.0  &        -        & 0.2$\pm$0.3\\
6   & 2023-12-02  & H$\gamma$ & 280.74595 & 280.76083 & 280.76083 & 21.4  & 21.4 & 0.0  &        -        & 0.6$\pm$0.4\\
    & 2023-12-02  & g-band    & 280.75580 & 280.75664 & 280.76598 & 10.0  & 1.2  & 13.4 &     2.5         & 25.3$\pm$1.7\\
7   & 2023-12-06  & H$\alpha$ & 284.59308 & 284.60098 & 284.60890 & 22.8  & 11.4 & 11.4 &        -        & 0.2$\pm$0.1\\
7   & 2023-12-06  & H$\beta$  & 284.59308 & 284.60098 & 284.60890 & 22.8  & 11.4 & 11.4 &        -        & 0.3$\pm$0.1\\
7   & 2023-12-06  & H$\gamma$ & 284.59308 & 284.60098 & 284.60890 & 22.8  & 11.4 & 11.4 &        -        & 0.2$\pm$0.1\\
    & 2023-12-06  & g-band    & 284.59457 & 284.59596 & 284.60715 & 20.0  & 2.0  & 16.1 &     4.0         & 17.9$\pm$3.5\\
8   & 2023-12-11  & H$\alpha$ & 289.63318 & 289.64109 & 289.70438 & 102.5 & 11.4 & 91.1 &        -        & 0.8$\pm$0.2\\
8   & 2023-12-11  & H$\beta$  & 289.63318 & 289.64109 & 289.70042 & 96.8  & 11.4 & 85.4 &        -        & 0.9$\pm$0.3\\
8   & 2023-12-11  & H$\gamma$ & 289.63318 & 289.64109 & 289.70042 & 96.8  & 11.4 & 85.4 &        -        & 1.1$\pm$0.4\\
    & 2023-12-11  & g-band    & 289.63440 & 289.63519 & 289.65576 & 30.0  & 1.1  & 29.6 &      5.7        & 27.6$\pm$2.5\\
9   & 2024-01-10  & H$\alpha$ & 319.65541 & 319.67123 & 319.70443 & 70.6  & 22.8 & 47.8 &        -        & 0.8$\pm$0.2\\
9   & 2024-01-10  & H$\beta$  & 319.65937 & 319.67518 & 319.70443 & 64.9  & 22.8 & 42.1 &        -        & 1.0$\pm$0.2\\
9   & 2024-01-10  & H$\gamma$ & 319.65937 & 319.67123 & 319.70443 & 64.9  & 17.1 & 47.8 &        -        & 1.2$\pm$0.3\\
   
\hline
   
	\end{tabular}
\end{table*}

\begin{table*}
	\centering
	\contcaption{}
		\begin{tabular}{lcccccccccc}
		\hline
  
Flare &   night    &   line    &   flare start      &     flare peak     &    flare stop       &  t$_{\mathrm{flare}}$  & t$_{\mathrm{imp}}$  & t$_{\mathrm{grad}}$ & t$_{\mathrm{decay}}$ &        E      \\
      &            &           &                    &                    &                     &                        &                     &                     &                      &             \\
 no.  &    [UT]    &           & [JD-2460000]       &    [JD-2460000]    &                     &          [min ]        &        [min]        &         [min]       &      [min]           &       [10$^{31}$~erg]\\ 
  \hline 
   
  weak flares &&&&&&&&&& \\
  \hline

10 & 2023-10-26 & H$\alpha$ & 243.81133 & 243.82334 & 243.87951 & 98.2 & 17.3  & 80.9 &        -        &  0.5$\pm$0.3\\
11 & 2023-11-06 & H$\alpha$ & 254.60902 & 254.63309 & 254.66517 & 80.9 & 34.7  & 46.2 &        -        &  0.9$\pm$0.2\\
11 & 2023-11-06 & H$\beta$  & 254.60501 & 254.63309 & 254.66517 & 86.6 & 40.4  & 46.2 &        -        & 1.0$\pm$0.3\\
11 & 2023-11-06 & H$\gamma$ & 254.61703 & 254.62907 & 254.66116 & 63.6 & 17.3  & 46.2 &        -        & 1.0$\pm$0.4\\
12 & 2023-11-19 & H$\alpha$ & 267.69447 & 267.73797 & 267.74193 & 68.3 & 62.6  & 5.7  &        -        &  2.5$\pm$0.5\\
12 & 2023-11-19 & H$\beta$  & 267.70238 & 267.73797 & 267.74193 & 56.9 & 51.3  & 5.7  &        -        & 1.8$\pm$0.9\\
12 & 2023-11-19 & H$\gamma$ & 267.70238 & 267.73797 & 267.74193 & 56.9 & 51.3  & 5.7  &        -        & 2.6$\pm$2.6\\
13 & 2023-11-23 & H$\alpha$ & 271.75402 & 271.76193 & 271.76193 & 11.4 & 11.4  & 0.0  &        -        &  0.1$\pm$0.2\\
13 & 2023-11-23 & H$\beta$  & 271.75402 & 271.76193 & 271.76193 & 11.4 & 11.4  & 0.0  &        -        & 0.1$\pm$0.2\\
13 & 2023-11-23 & H$\gamma$ & 271.75797 & 271.76193 & 271.76193 & 5.7  & 5.7   & 0.0  &        -        & 0.1$\pm$0.2\\
14 & 2023-11-24 & H$\alpha$ & 272.62342 & 272.63530 & 272.70722 & 120.7& 17.1  & 103.6&        -        & 1.6$\pm$2.1\\
14 & 2023-11-24 & H$\alpha$ & 272.70722 & 272.71909 & 272.74678 & 57.  & 17.1  & 39.9 &        -        & 0.3$\pm$1.5\\
15 & 2023-11-28 & H$\alpha$ & 276.61620 & 276.64009 & 276.66786 & 74.4 & 34.4  & 40.0 &        -        &  0.4$\pm$0.2\\
16 & 2023-12-02 & H$\alpha$ & 280.56536 & 280.58513 & 280.70879 & 206.5& 28.5  & 178.1&        -        & 1.6$\pm$0.2\\
17 & 2023-12-05 & H$\alpha$ & 283.71233 & 283.73213 & 283.73213 & 28.5 & 28.5  & 0.0  &        -        &  0.2$\pm$0.2\\
17 & 2023-12-05 & H$\beta$  & 283.71629 & 283.73213 & 283.73213 & 22.8 & 22.8  & 0.0  &        -        & 0.9$\pm$0.3\\
17 & 2023-12-05 & H$\gamma$ & 283.69651 & 283.73213 & 283.73213 & 51.3 & 51.3  & 0.0  &        -        & 2.8$\pm$1.0\\
18 & 2023-12-21 & H$\alpha$ & 299.71570 & 299.72361 & 299.75524 & 56.9 & 11.4  & 45.6 &        -        &  0.70$\pm$0.2\\
18 & 2023-12-21 & H$\beta$  & 299.71570 & 299.72361 & 299.75524 & 56.9 & 11.4  & 45.6 &        -        & 1.4$\pm$0.2\\
18 & 2023-12-21 & H$\gamma$ & 299.71570 & 299.72756 & 299.75524 & 56.9 & 17.1  & 39.9 &        -        & 1.9$\pm$0.3\\
19 & 2023-12-23 & H$\alpha$ & 301.58982 & 301.64590 & 301.69941 & 157.8& 80.8  & 77.0 &        -        &  1.8$\pm$0.3\\
19 & 2023-12-23 & H$\beta$  & 301.60131 & 301.64590 & 301.68754 & 124.2& 64.2  & 60.0 &        -        & 1.2$\pm$0.5\\
19 & 2023-12-23 & H$\gamma$ & 301.60131 & 301.64590 & 301.67964 & 112.8& 64.2  & 48.6 &        -        & 1.5$\pm$0.9\\
20 & 2024-01-05 & H$\alpha$ & 314.68098 & 314.69284 & 314.72052 & 56.9 & 17.1  & 39.9 &        -        &  0.3$\pm$0.1\\
20 & 2024-01-05 & H$\beta$  & 314.68098 & 314.69284 & 314.70866 & 39.9 & 17.1  & 22.8 &        -        & 0.2$\pm$0.1\\
20 & 2024-01-05 & H$\gamma$ & 314.68098 & 314.68889 & 314.71262 & 45.6 & 11.4  & 34.2 &        -        & 0.4$\pm$0.2\\
21 & 2024-01-07 & H$\alpha$ & 316.63532 & 316.70752 & 316.71553 & 115.5& 104.0 & 11.5 &        -        &  2.3$\pm$0.4\\
21 & 2024-01-07 & H$\beta$  & 316.63934 & 316.71553 & 316.71553 & 109.7& 109.7 & 0.0  &        -        & 1.8$\pm$0.4\\
21 & 2024-01-07 & H$\gamma$ & 316.65141 & 316.71153 & 316.71553 & 92.3 & 86.6  & 5.8  &        -        & 2.3$\pm$0.6\\
22 & 2024-01-08 & H$\alpha$ & 317.62251 & 317.62952 & 317.65404 & 45.4 & 10.1  & 35.3 &        -        &  0.4$\pm$0.2\\

   \hline
  gradual phase only &&&&&&&&&& \\
  \hline

23 & 2023-10-28 & H$\alpha$ & 245.62941 & 245.62941 & 245.73166 & 147.300 & 0.0 & 147.300 &        -        &  1.3$\pm$0.5\\
23 & 2023-10-28 & H$\beta$  & 245.62941 & 245.62941 & 245.73166 & 147.300 & 0.0 & 147.300 &        -        & 1.4$\pm$0.7\\
24 & 2023-11-13 & H$\alpha$ & 261.61812 & 261.61812 & 261.67095 & 76.1000 & 0.0 & 76.1000 &        -        &  0.3$\pm$0.7\\
25 & 2023-11-22 & H$\alpha$ & 270.56114 & 270.56114 & 270.73314 & 247.700 & 0.0 & 247.700 &        -        &  3.8$\pm$0.4\\
25 & 2023-11-22 & H$\beta$  & 270.56114 & 270.56114 & 270.72918 & 242.000 & 0.0 & 242.000 &        -        & 2.7$\pm$0.4\\
25 & 2023-11-22 & H$\gamma$ & 270.56114 & 270.56114 & 270.71732 & 224.900 & 0.0 & 224.900 &        -        & 2.1$\pm$0.6\\
26 & 2023-11-30 & H$\alpha$ & 278.61772 & 278.61772 & 278.70522 & 126.000 & 0.0 & 126.000 &        -        &  0.7$\pm$0.3\\
26 & 2023-11-30 & H$\beta$  & 278.61772 & 278.61772 & 278.70125 & 120.300 & 0.0 & 120.300 &        -        & 1.2$\pm$0.3\\
26 & 2023-11-30 & H$\gamma$ & 278.61772 & 278.61772 & 278.69332 & 108.900 & 0.0 & 108.900 &        -        & 1.1$\pm$0.5\\
27 & 2023-12-12 & H$\alpha$ & 290.54803 & 290.54803 & 290.68855 & 202.400 & 0.0 & 202.400 &        -        &  1.7$\pm$0.7\\
28 & 2023-12-16	& H$\alpha$ & 294.53075 & 294.53075 & 294.66541 & 193.900 & 0.0 & 193.900 &        -        & 1.3$\pm$0.7\\
29 & 2023-12-19 & H$\alpha$ & 297.54660 & 297.54660 & 297.70220 & 224.100 & 0.0 & 224.100 &        -        &  2.4$\pm$0.4\\
29 & 2023-12-19 & H$\beta$  & 297.54660 & 297.54660 & 297.68620 & 201.000 & 0.0 & 201.000 &        -        & 1.6$\pm$0.4\\
29 & 2023-12-19 & H$\gamma$ & 297.54660 & 297.54660 & 297.66642 & 172.500 & 0.0 & 172.500 &        -        & 1.9$\pm$0.6\\
30 & 2024-01-06 & H$\alpha$ & 315.58144 & 315.58144 & 315.63433 & 76.2000 & 0.0 & 76.2000 &        -        &  0.7$\pm$0.2\\
30 & 2024-01-06 & H$\beta$  & 315.58144 & 315.58144 & 315.63033 & 70.4000 & 0.0 & 70.4000 &        -        & 0.5$\pm$0.2\\
30 & 2024-01-06 & H$\gamma$ & 315.58144 & 315.58144 & 315.64234 & 87.7000 & 0.0 & 87.7000 &        -        & 0.7$\pm$0.3\\
31 & 2024-01-11 & H$\alpha$ & 320.54135 & 320.54135 & 320.62045 & 113.900 & 0.0 & 113.900 &        -        &  1.2$\pm$0.2\\
31 & 2024-01-11 & H$\beta$  & 320.54135 & 320.54135 & 320.62836 & 125.300 & 0.0 & 125.300 &        -        & 1.9$\pm$0.3\\
31 & 2024-01-11 & H$\gamma$ & 320.54135 & 320.54135 & 320.62836 & 125.300 & 0.0 & 125.300 &        -        & 2.4$\pm$0.4\\

 \hline

	\end{tabular}
\end{table*}

\clearpage


\section{H$\alpha$ spectral line variability caused by the binarity of CC~Eri}
\label{Hamorph}
Starting with the ''distinct flares`` and flare no.~1. When the impulsive phase occurs the H$\alpha$ profiles are already merged and the M-dwarf component lies in the red wing of the merged line profile. As we do not see any red wing enhancement but only a line core enhancement we can assign this flare to the K-dwarf component.\\
For flare no.~2, when the impulsive phase occurs one can see both components separating with the K-dwarf component on the blue and the M-dwarf component on the red side (see Fig.~\ref{fig:flaretimeseries}). During the impulsive phase the M-dwarf component shows extra emission as well as an extra emission occurring on the blue side of the K- and M-dwarf profile. In the peak as well as in the gradual flare phase the K-dwarf component is enhanced. \\
For flare no.~3 one can see both components clearly separated in the flare phase with the K-dwarf component on the red side and the M-dwarf component on the blue side. The K-dwarf component shows excess emissions whereas the M-dwarf component stays without any excess emissions.\\ 
For flare no.~4 the M-dwarf component starts to merge with the K-dwarf component (from the blue side) but stays during the impulsive flare phase on the blue side of the merged profile. As the flare evolves further to the peak, the K-dwarf shows clear excess emission. If the M-dwarf reveals also excess emission can not be excluded but certainly not as strong as the K-dwarf. As the flare decays we still see an enhanced peak of the merged H$\alpha$ profile which is clearly dominated by the K-dwarf. The M-dwarf component has moved already closer into the K-dwarf component profile but still staying in the blue wing of the merged profile.\\ 
For flare no.~5 we see a separated H$\alpha$ profile with the K-dwarf component at the red and the M-dwarf component on the blue side. The flare has only a rise phase and during this phase one can see only excess emission originating from the K-dwarf component.\\
Flare no.~6 also shows only a rise phase at the end of the H$\alpha$ light curve consisting of two spectra only. The H$\alpha$ profile at that time is merged with the M-dwarf component being located at the red side of the merged profile (M-dwarf component moving from red to blue). Both spectra show the line core enhanced and in one spectrum also a weak enhancement in the spectral region where the M-dwarf component is.\\  
The flare evolution of flare no.~7 was also not fully captured. Here we possibly see the peak only, with two preceeding and one successive spectrum. The H$\alpha$ profile is merged with the M-dwarf component being located at the red side. Only the peak spectrum shows a clear excess emission appearing at the line core (K-dwarf) but also a weaker excess emission at the spectral location where the M-dwarf component is.\\ 
Flare no.~8 shows a separated profile during the impulsive as well as during the gradual flare phase. The K-dwarf component appears on the blue side and the M-dwarf component on the red side in the spectral profile. We clearly see excess emission originating from the K-dwarf component only.\\ 
Flare no.~9 shows a separated H$\alpha$ profile at the beginning of the light curve. As time evolves the M-dwarf component being located at the red side of the K-dwarf component starts to merge with the K-component. During the impulsive flare phase both profiles have already started to merge. One can still clearly see an excess emission of the M-dwarf but one spectrum later also excess emission from the K-dwarf is visible.\\
For the ''`weak flares``we continue with flare no.~10. During this flare the H$\alpha$ profile is separated. The K-dwarf component is located on the red and the M-dwarf component on the blue side. We see very subtle excess emission originating from both components.\\
Flare no.~11 also shows both components well separated,  with the K-dwarf component on the red and the M-dwarf component on the blue side. One can see clear excess emission originating from the K-dwarf. \\
For flare no.~12 we have captured a slow rise phase only. The H$\alpha$ line profile is merged, but starts to separate during the rise phase (M-dwarf component moving to the red). Excess emission is evident in the K-dwarf component as a well as a distinct additional blue extra emission occurring during the last six spectra which are not related to the M-dwarf component.\\
For flare no. 13 we have captured only the first two spectra of the impulsive phase. During the two spectra the H$\alpha$ profile is separated  with the K-dwarf component on the blue and the M-dwarf component on the red side. One can see that the excess emission causing the rise in flux originates from the K-star component only.\\ 
For flare no.~14 the H$\alpha$ profile is separated,  with the K-dwarf component on the blue and the M-dwarf component on the red side. During the impulsive flare phase we see excess emission from both stellar components.\\
Also for flare no.~15 the H$\alpha$ profile is separated,  with the K-dwarf component on the red and the M-dwarf component on the blue side. We see excess emission from both stellar components during the impulsive flare phase.\\
For flare no.~16 the spectral profile is well separated during the impulsive flare phase  with the K-dwarf component on the blue and the M-dwarf component on the red side. Excess emission is visible in both components.\\ 
For flare no.~17 we have captured a part of the impulsive flare phase only, with the K-dwarf component on the red and the M-dwarf component on the blue side. During the impulsive flare phase the H$\alpha$ profile is separated and we see excess emission  from the M-dwarf component only. \\
For flare no.~18 the H$\alpha$ profile is merged but starts to separate during the impulsive phase of the flare (M-dwarf component moves to the blue). We see excess emission from the K-dwarf component. \\
For flare no~19 the H$\alpha$ profile is separated during the impulsive phase but starts to merge during the gradual phase (M-dwarf component moves to the red). During the impulsive flare phase we see extra emission from the K-dwarf component. Possible weak contribution from the M-dwarf component may be existent.\\
For flare no.~20  the H$\alpha$ profile is separated,  with the K-dwarf component on the blue and the M-dwarf component on the red side. We see subtle excess emission from the K-dwarf component.\\
For flare no.~21 the H$\alpha$ both profiles are merged, with the M-dwarf component being located in the red wing of the merged profile. The flare shows a long impulsive phase during which the M-dwarf component shifts into the  K-dwarf component (from red to blue). We see excess emission in the core of the merged H$\alpha$ profile which we attribute to the K-dwarf component. As the position of the M-dwarf component is in the red wing of the merged profile it is not possible to completely exclude a contribution from the M-dwarf component in this case.\\
For flare no.~22 the H$\alpha$ profile is merged, but starts to separate (M-dwarf component moving from blue to red) during the impulsive flare phase. For this flare we can not evaluate from which component the flare may stem from.\\
For the ``gradual phase only'' flares we continue with flare no.~23 which shows a merged H$\alpha$ spectrum during the gradual flare phase. We see excess emission in the core of the merged profile. In this case we can not distinguish if the flare originates from the M- or K-dwarf component.\\
For flare no.~24 the H$\alpha$ profile is separated with the K-dwarf component on the blue and the M-dwarf component on the red side. We see excess emission from the K-dwarf component only. \\
For flare no.~25 the H$\alpha$ profile is merging (M-dwarf from blue to red). Excess emission comes from the merged line core and the blue wing of the merged profile. Therefore we suggest that this flare was originating from both components.\\ 
For flare no.~26 the H$\alpha$ profile is separating with the K-dwarf component on the blue and the M-dwarf component on the red side. Excess emission is seen in the K-dwarf component.\\
For flare.~27 the H$\alpha$ profile is separated with the K-dwarf component on the red and the M-dwarf component on the blue side. Here excess emission is originating from the M-dwarf.\\
For flare.~28 the H$\alpha$ profile is separated with the K-dwarf component on the blue and the M-dwarf component on the red side. Excess emission is originating from the K-dwarf. A minor contribution from the M-dwarf component can not be excluded.\\
For flare no.~29 the H$\alpha$ profile is again separated, with the K-dwarf component on the blue and the M-dwarf component on the red side. Excess emission is also originating from the K-dwarf.\\
For flare no.~30 the H$\alpha$ profile is again separated, with the K-dwarf component on the red and the M-dwarf component on the blue side. Excess emission is originating again from the K-dwarf.\\ 
For the last flare in the sample, which is flare no.~31, we see a merged H$\alpha$ profile and the M-dwarf component is located in the blue wing (moving from blue to red) of the merged profile. We see excess emission originating from the blue wing of the merged profile. In this case we can not assign the excess emission to a component as we also see weaker excess emission also from the K-dwarf.\\

\section{Flares and their asymmetries}
\label{asymap}
In the following we describe every asymmetry in the Balmer lines we have wittnessed in the analysis:\\
\textbf{Flare no.~1 (2023-10-27)}: During this flare we detect an enhancement in the blue wing of H$\alpha$ in the impulsive phase of the flare. The enhancement lasts until one spectrum after the peak. The mean projected bulk velocity is -115.1$\pm$35.2 and the projected maximum velocity is -167.0~km~s$^{-1}$. In H$\beta$ and  H$\gamma$ also blue wing enhancements are evident but only in two spectra each. We can not comment on the red wing of H$\alpha$ during the flare as the M-star component is located on the red side of the K-star component. \\
\textbf{Flare no.~2 (2023-11-08)}: This flare shows also a blue wing enhancement during the impulsive flare phase. The blue wing enhancement starts in the first spectrum of the impulsive phase and lasts until four spectra after the peak. The total duration of the blue wing enhancement is 51~minutes. The projected bulk velocity is -86.2$\pm$9.4 and the projected maximum velocity is -161.2~km~s$^{-1}$. H$\beta$ and H$\gamma$ show asymmetries also after the peak as well. Also here we can not comment on the red side of H$\alpha$ as the M-star component is located on the red side of the K-star component.\\
\textbf{Flare no.~3 (2023-11-12)}: This flare exhibits marginal red wing enhancements starting one spectrum before the flare peak and lasting until one spectrum after the flare peak. We deduce a bulk velocity of 148.8$\pm$6.9 and a projected maximum velocity of 160.2~km~s$^{-1}$. In H$\beta$ we also see those subtle red wing enhancements, but not in H$\gamma$.\\
\textbf{Flare no.~4 (2023-11-14)}: This flare exhibits during its impulsive phase a red wing enhancement which increases in velocity during the gradual phase and lasts until the last spectrum of the series. In H$\alpha$ the projected bulk velocity is 137.5$\pm$14.7 and the projected maximum velocity is 279.5~km~s$^{-1}$. H$\beta$ and also H$\gamma$ show a similar behaviour.\\
\textbf{Flare no.~7 (2023-12-06)}: This flare reveals a fast blue wing enhancement in one spectrum of the impulsive phase. The projected bulk velocity is -213.5$\pm$49.3 and the maximum velocity reaches -842.1~km~s$^{-1}$. Also in H$\beta$ and H$\gamma$ these blue-wing enhancements can be seen.\\
\textbf{Flare no.~8 (2023-12-11)}: In this flare, blue wing enhancements are evident in two spectra during the impulsive phase. We deduce projected bulk and maximum velocities of -143.9$\pm$50.0 and -213.3~km~s$^{-1}$, respectively. In H$\beta$ and H$\gamma$ those are hardly visible.\\
\textbf{Flare no.~9 (2024-01-10)}: This flare reveals a blue wing enhancement in one spectrum during the impulsive phase too subtle to be fitted. A similar enhancement is visible in H$\gamma$ but not in H$\beta$.\\
\textbf{Flare no.~11 (2023-11-06)}: Starting from the flare peak and lasting few spectra into the gradual phase, subtle red wing enhancements can be seen. The enhancements can not be fitted with gaussian functions to determine their projected velocities. These subtle enhancements are also visible in H$\beta$ and H$\gamma$.\\
\textbf{Flare no.~12 (2023-11-19)}: In this spectral time series we see a rise in H$\alpha$ flux until the end of the time series lasting for 57~minutes. This rise is partly caused by blue wing enhancements which increase towards the end of the time series. The projected bulk and maximum velocities are -74.7$\pm$3.7 and -178.6~km~s$^{-1}$, respectively. Also in H$\beta$ we see blue wing enhancements but not that distinct as in H$\alpha$.\\
\textbf{Flare no.~14 (2023-11-24)}: In the impulsive phase of this flare subtle H$\alpha$ blue wing enhancements are seen in two spectra. The enhancements can not be fitted with gaussian functions to determine their projected velocities. These subtle enhancements are also visible in H$\gamma$ but not in H$\beta$.\\
\textbf{Flare no.~15 (2023-11-28)}: Subtle red wing enhancements are seen in the peak of this flare. The enhancements can not be fitted with gaussian functions to determine their projected velocities. These subtle enhancements are also visible in H$\beta$ but not in H$\gamma$.\\
\textbf{Flare no.~17 (2023-12-05)}: In this spectral time series we see a rise in H$\alpha$ flux until the end of the time series. Here the line core of the well separated M-star component is enhanced as well as the blue wing of the K-star component. One can also see a subtle excess emission in the K-star component. If the asymmetry belongs to the M-star component may be reasonable and the excess emisson in the K-star component might be also caused by the blue asymmetry. If so then the projected velocity which we determined is higher. We deduced projected bulk and maximum velocities of -74.7$\pm$13.7 and -143.9~km~s$^{-1}$ (assuming that the blue wing asymmetry belongs to the K-star component ), respectively. The blue wing enhancements are not existent in H$\beta$ or H$\gamma$.\\
\textbf{Flare no.~18 (2023-12-21)}: In this flare we find red wing enhancements in the gradual phase of the flare with projected bulk and maximum velocities of 51.6$\pm$3.4 and 103.1~km~s$^{-1}$, respectively. Red wing enhancements in the gradual phase are also seen for H$\beta$.\\
\textbf{Flare no.~21 (2024-01-07)}: At the end of this spectral time series we see a rise until the end of the time series lasting for 29~minutes. During the rise phase we see subtle blue wing enhancements.  The enhancements can not be fitted with gaussian functions to determine their projected velocities. These subtle enhancements are also visible in H$\beta$ and H$\gamma$.\\
\textbf{Flare no.~22 (2024-01-08)}: During the impulsive phase we see subtle blue wing enhancements too weak to be fitted. Also in H$\beta$ those are visible but not in H$\gamma$. \\
\textbf{Flare no.~25 (2023-11-22)}: Red wing enhancements in the gradual phase of this flare with bulk and maximum velocities of 165.9$pm$6.6 and 228.5~km~s$^{-1}$, respectively, are seen. Those are not detectable in H$\beta$ or H$\gamma$.\\
\textbf{Flare no.~30 (2023-01-06)}: Red wing enhancements in the gradual phase of this flare with bulk and maximum velocities of 57.3$\pm$5.8 and 137.3~km~s$^{-1}$, respectively, are seen. Those are seen in H$\beta$, but not in H$\gamma$.\\
\textbf{Flare no.~31 (2023-01-11)}: Blue wing enhancements in the impulsive phase of this flare are detected with bulk and maximum velocities of -120.9$\pm$11.0 and -190.3~km~s$^{-1}$. Those are seen also in H$\beta$ and H$\gamma$.\\

\begin{figure*}
 \includegraphics[width=17cm]{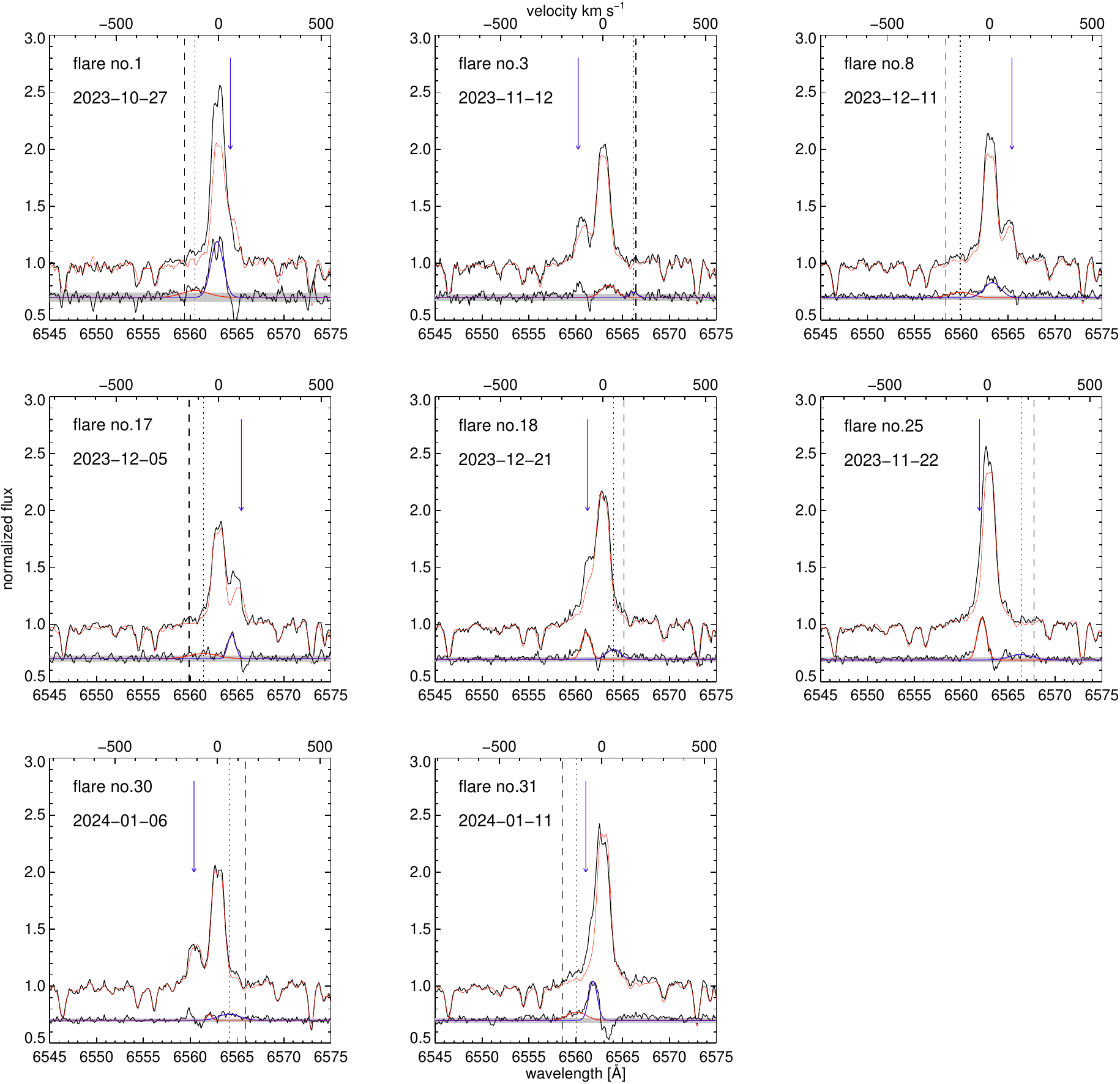}
  \caption{H$\alpha$ spectral line asymmetries detected in flares 1, 3, 8, 17, 18, 25, 30, and 31. In every subpanel we plot the spectrum showing the largest bulk velocity per night (solid black line). Overplotted is a pre- or post-flare spectrum (red solid line). Additionally we plot also in every subpanel below the actual spectrum the residual spectrum (shifted to a value of 0.7). The residual spectrum is overplotted with two gaussians accounting for the blue-or red asymmetries. Furthermore we have indicated the Doppler shift caused by the bulk velocities with a vertical dotted line \textbf{and the one caused by the maximum velocity with a vertical dashed line}. Also overplotted is a blue arrow indicating the position of the M-star component. }
 \label{fig:asymapp}
\end{figure*}

\begin{figure*}
 \includegraphics[width=7cm]{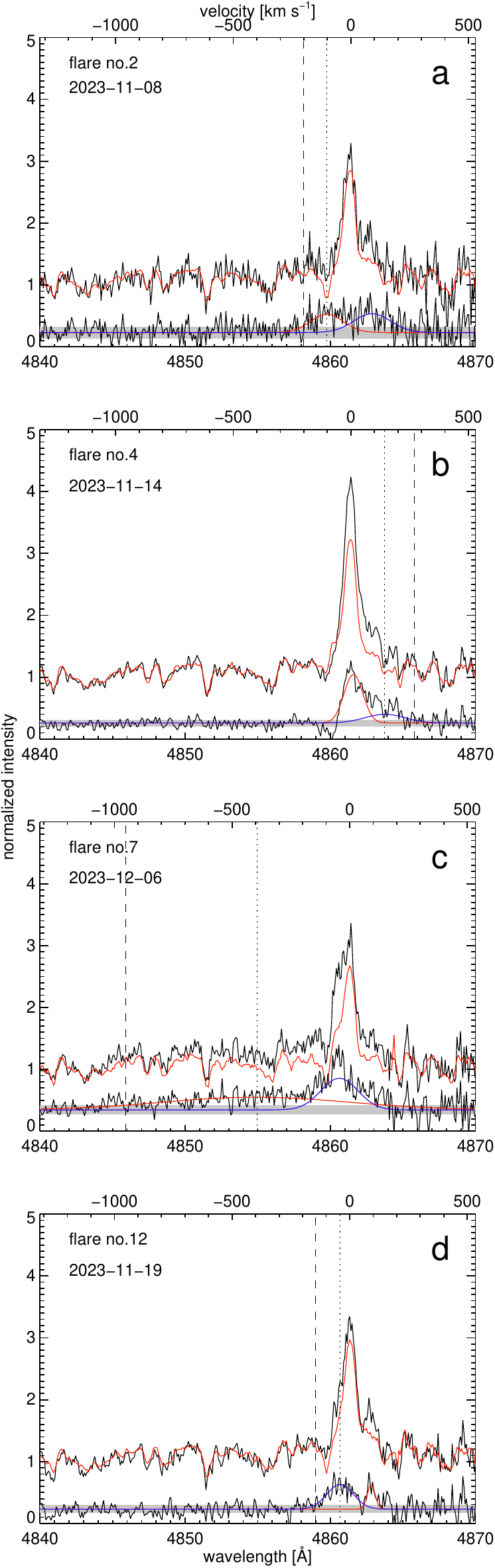}
  \caption{H$\beta$ spectral line asymmetries detected in flares 2 (a), 4 (b), 7 (c), and 12 (d, same flares as shown in in Fig.~\ref{fig:allasym}). In every subpanel we plot the spectrum showing the largest bulk velocity per night (solid black line). Overplotted is a pre- or post-flare spectrum (red solid line). Additionally we plot also in every subpanel below the actual spectrum the residual spectrum (shifted to a value of 0.25). The residual spectrum is overplotted with two gaussians accounting for the blue-or red asymmetries. Furthermore we have indicated the Doppler shift caused by the bulk velocities with a vertical dotted line and the one caused by the maximum velocity with a vertical dashed line. }
 \label{fig:asymapphbeta}
\end{figure*}

\begin{table*}
	\centering
	\caption{Flare events during which H$\alpha$ blue- or red asymmetries have been detected. The first column gives the flare number, the second the night in which the flare occurred, the third column the spectral line on which wing the asymmetry has been detected, the fourth column the start of the asymmetry in units of JD-2460000, the fifth column the end of the asymmetry, the sixth column the duration of the asymmetry in minutes, the seventh column the bulk velocity of the asymmetry, i.e. the projected velocity determined from the center of the fitted gaussian, and finally in the eighth column the maximum velocity, i.e. the projected velocity where the fitted gaussian merges with the background ($>$5\% above background). If we can not fit a gaussian, as the asymmetry is too subtle but obvious by eye we write ``+'' for a red asymmetry and ``-'' for a blue asymmetry.}
	\label{tab:C1}
	\begin{tabular}{lcccccccccc}
		\hline
 Flare   &   night      &   asymmetry start  &    asymmetry stop    &    t$_{\mathrm{asym}}$     &     v$_{\mathrm{bulk}}$    &      v$_{\mathrm{max}}$       \\
         &              &                    &                      &                            &                            &                                \\
   no.   &    [UT]      &    [JD-2460000]    &     [JD-2460000]     &          [min ]            &         [km~s$^{-1}$]       &          [km~s$^{-1}$]          \\ 
  \hline      
 
  distinct flares &&&&&&&&&& \\

  \hline

1  & 2023-10-27 & 244.80928 & 244.82130 & 17.3    & -115.1$\pm$35.2  &  -167.0   \\ 
2  & 2023-11-08 & 256.74415 & 256.77976 & 51.3    & -86.2$\pm$9.4    &  -161.2   \\ 
3  & 2023-11-12 & 260.79404 & 260.79800 &  5.7    &  148.8$\pm$6.9   &   160.2   \\ 
4  & 2023-11-14 & 262.70875 & 262.83929 & 188.0   &  137.5$\pm$14.7  &   279.5   \\ 
5  & 2023-11-17 &     0     &     0     &   0     &        0         &      0    \\ 
6  & 2023-12-02 &     0     &     0     &   0     &        0         &      0    \\ 
7  & 2023-12-06 & 284.60098 & 284.60494 &   5.7   &  -213.5$\pm$49.3 &  -842.1   \\ 
8  & 2023-12-11 & 289.64109 & 289.64109 &   5.0   &  -143.9$\pm$50.0     &  -213.3   \\ 
9  & 2024-01-10 & 319.67123 & 319.67914 &   11.4  &        -         &     -     \\ 

  \hline 
   
  weak flares &&&&&&&&&& \\
  \hline
10 & 2023-10-26 &     0     &     0     &     0   &        0         &    0      \\ 
11 & 2023-11-06 & 254.63309 & 254.64511 &   17.3  &        +         &    +      \\ 
12 & 2023-11-19 & 267.70634 & 267.74588 &   56.9  &    -74.7$\pm$3.7 &  -178.6   \\   
13 & 2023-11-23 &     0     &     0     &     0   &        0         &    0      \\   
14 & 2023-11-24 & 272.64322 & 272.65580 &   18.1  &        -         &    -      \\   
15 & 2023-11-28 & 276.64009 & 276.64803 &   11.4  &        +         &    +      \\   
16 & 2023-12-02 &     0     &     0     &     0   &        0         &    0      \\   
17 & 2023-12-05 & 283.64906 & 283.66092 &   17.1  &   -74.7$\pm$13.7 &   -143.9  \\    
18 & 2023-12-21 & 299.71965 & 299.72361 &    5.7  &     51.6$\pm$3.4 &   103.1   \\   
19 & 2023-12-23 &     0     &     0     &     0   &        0         &    0      \\   
20 & 2024-01-05 &     0     &     0     &     0   &        0         &    0      \\   
21 & 2024-01-07 & 316.69151 & 316.71153 &    28.8 &  - & -\\   
22 & 2024-01-08 & 317.62952 & 317.63303 &    15.1 &  - & -\\   

   \hline
  decaying tail &&&&&&&&&& \\
  \hline
23 & 2023-10-28 &     0     &     0     &     0   &        0         &     0     \\ 
24 & 2023-11-13 &     0     &     0     &     0   &        0         &     0     \\ 
25 & 2023-11-22 & 270.56510 & 270.61256 &   68.3  &    165.9$\pm$6.6 &    228.5  \\  
26 & 2023-11-30 &     0     &     0     &     0   &        0         &     0     \\
27 & 2023-12-12 &     0     &     0     &     0   &        0         &     0     \\
28 & 2023-12-16 &     0     &     0     &     0   &        0         &     0     \\
29 & 2023-12-19 &     0     &     0     &     0   &        0         &     0     \\
30 & 2024-01-06 & 315.58630 & 315.60231 &   23.1  &    57.3$\pm$5.8  &   137.3   \\ 
31 & 2024-01-11 & 320.54530 & 320.54927 &   5.7   &  -120.9$\pm$11.0 &  -190.3  \\ 
  \hline

	\end{tabular}
\end{table*}

\clearpage

\section{Flares and their spectral line emissions}

\begin{landscape}
\begin{table}
\centering
\tiny
\caption{Line fluxes of spectral lines which showed excess emission during the distinctive flares from table~\ref{tab:B1}. Shown are impulsive and gradual phases, where available.}
\label{tab:D1}
\begin{tabular}{lcccccccccccccccc}
		\hline

	&	2023-10-27	&	2023-10-27	&	2023-11-08	&	2023-11-08	&	2023-11-12	&	2023-11-12	&	2023-11-14	&	2023-11-14	&	2023-11-17	&	2023-12-02	&	2023-12-06	&	2023-12-06	&	2023-12-11	&	2023-12-11	&	2024-01-10	&	2024-01-10	\\
	&	impulsive	&	gradual	&	impulsive	&	gradual	&	impulsive	&	gradual	&	impulsive	&	gradual	&	impulsive	&	impulsive	&	impulsive	&	gradual	&	impulsive	&	gradual	&	impulsive	&	gradual	\\
    &	flare 1	&	flare 1	&	flare 2	&	flare 2	&	flare 3	&	flare 3	&	flare 4	&	flare 4	&	flare 5	&	flare 6	&	flare 7	&	flare 7	&	flare 8	&	flare 8	&	flare 9	&	flare 9	\\
    \hline
CaI7148.15	&	-	&	-	&	-	&	1.47$\pm$0.44	&	-	&	-	&	0.29$\pm$0.16	&	0.36$\pm$0.48	&	-	&	-	&	0.60$\pm$0.26	&	0.55$\pm$0.25	&	-	&	-	&	0.00$\pm$0.29	&	-	\\
HeI6678.15	&	0.68$\pm$0.16	&	0.84$\pm$0.34	&	0.70$\pm$0.25	&	-	&	-	&	-	&	1.49$\pm$0.16	&	0.90$\pm$0.19	&	-	&	0.54$\pm$0.15	&	1.27$\pm$0.23	&	0.46$\pm$0.23	&	-	&	-	&	-	&	-	\\
H6562.79	&	8.83$\pm$0.48	&	8.26$\pm$0.79	&	7.86$\pm$0.62	&	2.15$\pm$0.79	&	3.61$\pm$0.35	&	1.64$\pm$0.27	&	14.79$\pm$0.28	&	21.15$\pm$0.43	&	5.00$\pm$0.26	&	6.64$\pm$0.56	&	23.98$\pm$0.68	&	18.83$\pm$0.55	&	5.22$\pm$0.63	&	1.51$\pm$0.49	&	4.92$\pm$0.43	&	5.44$\pm$0.26	\\
NaDI5895.92	&	0.56$\pm$0.17	&	0.80$\pm$0.39	&	1.81$\pm$0.22	&	0.69$\pm$0.18	&	0.77$\pm$0.16	&	0.48$\pm$0.11	&	2.13$\pm$0.12	&	1.11$\pm$0.17	&	0.76$\pm$0.13	&	0.83$\pm$0.16	&	0.83$\pm$0.27	&	-	&	1.17$\pm$0.23	&	-	&	0.26$\pm$0.15	&	0.21$\pm$0.13	\\
NaDII5889.95	&	0.69$\pm$0.17	&	1.18$\pm$0.39	&	2.19$\pm$0.22	&	0.92$\pm$0.18	&	0.82$\pm$0.16	&	0.71$\pm$0.11	&	2.44$\pm$0.12	&	1.37$\pm$0.17	&	0.93$\pm$0.13	&	0.98$\pm$0.16	&	1.02$\pm$0.27	&	-	&	1.14$\pm$0.23	&	-	&	0.32$\pm$0.15	&	0.45$\pm$0.13	\\
HeI5876.65	&	0.52$\pm$0.17	&	0.29$\pm$0.39	&	5.15$\pm$0.22	&	-	&	-	&	-	&	3.97$\pm$0.12	&	3.99$\pm$0.16	&	2.53$\pm$0.12	&	0.69$\pm$0.16	&	2.82$\pm$0.33	&	-	&	0.12$\pm$0.23	&	-	&	0.77$\pm$0.15	&	0.92$\pm$0.13	\\
MgI5528.40	&	-	&	-	&	0.22$\pm$0.17	&	-	&	-	&	-	&	0.28$\pm$0.10	&	-	&	-	&	0.16$\pm$0.15	&	-	&	-	&	0.08$\pm$0.20	&	-	&	-	&	-	\\
FeI5455.61	&	-	&	-	&	0.54$\pm$0.14	&	-	&	-	&	-	&	0.40$\pm$0.10	&	-	&	-	&	0.19$\pm$0.12	&	-	&	-	&	0.19$\pm$0.15	&	-	&	0.09$\pm$0.08	&	-	\\
FeI5446.92	&	-	&	-	&	0.36$\pm$0.15	&	-	&	-	&	-	&	0.47$\pm$0.10	&	-	&	-	&	0.19$\pm$0.12	&	-	&	-	&	0.26$\pm$0.15	&	-	&	-	&	-	\\
FeI5429.70	&	-	&	-	&	0.72$\pm$0.13	&	-	&	-	&	-	&	0.47$\pm$0.09	&	-	&	-	&	0.06$\pm$0.10	&	-	&	-	&	-	&	-	&	-	&	-	\\
CrI5409.79	&	-	&	-	&	-	&	-	&	-	&	-	&	0.18$\pm$0.09	&	-	&	-	&	-	&	-	&	-	&	-	&	-	&	-	&	-	\\
FeI5405.77	&	-	&	-	&	0.31$\pm$0.14	&	-	&	-	&	-	&	0.34$\pm$0.10	&	-	&	-	&	0.25$\pm$0.11	&	-	&	-	&	-	&	-	&	-	&	-	\\
FeI5397.13	&	-	&	-	&	0.41$\pm$0.15	&	-	&	-	&	-	&	0.49$\pm$0.11	&	-	&	-	&	0.37$\pm$0.12	&	-	&	-	&	-	&	-	&	-	&	-	\\
FeI5371.87	&	-	&	-	&	0.52$\pm$0.12	&	-	&	-	&	-	&	0.50$\pm$0.09	&	-	&	-	&	0.23$\pm$0.10	&	-	&	-	&	-	&	-	&	-	&	-	\\
FeI5328.04	&	-	&	-	&	1.07$\pm$0.19	&	0.41$\pm$0.14	&	-	&	-	&	0.92$\pm$0.15	&	-	&	-	&	0.48$\pm$0.14	&	-	&	-	&	-	&	-	&	-	&	-	\\
FeII5316.61	&	-	&	-	&	0.28$\pm$0.17	&	-	&	-	&	-	&	0.82$\pm$0.14	&	0.53$\pm$0.11	&	-	&	0.07$\pm$0.13	&	0.24$\pm$0.18	&	-	&	0.01$\pm$0.17	&	-	&	-	&	-	\\
FeII5275.99	&	-	&	-	&	0.33$\pm$0.19	&	-	&	-	&	-	&	0.65$\pm$0.17	&	0.24$\pm$0.13	&	-	&	-	&	-	&	-	&	0.19$\pm$0.20	&	-	&	-	&	-	\\
FeI5269.54	&	-	&	-	&	1.02$\pm$0.22	&	0.28$\pm$0.16	&	-	&	-	&	1.26$\pm$0.19	&	0.40$\pm$0.15	&	-	&	0.57$\pm$0.17	&	-	&	-	&	-	&	-	&	-	&	-	\\
FeII5234.62	&	-	&	-	&	-	&	-	&	-	&	-	&	0.42$\pm$0.14	&	-	&	-	&	-	&	-	&	-	&	-	&	-	&	-	&	-	\\
FeI5227.19	&	-	&	-	&	0.69$\pm$0.13	&	-	&	-	&	-	&	0.90$\pm$0.12	&	-	&	-	&	0.27$\pm$0.10	&	-	&	-	&	0.30$\pm$0.13	&	-	&	0.00$\pm$0.08	&	-	\\
CrI5208.44	&	-	&	-	&	0.66$\pm$0.24	&	-	&	-	&	-	&	0.53$\pm$0.28	&	0.18$\pm$0.16	&	-	&	0.21$\pm$0.16	&	-	&	-	&	-	&	-	&	-	&	-	\\
MgI5183.60	&	0.27$\pm$0.12	&	-	&	0.90$\pm$0.18	&	-	&	0.26$\pm$0.10	&	-	&	1.06$\pm$0.21	&	0.39$\pm$0.12	&	0.23$\pm$0.08	&	0.59$\pm$0.12	&	0.22$\pm$0.17	&	0.19$\pm$0.15	&	0.50$\pm$0.15	&	-	&	0.08$\pm$0.09	&	0.06$\pm$0.07	\\
MgI5172.68	&	0.34$\pm$0.15	&	-	&	1.03$\pm$0.21	&	0.19$\pm$0.12	&	0.24$\pm$0.12	&	-	&	1.46$\pm$0.24	&	0.63$\pm$0.14	&	0.29$\pm$0.09	&	0.45$\pm$0.14	&	0.49$\pm$0.20	&	0.18$\pm$0.18	&	0.57$\pm$0.18	&	-	&	0.30$\pm$0.10	&	0.25$\pm$0.08	\\
FeII5169.03	&	0.24$\pm$0.10	&	-	&	0.70$\pm$0.14	&	0.27$\pm$0.08	&	-	&	0.21$\pm$0.05	&	1.40$\pm$0.16	&	0.73$\pm$0.09	&	0.14$\pm$0.06	&	0.36$\pm$0.09	&	0.06$\pm$0.13	&	0.36$\pm$0.12	&	0.27$\pm$0.12	&	-	&	0.32$\pm$0.07	&	0.20$\pm$0.06	\\
MgI5167.32	&	0.50$\pm$0.10	&	-	&	0.97$\pm$0.15	&	0.44$\pm$0.09	&	-	&	-	&	1.08$\pm$0.17	&	0.44$\pm$0.10	&	0.56$\pm$0.07	&	0.19$\pm$0.10	&	0.44$\pm$0.14	&	0.75$\pm$0.12	&	0.36$\pm$0.13	&	-	&	-	&	-	\\
FeII5018.43	&	0.38$\pm$0.16	&	-	&	0.59$\pm$0.16	&	0.23$\pm$0.13	&	0.20$\pm$0.13	&	-	&	1.03$\pm$0.14	&	0.53$\pm$0.11	&	-	&	0.23$\pm$0.13	&	0.38$\pm$0.20	&	-	&	-	&	-	&	0.22$\pm$0.10	&	-	\\
HeI5015.68	&	-	&	-	&	0.52$\pm$0.16	&	-	&	0.38$\pm$0.13	&	-	&	0.63$\pm$0.13	&	0.32$\pm$0.11	&	-	&	-	&	-	&	-	&	0.04$\pm$0.17	&	-	&	0.06$\pm$0.10	&	-	\\
FeI4957.30	&	-	&	-	&	0.81$\pm$0.16	&	-	&	-	&	-	&	0.54$\pm$0.13	&	-	&	-	&	0.44$\pm$0.12	&	-	&	-	&	0.33$\pm$0.16	&	-	&	0.22$\pm$0.09	&	-	\\
FeII4923.92	&	0.17$\pm$0.11	&	-	&	0.40$\pm$0.12	&	0.34$\pm$0.08	&	-	&	-	&	1.19$\pm$0.10	&	0.88$\pm$0.08	&	-	&	0.29$\pm$0.09	&	-	&	-	&	0.08$\pm$0.12	&	-	&	-	&	-	\\
FeI4891.50	&	-	&	-	&	0.10$\pm$0.08	&	-	&	-	&	-	&	0.17$\pm$0.04	&	-	&	-	&	-	&	-	&	-	&	-	&	-	&	-	&	-	\\
FeI4890.76	&	-	&	-	&	0.26$\pm$0.08	&	-	&	-	&	-	&	0.19$\pm$0.04	&	-	&	-	&	-	&	-	&	-	&	-	&	-	&	-	&	-	\\
H4861.35	&	9.73$\pm$0.27	&	10.35$\pm$0.60	&	18.36$\pm$0.30	&	8.09$\pm$0.23	&	4.57$\pm$0.24	&	2.17$\pm$0.17	&	21.41$\pm$0.17	&	14.11$\pm$0.17	&	7.95$\pm$0.17	&	8.42$\pm$0.28	&	16.91$\pm$0.41	&	8.85$\pm$0.37	&	5.46$\pm$0.35	&	-	&	6.97$\pm$0.21	&	4.84$\pm$0.16	\\
MnI4783.42	&	-	&	-	&	-	&	-	&	-	&	-	&	0.30$\pm$0.09	&	-	&	-	&	-	&	-	&	-	&	-	&	-	&	-	&	-	\\
MgI4702.98	&	-	&	-	&	0.28$\pm$0.15	&	-	&	-	&	-	&	0.21$\pm$0.09	&	-	&	-	&	-	&	-	&	-	&	-	&	-	&	0.26$\pm$0.10	&	-	\\
HeII4686.68	&	-	&	-	&	0.62$\pm$0.17	&	-	&	-	&	0.50$\pm$0.11	&	0.65$\pm$0.10	&	0.45$\pm$0.10	&	-	&	0.22$\pm$0.14	&	0.80$\pm$0.24	&	-	&	-	&	-	&	-	&	-	\\
TiI4629.34	&	-	&	-	&	-	&	-	&	-	&	-	&	0.48$\pm$0.10	&	0.37$\pm$0.10	&	-	&	-	&	-	&	-	&	-	&	-	&	-	&	-	\\
FeII4583.83	&	-	&	-	&	-	&	-	&	-	&	-	&	0.98$\pm$0.17	&	0.75$\pm$0.15	&	-	&	-	&	-	&	-	&	-	&	-	&	-	&	-	\\
FeII4555.8	&	-	&	-	&	-	&	-	&	-	&	-	&	0.54$\pm$0.17	&	0.18$\pm$0.14	&	-	&	-	&	-	&	-	&	-	&	-	&	-	&	-	\\
FeII4549.47	&	-	&	-	&	0.50$\pm$0.15	&	-	&	-	&	-	&	0.69$\pm$0.14	&	0.32$\pm$0.11	&	-	&	-	&	-	&	-	&	-	&	-	&	-	&	-	\\
TiI4522.80	&	-	&	-	&	0.23$\pm$0.15	&	-	&	-	&	-	&	0.34$\pm$0.10	&	0.16$\pm$0.08	&	-	&	-	&	0.03$\pm$0.19	&	-	&	0.52$\pm$0.16	&	-	&	-	&	-	\\
FeI4482.17	&	-	&	-	&	0.30$\pm$0.18	&	-	&	-	&	-	&	0.48$\pm$0.13	&	0.26$\pm$0.09	&	-	&	-	&	-	&	-	&	0.34$\pm$0.18	&	-	&	0.23$\pm$0.11	&	-	\\
MgII4481.33	&	-	&	-	&	0.57$\pm$0.19	&	-	&	-	&	-	&	0.55$\pm$0.13	&	-	&	-	&	-	&	-	&	-	&	-	&	-	&	-	&	-	\\
HeI4471.48	&	0.63$\pm$0.24	&	-	&	1.13$\pm$0.24	&	-	&	-	&	-	&	1.32$\pm$0.17	&	0.64$\pm$0.12	&	-	&	-	&	-	&	-	&	-	&	-	&	0.26$\pm$0.15	&	-	\\
FeI4459.12	&	-	&	-	&	-	&	-	&	-	&	-	&	0.70$\pm$0.18	&	-	&	-	&	-	&	-	&	-	&	-	&	-	&	-	&	-	\\
CaI4456.61	&	-	&	-	&	-	&	-	&	-	&	-	&	0.39$\pm$0.14	&	-	&	-	&	-	&	-	&	-	&	-	&	-	&	-	&	-	\\
TiII4450.49	&	-	&	-	&	-	&	-	&	-	&	-	&	0.10$\pm$0.12	&	0.11$\pm$0.11	&	-	&	-	&	-	&	-	&	-	&	-	&	-	&	-	\\
TiI4443.80	&	-	&	-	&	-	&	-	&	-	&	-	&	0.29$\pm$0.13	&	-	&	-	&	-	&	-	&	-	&	-	&	-	&	-	&	-	\\
FeI4427.31	&	-	&	-	&	0.15$\pm$0.26	&	-	&	-	&	-	&	0.32$\pm$0.13	&	0.15$\pm$0.12	&	-	&	-	&	-	&	-	&	-	&	-	&	-	&	-	\\
FeI4415.13	&	-	&	-	&	0.49$\pm$0.29	&	-	&	-	&	-	&	0.72$\pm$0.15	&	-	&	-	&	-	&	-	&	-	&	-	&	-	&	-	&	-	\\
FeI4405.75	&	-	&	-	&	-	&	-	&	-	&	-	&	0.41$\pm$0.11	&	-	&	-	&	-	&	-	&	-	&	-	&	-	&	-	&	-	\\
TiII4395.03	&	-	&	-	&	-	&	-	&	-	&	-	&	0.13$\pm$0.10	&	-	&	-	&	-	&	-	&	-	&	-	&	-	&	-	&	-	\\
FeI4384.68	&	-	&	-	&	-	&	-	&	-	&	-	&	0.19$\pm$0.06	&	-	&	-	&	-	&	-	&	-	&	-	&	-	&	-	&	-	\\
FeI4383.55	&	-	&	-	&	0.22$\pm$0.15	&	-	&	-	&	-	&	0.24$\pm$0.06	&	-	&	-	&	-	&	-	&	-	&	-	&	-	&	-	&	-	\\
FeI4375.93	&	-	&	-	&	-	&	-	&	-	&	-	&	0.16$\pm$0.06	&	-	&	-	&	-	&	-	&	-	&	-	&	-	&	-	&	-	\\
H4340.47	&	-	&	-	&	6.66$\pm$0.31	&	2.08$\pm$0.19	&	0.37$\pm$0.28	&	-	&	8.29$\pm$0.21	&	4.66$\pm$0.13	&	1.47$\pm$0.17	&	3.54$\pm$0.20	&	6.91$\pm$0.39	&	5.08$\pm$0.45	&	-	&	-	&	2.36$\pm$0.24	&	1.43$\pm$0.17	\\
FeII4303.17	&	-	&	-	&	-	&	-	&	-	&	-	&	0.29$\pm$0.27	&	-	&	-	&	-	&	-	&	-	&	-	&	-	&	-	&	-	\\
CrI4274.80	&	-	&	-	&	-	&	-	&	-	&	-	&	1.11$\pm$0.30	&	-	&	-	&	-	&	-	&	-	&	-	&	-	&	-	&	-	\\
FeI4271.65	&	-	&	-	&	-	&	-	&	-	&	-	&	1.33$\pm$0.22	&	-	&	-	&	-	&	-	&	-	&	-	&	-	&	-	&	-	\\
FeII4233.17	&	-	&	-	&	-	&	-	&	-	&	-	&	0.41$\pm$0.05	&	-	&	-	&	-	&	-	&	-	&	-	&	-	&	-	&	-	\\
CaI4226.73	&	-	&	-	&	-	&	-	&	-	&	-	&	0.48$\pm$0.08	&	-	&	-	&	-	&	-	&	-	&	-	&	-	&	-	&	-	\\
FeII4178.86	&	-	&	-	&	-	&	-	&	-	&	-	&	0.19$\pm$0.22	&	-	&	-	&	-	&	-	&	-	&	-	&	-	&	-	&	-	\\
FeII4173.45	&	-	&	-	&	-	&	-	&	-	&	-	&	0.59$\pm$0.38	&	-	&	-	&	-	&	-	&	-	&	-	&	-	&	-	&	-	\\
H4101.73	&	15.51$\pm$3.47	&	25.36$\pm$14.85	&	-	&	-	&	-	&	-	&	33.18$\pm$2.15	&	-	&	-	&	-	&	-	&	-	&	-	&	-	&	-	&	-	\\

\hline
\end{tabular}
\end{table}
\begin{scriptsize}
\end{scriptsize}
\end{landscape}

\clearpage

\begin{landscape}
\begin{table}
\centering
\tiny
\caption{Line fluxes of spectral lines which showed excess emission during the weak flares from table~\ref{tab:B1}. Shown are impulsive and gradual phases, where available.}
\label{tab:D2}
\begin{tabular}{lccccccccccccccccc}
		\hline

		&	2023-10-26	&	2023-10-26	&	2023-11-06	&	2023-11-06	&	2023-11-19	&	2023-11-23	&	2023-11-24	&	2023-11-24	&	2023-11-28	&	2023-11-28	&	2023-12-02	&	2023-12-02	&	2023-12-05	&	2023-12-21	&	2023-12-21	&	2023-12-23	&	2023-12-23 \\
	        &	impulsive	&	 gradual	&	impulsive	&	gradual	    &	impulsive	&	impulsive	&	impulsive	&	gradual	    &	impulsive	&	gradual	    &	impulsive	&	gradual	    &	impulsive	&	impulsive	&	gradual	&	impulsive	&	gradual \\
        	&	flare 10	&    flare 10	&	flare 11	&	flare 11	&	flare 12	&	flare 13	&	flare 14	&	flare 14	&	flare 15	&	flare 15	&	flare 16	&	flare 16	&	flare 17	&	flare 18	&  flare 18 &   flare 19    & flare 19 \\
\hline
CaI7148.15	&	     -   	&	     -    	&	     -   	&	     -    	&	     -   	&	    - 	    &      	 -   	&	     -	    &	    -	    &	    -	    &	     -      &	    -	    & 0.73$\pm$0.36	&	    -      	&	    -      	&	     -	    &	- \\
HeI6678.15	&	     -   	&	     -  	&	     -	    &	     -	    &	     -   	&	    -	    &	     -   	&	     -   	&	    -   	&	    -   	&        -   	&	    -    	& 0.34$\pm$0.15	& 0.32$\pm$0.13	& 0.27$\pm$0.11	&	     -	    &	- \\
H6563	    & 3.00$\pm$0.31	& 1.52$\pm$0.28	& 3.10$\pm$0.27	& 0.86$\pm$0.34	& 4.75$\pm$0.52	& 1.25$\pm$0.45	& 5.64$\pm$0.49	& 0.62$\pm$0.52	& 3.25$\pm$0.30	& 2.72$\pm$0.33	& 2.56$\pm$0.39	& 1.38$\pm$0.36	& 3.61$\pm$0.65	& 4.34$\pm$0.23	& 6.81$\pm$0.23	& 5.00$\pm$0.22	& 7.10$\pm$0.26 \\
NaDI5895.92	&	     -	    &	     -	    &	-	&	-	&	0.79$\pm$0.31	&	-	&	-	&	-	&	-	&	-	&	-	&	-	&	-	&	0.54$\pm$0.16	&	0.65$\pm$0.13	&	0.09$\pm$0.14	&	0.16$\pm$0.14 \\
NaDII5889.95 &	     -	    &	     -	    &	-	&	-	&	-	&	-	&	-	&	-	&	-	&	-	&	-	&	-	&	-	&	0.50$\pm$0.16	&	0.77$\pm$0.13	&	0.32$\pm$0.14	&	0.47$\pm$0.14 \\
HeI5876.65	&	     -	    &	     -	    &	-	&	-	&	-	&	-	&	-	&	-	&	-	&	-	&	-	&	-	&	0.65$\pm$0.19	&	0.95$\pm$0.16	&	0.73$\pm$0.13	&	0.30$\pm$0.14	&	0.74$\pm$0.14 \\
MgI5528.40	&	     -	    &	     -	    &	-	&	-	&	-	&	-	&	-	&	-	&	-	&	-	&	-   &	-	&	-	&	-	&	-	&	-	&	- \\
FeI5455.61	&	     -	    &	     -	    &	-	&	-	&	-	&	-	&	-	&	-	&	-	&	-	&	-	&	-	&	-	&	-	&	-	&	-	&	- \\
FeI5446.92	&	     -	    &	     -	    &	-	&	-	&	-	&	-	&	-	&	-	&	-	&	-	&	-	&	-	&	-	&	-	&	-	&	-	&	- \\ 
FeI5429.70	&	     -	    &	     -	    &	-	&	-	&	-	&	-	&	-	&	-	&	-	&	-	&	-	&	-	&	-	&	-	&	-	&	-	&	- \\
CrI5409.79	&	     -	    &	     -	    &	-	&	-	&	-	&	-	&	-	&	-	&	-	&	-	&	-	&	-	&	-	&	-	&	-	&	-	&	- \\
FeI5405.77	&	     -	    &	     -	    &	-	&	-	&	-	&	-	&	-	&	-	&	-	&	-	&	-	&	-	&	-	&	-	&	-	&	-	&	- \\
FeI5397.13	&	     -	    &	     -	    &	-	&	-	&	-	&	-	&	-	&	-	&	-	&	-	&	-	&	-	&	-	&	-	&	-	&	-	&	- \\
FeI5371.87	&	     -	    &	     -	    &	-	&	-	&	-	&	-	&	-	&	-	&	-	&	-	&	-	&	-	&	-	&	-	&	-	&	-	&	- \\
FeI5328.04	&	     -	    &	     -	    &	-	&	-	&	-	&	-	&	-	&	-	&	-	&	-	&	-	&	-	&	-	&	-	&	-	&	-	&	- \\
FeII5316.61	&	     -	    &	     -	    &	-	&	-	&	-	&	-	&	-	&	-	&	-	&	-	&	-	&	-	&	-	&	-	&	-	&	-	&	- \\
FeII5275.99	&	     -	    &	     -	    &	-	&	-	&	-	&	-	&	-	&	-	&	-	&	-	&	-	&	-	&	-	&	-	&	-	&	-	&	- \\
FeI5269.54	&	     -	    &	     -	    &	-	&	-	&	-	&	-	&	-	&	-	&	-	&	-	&	-	&	-	&	-	&	-	&	-	&	-	&	- \\
FeII5234.62	&	     -	    &	     -	    &	-	&	-	&	-	&	-	&	-	&	-	&	-	&	-	&	-	&	-	&	-	&	-	&	-	&	-	&	- \\
FeI5227.19	&	     -	    &	     -	    &	-	&	-	&	-	&	-	&	-	&	-	&	-	&	-	&	-	&	-	&	-	&	-	&	-	&	-	&	- \\
CrI5208.44	&	     -	    &	     -	    &	-	&	-	&	-	&	-	&	-	&	-	&	-	&	-	&	-	&	-	&	-	&	-	&	-	&	-	&	- \\
MgI5183.60	&	     -	    &	     -	    &	-	&	-	&	-	&	-	&	-	&	-	&	-	&	-	&	-	&	-	&	-	&	0.19$\pm$0.09	&	0.29$\pm$0.09	&	-	&	- \\
MgI5172.68	&	     -	    &	     -	    &	-	&	-	&	-	&	-	&	-	&	-	&	-	&	-	&	-	&	-	&	-	&	-	&	0.20$\pm$0.10	&	-	&	- \\
FeII5169.03	&	     -	    &	     -	    &	-	&	-	&	-	&	-	&	-	&	-	&	-	&	-	&	-	&	-	&	-	&	0.12$\pm$0.07	&	0.43$\pm$0.07	&	-	&	- \\
MgI5167.32	&	     -	    &	     -	    &	-	&	-	&	-	&	-	&	-	&	-	&	-	&	-	&	-   &	-	&	-	&	0.25$\pm$0.08	&	0.17$\pm$0.07	&	-	&	- \\
FeII5018.43	&	     -	    &	     -	    &	-	&	-	&	-	&	-	&	-	&	-	&	-	&	-	&	-	&	-	&	-	&	-	&	-	&	-	&	- \\
HeI5015.68	&	     -	    &	     -	    &	-	&	-	&	-	&	-	&	-	&	-	&	-	&	-	&	-	&	-	&	-	&	-	&	-	&	-	&	- \\
FeI4957.30	&	     -	    &	     -	    &	-	&	-	&	-	&	-	&	-	&	-	&	-	&	-	&	-	&	-	&	-	&	-	&	-	&	-	&	- \\
FeII4923.92	&	     -	    &	     -	    &	-	&	-	&	-	&	-	&	-	&	-	&	-	&	-	&	-	&	-	&	-	&	-	&	-	&	-	&	- \\
FeI4891.50	&	     -	    &	     -	    &	-	&	-	&	-	&	-	&	-	&	-	&	-	&	-	&	-	&	-	&	-	&	-	&	-	&	-	&	- \\
FeI4890.76	&	     -	    &	     -	    &	-	&	-	&	-	&	-	&	-	&	-	&	-	&	-	&	-	&	-	&	-	&	-	&	-	&	-	&	- \\
H4861	&	0.89$\pm$0.27	&	     -	    &	2.13$\pm$0.23	&	2.48$\pm$0.33	&	-	&	-	&	-	&	-	&	1.62$\pm$0.23	&	-	&	-	&	-	&	3.41$\pm$0.27	&	5.98$\pm$0.23	&	6.29$\pm$0.18	&	1.15$\pm$0.20	&	3.08$\pm$0.20 \\
MnI4783.42	&	     -	    &	     -	    &	-	&	-	&	-	&	-	&	-	&	-	&	-	&	-	&	-	&	-	&	-	&	-	&	-	&	-	&	- \\
MgI4702.98	&	     -	    &	     -	    &	-	&	-	&	-	&	-	&	-	&	-	&	-	&	-	&	-	&	-	&	-	&	-	&	-	&	-	&	- \\
HeII4686.68	&	     -	    &	     -	    &	-	&	-	&	-	&	-	&	-	&	-	&	-	&	-	&	-	&	-	&	-	&	-	&	-	&	-	&	- \\
TiI4629.34	&	     -	    &	     -	    &	-	&	-	&	-	&	-	&	-	&	-	&	-	&	-	&	-	&	-	&	-	&	-	&	-	&	-	&	- \\
FeII4583.83	&	     -	    &	     -	    &	-	&	-	&	-	&	-	&	-	&	-	&	-	&	-	&	-	&	-	&	-	&	-	&	-	&	-	&	- \\
FeII4555.8	&	     -	    &	     -	    &	-	&	-	&	-	&	-	&	-	&	-	&	-	&	-	&	-	&	-	&	-	&	-	&	-	&	-	&	- \\
FeII4549.47	&	     -	    &	     -	    &	-	&	-	&	-	&	-	&	-	&	-	&	-	&	-	&	-	&	-	&	-	&	-	&	-	&	-	&	- \\
TiI4522.80	&	     -	    &	     -	    &	-	&	-	&	-	&	-	&	-	&	-	&	-	&	-	&	-	&	-	&	-	&	-	&	-	&	-	&	- \\
FeI4482.17	&	     -	    &	     -	    &	-	&	-	&	-	&	-	&	-	&	-	&	-	&	-	&	-	&	-	&	-	&	-	&	-	&	-	&	- \\
MgII4481.33	&	     -	    &	     -	    &	-	&	-	&	-	&	-	&	-	&	-	&	-	&	-	&	-	&	-	&	-	&	-	&	-	&	-	&	- \\
HeI4471.48	&	     -	    &	     -	    &	-	&	-	&	-	&	-	&	-	&	-	&	-	&	-	&	-	&	-	&	-	&	-	&	-	&	-	&	- \\
FeI4459.12	&	     -	    &	     -	    &	-	&	-	&	-	&	-	&	-	&	-	&	-	&	-	&	-	&	-	&	-	&	-	&	-	&	-	&	- \\
CaI4456.61	&	     -	    &	     -	    &	-	&	-	&	-	&	-	&	-	&	-	&	-	&	-	&	-	&	-	&	-	&	-	&	-	&	-	&	- \\
TiII4450.49	&	     -	    &	     -	    &	-	&	-	&	-	&	-	&	-	&	-	&	-	&	-	&	-	&	-	&	-	&	-	&	-	&	-	&	- \\
TiI4443.80	&	     -	    &	     -	    &	-	&	-	&	-	&	-	&	-	&	-	&	-	&	-	&	-	&	-	&	-	&	-	&	-	&	-	&	- \\
FeI4427.31	&	     -	    &	     -	    &	-	&	-	&	-	&	-	&	-	&	-	&	-	&	-	&	-	&	-	&	-	&	-	&	-	&	-	&	- \\
FeI4415.13	&	     -	    &	     -	    &	-	&	-	&	-	&	-	&	-	&	-	&	-	&	-	&	-	&	-	&	-	&	-	&	-	&	-	&	- \\
FeI4405.75	&	     -	    &	     -	    &	-	&	-	&	-	&	-	&	-	&	-	&	-	&	-	&	-	&	-	&	-	&	-	&	-	&	-	&	- \\
TiII4395.03	&	     -	    &	     -	    &	-	&	-	&	-	&	-	&	-	&	-	&	-	&	-	&	-	&	-	&	-	&	-	&	-	&	-	&	- \\
FeI4384.68	&	     -	    &	     -	    &	-	&	-	&	-	&	-	&	-	&	-	&	-	&	-	&	-	&	-	&	-	&	-	&	-	&	-	&	- \\
FeI4383.55	&	     -	    &	     -	    &	-	&	-	&	-	&	-	&	-	&	-	&	-	&	-	&	-	&	-	&	-	&	-	&	-	&	-	&	- \\
FeI4375.93	&	     -	    &	     -	    &	-	&	-	&	-	&	-	&	-	&	-	&	-	&	-	&	-	&	-	&	-	&	-	&	-	&	-	&	- \\
H4340	    &	     -	    &	     -	    &	-	&	-	&	-	&	-	&	-	&	-	&	1.02$\pm$0.22	&	-	&	-	&	-	&	1.35$\pm$0.32	&	2.70$\pm$0.24	&	2.81$\pm$0.17	&	1.17$\pm$0.21	&	1.10$\pm$0.21 \\
FeII4303.17	&	     -	    &	     -	    &	-	&	-	&	-	&	-	&	-	&	-	&	-	&	-	&	-	&	-	&	-	&	-	&	-	&	-	&	- \\
CrI4274.80	&	     -	    &	     -	    &	-	&	-	&	-	&	-	&	-	&	-	&	-	&	-	&	-	&	-	&	-	&	-	&	-	&	-	&	- \\
FeI4271.65	&	     -	    &	     -	    &	-	&	-	&	-	&	-	&	-	&	-	&	-	&	-	&	-	&	-	&	-	&	-	&	-	&	-	&	- \\
FeII4233.17	&	     -	    &	     -	    &	-	&	-	&	-	&	-	&	-	&	-	&	-	&	-	&	-	&	-	&	-	&	-	&	-	&	-	&	- \\
CaI4226.73	&	     -	    &	     -	    &	-	&	-	&	-	&	-	&	-	&	-	&	-	&	-	&	-	&	-	&	-	&	-	&	-	&	-	&	- \\
FeII4178.86	&	     -	    &	     -	    &	-	&	-	&	-	&	-	&	-	&	-	&	-	&	-	&	-	&	-	&	-	&	-	&	-	&	-	&	- \\
FeII4173.45	&	     -	    &	     -	    &	-	&	-	&	-	&	-	&	-	&	-	&	-	&	-	&	-	&	-	&	-	&	-	&	-	&	-	&	- \\
H4101	    &        -	    &	     -	    &	-	&	-	&	-	&	-	&	-	&	-	&	-	&	-	&	-	&	-	&	-	&	-	&	-	&	-	&	- \\
	
    \hline
\end{tabular}
\end{table}
\begin{scriptsize}
\end{scriptsize}
\end{landscape}

\clearpage

\begin{landscape}
\begin{table}
\centering
\tiny
	\contcaption{}
\begin{tabular}{lccccc}
		\hline

	        &	  2024-01-05   &   2024-01-05	&	 2024-01-07	&	2024-01-08	&	2024-01-08	\\
	        &	  impulsive	   &	gradual	    &	 impulsive	&	impulsive	&	gradual	\\
            &	  flare 20     &	flare 20	&	 flare 21	&	flare 22	&	flare 22	\\
            
            \hline
CaI7148.15	&	1.64$\pm$0.40  & 1.53$\pm$0.43	&	      -	    & 0.28$\pm$0.35	&	0.67$\pm$0.28	\\
HeI6678.15	&	      -        &	   -    	&	      -  	& 0.23$\pm$0.25	&	       -	\\
H6562.79	&	1.05$\pm$0.29  & 0.72$\pm$0.30	&	3.96$\pm$0.56	&	3.46$\pm$0.51	&	0.13$\pm$0.49	\\
NaDI5895.92	&	0.32$\pm$0.13  &	-	&	0.46$\pm$0.19	&	-	&	-	\\
NaDII5889.95 &	0.18$\pm$0.13  &	-	&	0.55$\pm$0.19	&	-	&	-	\\
HeI5876.65	&	0.64$\pm$0.13  & 0.64$\pm$0.14	&	1.50$\pm$0.19	&	1.05$\pm$0.30	&	1.35$\pm$0.23	\\
MgI5528.40	&	-	&	-	&	-	&	-	&	-	\\
FeI5455.61	&	-	&	-	&	-	&	-	&	-	\\
FeI5446.92	&	-	&	-	&	-	&	-	&	-	\\
FeI5429.70	&	-	&	-	&	-	&	-	&	-	\\
CrI5409.79	&	-	&	-	&	-	&	-	&	-	\\
FeI5405.77	&	-	&	-	&	-	&	-	&	-	\\
FeI5397.13	&	-	&	-	&	-	&	-	&	-	\\
FeI5371.87	&	-	&	-	&	-	&	-	&	-	\\
FeI5328.04	&	-	&	-	&	-	&	-	&	-	\\
FeII5316.61	&	-	&	-	&	-	&	-	&	-	\\
FeII5275.99	&	-	&	-	&	-	&	-	&	-	\\
FeI5269.54	&	-	&	-	&	-	&	-	&	-	\\
FeII5234.62	&	-	&	-	&	-	&	-	&	-	\\
FeI5227.19	&	-	&	-	&	-	&	-	&	-	\\
CrI5208.44	&	-	&	-	&	-	&	-	&	-	\\
MgI5183.60	&	-	&	-	&	-	&	-	&	-	\\
MgI5172.68	&	-	&	-	&	-	&	-	&	-	\\
FeII5169.03	&	-	&	-	&	0.39$\pm$0.07	&	-	&	-	\\
MgI5167.32	&	-	&	-	&	-	&	-	&	-	\\
FeII5018.43	&	-	&	-	&	0.29$\pm$0.11	&	-	&	-	\\
HeI5015.68	&	-	&	-	&	-	&	-	&	-	\\
FeI4957.30	&	-	&	-	&	-	&	-	&	-	\\
FeII4923.92	&	-	&	-	&	-	&	-	&	-	\\
FeI4891.50	&	-	&	-	&	-	&	-	&	-	\\
FeI4890.76	&	-	&	-	&	-	&	-	&	-	\\
H4861.35	&	-	&	-	&	5.08$\pm$0.21	&	-	&	-	\\
MnI4783.42	&	-	&	-	&	-	&	-	&	-	\\
MgI4702.98	&	-	&	-	&	-	&	-	&	-	\\
HeII4686.68	&	-	&	-	&	-	&	-	&	-	\\
TiI4629.34	&	-	&	-	&	-	&	-	&	-	\\
FeII4583.83	&	-	&	-	&	-	&	-	&	-	\\
FeII4555.8	&	-	&	-	&	-	&	-	&	-	\\
FeII4549.47	&	-	&	-	&	-	&	-	&	-	\\
TiI4522.80	&	-	&	-	&	-	&	-	&	-	\\
FeI4482.17	&	-	&	-	&	-	&	-	&	-	\\
MgII4481.33	&	-	&	-	&	-	&	-	&	-	\\
HeI4471.48	&	-	&	-	&	-	&	-	&	-	\\
FeI4459.12	&	-	&	-	&	-	&	-	&	-	\\
CaI4456.61	&	-	&	-	&	-	&	-	&	-	\\
TiII4450.49	&	-	&	-	&	-	&	-	&	-	\\
TiI4443.80	&	-	&	-	&	-	&	-	&	-	\\
FeI4427.31	&	-	&	-	&	-	&	-	&	-	\\
FeI4415.13	&	-	&	-	&	-	&	-	&	-	\\
FeI4405.75	&	-	&	-	&	-	&	-	&	-	\\
TiII4395.03	&	-	&	-	&	-	&	-	&	-	\\
FeI4384.68	&	-	&	-	&	-	&	-	&	-	\\
FeI4383.55	&	-	&	-	&	-	&	-	&	-	\\
FeI4375.93	&	-	&	-	&	-	&	-	&	-	\\
H4340.47	&	-	&	-	&	1.77$\pm$0.20	&	-	&	-	\\
FeII4303.17	&	-	&	-	&	-	&	-	&	-	\\
CrI4274.80	&	-	&	-	&	-	&	-	&	-	\\
FeI4271.65	&	-	&	-	&	-	&	-	&	-	\\
FeII4233.17	&	-	&	-	&	-	&	-	&	-	\\
CaI4226.73	&	-	&	-	&	-	&	-	&	-	\\
FeII4178.86	&	-	&	-	&	-	&	-	&	-	\\
FeII4173.45	&	-	&	-	&	-	&	-	&	-	\\
H4101.73	&	-	&	-	&	-	&	-	&	-	\\

\hline
\end{tabular}
\end{table}
\begin{scriptsize}
\end{scriptsize}
\end{landscape}

\clearpage

\begin{landscape}
\begin{table}
\centering
\tiny
\caption{Line fluxes of spectral lines which showed excess emission during the flares from table~\ref{tab:B1} showing decaying tails only.}
\label{tab:D3}
\begin{tabular}{lccccccccc}
		\hline

	          &	  2023-10-28   &   2023-11-13   &   2023-11-22   &   2023-11-30   &  2023-12-12  &   2023-12-16   &   2023-12-19   &   2024-01-06   &   2024-01-11	\\
	        &    gradual	 &	  gradual	  &	   gradual     &	gradual	    &    gradual   &	 gradual	 &	   gradual	  &	   gradual	   &	gradual	\\
            &    flare 23	 &	  flare 24	  &	   flare 25    &	flare 26	&    flare 27  &	 flare 28	 &	   flare 29	  &	   flare 30	   &	flare 31	\\
            \hline
CaI7148.15	&	     -	     &	     -        &	       -	   &	      -	    &      -       & 1.24$\pm$0.31	 &	       -	  &	       -	   &	    -	\\
HeI6678.15	&	     -	     &     	 -     	  &	       -	   &	      -	    &     -       &	     -	     &	       -	  &	       -	   &	0.09$\pm$0.10	\\
H6562.79	& 6.85$\pm$0.66	 & 9.85$\pm$0.71  &	8.70$\pm$0.53  & 3.48$\pm$0.42	&       4.64$\pm$0.63       & 3.18$\pm$0.35	 & 5.16$\pm$0.38  &	2.75$\pm$0.18  &  4.11$\pm$0.23	\\
NaDI5895.92	&	     -	     &       -	      &	0.26$\pm$0.13  &	      -	    &-&	     -	     &	0.00$\pm$0.12 &	0.29$\pm$0.12  &	0.41$\pm$0.11	\\
NaDII5889.95 &	     -	     &	     -	      &      	-	   &	      -	    &-&	     -	     &	       -	  &    	    -	   &	0.37$\pm$0.11	\\
HeI5876.65	&	     -	     &	     -	      &	        -	   &	      -	    &-&	     -	     &	       -	  &	        -	   &	0.05$\pm$0.11	\\
MgI5528.40	&	     -	     &	     -	      &	        -	   &	      -	    &-&	     -	     &	       -	  &	        -	   &	      -	\\
FeI5455.61	&	     -	     &	     -	      &     	-	   &	      -	    &-&	     -	     &	       -	  &	        -	   &	      -	\\
FeI5446.92	&	     -	     &	-	          &	-	           &	-	        &-&	-	&	-	&	-	&	-	\\
FeI5429.70	&	     -	     &	-	          &	-	           &	-	        &-&	-	&	-	&	-	&	0.01$\pm$0.06	\\
CrI5409.79	&	     -	     &	-	          &	-	           &	-	        &-&	-	&	-	&	-	&	-	\\
FeI5405.77	&      	 -	     &	-	          &	-	           &	-	        &-&	-	&	-	&	-	&	-	\\
FeI5397.13	&	-	         &	-	          &	-	           &	-	        &-&	-	&	-	&	-	&	0.08$\pm$0.06	\\
FeI5371.87	&	-	         &	-	          &	-	           &	-	        &-&	-	&	-	&	-	&	0.18$\pm$0.05	\\
FeI5328.04	&	-	         &	-	          &	-	           &	-	        &-&	-	&	-	&	-	&	-	\\
FeII5316.61	&	-	         &	-	          &	-	           &	-	        &-&	-	&	-	&	-	&	0.09$\pm$0.07	\\
FeII5275.99	&	-	         &	-	          &	-	           &	-	        &-&	-	&	-	&	-	&	-	\\
FeI5269.54	&	-	         &	-	          &	-	           &	-	        &-&	-	&	-	&	-	&	-	\\
FeII5234.62	&	-	         &	-	          &	-	           &	-	        &-&	-	&	-	&	-	&	-	\\
FeI5227.19	&	-	         &	-	          &	-	           &	-	        &-&	-	&	-	&	-	&	0.13$\pm$0.06	\\
CrI5208.44	&	-	         &	-	          &	-	           &	-	        &-&	-	&	-	&	-	&	-	\\
MgI5183.60	&	-	         &	-	          &	-	           &	-	        &-&	-	&	-	&	-	&	0.18$\pm$0.07	\\
MgI5172.68	&	-	         &	-	          &	-	           &	-	        &-&	-	&	-	&	-	&	0.29$\pm$0.09	\\
FeII5169.03	&	-	         &	-	          &	-	           &	-	        &-&	-	&	-	&	0.14$\pm$0.05	&	0.30$\pm$0.06	\\
MgI5167.32	&	-	         &	-	          &	-	           &	-	        &-&	-	&	-	&	-	&	0.17$\pm$0.06	\\
FeII5018.43	&	-	         &	-	          &	-	           &	-	        &-&	-	&	-	&	-	&	0.28$\pm$0.07	\\
HeI5015.68	&	-	         &	-	          &	-	           &	-	        &-&	-	&	-	&	-	&	-	\\
FeI4957.30	&	-	         &	-	          &	-	           &	-	        &-&	-	&	-	&	-	&	-	\\
FeII4923.92	&	-	         &	-	          &	-	           &	-	        &-&	0.26$\pm$0.06	&	-	&	0.06$\pm$0.06	&	0.11$\pm$0.05	\\
FeI4891.50	&	-	         &	-	          &	-	           &	-	        &-&	-	&	-	&	-	&	-	\\
FeI4890.76	&	-	         &	-	          &	-	           &	-	        &-&	-	&	-	&	-	&	-	\\
H4861.35	& 0.57$\pm$0.39	 &	-	          &	1.54$\pm$0.20  &	-	        &-&	2.88$\pm$0.17	&	2.70$\pm$0.16	&	2.00$\pm$0.16	&	5.88$\pm$0.14	\\
MnI4783.42	&	-	         &	-	          &	-	           &	-	        &-&	-	&	-	&	-	&	-	\\
MgI4702.98	&	-	         &	-	          &	-	           &	-	        &-&	-	&	0.40$\pm$0.07	&	-	&	0.01$\pm$0.07	\\
HeII4686.68	&	-	         &	-	          &	-	           & 0.20$\pm$0.12	&-&	-	&	-	&	-	&	-	\\
TiI4629.34	&	-	         &	-	          &	-	           &	-	        &-&	-	&	-	&	-	&	-	\\
FeII4583.83	&	-	         &	-	          &	-	           &	-	        &-& 	-	&	-	&	-	&	0.06$\pm$0.09	\\
FeII4555.8	&	-	         &	-	          &	-	           &	-	        &-&	-	&	-	&	-	&	-	\\
FeII4549.47	&	-	         &	-	          &	0.16$\pm$0.12  &	-	        &-&	-	&	0.12$\pm$0.07	&	-	&	0.22$\pm$0.07	\\
TiI4522.80	&	-	         &	-	          &	-	           &	-	        &-&	-	&	-	&	-	&	-	\\
FeI4482.17	&	-	         &	-	          &	-	           &	-	        &-&	-	&	-	&	-	&	-	\\
MgII4481.33	&	-	         &	-	          &	-	           &	-	        &-&	-	&	-	&	-	&	-	\\
HeI4471.48	&	-	         &	-	          &	-	           &	-	        &-&	-	&	-	&	-	&	0.09$\pm$0.10	\\
FeI4459.12	&	-	         &	-	          &	-	           &	-	        &-&	-	&	-	&	-	&	-	\\
CaI4456.61	&	-	         &	-	          &	-	           &	-	        &-&	-	&	-	&	-	&	-	\\
TiII4450.49	&	-	         &	-	          &	-	           &	-	        &-&	-	&	-	&	-	&	-	\\
TiI4443.80	&	-	         &	-	          &	-	           &	-	        &-&	-	&	-	&	-	&	-	\\
FeI4427.31	&	-	         &	-	          &	-	           &	-	        &-&	-	&	-	&	-	&	-	\\
FeI4415.13	&	-	         &	-	          &	-	           &	-	        &-&	-	&	-	&	-	&	-	\\
FeI4405.75	&	-	         &	-	          &	-	           &	-	        &-&	-	&	-	&	-	&	-	\\
TiII4395.03	&	-	         &	-	          &	-	           &	-	        &-& 	-	&	-	&	-	&	-	\\
FeI4384.68	&	-	         &	-	          &	-	           &	-	        &-&	-	&	-	&	-	&	-	\\
FeI4383.55	&	-	         &	-	          &	-	           &	-	        &-&	-	&	-	&	-	&	-	\\
FeI4375.93	&	-	         &	-	          &	-	           &	-	        &-&	-	&	-	&	-	&	-	\\
H4340.47	&	-	         &	-	          &	0.55$\pm$0.20  & 0.87$\pm$0.20	&-&	0.68$\pm$0.17	&	0.83$\pm$0.13	&	0.93$\pm$0.13	&	2.48$\pm$0.16	\\
FeII4303.17	&	-	         &	-	          &	-	           &	-	        &-&	-	&	-	&	-	&	-	\\
CrI4274.80	&	-	         &	-	          &	-	           &	-	        &-&	-	&	-	&	-	&	-	\\
FeI4271.65	&	-	         &	-	          &	-	           &	-	        &-&	-	&	-	&	-	&	-	\\
FeII4233.17	&	-	         &	-	          &	-	           &	-	        &-&	-	&	-	&	-	&	-	\\
CaI4226.73	&	-	         &	-	          &	-	           &	-	        &-&	-	&	-	&	-	&	-	\\
FeII4178.86	&	-	         &	-	          &	-	           &	-	        &-&	-	&	-	&	-	&	-	\\
FeII4173.45	&	-	         &	-	          &	-	           &	-	        &-&	-	&	-	&	-	&	-	\\
H4101.73	&	-	         &	-	          &  	-	       &	-	        &-&	-	&	-	&	-	&	-	\\

\hline
\end{tabular}
\end{table}
\begin{scriptsize}
\end{scriptsize}
\end{landscape}

\clearpage

\clearpage

\section{Flare EW light curves in H$\alpha$, H$\beta$,H$\gamma$, and g'-band}

\begin{figure*}
 \includegraphics[width=8cm]{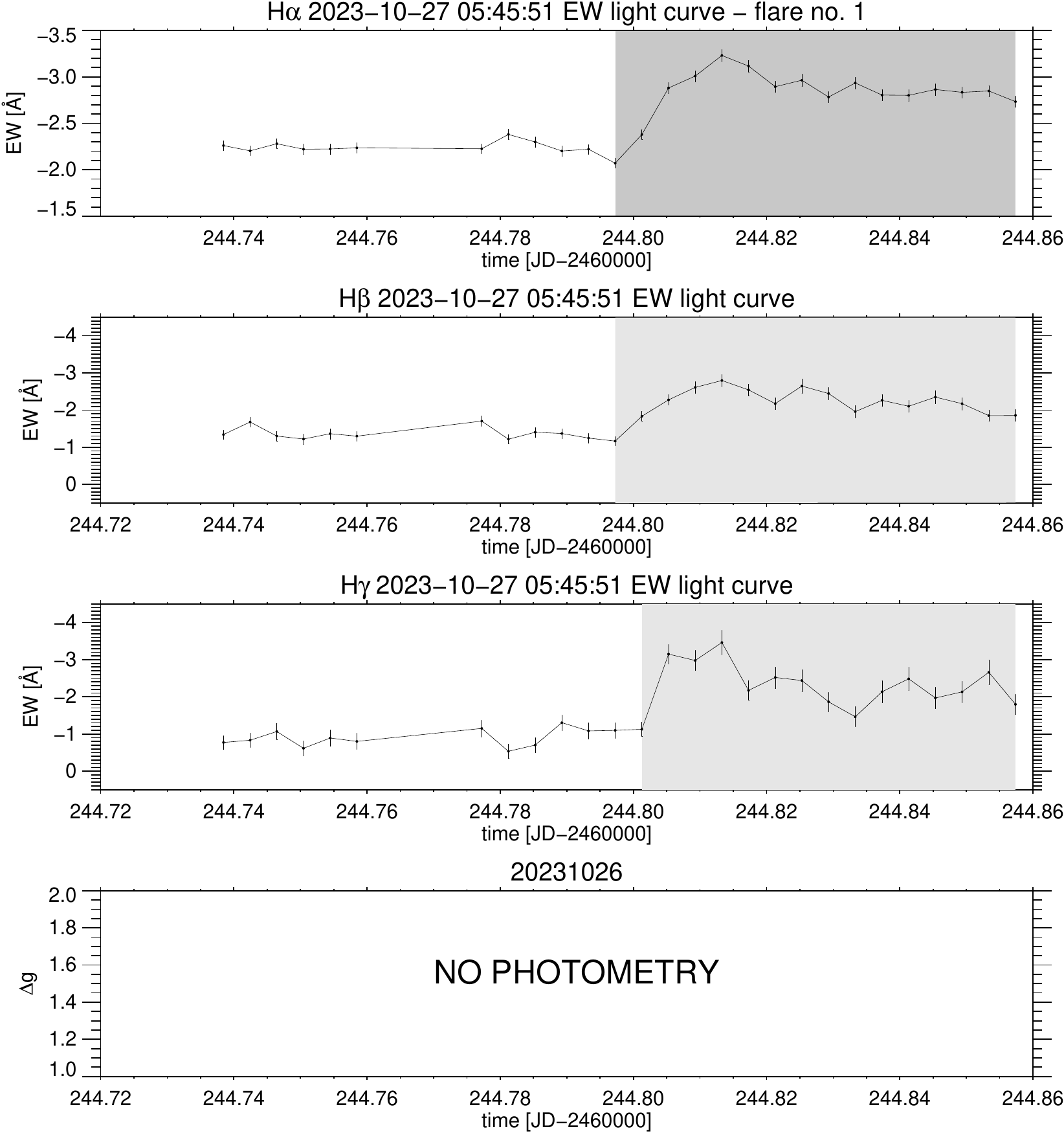}
 \hspace*{0.5cm}
 \includegraphics[width=8cm]{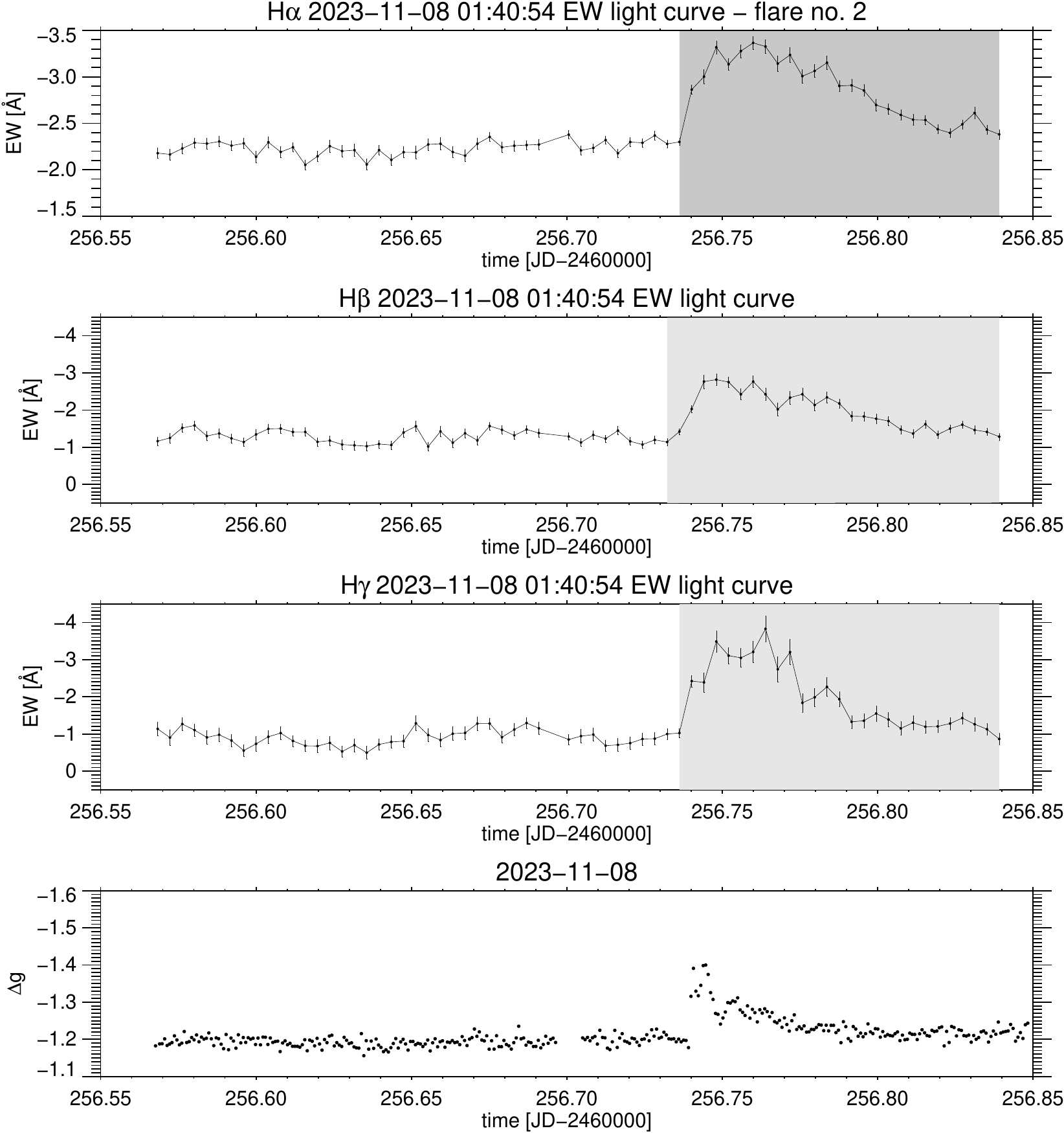}
 \\ 
 \vspace*{1.5cm} 
  \includegraphics[width=8cm]{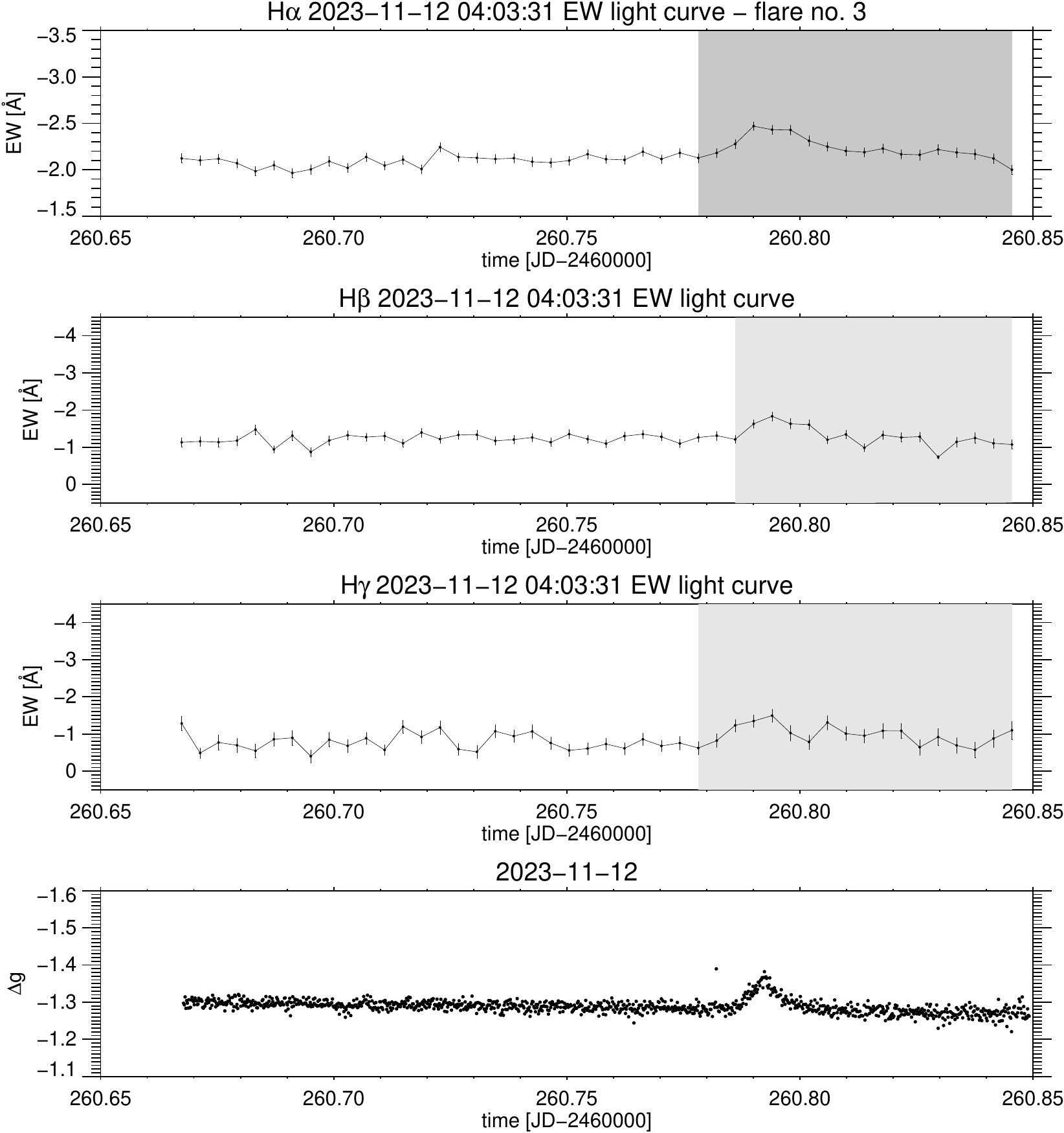}
 \hspace*{0.5cm} 
 \includegraphics[width=8cm]{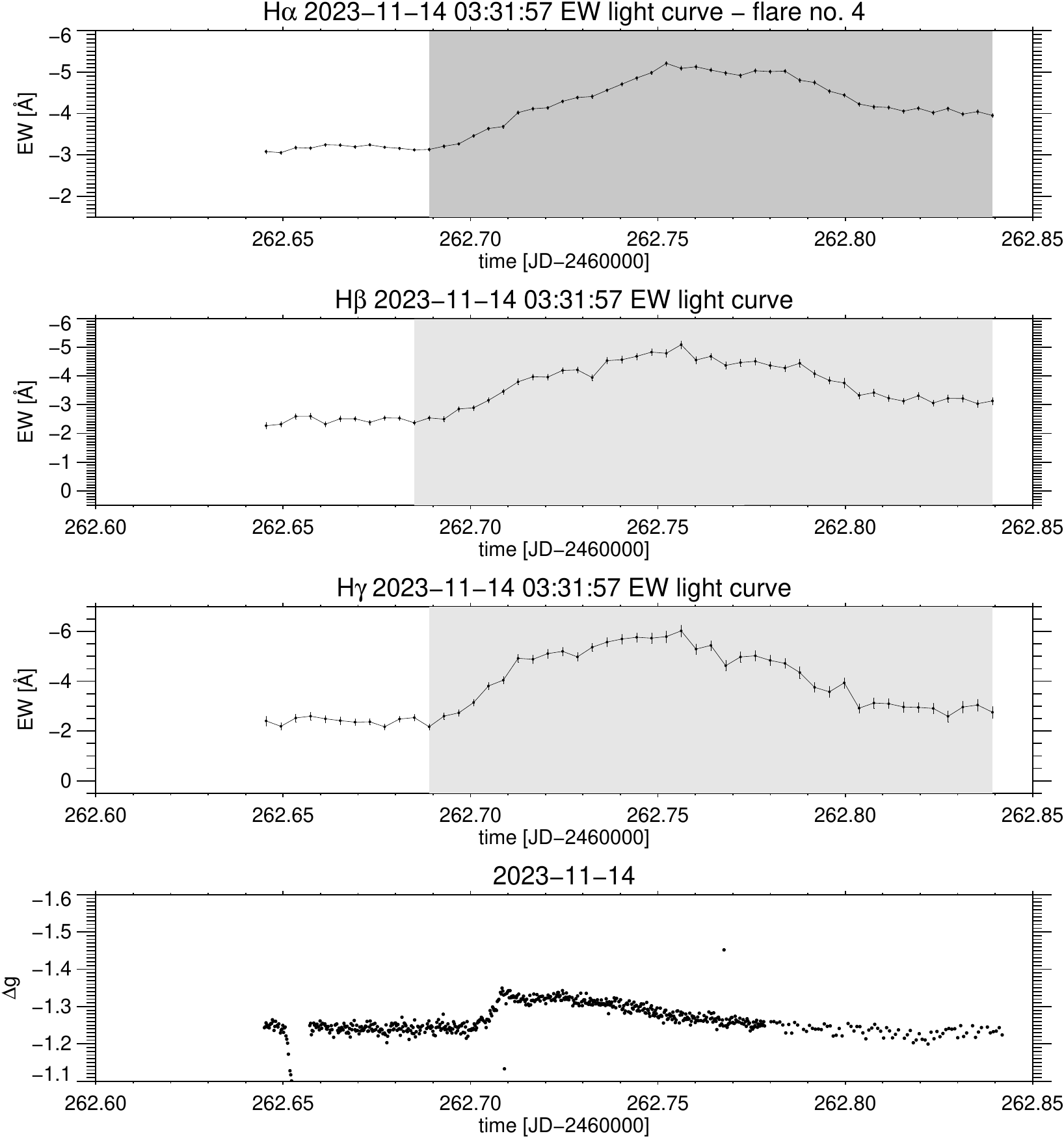}
 \caption{H$\alpha$, H$\beta$, H$\gamma$, and g'-band (from top to bottom) light curves of CC Eri in the nights of 2023-10-27 (upper left panel), 2023-11-08 (upper right panel), 2023-11-12 (lower left panel), and 2023-11-14 (lower right panel).}
 \label{fig:EWlc1}
\end{figure*}

\begin{figure*}
 \includegraphics[width=8cm]{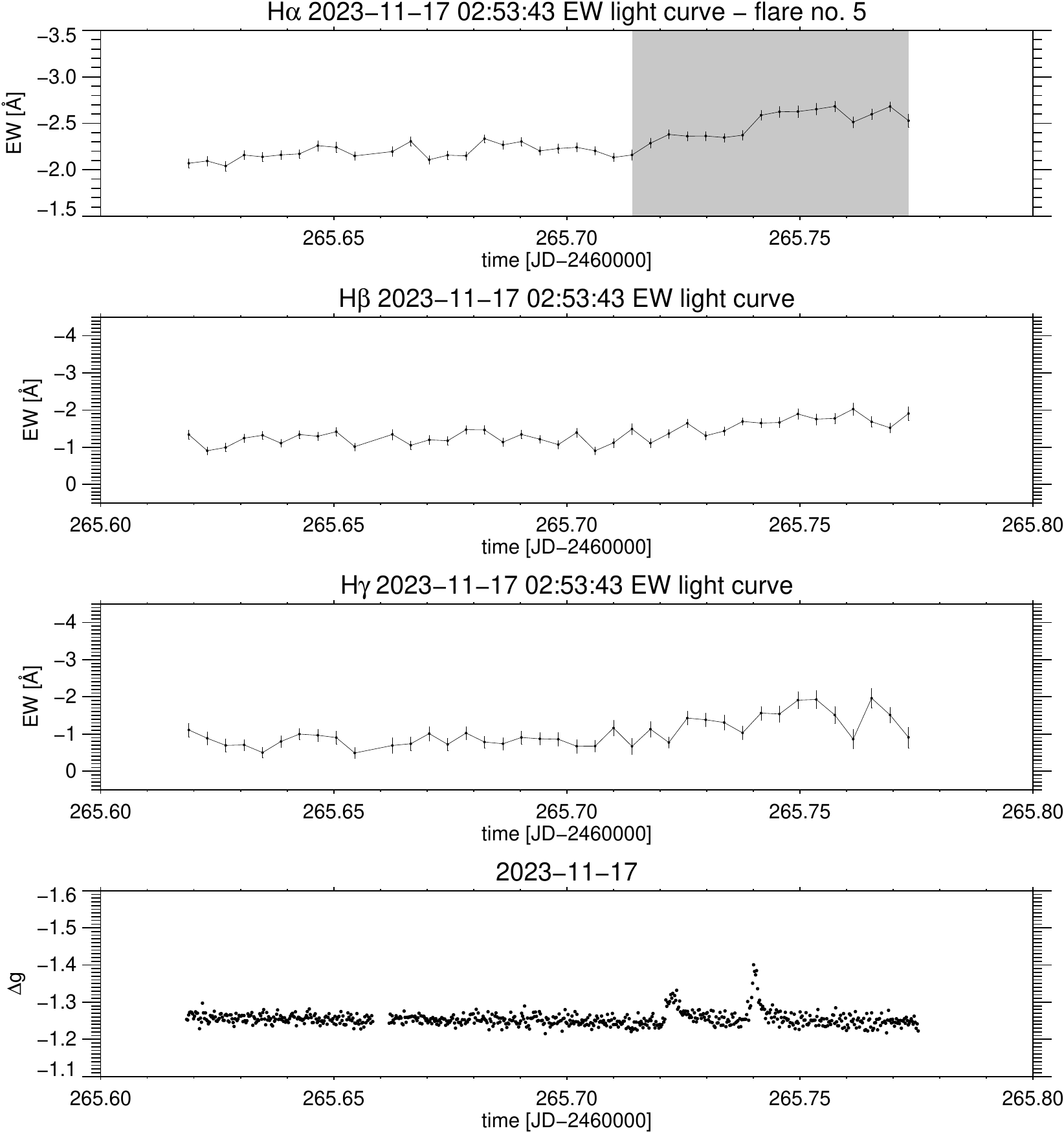}
 \hspace*{0.5cm}
 \includegraphics[width=8cm]{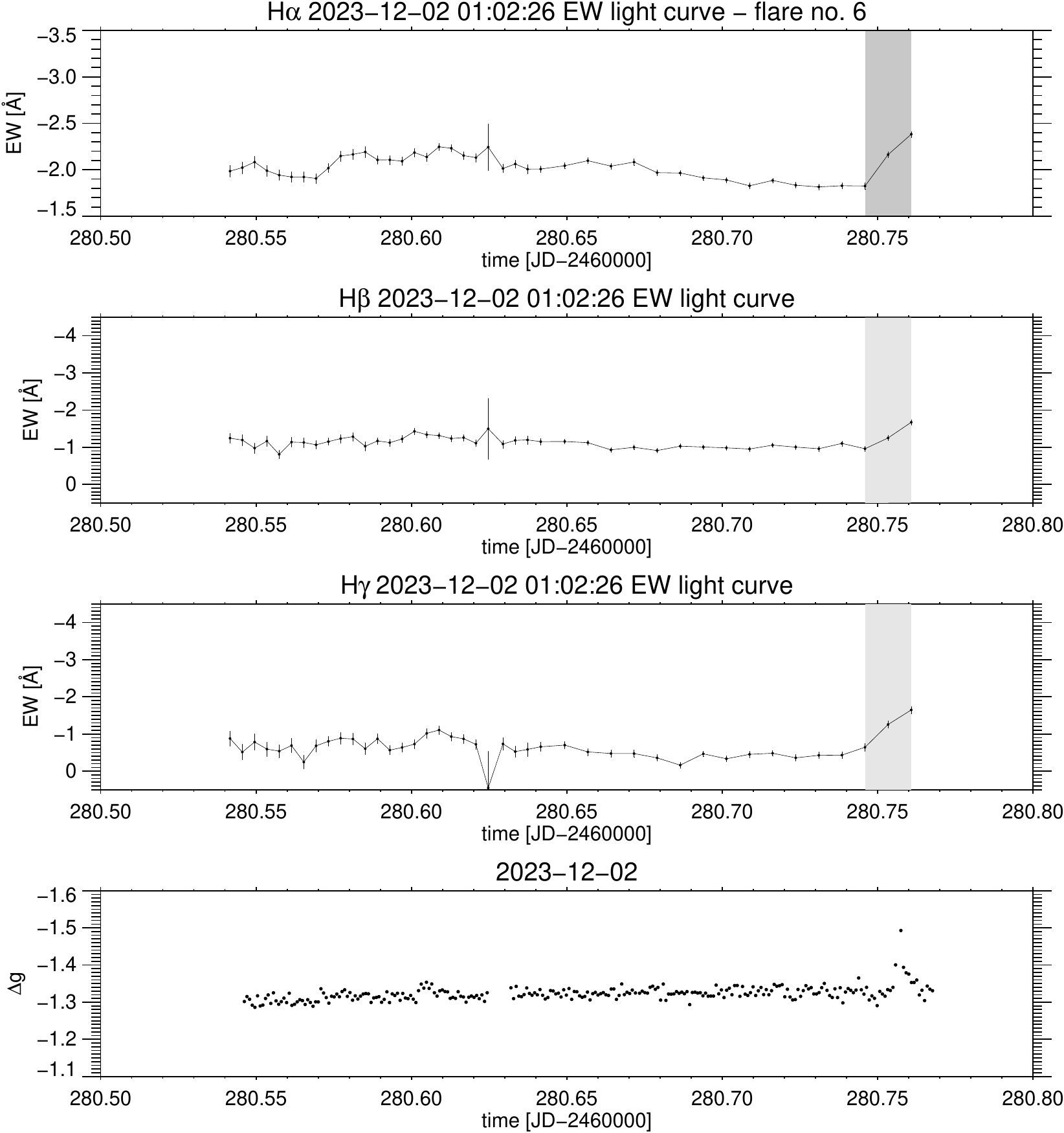}
 \\ 
 \vspace*{1.5cm} 
  \includegraphics[width=8cm]{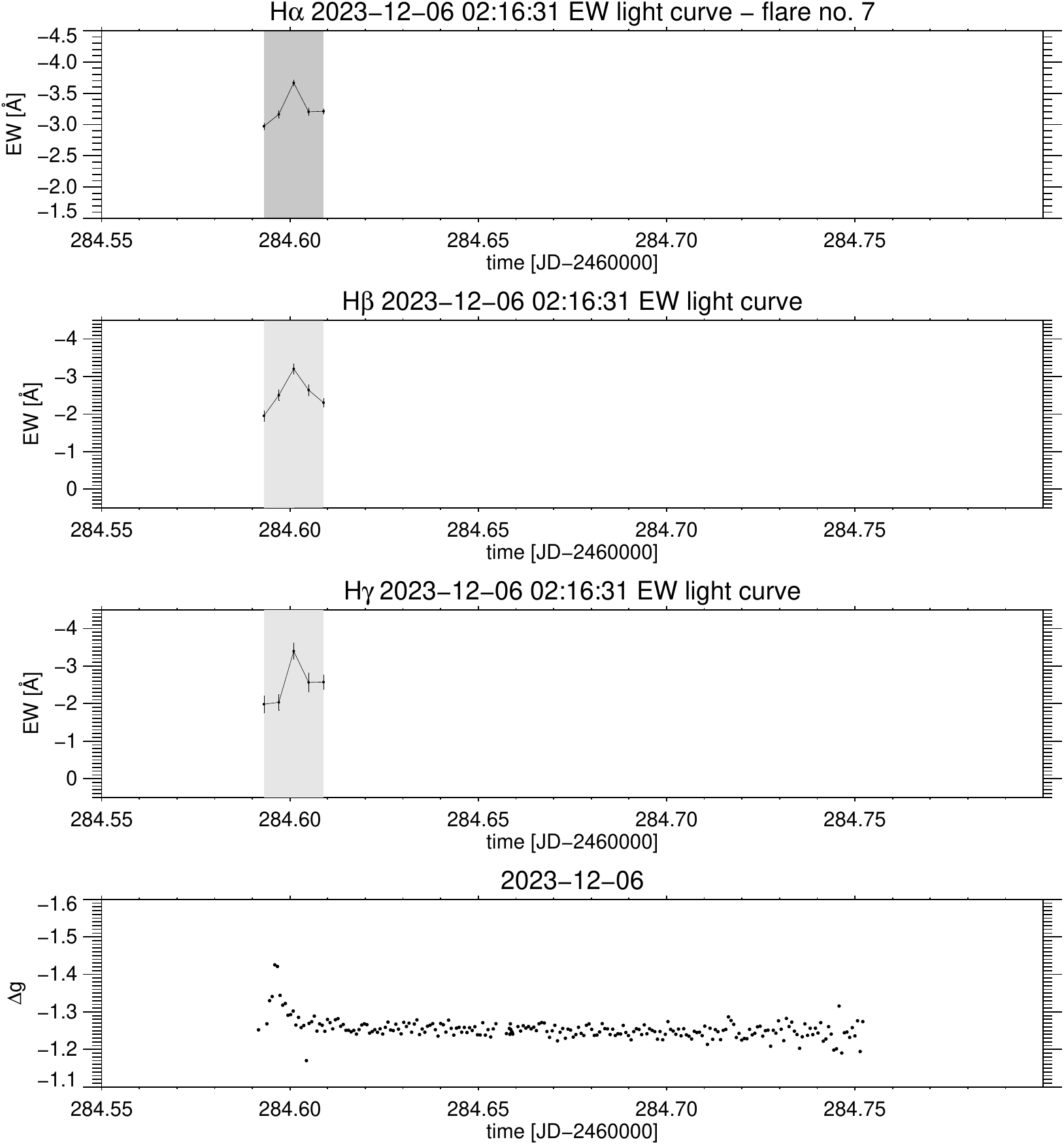}
 \hspace*{0.5cm} 
 \includegraphics[width=8cm]{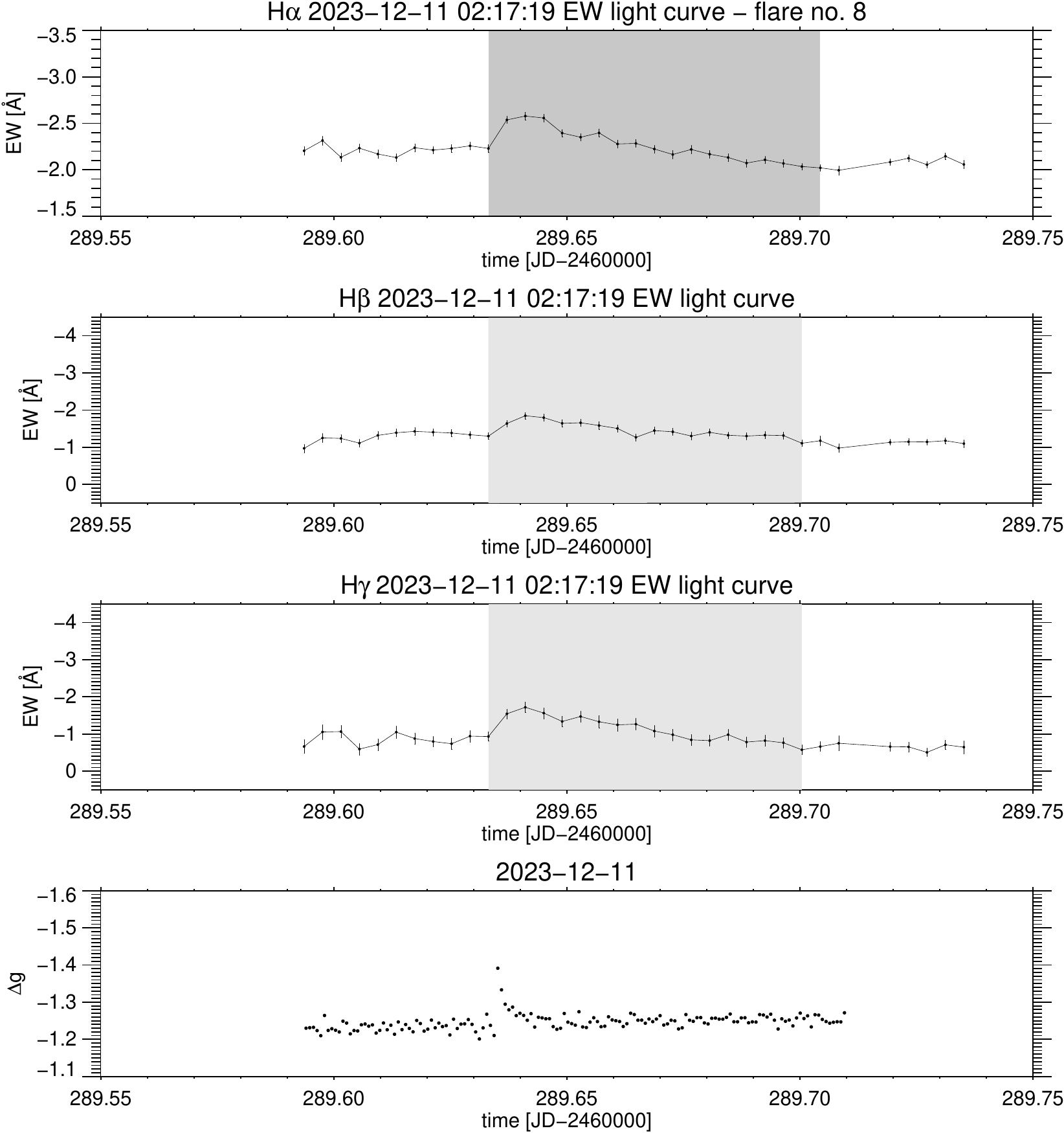}
 \caption{H$\alpha$, H$\beta$, H$\gamma$, and g'-band (from top to bottom) light curves of CC Eri in the nights of 2023-11-17 (upper left panel), 2023-12-02 (upper right panel), 2023-12-06 (lower left panel), and 2023-12-11 (lower right panel).}
 \label{fig:EWlc2}
\end{figure*}

\begin{figure*}
 \includegraphics[width=8cm]{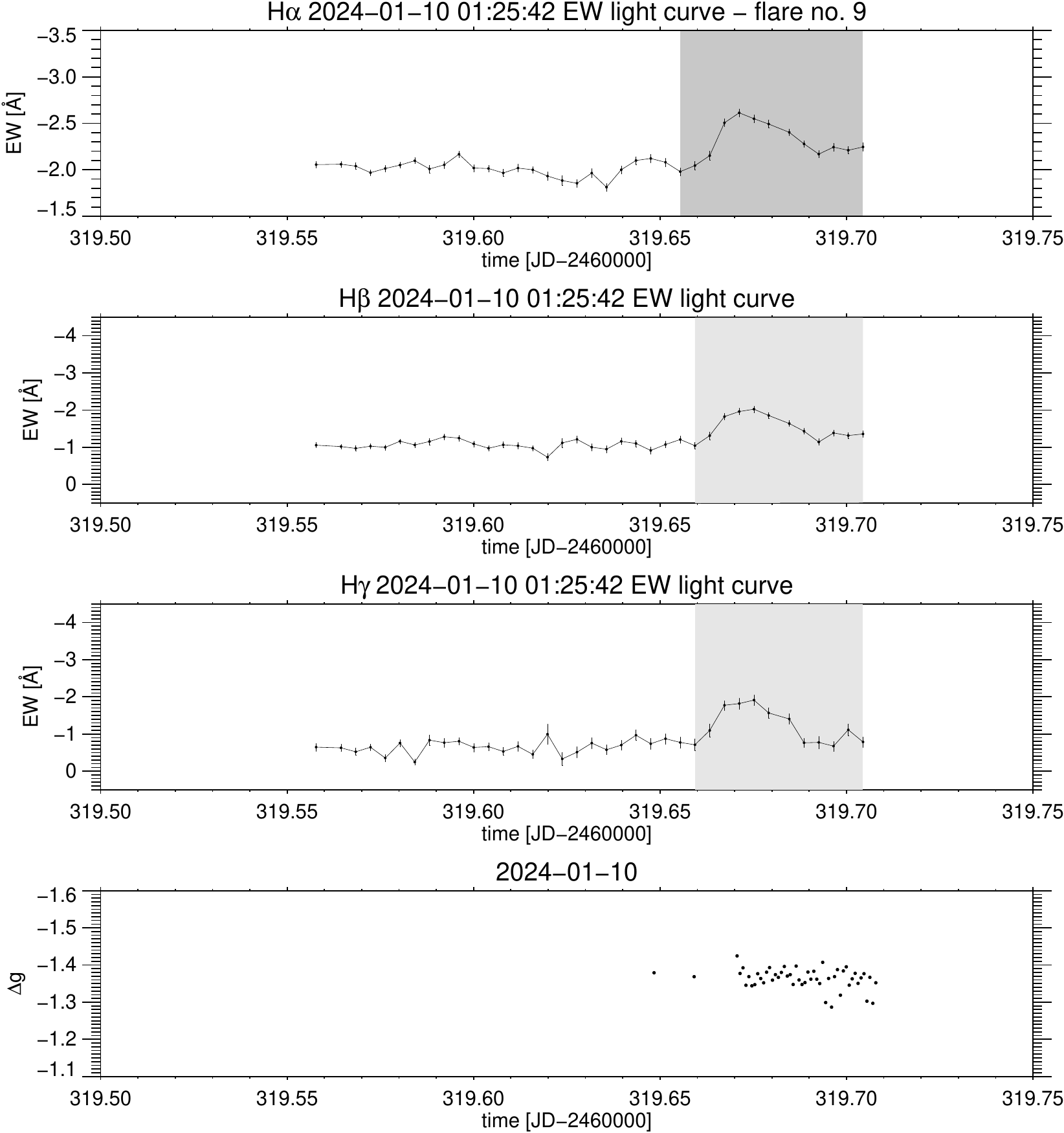}
 \hspace*{0.5cm}
 \includegraphics[width=8cm]{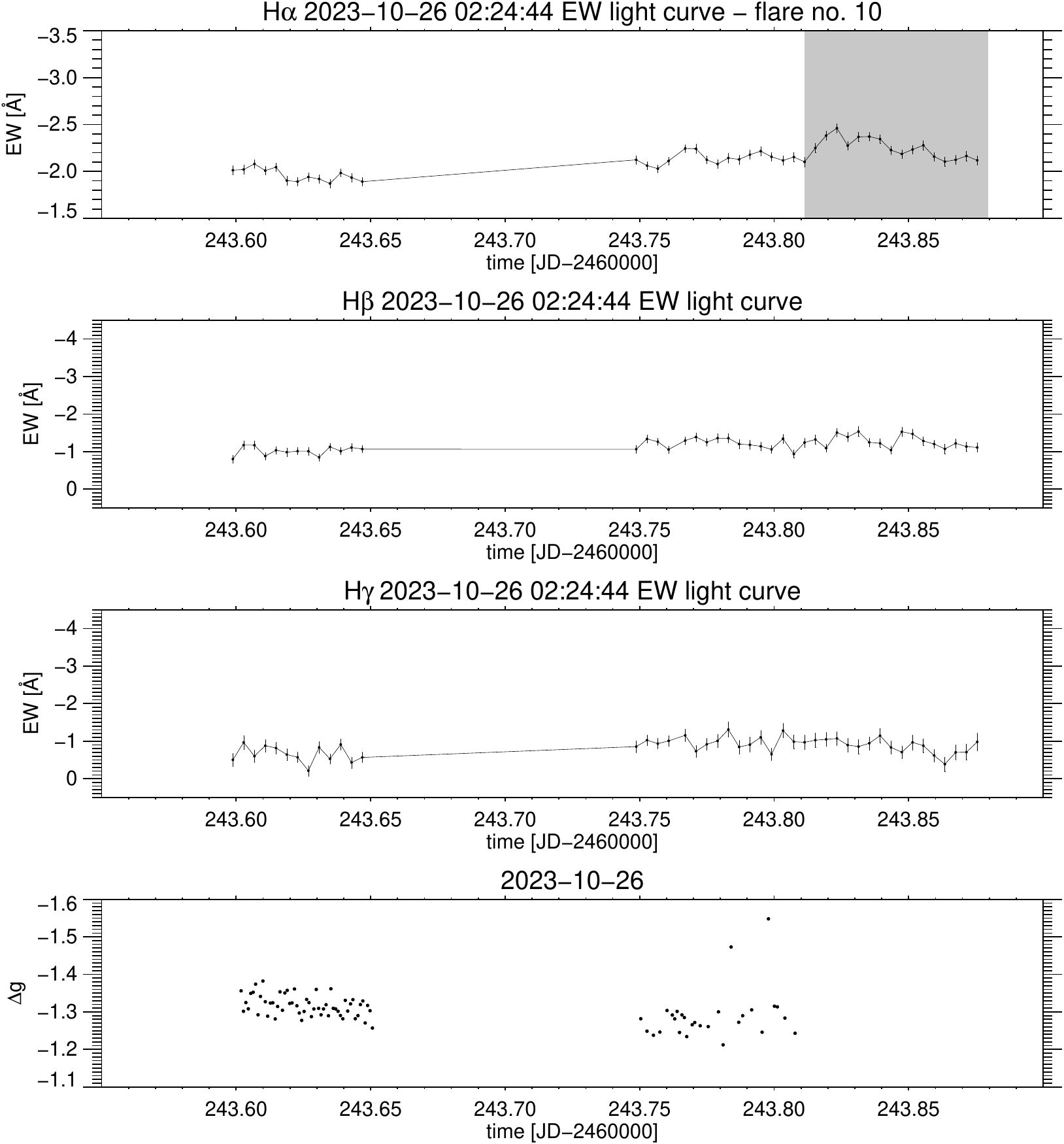}
 \\ 
 \vspace*{1.5cm} 
  \includegraphics[width=8cm]{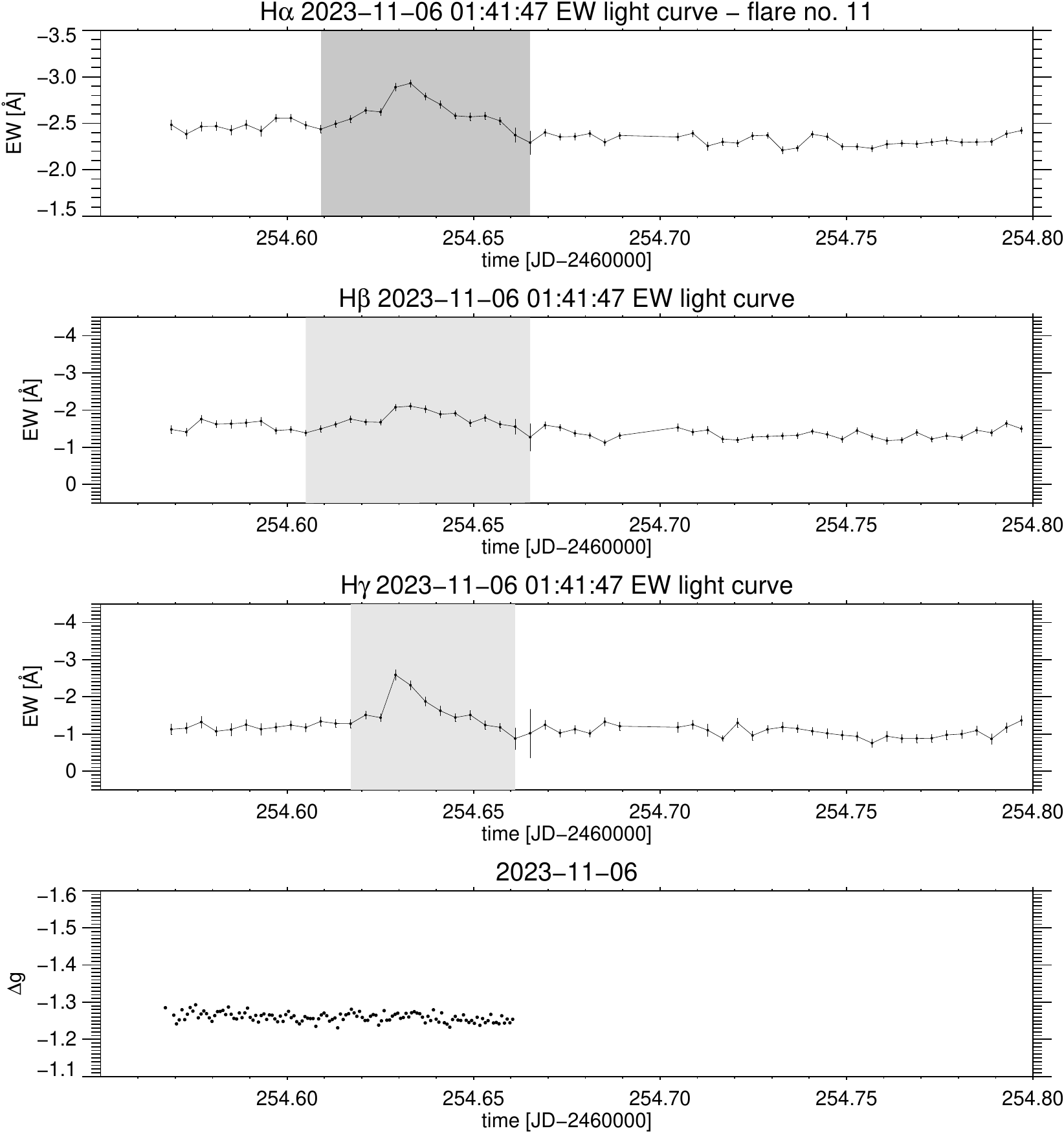}
 \hspace*{0.5cm} 
 \includegraphics[width=8cm]{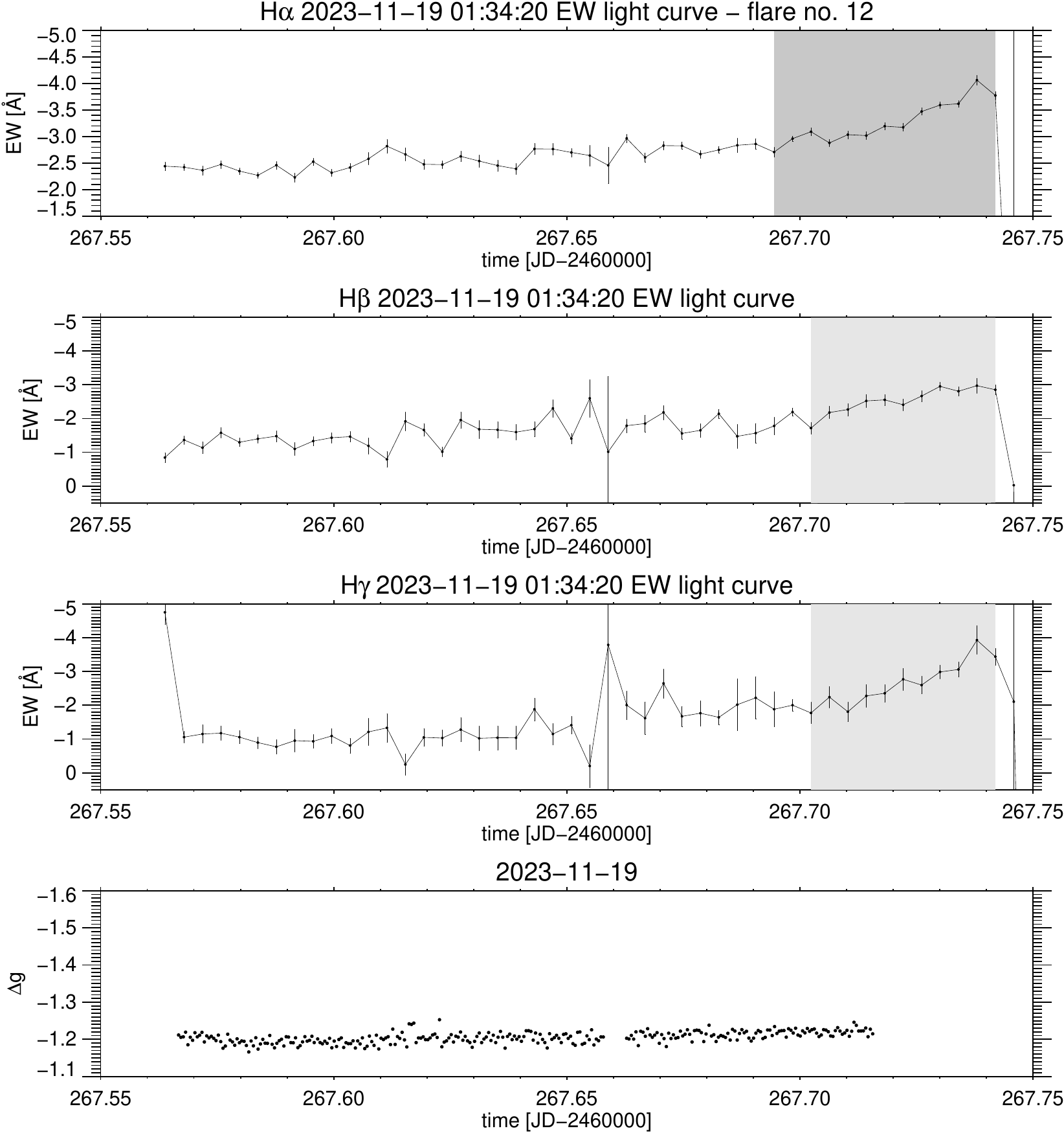}
 \caption{H$\alpha$, H$\beta$, H$\gamma$, and g'-band (from top to bottom) light curves of CC Eri in the nights of 2024-01-10 (upper left panel), 2023-10-26 (upper right panel), 2023-11-06 (lower left panel), and 2023-11-19 (lower right panel).}
 \label{fig:EWlc3}
\end{figure*}

\begin{figure*}
 \includegraphics[width=8cm]{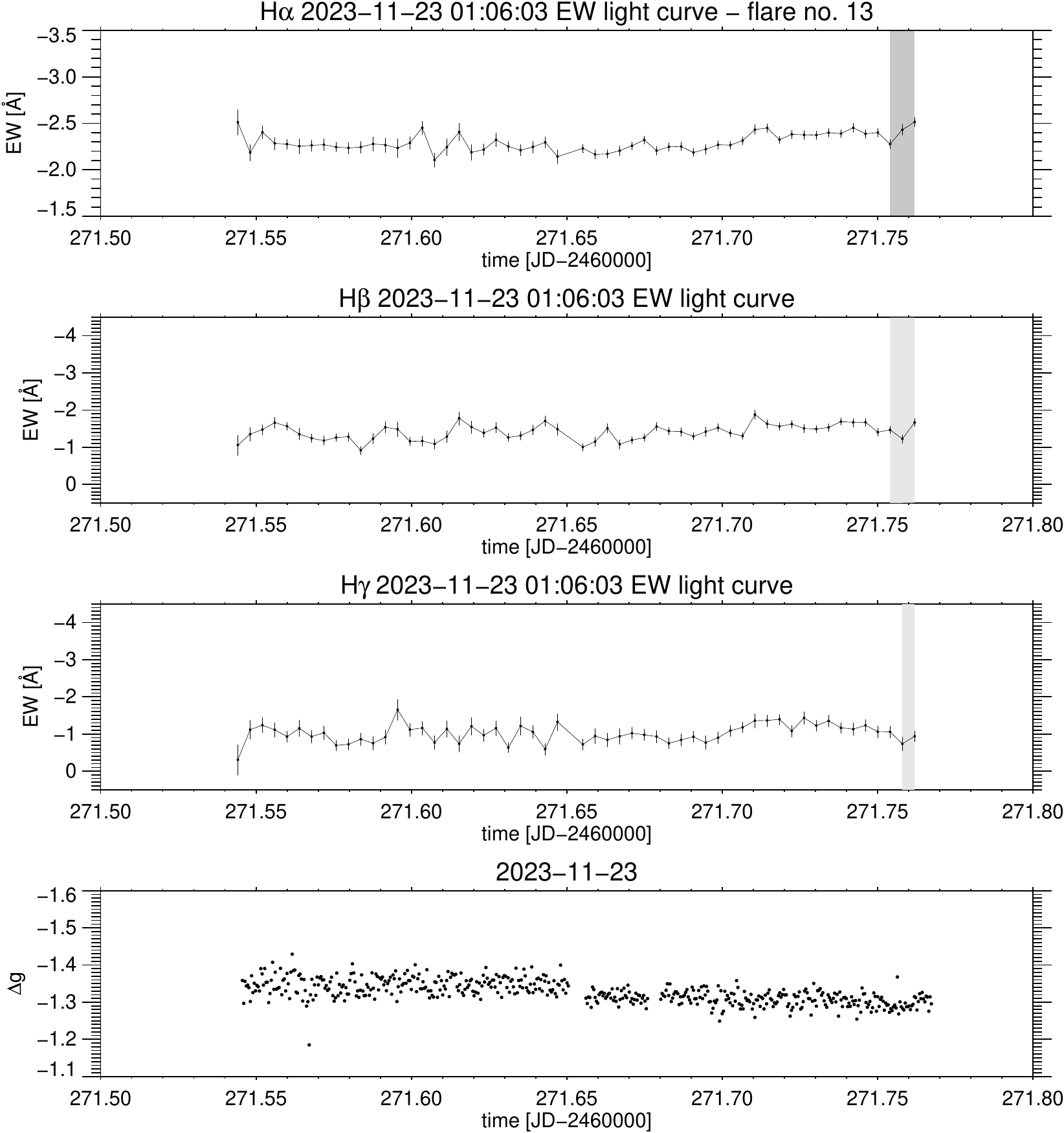}
 \hspace*{0.5cm}
 \includegraphics[width=8cm]{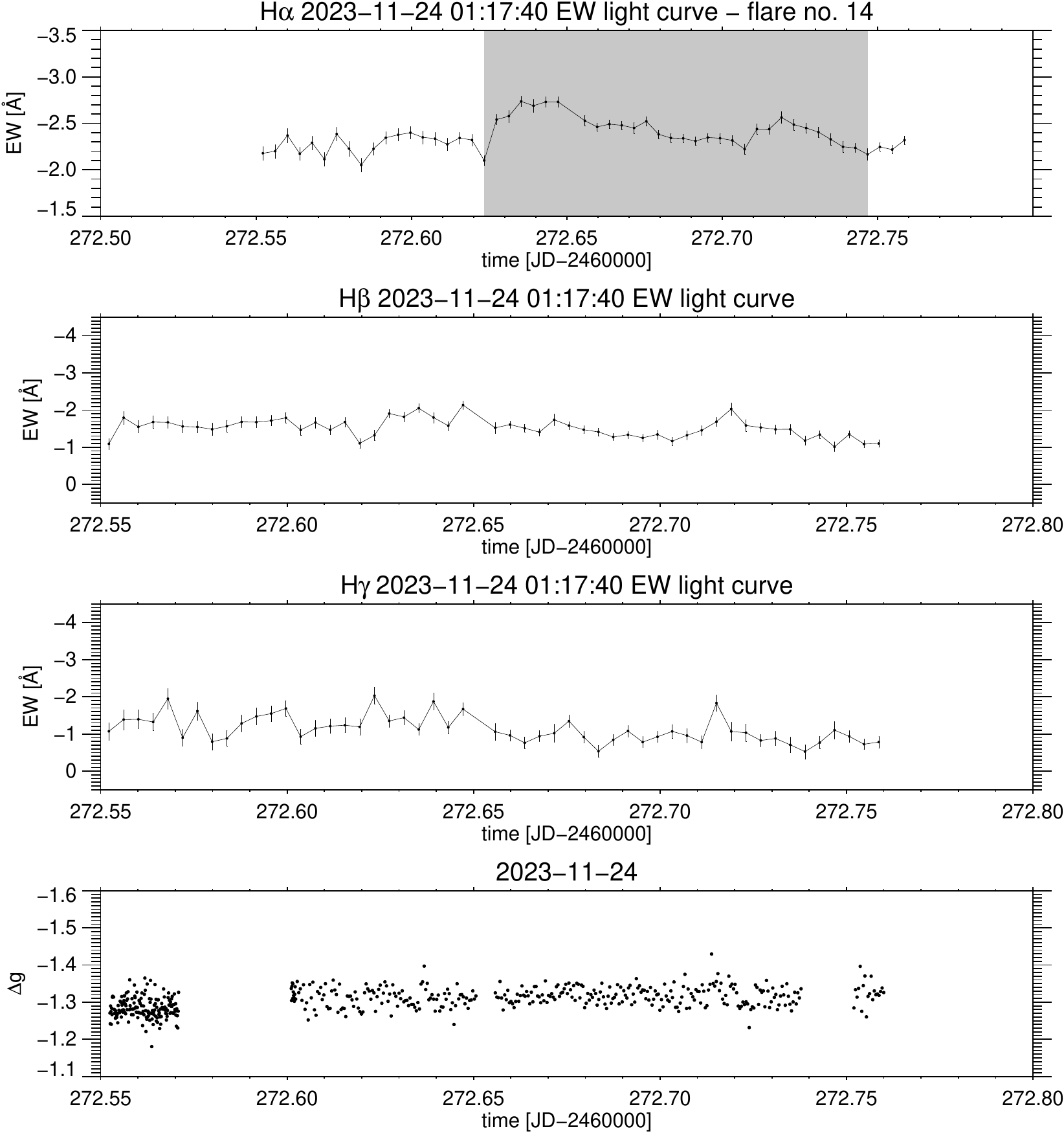}
 \\ 
 \vspace*{1.5cm} 
  \includegraphics[width=8cm]{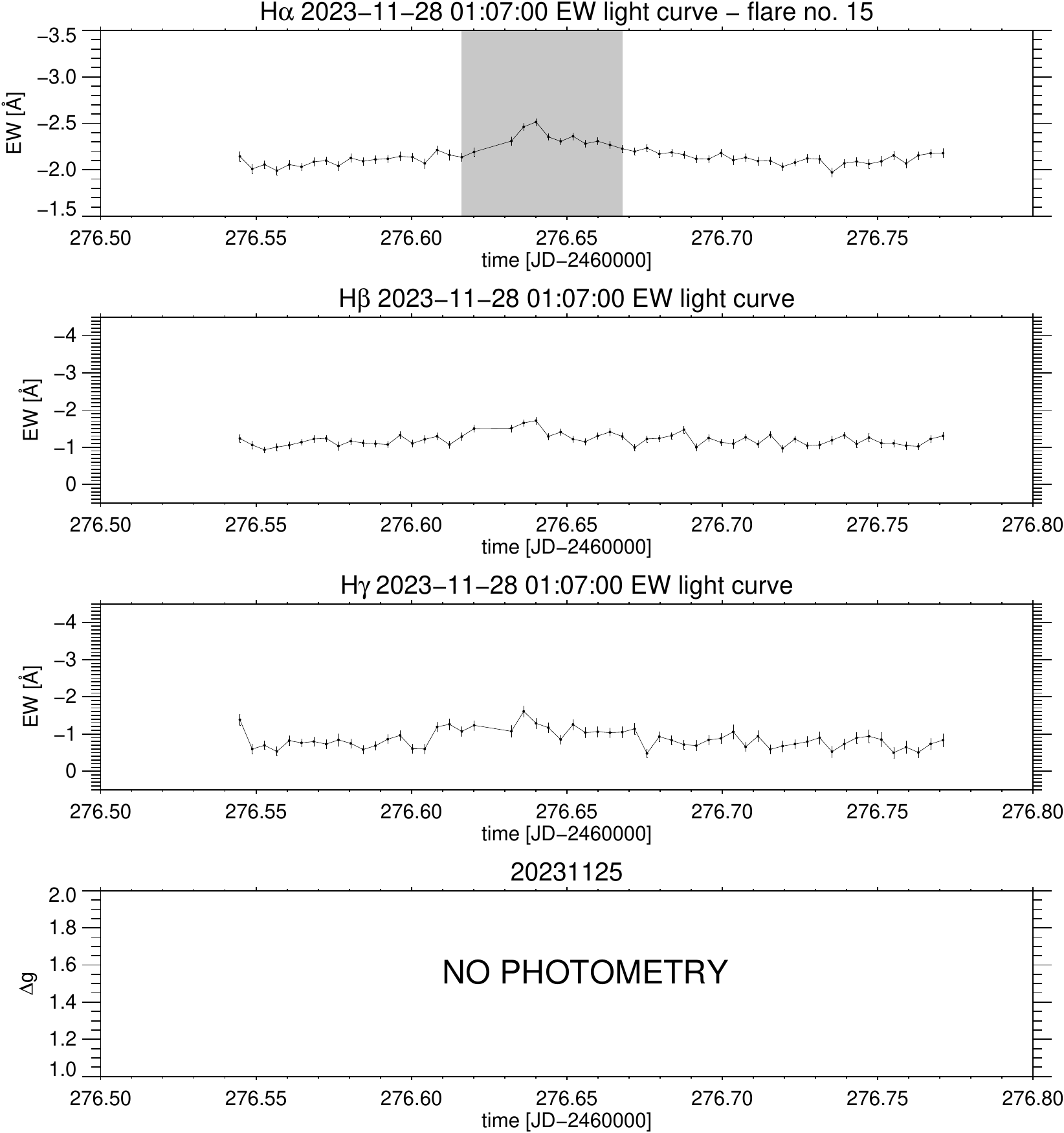}
 \hspace*{0.5cm} 
 \includegraphics[width=8cm]{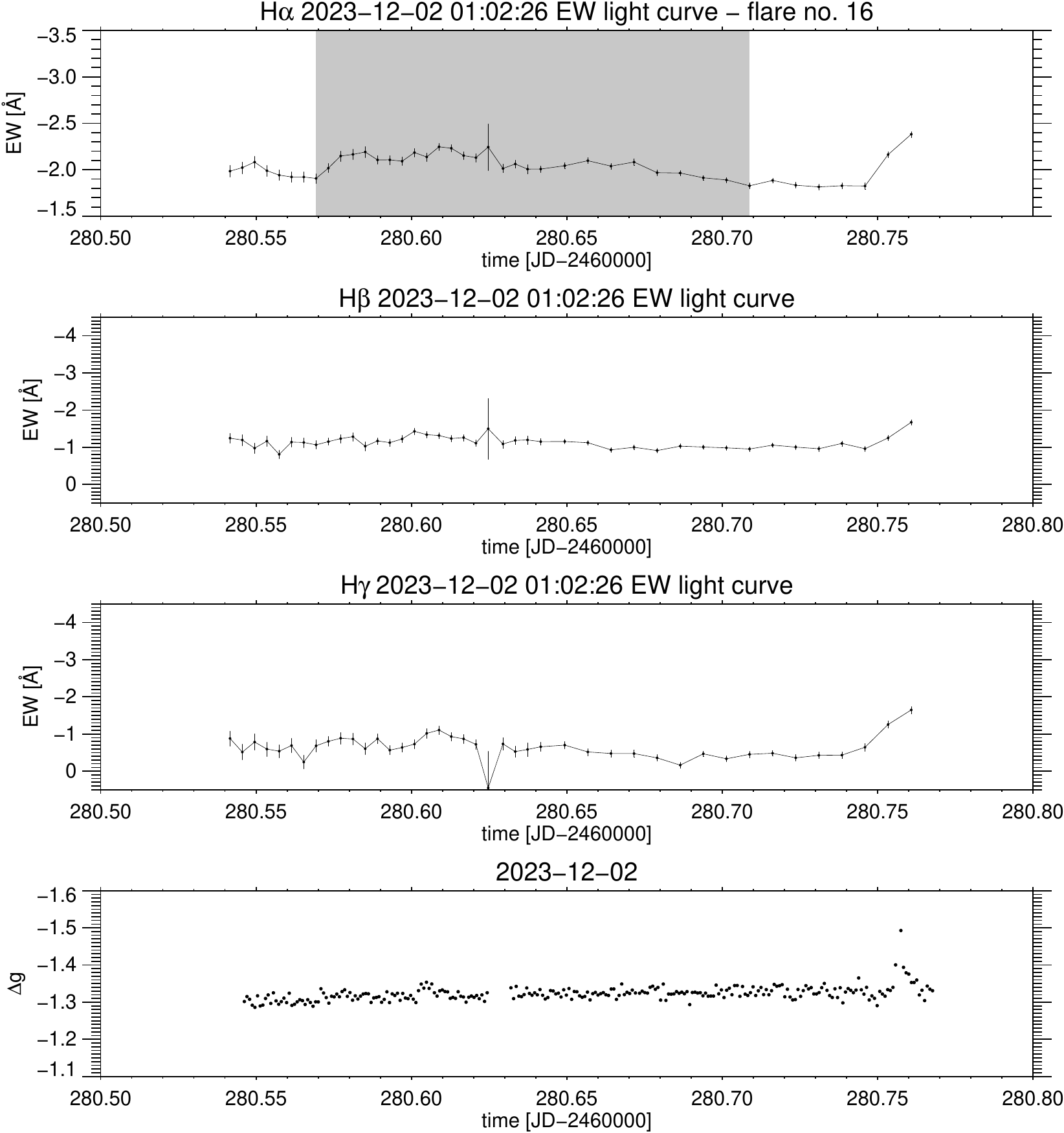}
 \caption{H$\alpha$, H$\beta$, H$\gamma$, and g'-band (from top to bottom) light curves of CC Eri in the nights of 2023-11-23 (upper left panel), 2023-11-24 (upper right panel), 2023-11-28 (lower left panel), and 2023-12-02 (lower right panel).}
 \label{fig:EWlc4}
\end{figure*}

\begin{figure*}
 \includegraphics[width=8cm]{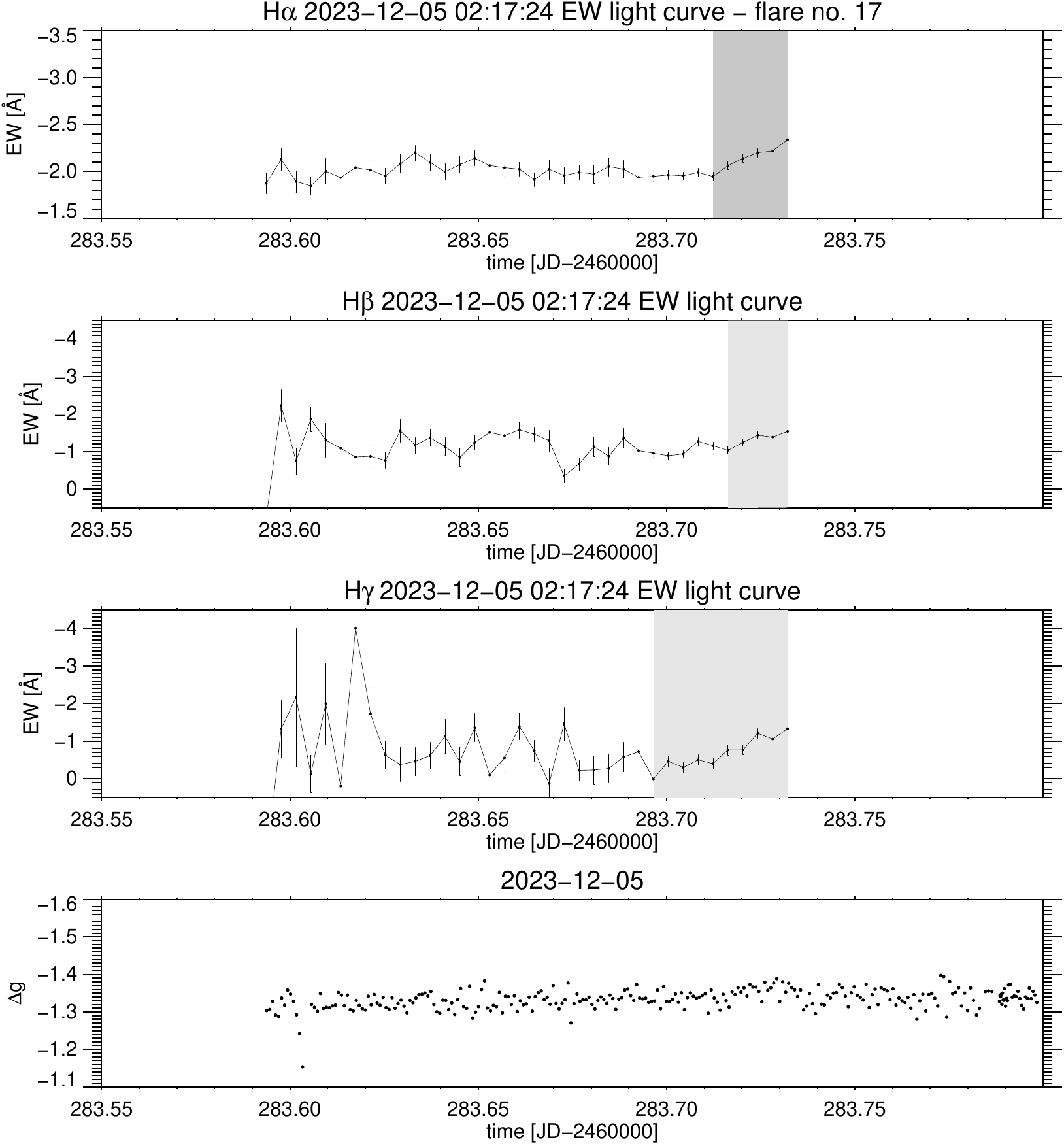}
 \hspace*{0.5cm}
 \includegraphics[width=8cm]{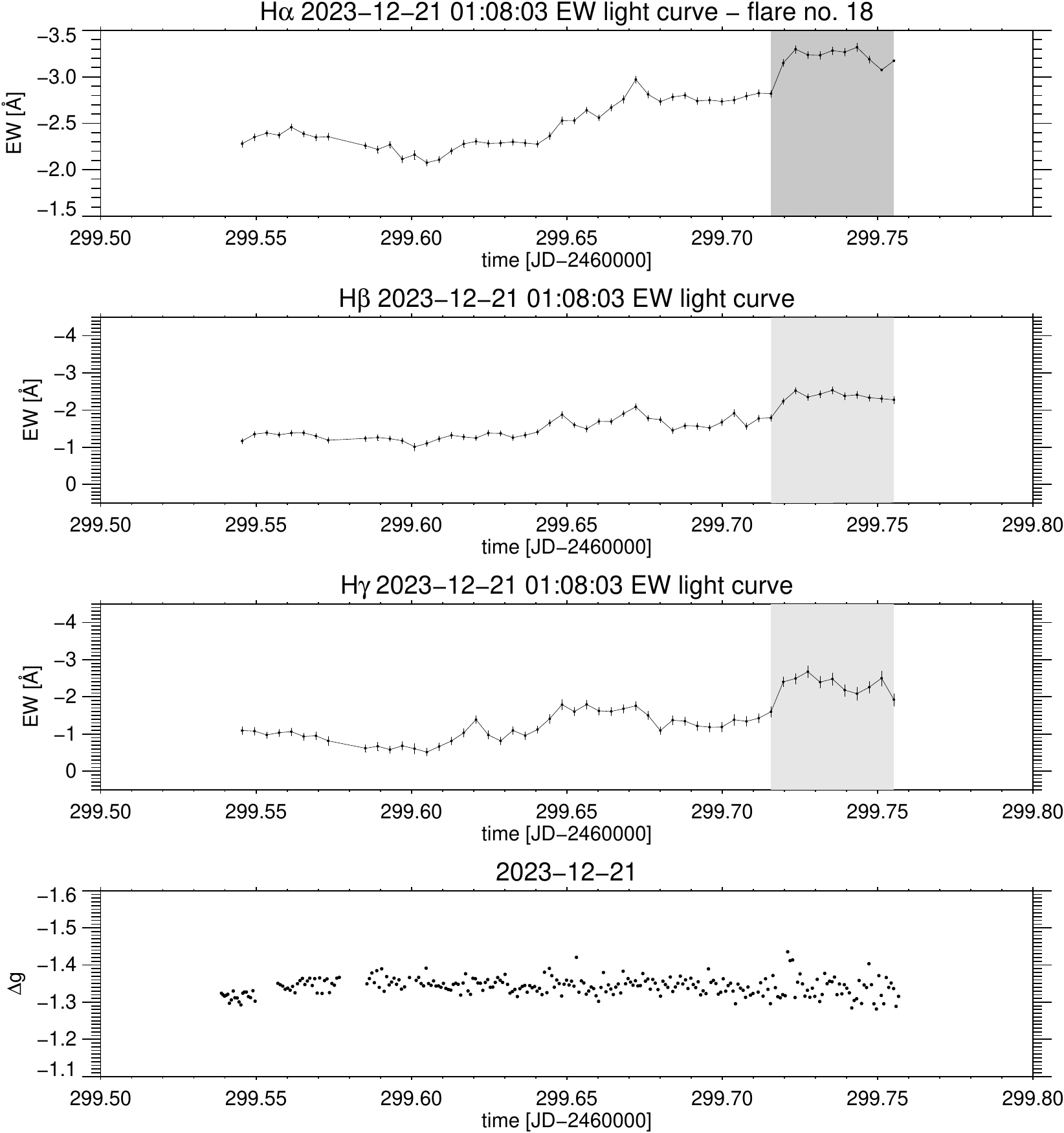}
 \\ 
 \vspace*{1.5cm} 
  \includegraphics[width=8cm]{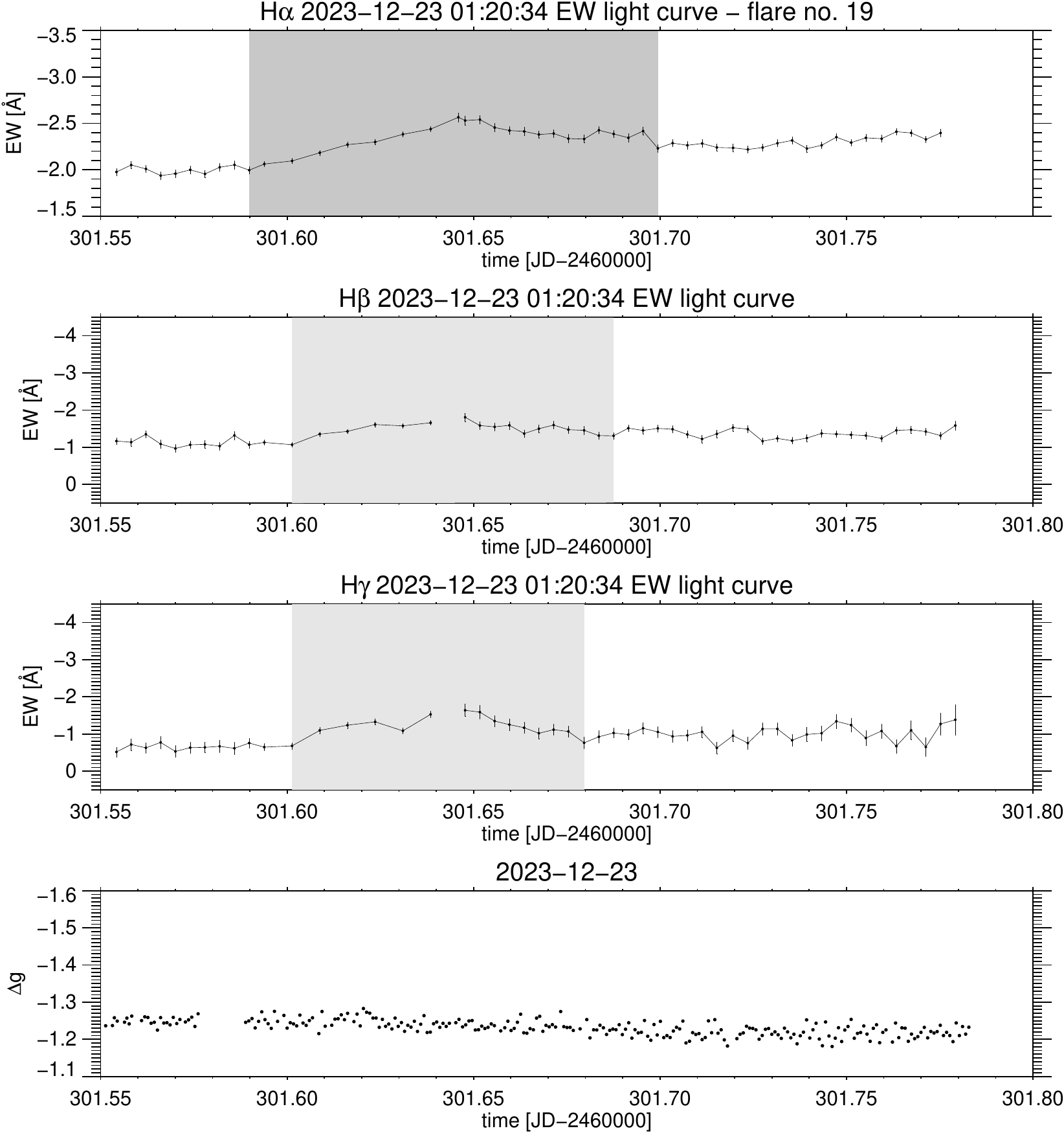}
 \hspace*{0.5cm} 
 \includegraphics[width=8cm]{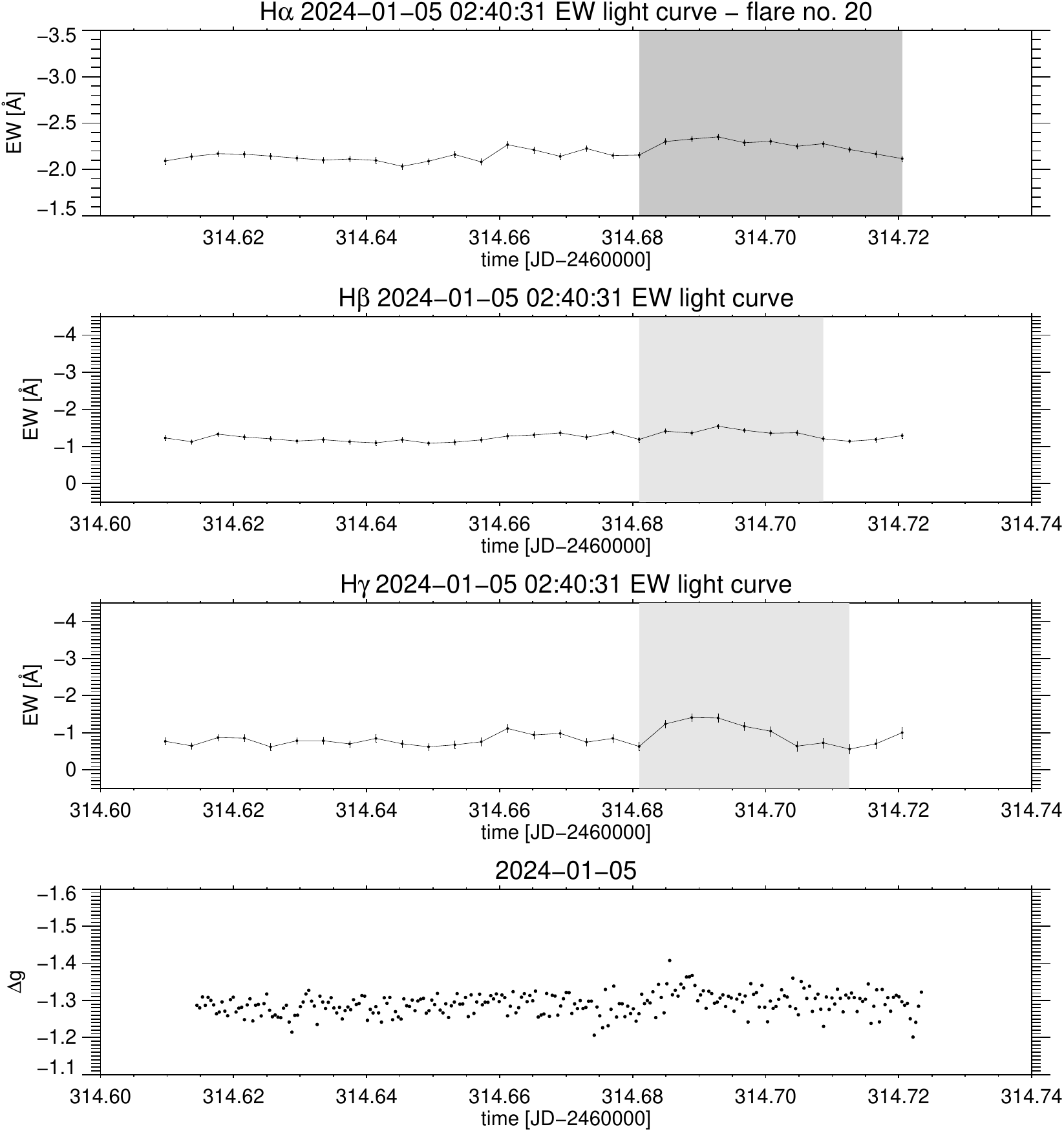}
 \caption{H$\alpha$, H$\beta$, H$\gamma$, and g'-band (from top to bottom) light curves of CC Eri in the nights of 2023-12-05 (upper left panel), 2023-12-21 (upper right panel), 2023-12-23 (lower left panel), and 2024-01-05 (lower right panel).}
 \label{fig:EWlc5}
\end{figure*}

\begin{figure*}
 \includegraphics[width=8cm]{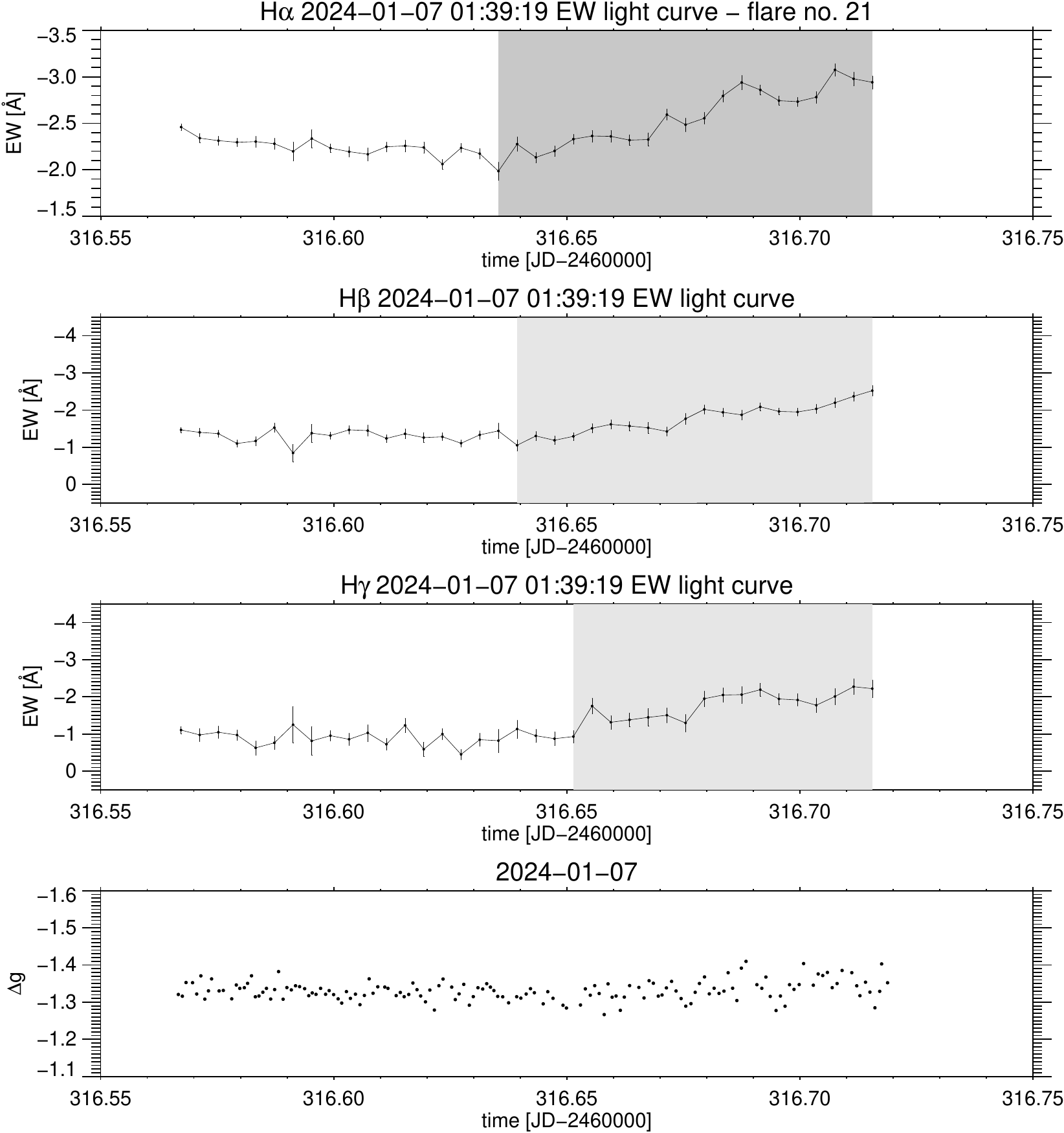}
 \hspace*{0.5cm}
 \includegraphics[width=8cm]{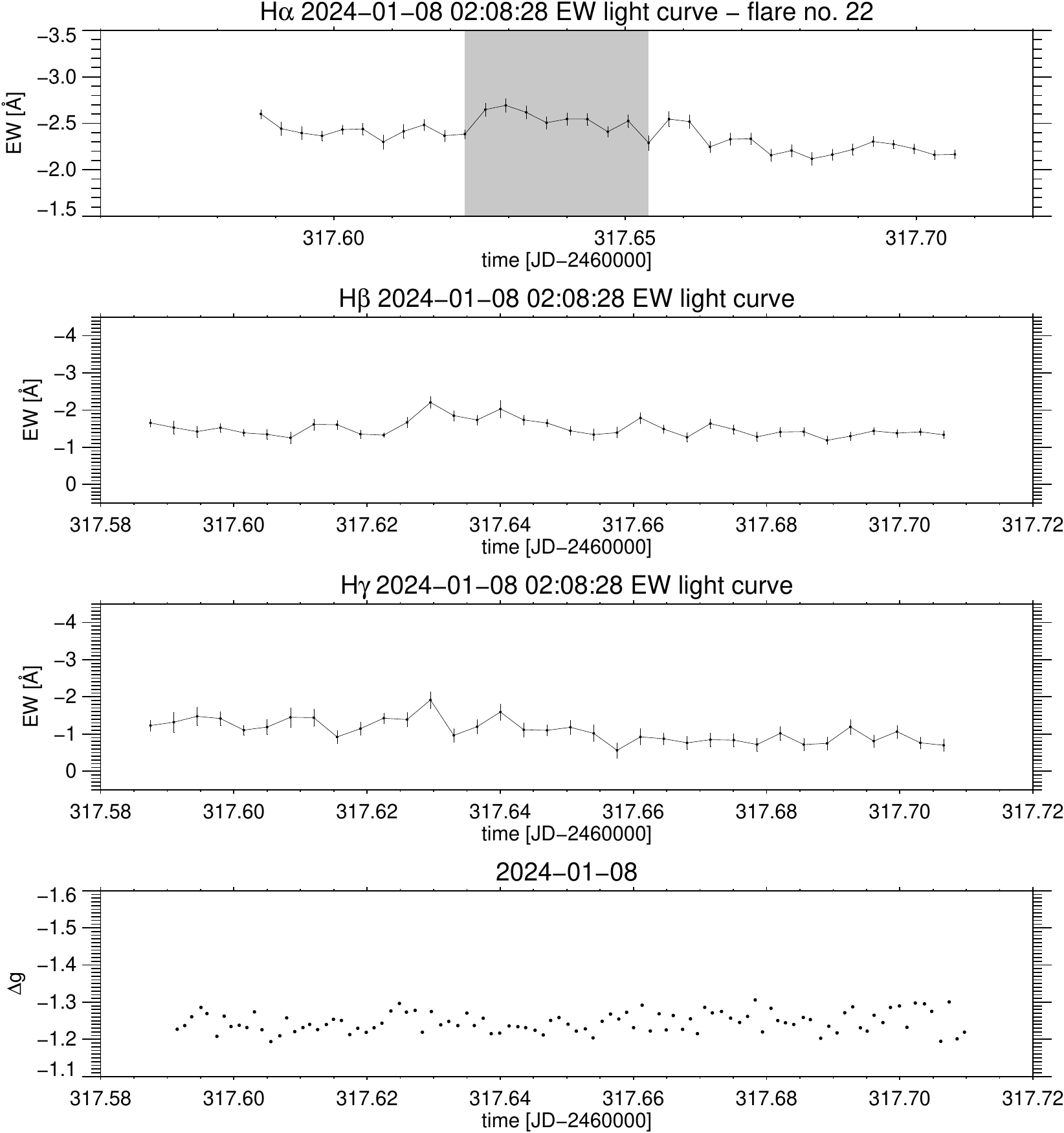}
 \\ 
 \vspace*{1.5cm} 
  \includegraphics[width=8cm]{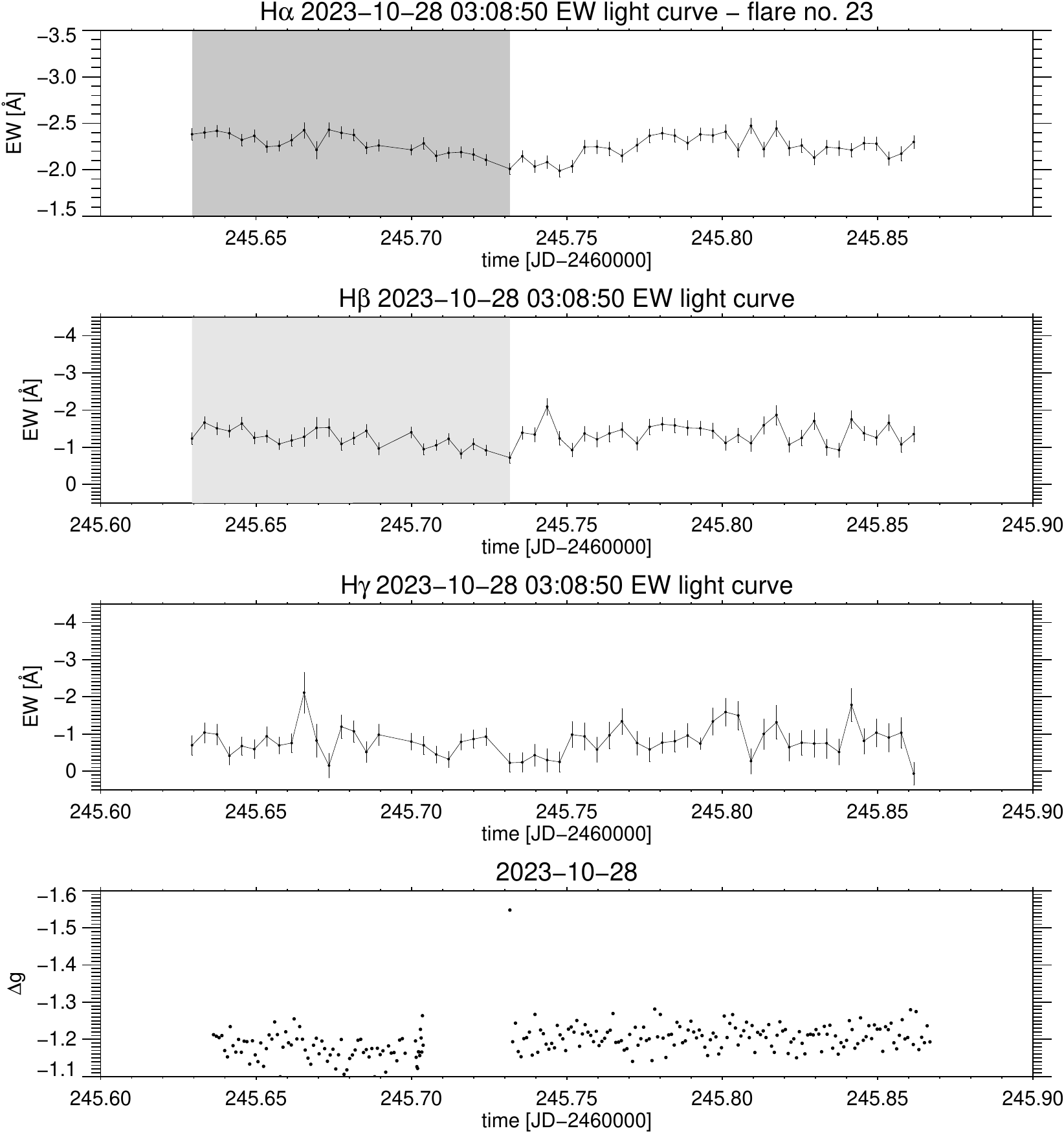}
 \hspace*{0.5cm} 
 \includegraphics[width=8cm]{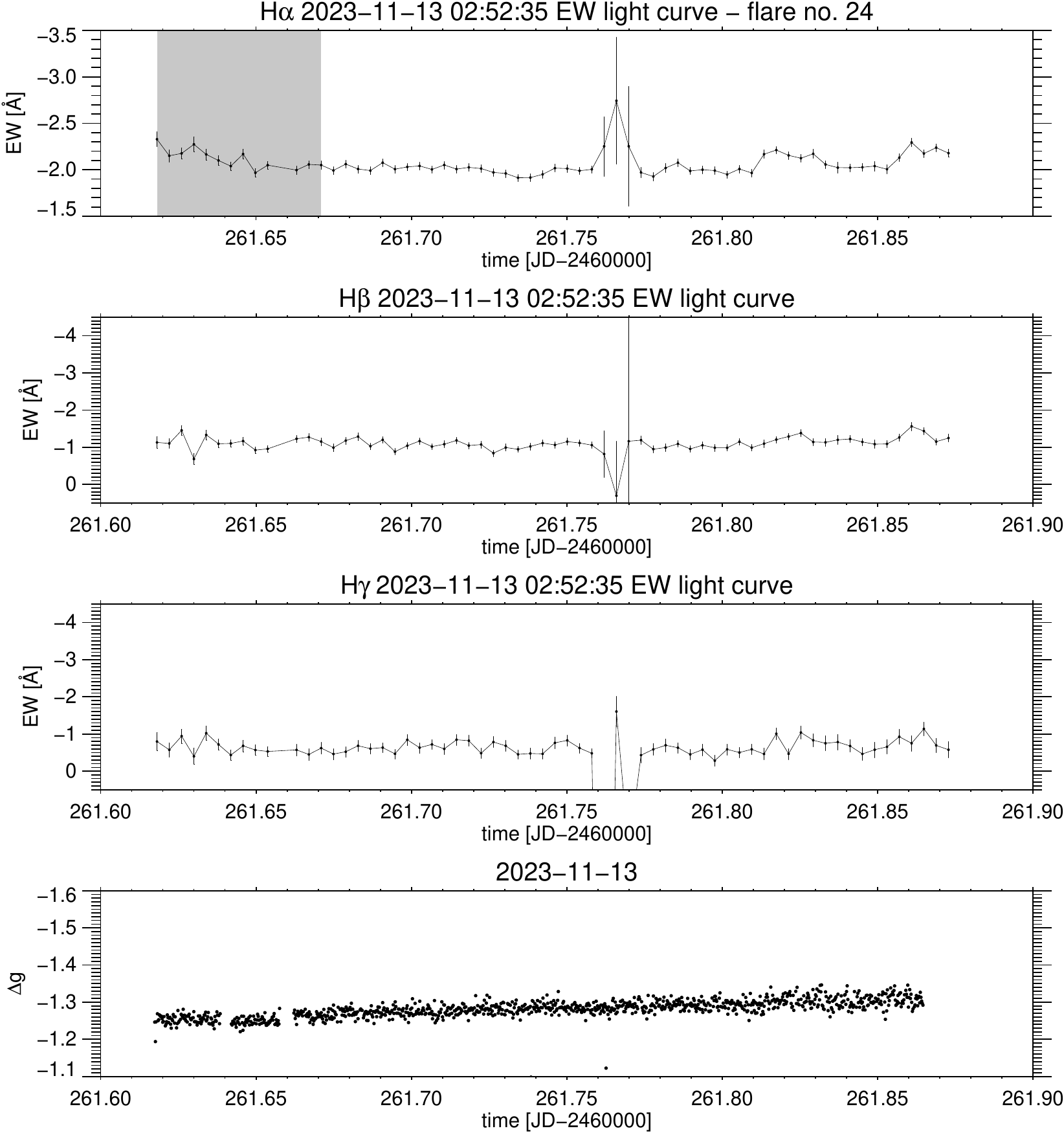}
 \caption{H$\alpha$, H$\beta$, H$\gamma$, and g'-band (from top to bottom) light curves of CC Eri in the nights of 2024-01-07 (upper left panel), 2024-01-08 (upper right panel), 2023-10-28 (lower left panel), and 2023-11-13 (lower right panel).}
 \label{fig:EWlc6}
\end{figure*}

\begin{figure*}
 \includegraphics[width=8cm]{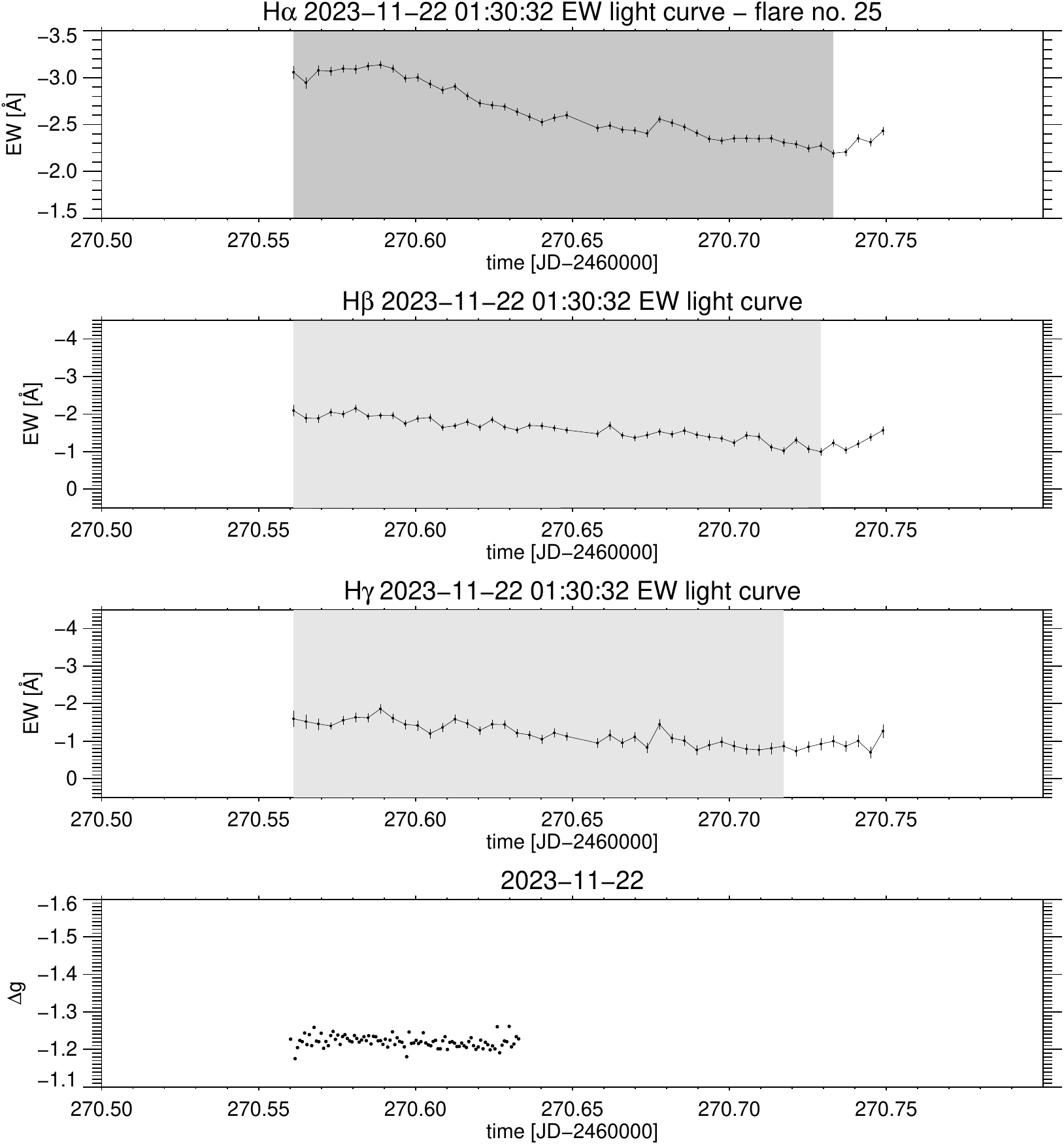}
 \hspace*{0.5cm}
 \includegraphics[width=8cm]{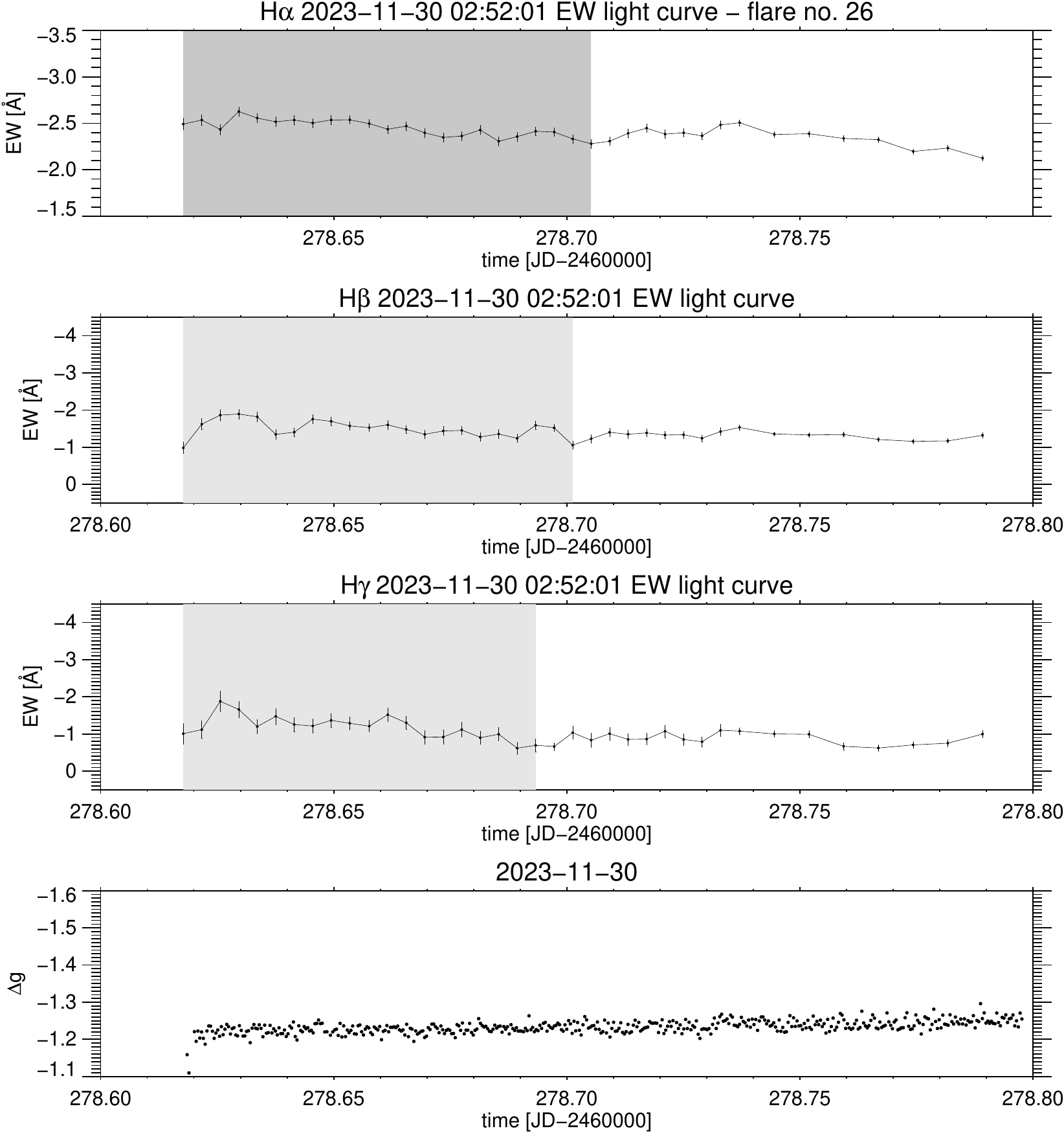}
 \\ 
 \vspace*{1.5cm} 
  \includegraphics[width=8cm]{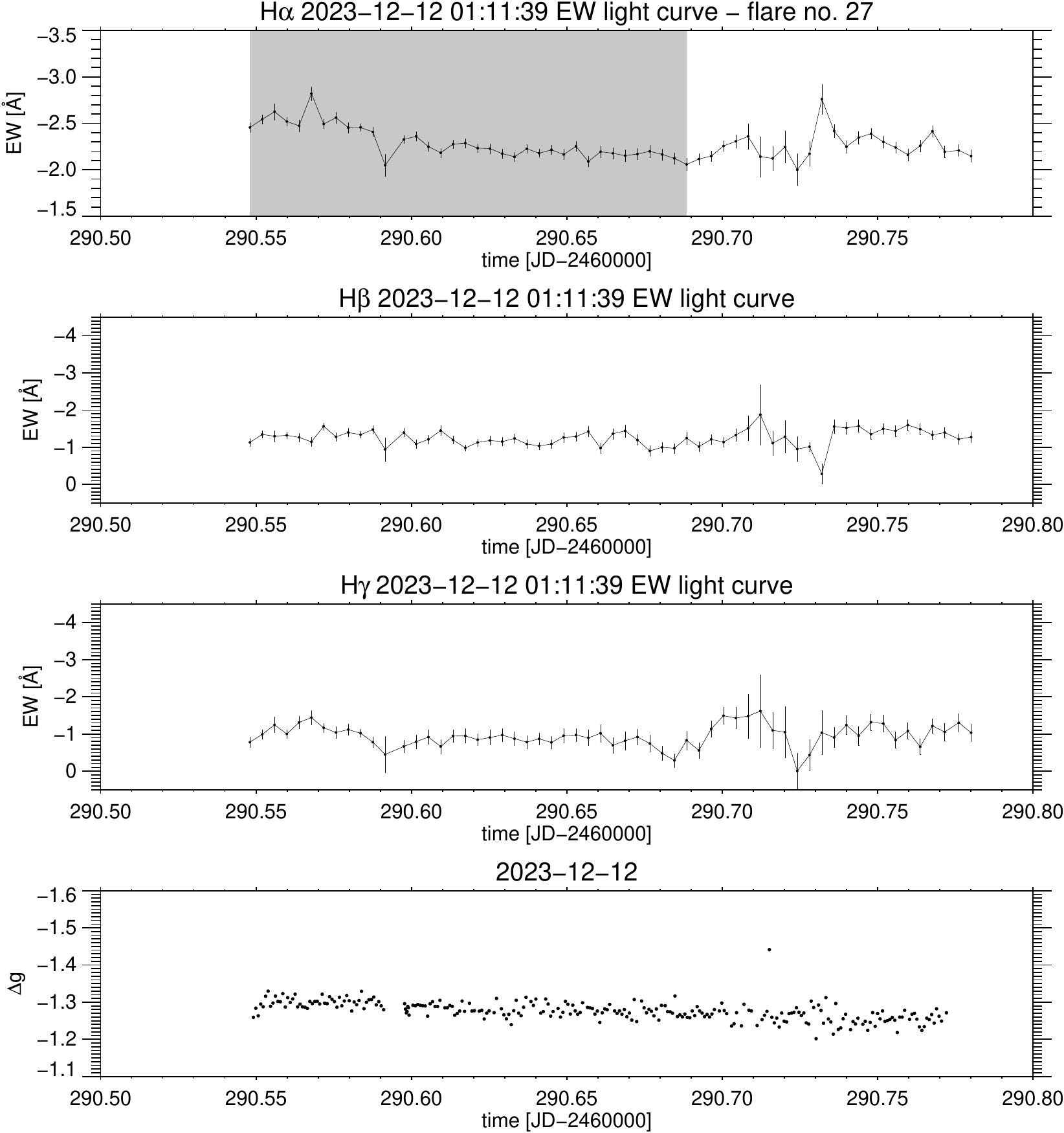}
 \hspace*{0.5cm} 
 \includegraphics[width=8cm]{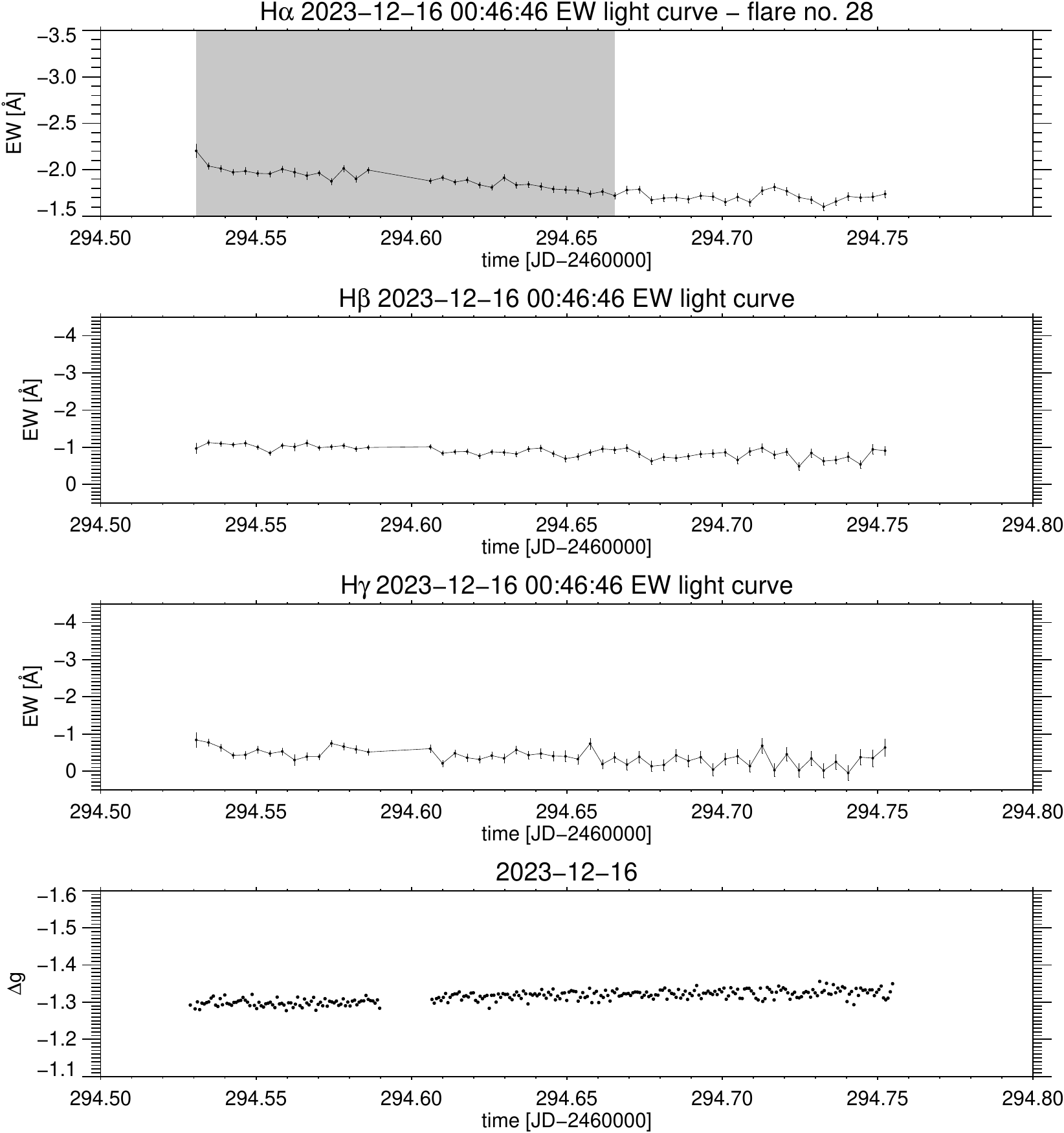}
 \caption{H$\alpha$, H$\beta$, H$\gamma$, and g'-band (from top to bottom) light curves of CC Eri in the nights of 2023-11-22 (upper left panel), 2023-11-30 (upper right panel), 2023-12-12 (lower left panel), and 2023-12-16 (lower right panel).}
 \label{fig:EWlc7}
\end{figure*}

\begin{figure*}
 \includegraphics[width=8cm]{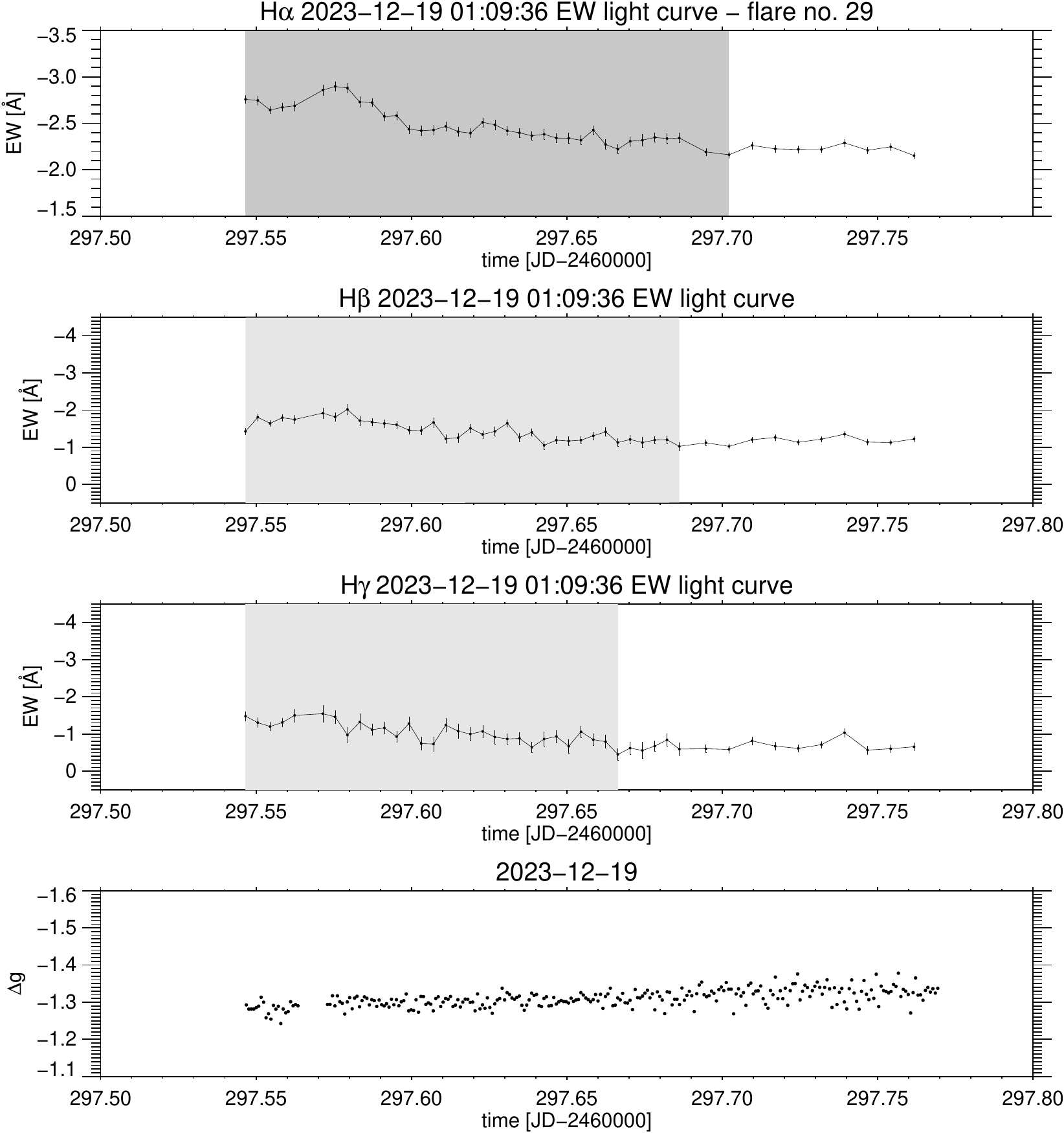}
 \hspace*{0.5cm}
 \includegraphics[width=8cm]{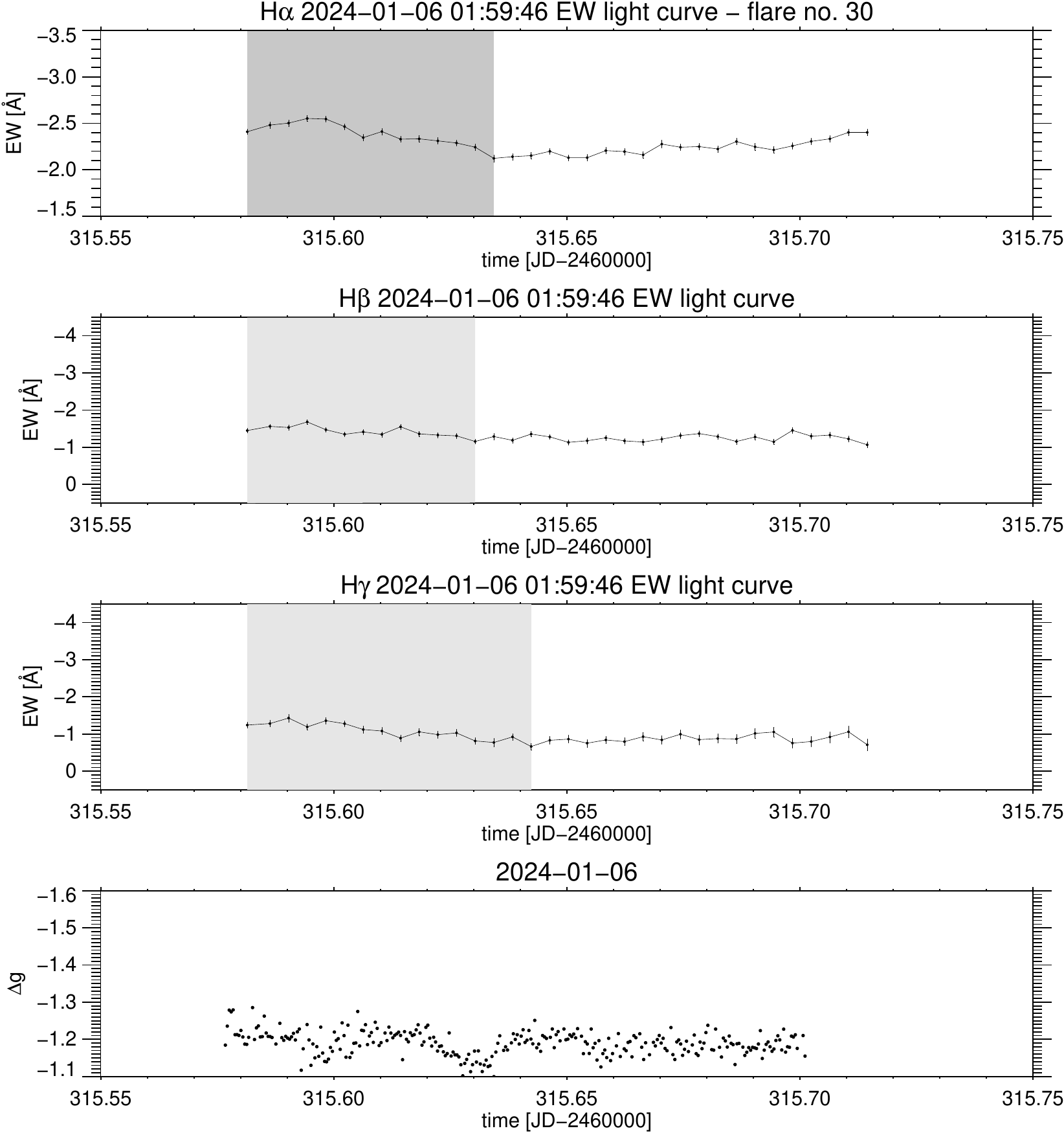}
 \\ 
  \vspace*{1.5cm} 
  \includegraphics[width=8cm]{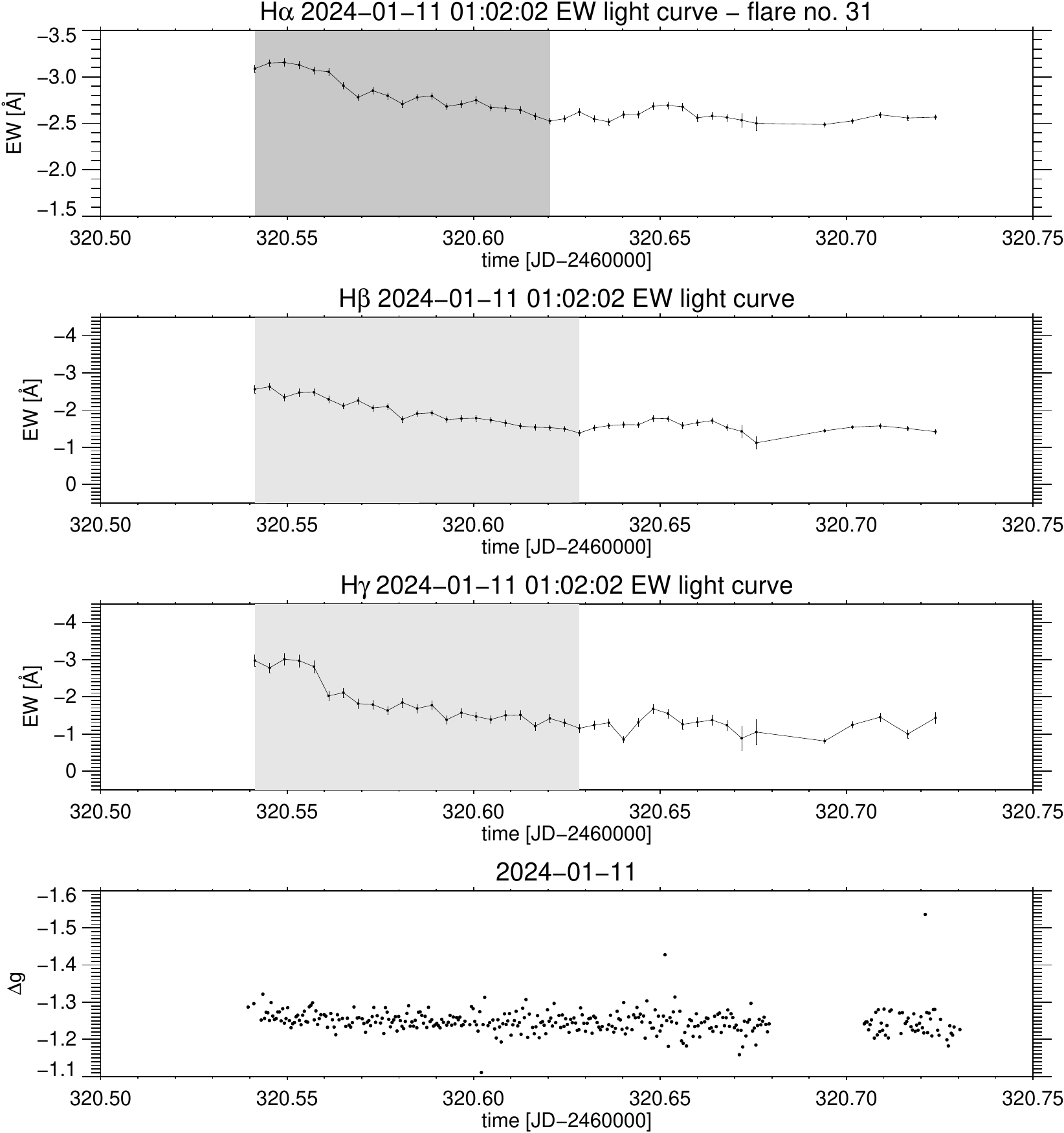}
 \caption{H$\alpha$, H$\beta$, H$\gamma$, and g'-band (from top to bottom) light curves of CC Eri in the nights of 2023-12-19 (upper left panel), 2024-01-06 (upper right panel), and 2024-01-11(lower panel).}
 \label{fig:EWlc8}
\end{figure*}

\newpage
 \begin{figure*}
 \includegraphics[width=5.5cm]{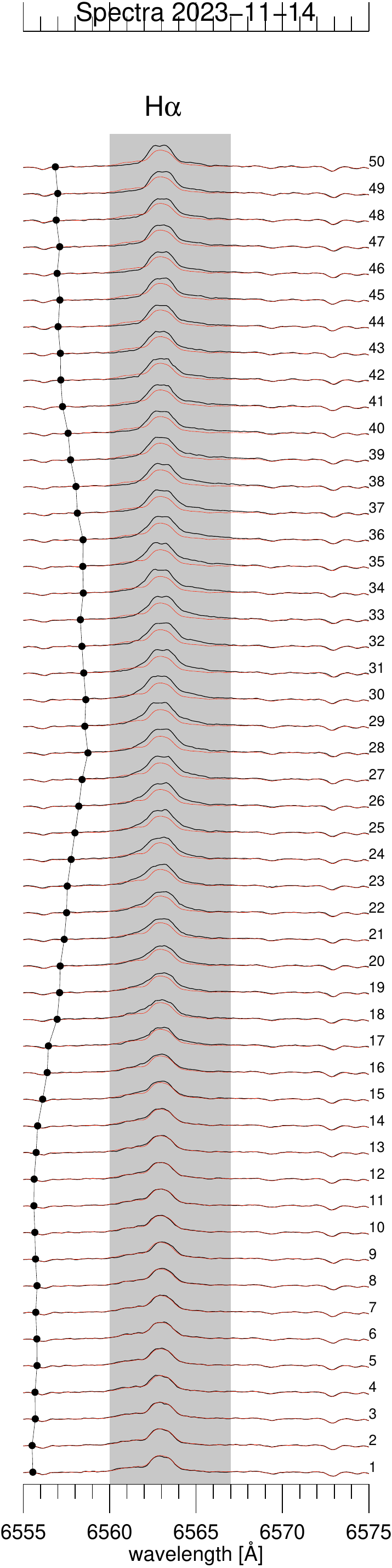}
 \includegraphics[width=5.21cm]{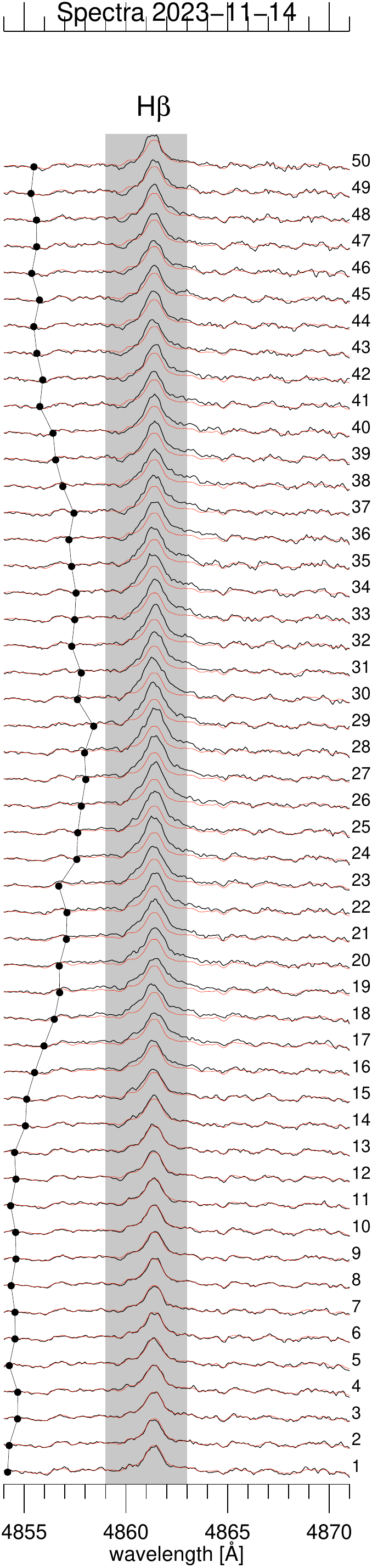}
 \includegraphics[width=5.5cm]{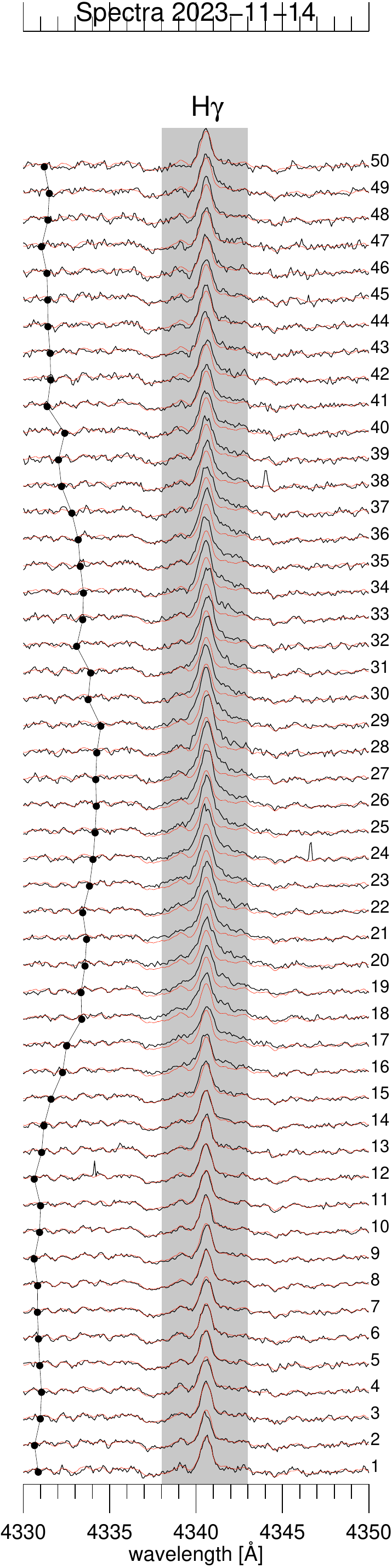}
 \caption{H$\alpha$, H$\beta$, and H$\gamma$ spectral time series of CC Eri in the night of 2023-11-14. In each panel the corresponding light curve is plotted vertically to identify corresponding spectra at each time step in the light curve. Black solid lines correspond to the actual spectrum and red solid lines correspond to an average spectrum. The grey shaded area forgrounds the spectral line of interest.}
 \label{fig:singlespectranalysis0}
\end{figure*}

\newpage

\newpage
 \begin{figure*}
 \includegraphics[width=4.00cm]{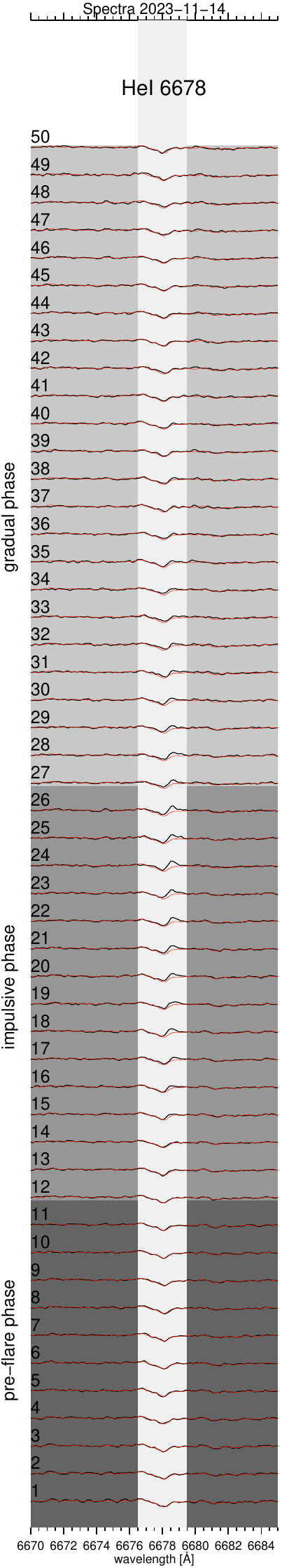}
 \includegraphics[width=4.02cm]{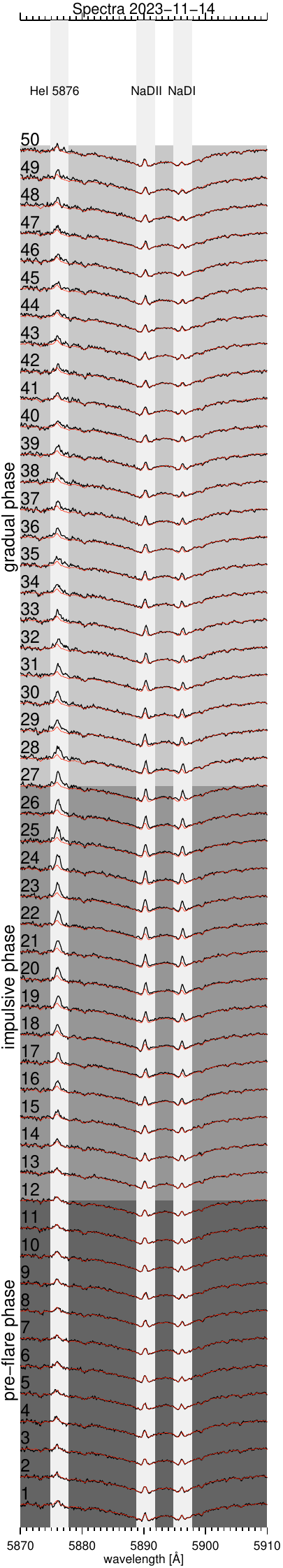}
 \includegraphics[width=4.1cm]{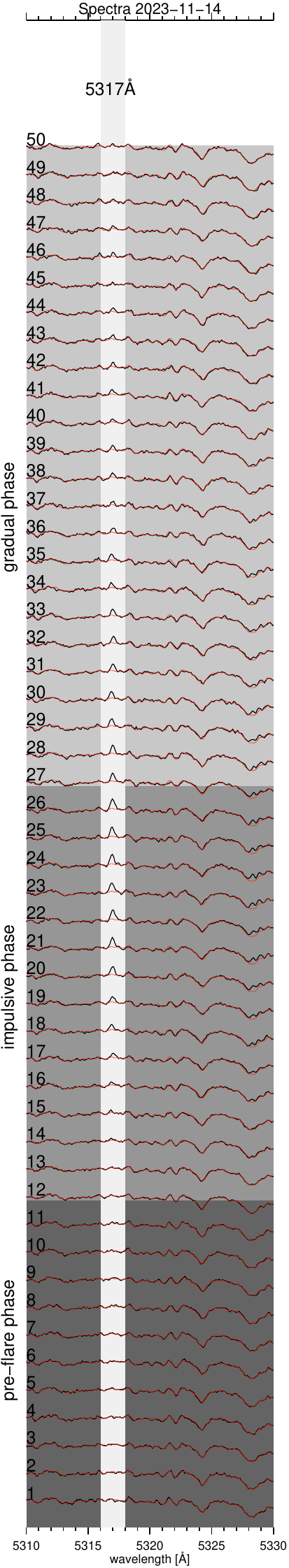}
 \includegraphics[width=4.cm]{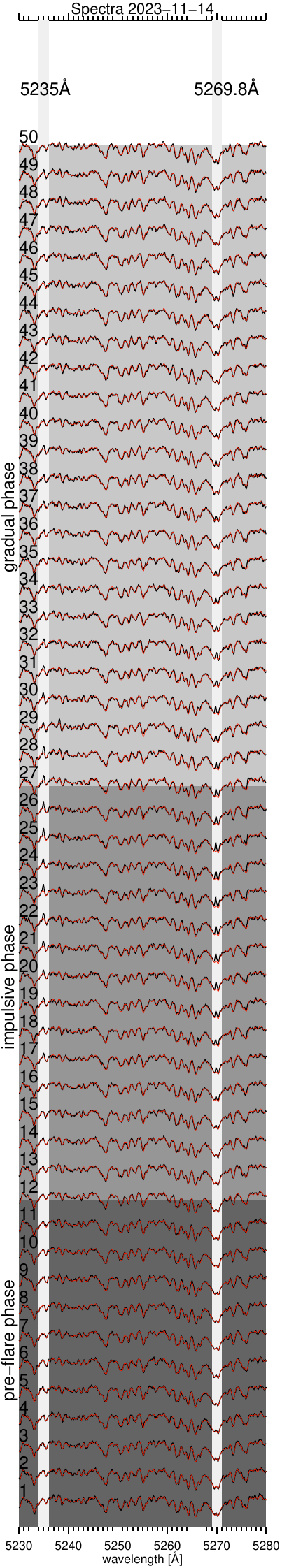}
 \caption{HeII (6678\AA), HeI D3 (5786\AA), NaDI+II (5896, 5890\AA),  5317\AA, 5235\AA, and FeI5270(\AA) spectral time series of CC~Eri in the night of 2023-11-14. The spectral time series are plotted on dark grey, grey and light grey backgrounds correpsonding to the pre-flare, impulsive, and gradual phases of the flare. Black solid lines correspond to the actual spectrum and red solid lines correspond to an average spectrum. The very light grey vertical background foregrounds the spectral line of interest.}
 \label{fig:singlespectranalysis1}
\end{figure*}

\newpage
\begin{figure*}
 \includegraphics[width=4.02cm]{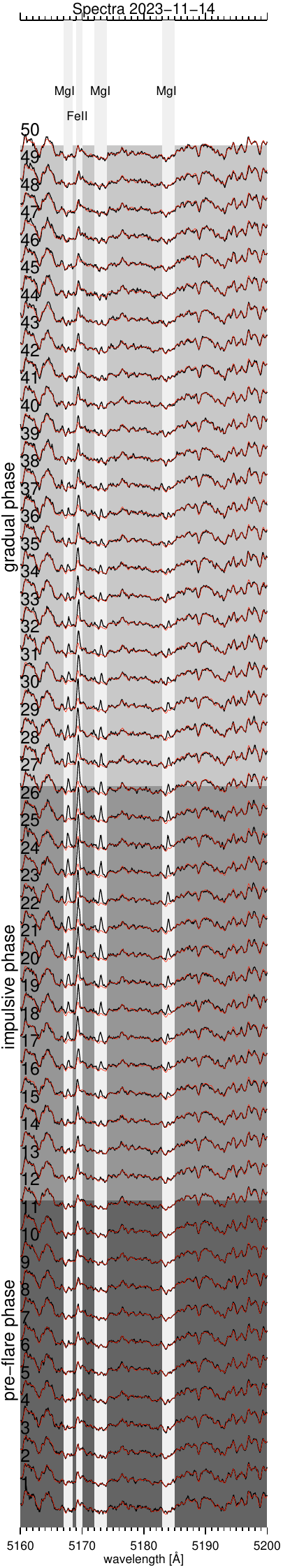}
 \includegraphics[width=4.10cm]{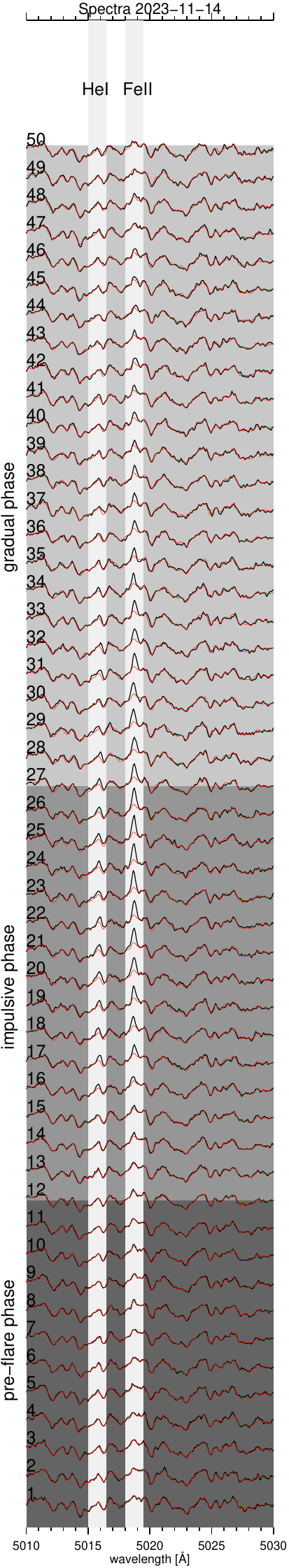}
 \includegraphics[width=4.165cm]{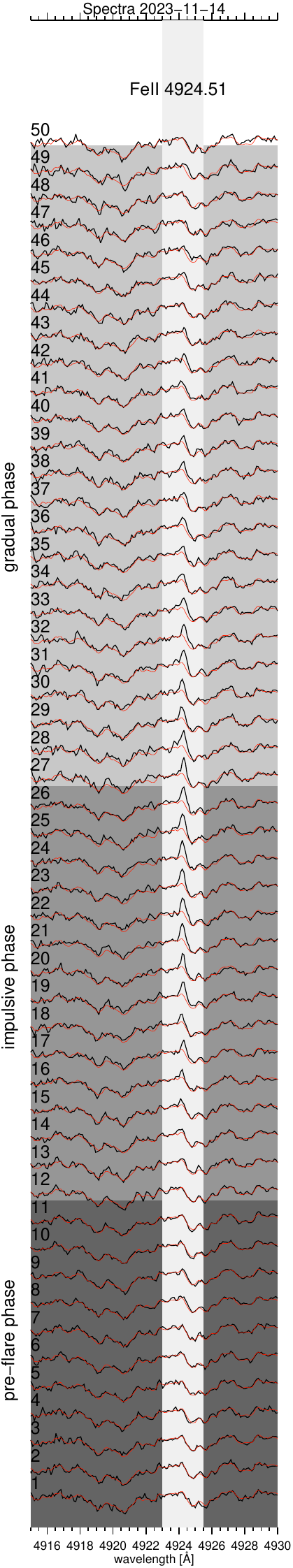}
 \includegraphics[width=4.00cm]{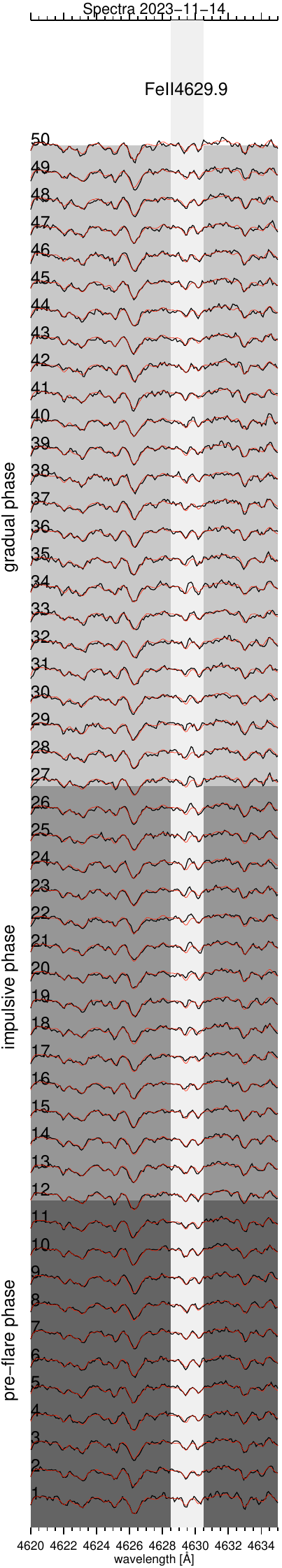}
 \caption{MgI triplet (5168\AA, 5173\AA, 5184\AA), FeII (5169\AA), HeI (5015.7\AA), FeII (5018\AA), FeII (4925\AA), and FeII(4629.9\AA) spectral time series of CC~Eri in the night of 2023-11-14. The spectral time series are plotted on dark grey, grey and light grey backgrounds correpsonding to the pre-flare, impulsive, and gradual phases of the flare. Black solid lines correspond to the actual spectrum and red solid lines correspond to an average spectrum. The very light grey vertical background foregrounds the spectral line of interest.}
  \label{fig:singlespectranalysis2}
\end{figure*}

\newpage
\begin{figure*}
 \includegraphics[width=4.18cm]{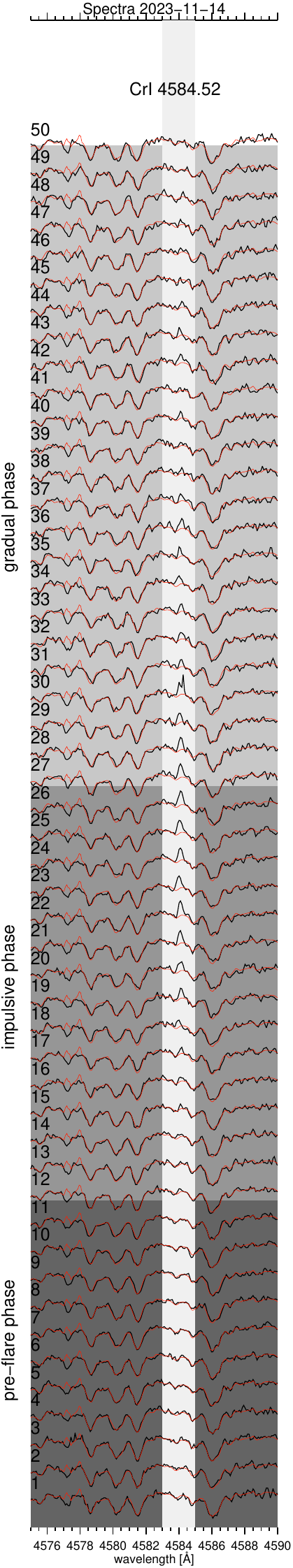}
 \includegraphics[width=4.15cm]{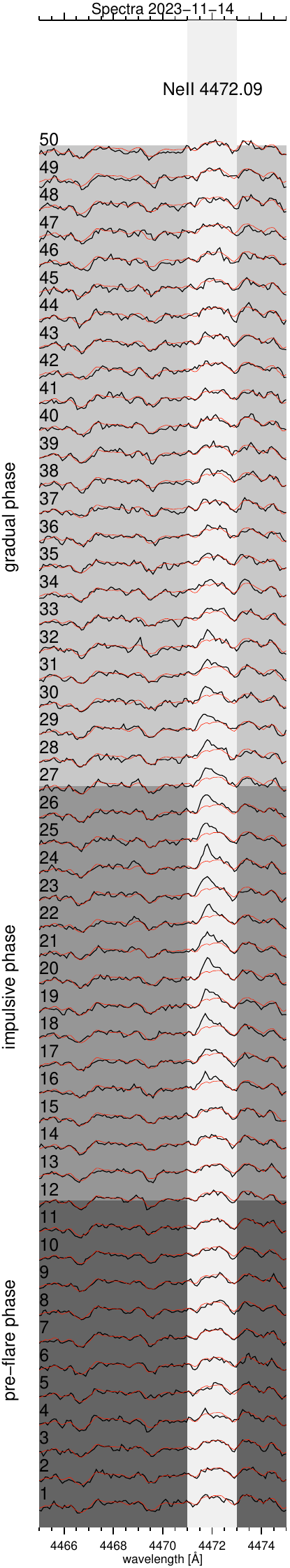}
 \caption{CrI (4585\AA) and NeII (4472\AA) spectral time series of CC~Eri in the night of 2023-11-14. The spectral time series are plotted on dark grey, grey and light grey backgrounds correpsonding to the pre-flare, impulsive, and gradual phases of the flare. Black solid lines correspond to the actual spectrum and red solid lines correspond to an average spectrum. The very light grey vertical background foregrounds the spectral line of interest.}
  \label{fig:singlespectranalysis3}
\end{figure*}

\bsp	
\label{lastpage}
\end{document}